\documentclass[aps,prr,twocolumn,superscriptaddress,longbibliography,floatfix]{revtex4-1}

\usepackage{graphicx}
\usepackage{bm}
\usepackage{amsmath}
\usepackage{amssymb}
\usepackage{mathtools}
\usepackage{amsfonts}
\usepackage[all]{xy}
\usepackage{xspace}
\usepackage{accents}
\usepackage{stmaryrd}
\usepackage{tabularx}
\usepackage{extarrows}
\usepackage{float}
\usepackage{multirow}
\usepackage{makecell}

\newcounter{fnnumber}

\newenvironment{smallarray}[1]
 {\null\,\vcenter\bgroup\small
  \renewcommand{\arraystretch}{0.7}%
  \arraycolsep=.13885em
  \hbox\bgroup$\array{@{}#1@{}}}
 {\endarray$\egroup\egroup\,\null}
\newlength{\dhatheight}
\newcommand{\quadcover}[1]{
    \settoheight{\dhatheight}{\ensuremath{\hat{#1}}}
    \addtolength{\dhatheight}{-0.33ex}
    \hat{\vphantom{\rule{1pt}{\dhatheight}}
    \smash{\hat{#1}}}
}
\newcommand{\minor}[1]{#1'}
\newcommand{\spincover}[1]{\tilde{#1}}
\newcommand{\twistcover}[1]{\hat{#1}}
\newcommand{\ket}[1]{{\left|#1\right\rangle}}
\newcommand{\tket}[1]{{|#1\rangle}}

\newcommand{\tmod}{{\,\mathrm{mod}\,}}
\newcommand{\Rep}{\mathrm{Rep}}
\newcommand{\CRep}{\mathbb{C}\Rep}
\newcommand{\RRep}{\mathbb{R}\Rep}
\newcommand{\gRep}{\Rep_{\ZZ_2}}

\newcommand{\gRRep}{\mathbb{R}\Rep_{\ZZ_2}}
\newcommand{\Irr}{\mathrm{Irr}}
\newcommand{\gIrr}{\mathrm{Irr}_{\ZZ_2}}
\newcommand{\inv}{\iota}

\newcommand{\Cl}{\mathrm{Cl}}
\newcommand{\CCl}{\mathbb{C}\mathrm{l}}
\newcommand{\rr}{\mathfrak{r}}
\newcommand{\cc}{\mathfrak{c}}
\newcommand{\ConjOp}{\kappa}
\newcommand{\parity}{\varpi}

\newcommand{\SSp}{\mathbb{S}}
\newcommand{\bSSp}[1]{{\bar{\Sigma}^{#1}\mathbb{S}}}
\newcommand{\TSp}{\mathbb{T}}
\newcommand{\bTSp}[1]{{\bar{\mathbb{T}}^{#1}}}
\newcommand{\trivg}{1}
\newcommand{\RR}{\mathbb{R}}
\newcommand{\CC}{\mathbb{C}}
\newcommand{\HH}{\mathbb{H}}
\newcommand{\ZZ}{\mathbb{Z}}

\newcommand{\NN}{\mathbb{N}}
\newcommand{\TR}{\Theta}
\newcommand{\PH}{\Xi}
\newcommand{\Chiral}{\Pi}
\newcommand{\ZT}{\mathbb{Z}_2^{T}}
\newcommand{\sC}{\mathcal{C}}
\newcommand{\sR}{\mathcal{R}}
\newcommand{\gotimes}{\hat\otimes}
\DeclareMathOperator{\Id}{Id}
\DeclareMathOperator{\Res}{Res}
\DeclareMathOperator{\Ind}{Ind}
\DeclareMathOperator{\Hom}{Hom}
\DeclareMathOperator{\Mod}{Mod}
\DeclareMathOperator{\End}{End}
\DeclareMathOperator{\Aut}{Aut}

\DeclareMathOperator{\Map}{Map}
\DeclareMathOperator{\Norm}{N}
\newcommand{\CMon}{\mathrm{CMon}}

\newcommand{\Vect}{\mathrm{Vect}}
\newcommand{\Sp}{\mathbf{Spct}} 
\newcommand{\Ab}{\mathbf{Ab}}
\newcommand{\grAb}{\mathbf{Ab}_{\ZZ_2}}
\newcommand{\spaces}{\mathbf{Sps}}
\newcommand{\Mat}[1]{\mathrm{M}_{#1\times #1}}
\newcommand{\tMat}[1]{\mathrm{M}_{#1\!\times #1}\!}
\newcommand{\ttMat}[1]{\mathrm{M}_{#1\!\times\! #1}\!}
\newcommand{\Under}[1]{{\underline{#1}}}
\newcommand{\AI}{\mathbf{AI}}
\newcommand{\fundAI}[1]{v_{#1}^1}
\newcommand{\BS}{{\mathbf{BS}_{\scriptscriptstyle\mathrm{TISC}}}}
\newcommand{\BSfull}{{\mathbf{BS}}}
\newcommand{\SI}{{\mathbf{SI}_{\scriptscriptstyle\mathrm{TISC}}}}
\newcommand{\SIfull}{{\mathbf{SI}}}
\newcommand{\Surf}{\mathbf{ASS}}

\newcommand{\BZ}{\mathcal{BZ}}

\newcommand{\KR}{{\mathrm{K}\mathbb{R}}}
\newcommand{\KO}{\mathrm{KO}}
\newcommand{\KU}{\mathrm{KU}}
\newcommand{\KC}{{\mathrm{K}\mathbb{C}}}

\newcommand{\rKR}{\widetilde{\mathrm{K}\mathbb{R}}}
\newcommand{\rKC}{\widetilde{\mathrm{K}\mathbb{C}}}
\newcommand{\trKU}{\widetilde{\mathrm{KU}}{}}
\newcommand{\trKR}{\widetilde{\mathrm{K}\mathbb{R}}{}}
\newcommand{\trKC}{\widetilde{\mathrm{K}\mathbb{C}}{}}
\newcommand{\KK}{{\mathrm{K}}}

\newcommand{\GL}{\mathrm{GL}}

\newcommand{\Pin}{\mathrm{Pin}}
\newcommand{\SPin}{\mathrm{Spin}}
\newcommand{\Ort}{\mathrm{O}}
\newcommand{\SOrt}{\mathrm{SO}}
\newcommand{\Unit}{\mathrm{U}}
\newcommand{\Simp}{\mathrm{Sp}}
\newcommand{\SUnit}{\mathrm{SU}}
\newcommand{\im}{{\mathrm{Im}}}
\newcommand{\conj}[1]{\bar{#1}}
\newcommand{\wideconj}[1]{\overline{#1}} 
\newcommand{\Ss}{S}
\newcommand{\bSs}{\bar{S}}
\newcommand{\Tt}{T}
\newcommand{\bTt}{\bar{T}}
\newcommand{\gp}{\mathrm{grp}}
\newcommand{\vx}{\mathbf{x}}
\newcommand{\vk}{\mathbf{k}}

\newcommand{\vb}{\mathbf{b}}
\newcommand{\va}{\mathbf{a}}
\newcommand{\sigmaO}{\sigma_0}
\newcommand{\lv}{\vb}
\newcommand{\dlv}{\va}
\newcommand{\hsm}{\vk}
\newcommand{\wyck}{\vx}
\newcommand{\wyckcoord}{x}
\newcommand{\hsmcoord}{k}

\newcommand{\uniwyck}{\mathcal{WP}}

\newcommand{\pt}{\mathrm{pt}}
\newcommand{\transpose}{{\mathrm{T}}}
\newcommand{\newg}[1]{#1'}
\newcommand{\pinelm}{a}

\newcommand{\rvs}{V}
\newcommand{\cvs}{U}

\newcommand{\Relatt}{\Lambda}
\newcommand{\Euc}[1]{\mathbb{E}^{#1}}
\newcommand{\HSM}{\mathcal{HSM}}
\newcommand{\wtf}[1]{#1^*}

\newcommand{\AIfunc}[3]{\nu_{#1}^{#2}(#3)}

\newcommand{\AItoK}{\mathfrak{ai}}
\newcommand{\defeq}{\overset{\scriptscriptstyle\mathrm{def}}{=}}
\newcommand{\arsim}{\overset{\sim}{\to}}
\newcommand{\oto}[1]{\xrightarrow{#1}}

\newcommand{\repgr}[1]{\varpi_{#1}}

\newcommand{\iR}{{\rho_\RR}}
\newcommand{\iC}{{\rho_\CC}}

\newcommand{\signrep}{\rho_{\mathrm{sign}}}

\newcommand{\Gwyck}{\uniwyck_G}

\newcommand{\proj}{P}

\newcommand{\cliff}[1]{e_{#1}}
\newcommand{\cliffP}[1]{e_{#1}}
\newcommand{\cliffM}[1]{e_{#1}}

\usepackage[
  colorlinks=true,
  urlcolor=blue,
  linkcolor=blue,
  citecolor=blue,
  bookmarks=false,
  pdftitle={Tenfold Topology of Crystals},
  pdfauthor={E. Cornfeld and S. Carmeli}
]{hyperref}

\newcommand{\mytitle}{Tenfold topology of crystals: \\ Unified classification of crystalline topological insulators and superconductors}

\begin{document}

\title{\mytitle}

\author{Eyal Cornfeld}
\affiliation{Department of Condensed Matter Physics, Weizmann Institute of Science, Rehovot 7610001, Israel}

\author{Shachar Carmeli}
\affiliation{Department of Mathematics and Computer Science, Weizmann Institute of Science, Rehovot 7610001, Israel}

\begin{abstract}
The celebrated tenfold-way of Altland-Zirnbauer symmetry classes discern any quantum system by its pattern of non-spatial symmetries. It lays at the core of the periodic table of topological insulators and superconductors which provided a complete classification of weakly-interacting electrons' non-crystalline topological phases for all symmetry classes.
Over recent years, a plethora of topological phenomena with diverse surface states has been discovered in crystalline materials.
In this paper, we obtain an exhaustive classification of topologically distinct groundstates as well as topological phases with anomalous surface states of crystalline topological insulators and superconductors for key space-groups, layer-groups, and rod-groups. This is done in a unified manner for the full tenfold-way of Altland-Zirnbauer non-spatial symmetry classes.
We establish a comprehensive paradigm that harnesses the modern mathematical framework of equivariant spectra; it allows us to obtain results applicable to generic topological classification problems.
In particular, this paradigm provides efficient computational tools that enable an inherently unified treatment of the full tenfold-way.
\end{abstract}

\maketitle

\section{Introduction}\label{sec:intro}
\begin{figure}[t]
\centering
\includegraphics[width=\linewidth]{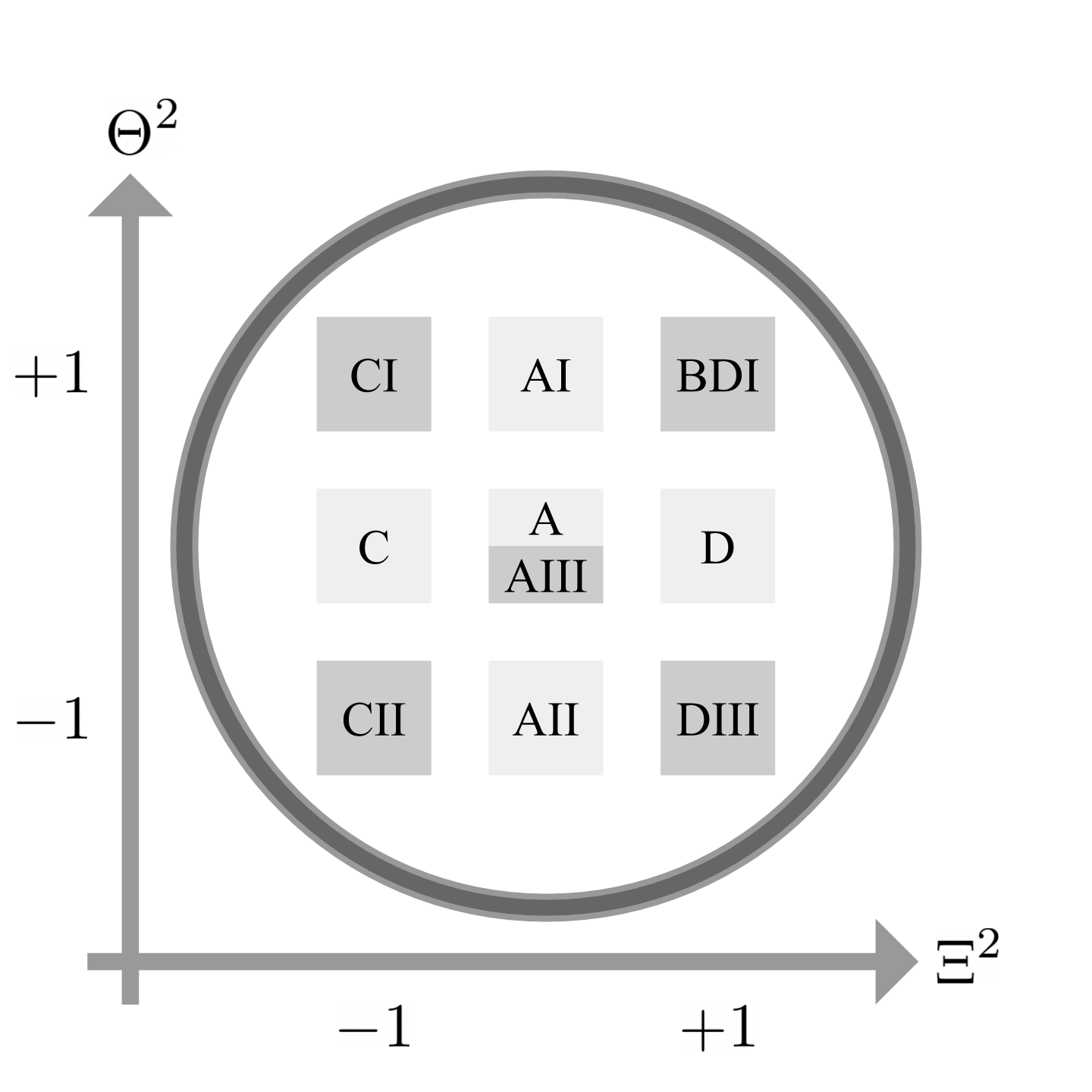}
\caption{The Bott clock. The ten Altland-Zirnbauer classes arranged by the tenfold-way pattern of non-spatial symmetries. These are the anti-unitary time-reversal symmetry, $\TR$, and the anti-unitary particle-hole anti-symmetry, $\PH$. The darkly shaded classes are invariant under the unitary sublattice/chiral anti-symmetry, $\Chiral=\TR\PH$. See Table~\ref{tab:per}.}
\label{fig:bott}
\end{figure}

The discovery of the periodic table of topological insulators and superconductors (TISC) and the tenfold-way has revolutionized our understanding of topological phases of quantum matter~\cite{Schnyder2008Classification,kitaev2009periodic,schnyder2009classification,ryu2010topological,Hasan2010Colloquium,moore2010birth,stone2010symmetries,Teo2010Topological,franz2013topological,witten2015three,Chiu2016Classification}. The table has enabled the classification of all weakly-interacting fermionic phases in systems with any Altland-Zirnbauer (AZ) symmetry class, which are the ten possible classes of non-spatial symmetries~\cite{Altland1997Nonstandard,heinzner2005symmetry,zirnbauer2010symmetry}; see Fig.~\ref{fig:bott} and Table~\ref{tab:per}. These classes are characterized by the presence or absence of internal symmetries of the system, such as time-reversal, particle-hole symmetry, spin rotations, and charge conservation. In the absence of spatial crystalline symmetries, the tenfold-way also provides a classification of ``strong" and ``weak" anomalous topological surface states due to the celebrated bulk-boundary correspondence, relating the topological properties of a gapped bulk to the properties of its anomalous surface theory~\cite{Kane2005Topological,Kane2005Quantum,FuKaneMele2007Topological,Moore2007Topological,Hsieh2009Observation,Roy2009Topological,Fu2007Topological,Konig2007Quantum,xia2009observation,Chen2009Experimental,franz2013topological,Khalaf2018Symmetry}.

The tenfold-way and the mutual relations amongst symmetry classes within it are thus at the physical foundation of our understanding of topological quantum phenomena. One of the insightful ways to derive and decipher the structure of the periodic table of TISC is to naturally encapsulate the tenfold-way within ``complex $\KK$-theory" and ``real $\KK$-theory", the former of which is sometimes referred to as `the simplest generalized cohomology theory'. Since our understanding of their importance, this and other cohomology theories have been used throughout physics to explore and understand various phenomena. Examples of such are complex and real cobordisms which have been used to classify the strongly-interacting invertible fermionic topological phases~\cite{freed2016reflection}.

In this paper, we utilize the mathematical notion of equivariant \emph{``spectra"} and the theory of equivariant stable homotopy, which further generalize the notion of a cohomology theory and unveils deeper relations within the symmetry classes of quantum systems; see Refs.~\onlinecite{may1996equivariant,elmendorf1997rings}. Similar notions had been surfacing in various areas of physics over the past decade, notable examples include: Kitaev~\cite{kitaev2011toward,kitaev2013topological,kitaev2015homotopy}, Freed and Hopkins~\cite{freed2014short,freed2016reflection}, Kapustin et~al.~\cite{kapustin2014symmetry,Kapustin2015Topological} and  Chen, Gu, Liu, and Wen~\cite{Chen2013Symmetry,Gu2014Symmetry}; for an overview see Xiong at Ref.~\onlinecite{Xiong2018Minimalist}.
Moreover, as the modern mathematical framework for algebra in spectra has also been building up over the past decade (a comprehensive treatment is provided in Ref.~\onlinecite{Lurie2017Higher}), we believe this modern formulation may shed light on various further new and exciting topics across physics. We move on to show how to use equivariant spectra in order to perform a unified analysis of crystalline TISC (CTISC) in all ten AZ symmetry classes, and obtain a complete classification of topological phases as well as anomalous surface states in various key crystalline space-group symmetries; see Table~\ref{tab:mainres} in Sec.~\ref{sec:main_results}.

\begin{table}[t]
\centering
\caption{Topological invariants, \(\pi_0(\sC_{q-d})\) and \(\pi_0(\sR_{q-d})\), 
for complex and real Altland-Zirnbauer symmetry classes, \(q\), in \(d\) spatial dimensions. See Fig.~\ref{fig:bott}.}\label{tab:per}
\begin{tabularx}{\linewidth}{llXXXXXX}
\hline\hline
$q$ & 
Classifying space & Class & $d=0$ & $d=1$ & $d=2$ & $d=3$
\\ \hline
0 & 
$\sC_0=\prod_{m}\tfrac{\Unit(n)}{\Unit(m)\times \Unit(n-m)}$
& A 	& $\ZZ$	& 0		& $\ZZ$	& 0								
\\ 
1 & 
$\sC_1=\Unit(n)$									
& AIII & 0		& $\ZZ$	& 0		& $\ZZ$									
\\ \hline
0 & 
$\sR_0=\prod_{m} \tfrac{\Ort(n)}{\Ort(m)\times \Ort(n-m)}$	
& AI 	& $\ZZ$	& 0 	& 0		& 0						
\\ 
1 & 
$\sR_1=\Ort(n)$									
& BDI 	& $\ZZ_2$	& $\ZZ$	& 0		& 0										
\\ 
2 & 
$\sR_2=\Ort(2n)/\Unit(n)$
& D 	& $\ZZ_2$	& $\ZZ_2$	& $\ZZ$	& 0							
\\ 
3 & 
$\sR_3=\Unit(2n)/\Simp(n)$								
& DIII & 0		& $\ZZ_2$	& $\ZZ_2$	& $\ZZ$							
\\
4 & 
$\sR_4=\prod_{m} \tfrac{\Simp(n)}{\Simp(m)\times \Simp(n-m)}$
& AII 	& $\ZZ$	& 0		& $\ZZ_2$	& $\ZZ_2$ 
\\ 
5 & 
$\sR_5=\Simp(n)$
& CII 	& 0		& $\ZZ$	& 0		& $\ZZ_2$									
\\ 
6 & 
$\sR_6=\Simp(n)/\Unit(n)$
& C 	& 0		& 0		& $\ZZ$	& 0						
\\ 
7 & 
$\sR_7=\Unit(n)/\Ort(n)$
& CI 	& 0		& 0		& 0		& $\ZZ$			
\\ \hline\hline
\end{tabularx}
\end{table}

Over the past several years, many naturally inquired the effects of spatial crystalline symmetries of CTISC on their topological classification~\cite{Fu2010Odd,Fu2011Topological,hsieh2012topological,dziawa2012topological,tanaka2012experimental,xu2012observation,Mong2010Antiferromagnetic,slager2013space,Chiu2013Classification,Benalcazar2014Classification,Varjas2015Bulk,Shiozaki2015Z2,Cho2015Topological,Yang2015Topological,wang2016hourglass,Ezawa2016Hourglass,Varjas2017Space,Yang2017Topological,Wieder2018Wallpaper,Bouhon2017Global,bouhon2017bulk,Kruthoff2017Topological,kruthoff2017topology,Cornfeld2019Classification,Song2019Topological}. A major advancement in that topic came with the formulation of symmetry indicators (SI) and topological quantum chemistry which enable the detection of topological phases from the crystalline symmetry properties of the band structure~\cite{Dong2016Classification,po2017symmetry,bradlyn2017topological,Watanabe2018Structure,Bradlyn2018Band,song2018quantitative,Cano2018Building,Vergniory2017Graph,Ono2018Unified,vergniory2018high,Khalaf2018Symmetry,Khalaf2018Higher,Geier2020Symmetry,Ono2020Refined,ono2020Z2}. These have contributed to the discovery and understanding of diverse topological phenomena such as higher-order TISC (HOTISC)~\cite{parameswaran2017topological,Benalcazar2017Quantized,Benalcazar2017Electric,Song2017d,Langbehn2017Reflection,Schindler2018Higher,schindler2018bismuth,xu2017topological,Shapourian2018Topological,lin2017topological,Ezawa2018Higher,Khalaf2018Higher,Geier2018Second,trifunovic2018higher,fang2017rotation,okuma2018topological}, ``fragile" TISC~\cite{bradlyn2017topological,Po2018Fragile,bouhon2018wilson,bradlyn2018disconnected}, obstructed atomic limits~\cite{bradlyn2017topological,bradlyn2018disconnected}, and boundary-obstructed TISC~\cite{khalaf2019boundary}.

The majority of classification efforts have naturally focused on time-reversal invariant topological insulators (TIs) with spin-orbit coupling, i.e., AZ class AII; see Fig.~\ref{fig:bott}. Nevertheless, our understanding of topological crystalline phenomena has since expanded to incorporate other AZ classes. However, it has been known that some CTISC are not detectable by their SI and that other topological invariants (such as Berry phases) are needed to discern them, this has been dubbed surface state ambiguity. Particularly, for three AZ classes, AIII, DIII, and CI, no gapped topology is detectable by SI~\footnote{In general, this does not hold for the extensions discussed in Sec.~\ref{sec:mag}; see, e.g., Ref.~\onlinecite{Geier2020Symmetry}.};
\setcounter{fnnumber}{\thefootnote}
see discussion in Sec.~\ref{sec:no-SI}. Noticeably, over the past couple of years, two independent works by Khalaf et~al.~\cite{Khalaf2018Symmetry} and Song et~al.~\cite{Song2019Topological} have presented a classification of anomalous surface states for AZ class AII using different methods.

In this paper, we show that the topological phases of CTISC are all manifestations of an underlying spectrum for each space group symmetry. We first derive these spectra and harness their properties to gain a unified description of the $\KK$-theory and the topological invariants amongst all ten AZ symmetry classes. We further use spectra to gain an understanding of atomic insulators (AI) and thus of anomalous surface states which may be regarded as their complements.

Before reading on, we wish to draw the readers' attention to two key points. First, one should note that equivariant spectra are not to be confused with the related notion of ``spectral sequences", which have also been used to describe CTISC phenomena~\cite{shiozaki2018atiyah,shiozaki2018generalized,okuma2018topological,stehouwer2018classification}. Second, as we discuss in Sec.~\ref{sec:ext}, there are many possible extensions of our analysis. Prominently, one may study either the so-called ``magnetic space-groups"~\cite{Zhang2015Topological,Watanabe2018Structure,okuma2018topological,shiozaki2019classification,ono2020Z2} or superconductors where the particle-hole anti-symmetry does not commute with the crystalline symmetry~\cite{Geier2020Symmetry,Ono2020Refined,ono2020Z2}. Herein, we focus on the case where the anti-unitaries commute with the crystalline symmetries, leaving further extensions for future works.

The rest of the paper is organized as follows: In Sec.~\ref{sec:over}, we overview the classification problem of CTISC and present the essence of the equivariant spectra paradigm. In Sec.~\ref{sec:main_results}, we present our main classification results and outline the main methods used to derive them. In Sec.~\ref{sec:detailed}, we take a ``hands-on" approach and provide detailed derivation via a thorough study of a pedagogical example. We conclude in Sec.~\ref{sec:discussion}.

\section{Physical overview}\label{sec:over}
\begin{figure*}[t]
\centering
\includegraphics[width=\linewidth]{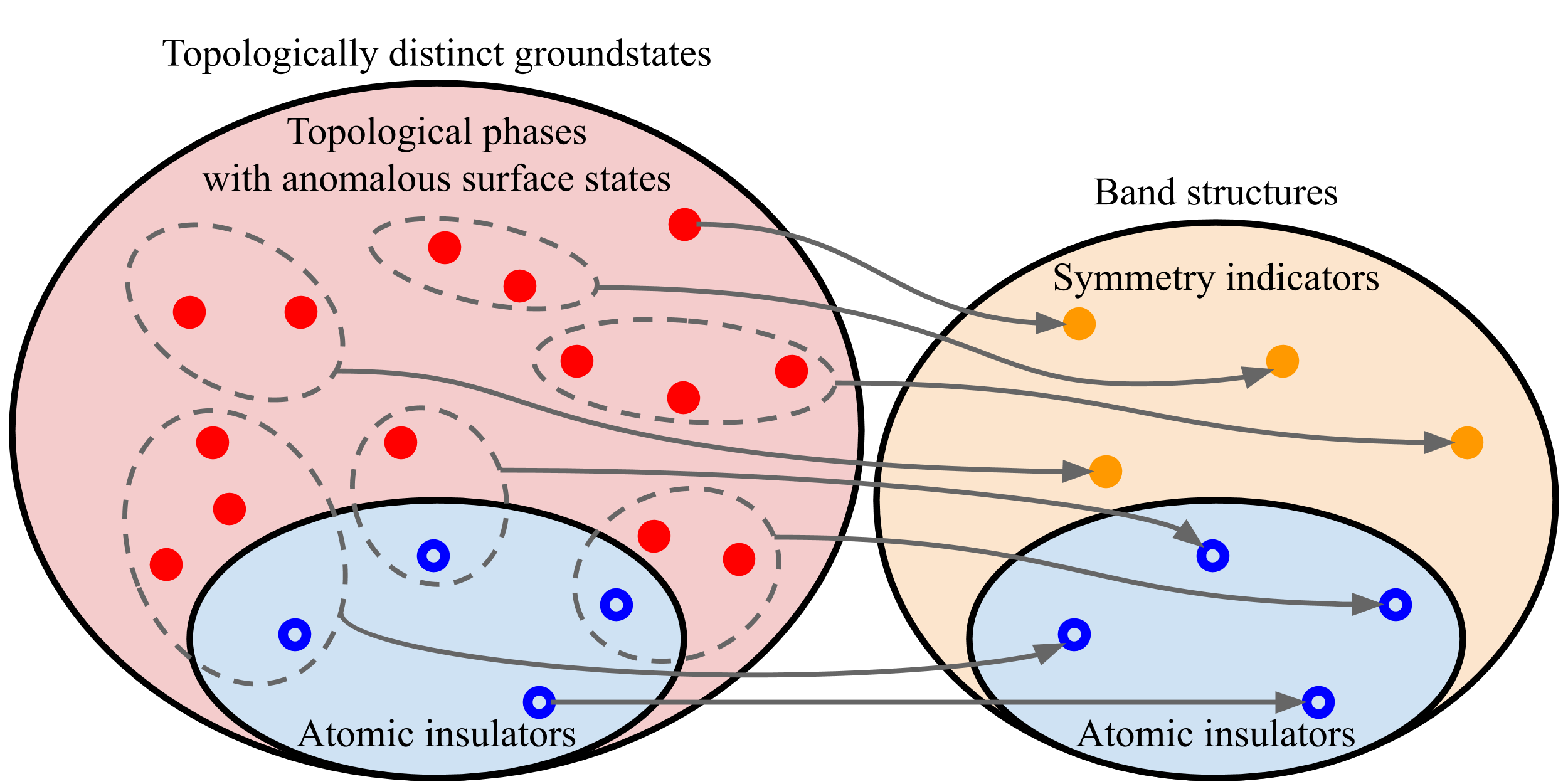}
\caption{Pictorial depiction of Eq.~\eqref{eq:main-diag}. Topological phases with anomalous surface states are topologically distinct groundstates modulo atomic insulators. The symmetry indicators of insulators and superconductors capture partial knowledge of the full classification.}
\label{fig:main-diag}
\end{figure*}

\subsection{Crystalline topological insulators and superconductors}\label{sec:intro_BS}

The classification of all quantum states protected by symmetries of the system is one of the main challenges facing the condensed matter physics community. Even in the absence of any crystalline symmetries, there is a plethora of topological phenomena including many exotic states of matter~\cite{Wen2019Choreographed}. In this work, we focus our efforts on the classification of topological states of weakly-interacting fermions. Their symmetries in general are split into internal non-spatial symmetries and spatial symmetries such as crystalline symmetries.
Before moving on, we note that in the absence of any crystalline symmetries, major progress was recently made in the classification of ``invertible" topological phases of \emph{strongly-interacting} fermions~\cite{freed2016reflection}; in Sec.~\ref{sec:cobord}, we discuss how this may be expanded using the formalism described in this paper.

In presence of crystalline symmetries, there is already a vast phenomenology of topological phases including HOTISC, fragile TISC, obstructed atomic limits, and boundary-obstructed TISC.
One of the remarkable properties of the TISC in absence of crystalline symmetries is that they all fit within the same systematic periodic table of TISC; see Table~\ref{tab:per}. These diverse phases include: integer quantum Hall (IQH) phases~\cite{Thouless1982Quantized,Avron1983Homotopy}; time-reversal symmetric quantum spin-Hall phases and strong/weak TIs~\cite{Hasan2010Colloquium,Qi2011Topological,Kane2005Quantum,Kane2005Topological,Konig2007Quantum,FuKaneMele2007Topological,Moore2007Topological,Roy2009Topological}; topological Majorana bound states and $p_x+ip_y$ topological superconductors (TSCs)~\cite{kitaev2001unpaired,Read2000Paired};
and many others. It is thus extremely desirable to treat CTISC phenomena in a similar unified manner; this is what we are set to do.

One says two quantum states belong to the same topological phase if they can be adiabatically deformed into one another while respecting the symmetries of the system and without encountering a phase transition (i.e., equivariantly  homotopic). We focus on either insulators or superconductors where the groundstate is characterized by the existence of an energy gap. The many-body groundstate of a free fermion gapped system is given by the choice of filled valance bands and empty conduction bands.
There is a natural notion of band addition by stacking bands together. Moreover, in the study of band representations, one often also considers band subtraction. This is known as the ``stable" limit, within it, as we explain in Sec.~\ref{sec:intro_class}, the classification of all topologically distinct groundstates is given by (twisted) equivariant $\KK$-theory~\cite{freed2013twisted,Morimoto2013Topological,Shiozaki2014Topology,Hsieh2014CPT,Shiozaki2016Topology,Shiozaki2017Topological,Trifunovic2017Bott,shiozaki2018atiyah,Geier2018Second,trifunovic2018higher,okuma2018topological}.

In this paper, we provide forceful techniques that allow us to carry out explicit calculations of the full $\KK$-theoretic classification. However, easily attainable partial knowledge of the topology of a state is often given in terms of its SI. We believe it is useful to first explain the relations between $\KK$-theory and SI.

At each point in the Brillouin Zone (BZ), $\vk\in\BZ$, one can always decompose each band according to representations of the symmetry group preserving that point. This can be viewed as a quantum state of an effective 0-dimensional (0D) system at $\vk$. The topological invariants of each representation at each point are called ``band labels". However, the fact that these band labels originate from a state on some $d$-dimensional BZ naturally imposes certain compatibility conditions; for example, continuity requires the labels to remain constant along high-symmetry lines within the BZ. It is thus natural to study physical band structures, $\BS$, as band labels modulo physical compatibility relations. In order to avoid possible confusion, we note that there exists variability in the literature regarding which compatibility relations to impose in the definition of $\BSfull$, this may lead to an inclusion of (semi-)metallic states within $\BSfull$ (see, e.g., discussion in Refs.~\onlinecite{po2017symmetry,Ono2018Unified,Geier2020Symmetry}); we focus our attention on TISC and hence use the stricter definition such that $\BS$ only includes band labels corresponding to gapped systems, i.e., insulators and superconductors~\footnote{The \emph{insulating} band structures are often referred to as representation-enforced quantum band insulators~\cite{po2017symmetry}.}.

Each distinct class of topologically equivalent TISC groundstates is represented by a distinct member of $\KK$-theory. Since $\BS$ are topological invariants, the band labels of every member of such an equivalence class would be identical. Therefore, two quantum states with different band labels are necessarily topologically distinct, and there exists an onto map (epimorphism), $\KK\to\BS$. However, the converse is in general not true; there often exist two quantum states which have identical band labels but are nevertheless topologically distinct~\cite{Khalaf2018Symmetry,song2018quantitative}; see Fig.~\ref{fig:main-diag}.

A compelling feature of TISC is bulk-boundary correspondence and the existence of anomalous surface states. This persists in CTISC; the boundary of a gapped insulator or superconductor which is not an atomic insulator or superconductor (AI) hosts either strong, weak, or higher-order anomalous surface states. We emphasize that in order to keep with the conventional jargon we use $\AI$ to denote both atomic insulators and atomic superconductors. Therefore, in order to classify anomalous surface states one broadens the topological equivalence such that two quantum states are considered equivalent if they only differ by stacking AIs. The topological invariants of AIs within $\KK$-theory, $\AI\to\KK$, are thus quotiented out and the anomalous surface states, $\Surf$, are given by $\Surf=\KK/\AI$. By construction, this also nullifies fragile CTISC and obstructed atomic limits, the study of which can be done by solely inspecting the structure of $\AI$~\cite{bradlyn2017topological,Po2018Fragile,bradlyn2018disconnected}.

A similar construction is standardly carried for $\BS$ to obtain the symmetry indicators of TISC, $\SI=\BS/\AI$. As discussed above, generically, two quantum states which have identical SI may still have topologically distinct anomalous surface states~\cite{Khalaf2018Symmetry,song2018quantitative}. Therefore, SI serve as easy-to-compute useful topological invariants which approximate the full $\KK$-theoretic topological classification of anomalous surface states. The above discussion is neatly summarized by the following commutative diagram:
\begin{equation}\label{eq:main-diag}
\xymatrix{
\AI\ar[r] \ar@{=}[d] & \KK \ar@{>>}^-{\rotatebox[origin=c]{270}{\text{onto}}}[d] \ar@{>>}[r] & \KK/\AI=\Surf \ar@{>>}^-{\rotatebox[origin=c]{270}{\text{onto}}}[d]\\
\AI\ar[r] & \BS \ar@{>>}[r] & \BS/\AI=\SI
}
\end{equation}
This is depicted in Fig.~\ref{fig:main-diag}. As a means of better approximating the $\KK$-theory, one may add non-local invariants such as Berry phases to the SI~\cite{bradlyn2017topological,song2018quantitative}. However, to the knowledge of the authors, no proof is yet given that the currently known invariants capture all possible CTISC phases in all AZ symmetry classes encapsulated by $\KK$-theory.

In the following sections, we briefly provide the essential basics of the full $\KK$-theoretic topological classification and how spectra may be utilized to resolve it.

\subsection{Topological classification}\label{sec:intro_class}

For simplicity, we initiate our discussion of topological classification by considering systems with no symmetry (other than charge conservation).
As discussed above, the many-body groundstate of a free fermion gapped system, $\tket{\psi(\vk)}$, is given by the choice of filled valance bands and empty conduction bands. For a system with $n$ bands, one has to first specify the number of filled bands, $m=0,\ldots,n$. Next, one has to pick which bands are filled and which are empty; this is equivalent to specifying a basis, i.e., a $\Unit(n)$ matrix. Recall that two quantum states belong to the same topological phase if they can be adiabatically deformed into one another without closing the energy gap, therefore, the exact basis choices within either the valance or the conduction bands, i.e., $\Unit(m)$ and $\Unit(n-m)$, are redundant. Thus, the space of topologically distinct basis choices is denoted $\sC_0(n)=\prod_{m}\tfrac{\Unit(n)}{\Unit(m)\times \Unit(n-m)}$. The groundstate of a translationally invariant quantum system is given by continuously making these basis choices for every point in the BZ, $\vk\in\BZ$, i.e., picking a continuous map $\tket{\psi(\vk)}\in\Map(\BZ,\sC_0(n))$, where $\Map(X,Y)$ is the space of maps from $X$ to $Y$. Therefore, the set of topologically distinct groundstates is given by, $\pi_0(\Map(\BZ,\sC_0(n)))$.
Here, $\pi_0(Y)$ is the set of topologically connected components of a space $Y$. Intuitively, the addition of trivial filled or empty bands should not drastically affect the groundstate of a quantum system. This intuition manifests itself in the mathematical notion of stability. Particularly, for any $n,n'$ \emph{large enough}, one finds $\Map(X,\sC_0(n))\simeq\Map(X,\sC_0(n'))$; this is noted as the ``stable limit"~\footnote{When the total number of bands is sufficiently small, non-stable topology is possible, examples include Hopf insulators and superconductors~\cite{Moore2008Topological,Deng2013Hopf,Kennedy2016Topological,kennedy2016bott}. Nevertheless, most physical systems have a very large number of bands and the topology is well captured by the stable limit.},
\begin{equation}\label{eq:basic-stable-homotopy}
\left\{\substack{\textstyle\text{Topologically distinct}\\ \textstyle\text{stable groundstates}}\right\}=\pi_0(\Map(\BZ,\sC_0)),
\end{equation}
where $\sC_0\defeq\lim_{n\to\infty}\sC_0(n)$ is denoted the ``classifying space"; see Table~\ref{tab:per}.

Although in general, the set of topologically distinct groundstates is not required to have any algebraic structure, the stable limit always yields a set with an abelian group structure.
This observation is deeply related to an alternative route leading to the same classification. As discussed in Sec.~\ref{sec:intro_BS} when studying topological BS one often allows formal differences of quantum states. This in-fact enforces an abelian (additive) group structure on the space of states, which is the defining property of $\KK$-theory,
\begin{equation}\label{eq:basic-KU}
\left\{\substack{\textstyle\text{Topologically distinct}\\ \textstyle\text{stable groundstates}}\right\}=\KU^0(\BZ).
\end{equation}
Indeed, as explained in further detail in Appendix~\ref{sec:KRKC-theory}, one has $\KU^0(X)=\pi_0(\Map(X,\sC_0))$ for any topological space $X$. Henceforth, we focus our attention on stable groundstates.

In the following sections, we consider the effects of symmetries on this classification scheme, where the $\KK$-theoretic classification becomes hard to compute. This is also where the relations between $\KK$-theory and equivariant spectra become more apparent.

\subsubsection{Symmetry \& equivariance}\label{sec:intro_sym}
In the presence of symmetries, quantum states are only considered equivalent if they can be adiabatically deformed into one another without breaking the symmetries. One thus has to refine the analysis of the previous section. The group of topologically distinct groundstates is given by equivalence classes of maps from the BZ to the classifying space which respect the symmetries; such maps are called ``equivariant" maps.

The ten AZ symmetry classes correspond to the presence or absence of the non-spatial symmetries. These are the two antiunitary symmetries, time-reversal symmetry, $\TR$, particle-hole anti-symmetry, $\PH$, as well as the unitary sublattice/chiral anti-symmetry, $\Chiral$; see Fig.~\ref{fig:bott}. In addition to these, a crystalline space-group symmetry acts via its point-group, $G$, on the BZ. In this paper we focus on non-magnetic space-groups where the crystalline symmetries act independently of the non-spatial symmetries; see Sec.~\ref{sec:mag} for further discussion.

The simplest symmetry is sublattice/chiral symmetry, $\Chiral$, which exchanges the filled and empty bands and thus acts only on the classifying space. This restricts the classifying space to $\Unit(n)=\sC_1(n)\subset\sC_0(2n)$ which is fixed under the symmetry. The group of topologically distinct groundstates in the absence or presence of $\Chiral$ is therefore given by equivariant $\KK$-theory which corresponds to equivariant maps,
\begin{equation}\label{eq:KU-via-Cq}
\KU_G^{-q}(\BZ)=\pi_0(\Map_G(\BZ,\sC_q)).
\end{equation}
Here, as in Table~\ref{tab:per}, $q=0,1$ correspond to AZ classes A, AIII, respectively; these are denoted the two complex AZ symmetry classes.

The situation becomes rather more intricate when considering the presence of antiunitary symmetries. We thus reserve the detailed discussion to Appendix~\ref{sec:KRKC-theory}; for a nice overview see Ref.~\onlinecite{kennedy2016bott}. Both $\TR$ and $\PH$ represent an antiunitary, $\ZT$-symmetry and in-fact differ only by their action on $\sC_q$. There are eight possible antiunitary actions, each of which corresponds to choosing a different real structure on $\sC_q$. These are denoted the eight real AZ symmetry classes and they are classified by real equivariant $\KK$-theory,
\begin{equation}\label{eq:KR-intro}
\KR_G^{-q,0}(\BZ)=\pi_0(\Map_{G\times\ZT}(\BZ,\sC_q)).
\end{equation}
Here, as in Table~\ref{tab:per}, $q=0,\ldots,7$ correspond to AZ classes AI, BDI, D, DIII, AII, CII, C, CI, respectively. Note that $\sC_{q}$ and $\sC_{q\tmod 2}$ are isomorphic as topological spaces, but differ by the action of $\ZT$. This becomes more apparent in the following example:

Consider a space, $X$, which is invariant under the antiunitary $\ZT$-symmetry. This can be, for example, either some parameter space of the system or a space surrounding a defect~\cite{Teo2010Topological,Chiu2016Classification}. Similar to the sublattice/chiral case, the $\ZT$-symmetry now acts to restrict $\sC_q$ to its fixed-points space under the symmetry, $\sR_q\subset\sC_q$, such that
\begin{equation}
\KR_G^{-q,0}(X)=\pi_0(\Map_G(X,\sR_q)).
\end{equation}
The fixed-points spaces, $\sR_q$, are denoted the real classifying spaces and explicit expressions for them are given in Table~\ref{tab:per}.

Before moving on, we note that the notation ``$\ZZ_2^T$" for the antiunitary $\ZT$-symmetry originates from its manifestation in classes AII and AI as time-reversal symmetry. Therefore, even though we provide a unified classification for all ten AZ classes, which includes manifestations as particle-hole symmetry, we stick to this notation in order to comply with the accepted dogmas.

\subsubsection{Spectra - an introduction}\label{sec:introno_spec}
In general, \emph{equivariant} $\KK$-theory [Eq.~\eqref{eq:KR-intro}] is not easy to compute. In this paper, we utilize the notion of equivariant spectra and use it to obtain explicit results. In order to acquaint the reader with this powerful notion, we provide some properties of $\KK$-theory and spectra which provide some useful intuitions.

Consider a 1-sphere (circle), $X=\Ss^1$, in AZ class, $q$, which is invariant under $\ZT$. The topological classification, $\KR^{-q,0}(S^1)$, is given by maps from $S^1$ to $\sR_q$. When examining the space of maps, one notices that $\sR_q$ are not always path-connected (e.g., $\sR_1(n)=\Ort(n)$ has two distinct connectivity components discerned by the sign of the determinant). Therefore, one must first pick a base-point within one of the connected components and then a map from the 1-sphere starting in that base-point. The latter is the definition of the fundamental group, $\pi_1$, and hence
\begin{equation}\label{eq:pi0pi1}
\KR^{-q,0}(S^1)=\pi_0(\Map(S^1,\sR_q))\simeq\pi_0(\sR_q)\times\pi_1(\sR_q).
\end{equation}
The based loop-space of $Y$, denoted $\Omega Y$, is the space of all based loops within $Y$, such that
\begin{equation}
\pi_1(Y)\simeq\pi_0(\Omega Y).
\end{equation}
Remarkably, all the classifying spaces $\sC_q$ and $\sR_q$ are all infinite loop-spaces such that
\begin{align}
&\sC_{q}=\Omega\sC_{q-1}, &\sR_{q}=\Omega\sR_{q-1}.
\end{align}
Here, $q\in\ZZ$ and one has $\sR_q\simeq\sR_{q\tmod 8}$.
This unique property is a manifestation of Bott-periodicity~\cite{atiyah1964clifford,bott1969lectures} and is at the heart of the periodic table of TISC. This also brings us to spectra.

The notion of spectra has considerably evolved over the past decades, see Refs.~\onlinecite{adams1974stable,Lurie2017Higher}.
Classically, a spectrum is a sequence of spaces, $X_i$, which satisfy $X_i=\Omega X_{i+1}$. We have thus already encountered two examples of such spectra: 
First, the $\KU$ spectrum defined by the (2-periodic) sequence $X_i=\sC_{-i}$. Second, the so-called $\KO$ spectrum defined by the (8-periodic) sequence $X_i=\sR_{-i}$.

An equivariant spectrum is similarly classically defined as a sequence of spaces, $X_i=\Omega X_{i+1}$ with an action of a group $G$; these have to satisfy some compatibility relations discussed in Appendix~\ref{sec:spectra}. One of the main equivariant spectra of interest in this work is the $\KR$ spectrum. It is defined by the 8-periodic sequence $X_i=\sC_{-i}$ with the action of $\ZT$ such that the fixed-point spaces are given by $\sR_{-i}\subset\sC_{-i}$.

\begin{figure}[t]
\centering
\includegraphics[width=\linewidth]{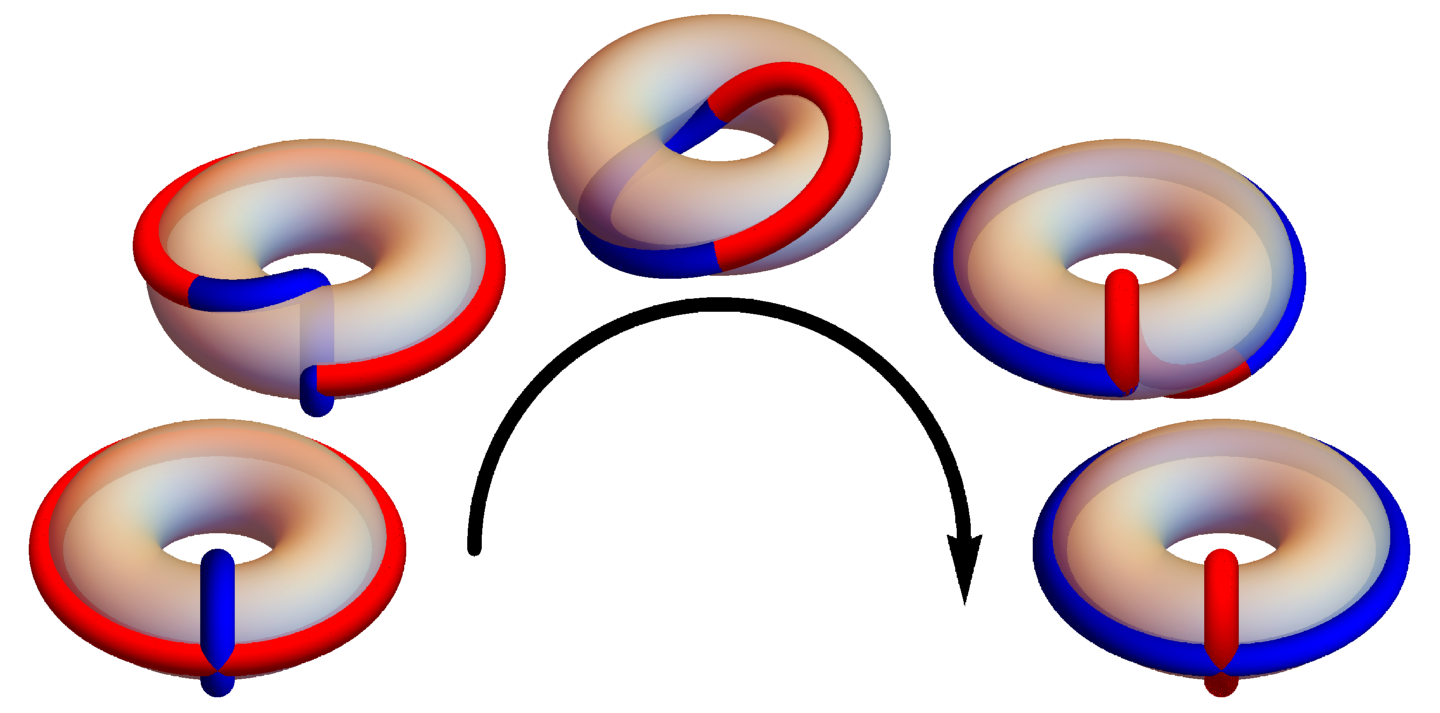}
\caption{Two concatenated loops on any topological space with a group structure always commute up to a homotopy. Here, such a homotopy is depicted for two loops (red and blue) on an abstract torus, $\Tt^2\simeq \Unit(1)\times\Unit(1)$; three intermediate stages of this homotopy are presented.}
\label{fig:loops}
\end{figure}

There is a natural internal group operation for loop-spaces given by concatenating two loops, where the inverse is given by reversing the orientation. Furthermore, since a spectrum consists of infinite loop-spaces one can show that this group operation must be commutative up to a continuous deformation, i.e., a homotopy; see Fig.~\ref{fig:loops}. This implies that every spectrum is endowed with an abelian additive group structure. Specifically, any two points $x,x'\in X_i$ may be added or subtracted such that $x\pm x'\in X_i$ is well-defined and unique up to homotopy.

As a consequence, many properties of abelian groups are also respected by spectra, this often poses a major simplification. In particular, the space of $G$-equivariant maps between two $G\times\ZT$-equivariant spectra $E,E'$ can be thought of as the ``abelian group" of $G$-equivariant homomorphisms $\Hom_G(E',E)$ which is by itself a $\ZT$-equivariant spectrum. Moreover, for any equivariant space, $X$, one can construct a spectrum $\SSp[X]$ such that~\footnote{A precise definition is provided in Appendix~\ref{sec:constructions-in-spectra}.}
\begin{equation} 
\Map_{G}(X,E)\approx\Hom_{G}(\SSp[X],E).
\end{equation}
This implies that the spaces of all quantum groundstates of a crystalline system for all real AZ symmetry classes can be expressed by one $\ZT$-equivariant spectrum,
\begin{equation}\label{eq:Hom-BZ-KR} 
\Hom_G(\SSp[\BZ],\KR_G).
\end{equation}
The complex AZ classes are similarly captured by an equivariant spectrum denoted $\KC_G$.
Furthermore, the atomic insulators also form a spectrum and therefore the map $\AI\to\KK$ whose cokernel yields the topological phases with ASS [see Eq.~\eqref{eq:main-diag}] also stems from a homomorphism of spectra.

This mindset of describing everything in terms of equivariant spectra allows us to perform algebraic abelian calculations that capture all AZ symmetry classes at once and yield explicit results we hereby present.

\section{Main results \& methods}\label{sec:main_results}
\begin{table*}[t]
\centering
\caption{Complete classification of topological phases with anomalous surface states of 3-dimensional crystalline topological insulators and superconductors for key point-groups (PG) and space-groups (SG) in all ten Altland-Zirnbauer symmetry classes. Each entry corresponds to the total group of all strong, weak, and higher-order surface states; more detailed results are provided in Appendix~\ref{app:tables}. Point-groups are given in Sch\"{o}nflies notation. Space-groups are given in Hermann-Mauguin notation and specified by their number as given in Ref.~\onlinecite{hahn1983international}.}\label{tab:mainres}
\renewcommand{\arraystretch}{1.05}
\begin{tabularx}{\textwidth}{XXXXXXXXXXX} 
\hline\hline
 PG  & $C_1$ & $C_i$ & $C_2$ &  & $C_s$ &  & $C_{2h}$ &  \\
 SG  & $\mathrm{P1}$ & $\mathrm{P\bar{1}}$ & $\mathrm{P2}$ & $\mathrm{C2}$ & $\mathrm{Pm}$ & $\mathrm{Cm}$ & $\mathrm{P2/m}$ & $\mathrm{C2/m}$ \\
 & \#1 & \#2 & \#3 & \#5 & \#6 & \#8 & \#10 & \#12 \\
\cline{1-9}
A & $\ZZ^{3}$ & $\ZZ^{3}\times\ZZ_{2}$ & $\ZZ$ & $\ZZ$ & $\ZZ^{3}$ & $\ZZ^{2}$ & $\ZZ^{3}$ & $\ZZ^{2}$ \\
AIII & $\ZZ^{4}$ & $0$ & $\ZZ^{6}$ & $\ZZ^{4}$ & $\ZZ^{6}$ & $\ZZ^{4}$ & $0$ & $0$ \\
\cline{1-9}
AI & $0$ & $0$ & $0$ & $0$ & $\ZZ^{2}$ & $\ZZ$ & $\ZZ^{2}$ & $\ZZ$ \\
BDI & $\ZZ^{3}$ & $0$ & $\ZZ$ & $\ZZ$ & $\ZZ^{2}$ & $\ZZ^{2}$ & $0$ & $0$ \\
D & $\ZZ^{3}\times\ZZ_{2}^{3}$ & $\ZZ^{3}$ & $\ZZ\times\ZZ_{2}^{3}$ & $\ZZ\times\ZZ_{2}^{2}$ & $\ZZ\times\ZZ_{2}$ & $\ZZ\times\ZZ_{2}$ & $\ZZ$ & $\ZZ$ \\
DIII & $\ZZ\times\ZZ_{2}^{6}$ & $0$ & $\ZZ^{5}\times\ZZ_{2}^{4}$ & $\ZZ^{3}\times\ZZ_{2}^{3}$ & $\ZZ^{4}\times\ZZ_{2}$ & $\ZZ^{2}\times\ZZ_{2}^{2}$ & $0$ & $0$ \\
AII & $\ZZ_{2}^{4}$ & $\ZZ_{2}^{3}\times\ZZ_{4}$ & $\ZZ_{2}^{5}$ & $\ZZ_{2}^{4}$ & $\ZZ^{2}\times\ZZ_{2}^{2}$ & $\ZZ\times\ZZ_{2}^{2}$ & $\ZZ^{2}\times\ZZ_{2}^{3}$ & $\ZZ\times\ZZ_{2}^{3}$ \\
CII & $\ZZ^{3}\times\ZZ_{2}$ & $\ZZ_{2}$ & $\ZZ\times\ZZ_{2}^{4}$ & $\ZZ\times\ZZ_{2}^{3}$ & $\ZZ^{2}\times\ZZ_{2}^{3}$ & $\ZZ^{2}\times\ZZ_{2}^{2}$ & $\ZZ_{2}^{4}$ & $\ZZ_{2}^{3}$ \\
C & $\ZZ^{3}$ & $\ZZ^{3}\times\ZZ_{2}$ & $\ZZ$ & $\ZZ$ & $\ZZ\times\ZZ_{2}$ & $\ZZ\times\ZZ_{2}$ & $\ZZ\times\ZZ_{2}$ & $\ZZ\times\ZZ_{2}$ \\
CI & $\ZZ$ & $0$ & $\ZZ^{5}$ & $\ZZ^{3}$ & $\ZZ^{4}$ & $\ZZ^{2}$ & $0$ & $0$ \\
\hline\hline
 PG  & $D_2$ &  & $C_{2v}$ &  &  & $D_{2h}$ &  \\
 SG  & $\mathrm{P222}$ & $\mathrm{C222}$ & $\mathrm{Pmm2}$ & $\mathrm{Cmm2}$ & $\mathrm{Amm2}$ & $\mathrm{Pmmm}$ & $\mathrm{Cmmm}$ \\
 & \#16 & \#21 & \#25 & \#35 & \#38 & \#47 & \#65 \\
\cline{1-8}
A & $0$ & $0$ & $\ZZ^{4}$ & $\ZZ^{2}$ & $\ZZ^{3}$ & $\ZZ^{6}$ & $\ZZ^{4}$ \\
AIII & $\ZZ^{13}$ & $\ZZ^{8}$ & $\ZZ^{5}$ & $\ZZ^{4}$ & $\ZZ^{4}$ & $0$ & $0$ \\
\cline{1-8}
AI & $0$ & $0$ & $\ZZ^{4}$ & $\ZZ^{2}$ & $\ZZ^{3}$ & $\ZZ^{6}$ & $\ZZ^{4}$ \\
BDI & $0$ & $0$ & $\ZZ$ & $\ZZ$ & $\ZZ$ & $0$ & $0$ \\
D & $0$ & $\ZZ_{2}$ & $0$ & $\ZZ_{2}$ & $0$ & $0$ & $0$ \\
DIII & $\ZZ^{13}$ & $\ZZ^{8}\times\ZZ_{2}$ & $\ZZ^{4}$ & $\ZZ^{3}\times\ZZ_{2}$ & $\ZZ^{3}$ & $0$ & $0$ \\
AII & $\ZZ_{2}^{6}$ & $\ZZ_{2}^{5}$ & $\ZZ^{4}\times\ZZ_{2}$ & $\ZZ^{2}\times\ZZ_{2}^{2}$ & $\ZZ^{3}\times\ZZ_{2}$ & $\ZZ^{6}$ & $\ZZ^{4}\times\ZZ_{2}$ \\
CII & $\ZZ_{2}^{6}$ & $\ZZ_{2}^{5}$ & $\ZZ\times\ZZ_{2}^{5}$ & $\ZZ\times\ZZ_{2}^{4}$ & $\ZZ\times\ZZ_{2}^{4}$ & $\ZZ_{2}^{6}$ & $\ZZ_{2}^{5}$ \\
C & $0$ & $0$ & $\ZZ_{2}^{5}$ & $\ZZ_{2}^{3}$ & $\ZZ_{2}^{3}$ & $\ZZ_{2}^{6}$ & $\ZZ_{2}^{4}$ \\
CI & $\ZZ^{13}$ & $\ZZ^{8}$ & $\ZZ^{4}\times\ZZ_{2}$ & $\ZZ^{3}\times\ZZ_{2}$ & $\ZZ^{3}$ & $0$ & $0$ \\
\hline\hline
 PG  & $C_4$ & $S_4$ & $C_{4h}$ & $D_4$ & $C_{4v}$ & $D_{2d}$ &  & $D_{4h}$ \\
 SG  & $\mathrm{P4}$ & $\mathrm{P\bar{4}}$ & $\mathrm{P4/m}$ & $\mathrm{P422}$ & $\mathrm{P4mm}$ & $\mathrm{P\bar{4}2m}$ & $\mathrm{P\bar{4}m2}$ & $\mathrm{P4/mmm}$ \\
 & \#75 & \#81 & \#83 & \#89 & \#99 & \#111 & \#115 & \#123 \\
\cline{1-9}
A & $\ZZ$ & $\ZZ\times\ZZ_{2}$ & $\ZZ^{3}$ & $0$ & $\ZZ^{3}$ & $\ZZ$ & $\ZZ^{2}$ & $\ZZ^{5}$ \\
AIII & $\ZZ^{9}$ & $\ZZ$ & $0$ & $\ZZ^{12}$ & $\ZZ^{6}$ & $\ZZ^{6}$ & $\ZZ^{4}$ & $0$ \\
\cline{1-9}
AI & $0$ & $0$ & $\ZZ^{2}$ & $0$ & $\ZZ^{3}$ & $\ZZ$ & $\ZZ^{2}$ & $\ZZ^{5}$ \\
BDI & $\ZZ^{3}$ & $0$ & $0$ & $0$ & $\ZZ^{3}$ & $0$ & $0$ & $0$ \\
D & $\ZZ\times\ZZ_{2}^{2}$ & $\ZZ\times\ZZ_{2}^{3}$ & $\ZZ$ & $0$ & $0$ & $0$ & $0$ & $0$ \\
DIII & $\ZZ^{6}\times\ZZ_{2}^{3}$ & $\ZZ\times\ZZ_{2}^{4}$ & $0$ & $\ZZ^{12}$ & $\ZZ^{3}$ & $\ZZ^{6}$ & $\ZZ^{4}$ & $0$ \\
AII & $\ZZ_{2}^{4}$ & $\ZZ_{2}^{2}\times\ZZ_{4}$ & $\ZZ^{2}\times\ZZ_{2}^{2}$ & $\ZZ_{2}^{5}$ & $\ZZ^{3}\times\ZZ_{2}$ & $\ZZ\times\ZZ_{2}^{3}$ & $\ZZ^{2}\times\ZZ_{2}^{2}$ & $\ZZ^{5}$ \\
CII & $\ZZ^{3}\times\ZZ_{2}^{3}$ & $\ZZ_{2}^{2}$ & $\ZZ_{2}^{3}$ & $\ZZ_{2}^{5}$ & $\ZZ^{3}\times\ZZ_{2}^{4}$ & $\ZZ_{2}^{4}$ & $\ZZ_{2}^{4}$ & $\ZZ_{2}^{5}$ \\
C & $\ZZ$ & $\ZZ\times\ZZ_{2}$ & $\ZZ\times\ZZ_{2}$ & $0$ & $\ZZ_{2}^{6}$ & $\ZZ_{2}$ & $\ZZ_{2}^{2}$ & $\ZZ_{2}^{5}$ \\
CI & $\ZZ^{6}$ & $\ZZ$ & $0$ & $\ZZ^{12}$ & $\ZZ^{3}\times\ZZ_{2}^{3}$ & $\ZZ^{6}$ & $\ZZ^{4}$ & $0$ \\
\hline\hline
 PG  & $C_3$ & $C_{3i}$ & $D_3$ & $C_{3v}$ & $D_{3d}$ & $T$ & $T_h$ & $O$ & $T_d$ & $O_h$ \\
 SG  & $\mathrm{R3}$ & $\mathrm{R\bar{3}}$ & $\mathrm{R32}$ & $\mathrm{R3m}$ & $\mathrm{R\bar{3}m}$ & $\mathrm{P23}$ & $\mathrm{Pm\bar{3}}$ & $\mathrm{P432}$ & $\mathrm{P\bar{4}3m}$ & $\mathrm{Pm\bar{3}m}$ \\
 & \#146 & \#148 & \#155 & \#160 & \#166 & \#195 & \#200 & \#207 & \#215 & \#221 \\
\cline{1-11}
A & $\ZZ$ & $\ZZ\times\ZZ_{2}$ & $0$ & $\ZZ$ & $\ZZ$ & $0$ & $\ZZ^{2}$ & $0$ & $\ZZ$ & $\ZZ^{3}$ \\
AIII & $\ZZ^{4}$ & $0$ & $\ZZ^{4}$ & $\ZZ^{4}$ & $0$ & $\ZZ^{7}$ & $0$ & $\ZZ^{9}$ & $\ZZ^{3}$ & $0$ \\
\cline{1-11}
AI & $0$ & $0$ & $0$ & $\ZZ$ & $\ZZ$ & $0$ & $\ZZ^{2}$ & $0$ & $\ZZ$ & $\ZZ^{3}$ \\
BDI & $\ZZ^{2}$ & $0$ & $0$ & $\ZZ^{2}$ & $0$ & $\ZZ$ & $0$ & $0$ & $\ZZ$ & $0$ \\
D & $\ZZ\times\ZZ_{2}$ & $\ZZ$ & $\ZZ_{2}$ & $0$ & $0$ & $0$ & $0$ & $0$ & $0$ & $0$ \\
DIII & $\ZZ^{2}\times\ZZ_{2}^{2}$ & $0$ & $\ZZ^{4}\times\ZZ_{2}$ & $\ZZ^{2}$ & $0$ & $\ZZ^{6}$ & $0$ & $\ZZ^{9}$ & $\ZZ^{2}$ & $0$ \\
AII & $\ZZ_{2}^{2}$ & $\ZZ_{2}\times\ZZ_{4}$ & $\ZZ_{2}^{3}$ & $\ZZ\times\ZZ_{2}$ & $\ZZ\times\ZZ_{2}^{2}$ & $\ZZ_{2}^{2}$ & $\ZZ^{2}$ & $\ZZ_{2}^{3}$ & $\ZZ\times\ZZ_{2}$ & $\ZZ^{3}$ \\
CII & $\ZZ^{2}\times\ZZ_{2}$ & $\ZZ_{2}$ & $\ZZ_{2}^{3}$ & $\ZZ^{2}\times\ZZ_{2}^{2}$ & $\ZZ_{2}^{3}$ & $\ZZ\times\ZZ_{2}^{2}$ & $\ZZ_{2}^{2}$ & $\ZZ_{2}^{3}$ & $\ZZ\times\ZZ_{2}^{2}$ & $\ZZ_{2}^{3}$ \\
C & $\ZZ$ & $\ZZ\times\ZZ_{2}$ & $0$ & $\ZZ_{2}^{2}$ & $\ZZ_{2}$ & $0$ & $\ZZ_{2}^{2}$ & $0$ & $\ZZ_{2}^{2}$ & $\ZZ_{2}^{3}$ \\
CI & $\ZZ^{2}$ & $0$ & $\ZZ^{4}$ & $\ZZ^{2}\times\ZZ_{2}$ & $0$ & $\ZZ^{6}$ & $0$ & $\ZZ^{9}$ & $\ZZ^{2}\times\ZZ_{2}$ & $0$ \\
\hline\hline
\end{tabularx}
\renewcommand{\arraystretch}{1}
\end{table*}

The formulation of topological physics in terms of equivariant spectra provides a useful tool for exploring hidden relations of topological states of quantum matter.
We hereby provide explicit quantitative results that demonstrate these qualitative ideas.

Concretely, we first use the properties of the spectra formalism to obtain the complete classification of topologically distinct quantum states of 3D CTISC in all AZ symmetry classes for key space-groups.
Second, we utilize other aspects of spectra to obtain an understating of the nature of AIs in the $\KK$-theory classification and provide a complete classification of topological phases with anomalous surface states for these space groups; see pictorial depiction in Fig.~\ref{fig:main-diag}(left).

The classification groups of topological phases are summarized in Table~\ref{tab:mainres}; detailed results including the full $\KK$-theory as well as 2-dimensional (2D) layer-groups and 1-dimensional (1D) rod groups are presented in Appendix~\ref{app:tables}.

The complete mathematical derivation will be provided in Appendix~\ref{sec:math_background}. Nevertheless, we hereby outline the essence of the methods used to obtain these results in Secs.~\ref{sec:distinct} and~\ref{sec:surf}.

\subsection{Topological classification and distinct states}\label{sec:distinct}

Similar to the discussion in Sec.~\ref{sec:introno_spec}, when classifying all equivariant maps from the BZ to the classifying spaces one must first specify a connected component of the classifying space and then focus on maps within that component. This entices the definition of the ``reduced" $\KK$-theory, $\rKR$,
\begin{equation}\label{eq:redK}
\KR_G(\BZ)\simeq\KR_G(\pt)\oplus\rKR_G(\BZ),
\end{equation}
where ``$\pt$" is the space comprised of a single point; cf.~Eq.~\eqref{eq:pi0pi1}.
This decomposition has a physical interpretation: Consider an AI originating from the trivial Wyckoff position [$\wyck=(0,0,0)$]. The groundstate corresponding to this AI has identical band structure at all points (momenta) in the BZ and hence is fully determined by $\KR_G(\pt)$.

The treatment of AIs at other Wyckoff positions is rather more complicated and is elucidated at Sec.~\ref{sec:surf}. However, in the \emph{absence} of crystalline symmetries the only Wyckoff position is the generic Wyckoff position, which therein gives the same contribution as the trivial position. Hence, it is very easy to distinguish between AIs which are classified by $\KR(\pt)$, and topological phases with anomalous surface states which are classified by $\rKR(\BZ)$.
For each AZ class, the AIs satisfy $\KR^{-q,0}(\pt)=\pi_0(\sR_q)$ and are easily read from the $d=0$ column in Table~\ref{tab:per}.

In general, determining the topological phases encapsulated by $\KR_G(\BZ)$ is a nontrivial task, whose solution we obtain using equivariant spectra. In order to acquaint the reader with this formalism, we first apply it to re-derive the classification of topological phases in the trivial space-group, $\mathrm{P1}$, see the first record of Table~\ref{tab:mainres}.
Historically, this was achieved by various methods such as the Baum-Connes isomorphism and the Poincar\'{e} duality~\cite{kitaev2009periodic}. However, as we explain below the spectra perspective is generalizable to non-trivial space-groups.

\subsubsection{First steps with spectra}\label{sec:P1}

\begin{figure}[t]
\centering
\includegraphics[width=\linewidth]{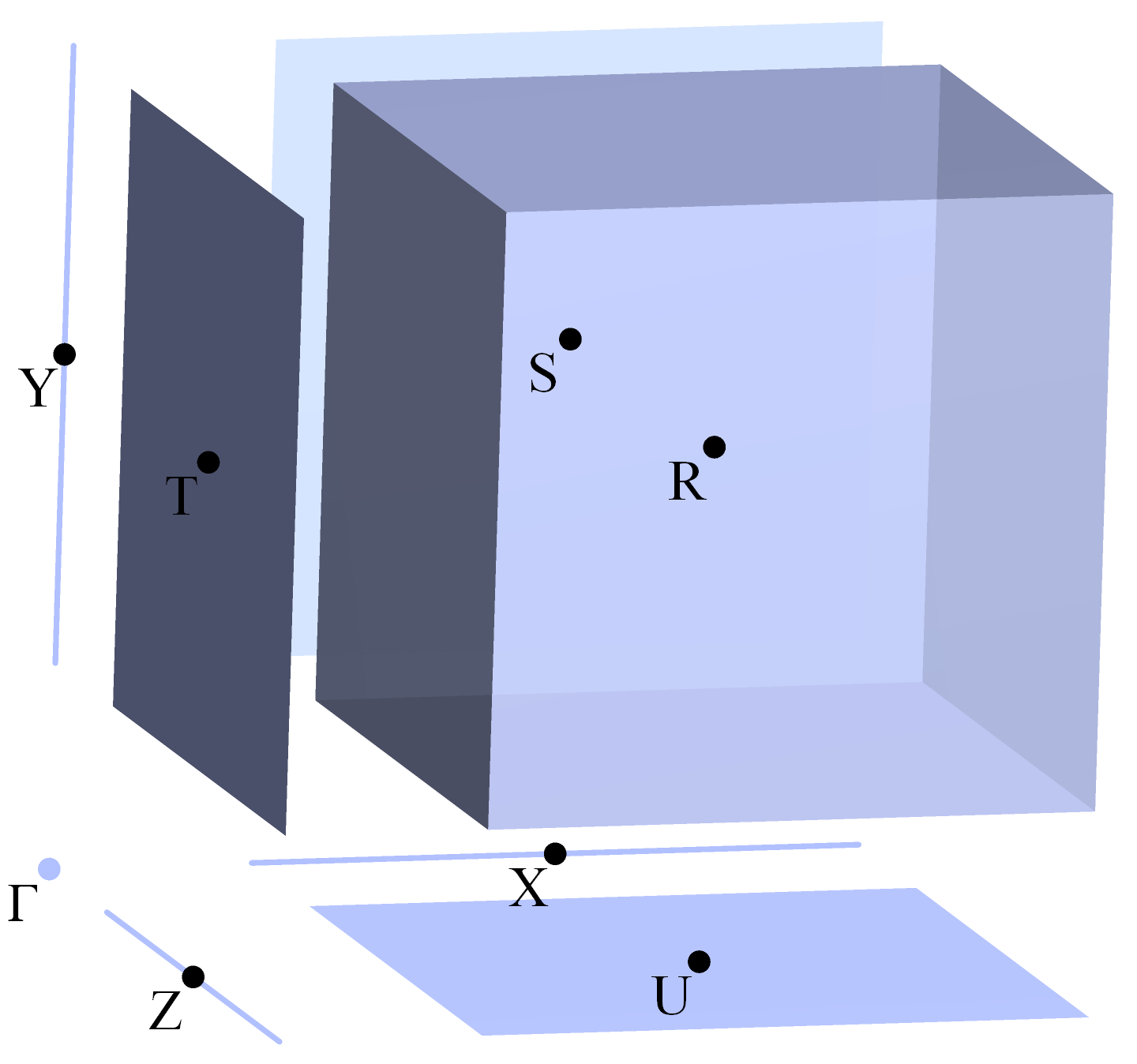}
\caption{An equivariant decomposition of the Brillouin zone torus into eight cells centered at high-symmetry momenta; see Eq.~\eqref{eq:TP1spec}.}
\label{fig:CWP1}
\end{figure}

Geometrically, the BZ of a 3D system is a 3-torus, $\BZ=\bTt^3=\bSs^1\times\bSs^1\times\bSs^1$, with an involutive action, i.e., $\vk\mapsto-\vk$, of the $\ZT$-symmetry on the momenta, $\vk\in\BZ$.
Instead of studying the BZ torus itself, we examine its ``free spectrum", $\bTSp{3} \defeq\SSp[\bTt^3]$, which has much simpler algebraic properties. In particular, $\SSp[X\times Y]\simeq\SSp[X]\otimes\SSp[Y]$, and hence $\bTSp{3}\simeq\SSp[\bSs^1]^{\otimes 3}$. Moreover, the free spectrum of the $\ZT$-equivariant $d$-sphere $\bSs^{d}$ may be constructed from the free spectrum of a point, $\SSp=\SSp[\pt]$,
\begin{equation}
\SSp[\bSs^d]\simeq\SSp\oplus\bSSp{d},
\end{equation}
and $\SSp$ is hence referred to as the ``sphere spectrum".
This decomposition provides the first hint for the usefulness of spectra, as the $\KK$-theory base-point decomposition [Eq.~\eqref{eq:redK}] is now evident already at the geometric level. Note, that the notation $\Sigma E\approx \Omega^{-1}E$ in spectra theory is analogous to (but is not to be confused with) the classical suspension, $\Sigma X$, of homotopy theory~\cite{hatcher2002algebraic}. The sphere spectrum satisfies $\bSSp{d_1}\otimes\bSSp{d_2}\simeq\bSSp{d_1+d_2}$; thus, gathering all the above, we obtain
\begin{align}\label{eq:TP1spec}
\bTSp{3}=\SSp[(\bSs^1)^3]&\simeq\SSp[\bSs^1]^{\otimes 3}\simeq(\SSp\oplus\bSSp{1})^{\otimes 3} \nonumber\\
&\simeq\SSp\oplus(\bSSp{1})^{\oplus 3}\oplus(\bSSp{2})^{\oplus 3}\oplus\bSSp{3}.
\end{align}
There is a very intuitive interpretation of this decomposition associated with some high-symmetry momenta (HSM): In Fig.~\ref{fig:CWP1} we see a decomposition of the 3-torus BZ as a $\ZT$-CW complex. In general a $G$-CW complex is a space composed of disjoint $d$-dimensional open cells such that each cell is mapped by the symmetry, $G$, to a cell of the same dimensionality. Here, we label the cells composing the BZ by the HSM at their centers. The 3-torus decomposes into eight cells, one 0D cell at the $\Gamma$ point, three 1D cells centered at the $X,Y,Z$ points, three 2D cells centered at the $S,T,U$ points, and one 3D cell centered at the $R$ point~\cite{bradley2010mathematical}. Each of these $d$-dimensional cells is associated with a $\bSSp{d}$ spectrum in Eq.~\eqref{eq:TP1spec}.

The distributivity of homomorphisms,
\begin{equation}
\Hom(E'\oplus E'',E)\simeq\Hom(E',E)\oplus\Hom(E'',E),
\end{equation}
together with the spectra to $\KK$-theory relation~\footnote{See discussion in Appendix~\ref{sec:KRKC-theory} and Appendix~\ref{sec:spectra}.},
$\pi_{q-p}(\Hom(\bSSp{d},\KR))\simeq\trKR^{p,q}(\bSs^{d})$,
immediately turns Eq.~\eqref{eq:TP1spec} to a $\KK$-theory decomposition,
\begin{multline}\label{eq:P1K}
\KR^{p,q}(\bTt^3)\simeq \KR^{p,q}(\pt) \\
\oplus \rKR^{p,q}(\bSs^1)^{\oplus 3}\oplus\rKR^{p,q}(\bSs^2)^{\oplus 3}\oplus\rKR^{p,q}(\bSs^3).
\end{multline}
Here, the $\KK$-theory of the spheres, $\bSs^d$, are much simpler objects and follow directly from Bott-periodicity~\cite{atiyah1964clifford,bott1969lectures,kitaev2009periodic},
\begin{equation}\label{eq:Bott-q}
\trKR^{-q,0}(\bSs^d)=\KR^{-q,-d}(\pt)\simeq\KR^{d-q,0}(\pt)\simeq\pi_0(\sR_{q-d}).
\end{equation}
These are easily read from Table~\ref{tab:per}.

For example, in AZ symmetry class AII, we have $q=4$ and thus find,
\begin{align}
\AI=\KR^{-4,0}(\pt)&\simeq\ZZ, \label{eq:P1AIdeg4} \\
\rKR^{-4,0}(\BZ)&\simeq 0^3\times\ZZ_2^3\times\ZZ_2,
\end{align}
cf.~$\mathrm{P1}$ in Table~\ref{tab:mainres}. Here, the $\ZZ$ atomic insulators correspond to the number of Kramer's pairs, while the $\ZZ_2^3$ correspond to the weak TIs and the last $\ZZ_2$ to the strong TI. The strong TISC are phases reflecting the topology of the bulk of a material and are indeed originate from the $R$-point cell; see Fig.~\ref{fig:CWP1}. The weak TISC phases, which are here associated with the $S$, $T$, and $U$-point cells, can be constructed by stacking 2D TISC along these directions.

We reemphasize that, in the absence of crystalline symmetries, there are more elementary ways to calculate the topological phases of the BZ torus. However, as we immediately show, the spectra perspective generalizes very well to the crystalline case.

\subsubsection{Crystalline symmetries \& equivariant spectra}\label{sec:symm-and-spec}

The key aspect revealing the structure of topological phases is that the free spectrum of the BZ torus itself decomposes into sphere spectra. This decomposition then implies that the classes of topologically distinct groundstates, represented by the $\KK$-theory of the BZ torus, decompose to classes within the $\KK$-theory of spheres which are much better understood. The equivariant spectra perspective reveals that this reasoning applies to crystalline systems as well.

In a previous work by one of the authors with A.~Chapman, see Ref.~\onlinecite{Cornfeld2019Classification}, an explicit classification of Dirac Hamiltonians invariant under any point-group symmetry, $G$, was provided. In our current settings, these classes of Dirac Hamiltonians correspond to topologically distinct groundstate classes associated with a $G$-equivariant sphere. In $\KK$-theoretic language, the classification by E.~C.~and A.~Chapman is provided by an \emph{isomorphism} between the $\KK$-theory of $G$-equivariant spheres and the non-equivariant $\KK$-theories of spheres corresponding to different representations of the point group $G$. Explicitly,
\begin{equation}\label{eq:CCiso}
\rKR_G^{p,q}(\bSs^d)\simeq\smashoperator{\bigoplus_{{\substack{\text{real}\\ \rho\in\gIrr^{\epsilon}(G)}}}}\rKR^{p-p_\rho,q-q_\rho}(\bSs^d) \oplus \smashoperator{\bigoplus_{{\substack{\text{complex}\\ \rho\in\gIrr^{\epsilon}(G)}}}}\rKC^{p-p_\rho,q}(\bSs^d).
\end{equation}
Here, $\gIrr^{\epsilon}(G)$ are all $\ZZ_2$-graded and $\epsilon$-twisted irreducible representations for suitable grading and twisting explicitly determined by the point-group, $G$; the degrees, $p_\rho,q_\rho$, are also explicitly determined. A proof of this isomorphism was given by M.~Karoubi in Ref.~\onlinecite{KAROUBI2002Equivariant}. For further details of this isomorphism see Appendix~\ref{app:iso}.

A question arises, whether the topologically distinct groundstates of crystalline systems, represented by $\KR_G(\bTt^3)$, can also be decomposed into equivariant spheres and thus, via the above isomorphism, to non-crystalline invariants.

We prove that this is indeed true for a class of space-groups presented in Table~\ref{tab:mainres}. The proof is given in Appendix~\ref{sec:signed-permutation} and becomes very straightforward once formulated in terms of equivariant spectra. Let us present the intuition behind this decomposition.

One of the key aspects of the non-crystalline decomposition in Sec.~\ref{sec:P1} was the decomposition of the BZ torus to spheres, $\bTt^3=\bSs^1\times\bSs^1\times\bSs^1$, where each sphere corresponds to one of the primitive reciprocal lattice vectors, $\vb_1,\vb_2,\vb_3$. However, a $G$ action may in general act on the BZ and map a primitive sphere into a non-primitive loop within the BZ. This corresponds to an equivalent but \emph{different} choice of primitive vectors. Let us consider the class of space-groups where there exists a choice of primitive vectors such that the point-group, $G$, respects this choice, i.e., for any $g\in G$ and any $\vb_i$ there exists $\vb_j$ such that $g(\vb_i)=\pm\vb_j$. This means that the point-group acts as a signed permutation on the primitive lattice vectors, and thus \emph{preserves} the direct product structure of the BZ torus; we dub this class of space-groups ``signed-permutation representations".

As proven in Appendix~\ref{sec:math_background}, we find that the free $G\times\ZT$-equivariant spectrum, $\bTSp{3}$, of the BZ torus of a signed-permutation representation decomposes into $G_\hsm$-equivariant sphere spectra,
\begin{equation}\label{eq:torus-G-decomposition}
\bTSp{3} \simeq \bigoplus\nolimits_{\hsm}\Ind_{G_\hsm}^{G}(\bSSp{d_\hsm}), 
\end{equation}
where, the ``little groups", $G_\hsm\subseteq G$ stabilize the HSM, $\hsm\in\BZ$, located at the \emph{centers} of $d_\hsm$-dimensional cells (points, lines, planes, and volumes) spanned by the primitive reciprocal lattice vectors~\footnote{A \emph{non-equivariant} homotopy-theory proof follows from Proposition~4I.1 in Ref.~\onlinecite{hatcher2002algebraic}.} For example, in a primitive cubic lattice, $\{\hsm\}$ are the $\Gamma,X,M,R$-points, while in a base-centered orthorhombic lattice, $\{\hsm\}$ are the $\Gamma,S,Z,R,Y,T$-points~\cite{bradley2010mathematical}.
This is analogous to the simpler $\ZT$-equivariant case discussed above; cf.~Eq.~\eqref{eq:TP1spec} and Fig.~\ref{fig:CWP1}. The induction from $G_\hsm$ to $G$ in Eq.~\eqref{eq:torus-G-decomposition}, which is a standard procedure in the study of band representations, here occurs already as a description of the BZ geometry.

As before, by using the spectra formulation, this decomposition which happens purely at the geometric level, immediately translates to the $\KK$-theory classification,
\begin{equation}\label{eq:KTtoKS}
\KR_G(\bTt^3) \simeq \bigoplus\nolimits_{\hsm}\rKR_{G_\hsm}(\bSs^{d_\hsm}),
\end{equation}
cf.~Eq.~\eqref{eq:P1K}; this $\KK$-theory result is similar to the assertion by Stehouwer et~al.~in Ref.~\onlinecite{stehouwer2018classification}. In order to complete the classification, one recalls that each of the summands in Eq.~\eqref{eq:KTtoKS} further decomposes according to the isomorphism of Ref.~\onlinecite{Cornfeld2019Classification}, see Eq.~\eqref{eq:CCiso}. We thus obtain our first quantitative result - an explicit formula for the full classification of all topologically distinct groundstates of all signed-permutation CTISC in all AZ symmetry classes; the classification tables are presented in Appendix~\ref{app:tables}.

The AIs at the trivial Wyckoff position are still captured by the $\Gamma$-point summand, $\rKR_{G_\Gamma}(\bSs^{d_\Gamma})=\rKR_{G}(\bSs^{0})\simeq\KR_G(\pt)$. However, the AIs at other Wyckoff positions are not so easily isolated. This is discussed in the following section.

\subsection{Topological phases and surface states}\label{sec:surf}
In order to obtain the full classification of topological phases with anomalous surface states one must quotient out the AIs from the full $\KK$-theory classification of topologically distinct groundstates obtained above; see pictorial depiction in Fig.~\ref{fig:main-diag}(left).

\subsubsection{Atomic insulators \& the \texorpdfstring{$\AItoK$}{AI to K-theory} map}\label{sec:AI}

An atomic insulator state, by definition, may always be adiabatically connected to a state with localized Wannier orbitals around specific Wyckoff positions.
Each Wyckoff position, $\wyck$, is invariant under a symmetry with point group $G_\wyck\subseteq G$. As such, the Wannier orbitals of the AIs at $\wyck$ must transform as representations, $\rho$, of $G_\wyck$.
In general, two AIs that differ either by their Wyckoff positions or by their representations of $G_\wyck$ cannot be continuously deformed into each other. They may hence correspond to different classes of topologically distinct groundstates and thus different elements of the $\KK$-theory. Therefore, the AI groundstates must be quotiented out in order to obtain the topological phases with anomalous surface states.

An AI formed by a Wannier orbital with a particular representation at a particular Wyckoff position is often referred to as an elementary band representation. Such elementary band representations form a generating set for the $\AI$ additive group.
This provides a concise description at AZ symmetry classes A, AI, and AII which form Dyson's threefold-way~\cite{Dyson1962Threefold,zirnbauer2010symmetry}, i.e.,
\begin{equation}\label{eq:AIDyson}
\AI\big|_{\substack{\text{AZ classes}\\\text{A,AI,AII}}} \simeq\bigoplus_\wyck\Rep(G_\wyck)\underset{\substack{\text{abelian}\\\text{groups}}}{\simeq}\bigoplus_\wyck\bigoplus_{\rho\in\Irr(G_\wyck)}\ZZ.
\end{equation}
Here, $\Irr(G)$ are the irreducible representations of $G$ which form the basis of the representation ring, $\Rep(G)$, consisting of all representations of $G$ with multiplication given by the tensor product of representations. This multiplicative structure would prove vital ahead.

Nevertheless, the generating set of irreducible representations is rather redundant for the description of the other AZ symmetry classes which unveil a deeper algebraic structure. The constraints imposed by the presence of either particle-hole or chiral symmetry (see Fig.~\ref{fig:bott}) imply that stacking identical AIs may be topologically equivalent to filling no Wannier orbitals. An AI at a particular Wyckoff position, $\wyck$, is a 0-dimensional system and is hence classified by the $\KK$-theory of a point, $\KR_{G_\wyck}(\pt)$. This automatically quotients-out the topological redundancies.

For example, the AIs at the generic Wyckoff position in space-group $\mathrm{P}1$ are given by $\KR^{-q,0}(\pt)$ at AZ symmetry class, $q$. The particle-hole and chiral symmetries reduce the $\ZZ$ AIs of Dyson's threefold-way (AZ classes A, AI, AII) to either $\ZZ_2$ or $0$ for the other AZ classes; see Table~\ref{tab:per}. This has been recently utilized in Refs.~\onlinecite{Geier2020Symmetry,Ono2020Refined,ono2020Z2} for the study of SI of superconducting systems.

The AIs in any space-group symmetry decompose according to the Wyckoff positions, 
\begin{equation}\label{eq:AIdef}
\AI=  \bigoplus\nolimits_\wyck\AI_{G_\wyck} = \bigoplus\nolimits_\wyck\KR_{G_\wyck}(\pt).
\end{equation}
For any AZ class, $q$, these are explicitly given by~\cite{Segal1968Equivariant}
\begin{multline}\label{eq:KRGpt}
\KR^{-q,0}_{G}(\pt) \simeq \smashoperator{\bigoplus_{{\substack{\text{real}\\ \rho\in\Irr(G)}}}}\KR^{-q,0}(\pt) \oplus \smashoperator{\bigoplus_{{\substack{\text{complex}\\ \rho\in\Irr(G)}}}}\KC^{-q,0}(\pt) \\
\oplus \smashoperator{\bigoplus_{{\substack{\text{quaternionic}\\ \rho\in\Irr(G)}}}}\KR^{-q-4,0}(\pt).
\end{multline}
This generalizes Eq.~\eqref{eq:AIDyson} to all AZ classes and indeed,
\begin{equation}\label{eq:AItoIrr}
\AI^{0,0}_{G}=\KR^{0,0}_{G}(\pt)\simeq\Rep(G)\underset{\substack{\text{abelian}\\\text{groups}}}{\simeq}\bigoplus_{\rho\in\Irr(G)}\ZZ.
\end{equation}
In order to gain an intuition for the shift of 4 for quaternionic representations in Eq.~\eqref{eq:KRGpt}, recall that a quaternionic representation may be expressed in terms of unit quaternions which form an $\SUnit(2)$ group. The four AZ classes, $q=2,3,4,5$ 
(D, DIII, AII, and CII), all have broken spin rotation $\SUnit(2)$ symmetry, while the other four AZ classes (C, CI, AI, and BDI) are all invariant; a quaternionic representation thus exchanges these two sets of classes. See Ref.~\onlinecite{kennedy2016bott} for further detail.

Before we can quotient out the AIs in order to find the topological phases with anomalous surface states, we must first find the map which evaluates the AIs as equivalence classes in the $\KK$-theory classification of topologically distinct groundstates,
\begin{equation}\label{eq:AItoK-diag}
\AI\xrightarrow{\AItoK} \KK\twoheadrightarrow\KK/\AI,
\end{equation}
cf.~Eq.~\eqref{eq:main-diag}. We denote this map by $\AItoK$. 

On the most basic level, we may treat this map as an abelian group-homomorphism, i.e., evaluating the AIs at a particular AZ symmetry class and returning the group element of their topological classification. However, much more structure is revealed when treating $\AItoK$ as a homomorphism on the level of module-spectra. We find that the contribution of any Wyckoff position to all AZ classes at once is captured by (at most) two integers per $\ZZ_2$-graded representation.

\subsubsection{Immediate implications for anomalous surface state}\label{sec:no-AI}
Before elaborating on the $\AItoK$ map, we note an immediate consequence of Eq.~\eqref{eq:KRGpt}. 
By explicitly plugging the abelian groups from the $d=0$ column in Table~\ref{tab:per}, one sees that $\AI=0$ for AZ classes AIII, DIII, and CI for all Wyckoff positions of any space-group symmetry. This implies that $\KK=\KK/\AI=\Surf$ and hence: \emph{In AZ symmetry classes \emph{AIII}, \emph{DIII}, and \emph{CI}, non-trivial topologically distinct groundstates of any weakly interacting fermionic crystalline system in any spatial dimension, all have anomalous surface states and correspond to non-trivial topological phases}~\footnotemark[\thefnnumber].

\subsubsection{Shiozaki's formula}\label{sec:shiozaki}
Consider the contribution of a single HSM summand in Eq.~\eqref{eq:KTtoKS}, classified by $\rKR_G(\bSs^d)$.
In a recent work, K.~Shiozki showed that AIs at the center of the point-group, $G$, can be realized as Dirac Hamiltonians with a spatially dependent ``hedgehog" mass-term; see Ref.~\onlinecite{shiozaki2019classification}. In $\KK$-theory language, this is captured by the identity $\KR_G(\pt)\simeq\rKR_G(\bSs^d\wedge\Ss^d)$, which is an equivariant version of Bott-periodicity. Here, $\bSs^d\wedge\Ss^d$ is a $G\times\ZT$-equivariant $2d$-dimensional sphere, such that the former $d$ coordinates describe the $\ZT$-odd momenta around the Dirac point and the latter $d$ coordinates describe the $\ZT$-even real-space dependence. The topological class of the groundstate of such a Dirac Hamiltonian may be readily found by neglecting the ``hedgehog" dependence of the mass-term.
This understanding enabled K.~Shiozaki to obtain an explicit formula for the $\AItoK$ map. In $\KK$-theory language, this stems from the (equivariant) embedding of the $d$-sphere within the $2d$-sphere,
\begin{equation}\label{eq:shiozaki}
\xymatrix{
\KR_G(\pt) \ar@{}[r]|-{\displaystyle\simeq} \ar@/^1.5pc/[rr]^\AItoK & \rKR_G(\bSs^d\wedge\Ss^d) \ar[r] & \rKR_G(\bSs^d).
}
\end{equation}
The isomorphism of Eq.~\eqref{eq:CCiso} provides an explicit basis for this equivariant $\KK$-theory map in terms of ($\ZZ_2$-graded) representation theory. Results for all magnetic and non-magnetic 3D point-groups are provided in Ref.~\onlinecite{shiozaki2019classification}.

For the signed-permutation representations, discussed in Sec.~\ref{sec:symm-and-spec}, we find that the AIs contribution to a HSM component, $\rKR_{G_\hsm}(\bSs^{d_\hsm})\subset\KR_G(\BZ)$ may be distilled (see Appendix~\ref{sec:math-AI}) to the AIs at the Wyckoff position, $\wyck$, reciprocal to $\hsm$. For example, in a primitive cubic lattice, the reciprocal to the $R$-point, $\hsm=(\pi,\pi,\pi)$, is the Wyckoff position, $\wyck=(\frac{1}{2},\frac{1}{2},\frac{1}{2})$. These are the AIs at the center of the point group, and one has $G_\wyck\simeq G_\hsm$~\footnote{Note, that in general, this is not a one-to-one correspondence as the same Wyckoff position may be reciprocal to many HSM if they lie on the same high-symmetry line/surface/volume.}. This implies that Shiozaki's formula may be utilized to compute the full $\AItoK$ map using our spectra decomposition, Eq.~\eqref{eq:KTtoKS}.

Let us thus focus on a particular HSM with a point-group, $G$, for which the $\AItoK$ map reduces to Eq.~\eqref{eq:shiozaki}.

\subsubsection{Ring-spectra \& multiplication of quantum states}\label{sec:mult}

A hint of a multiplicative structure already appeared with the identification $\KR_G^{0,0}(\pt)\simeq\Rep(G)$ [Eq.~\eqref{eq:AItoIrr}] with multiplication given by the tensor product of representations. For example, consider the $C_3$ group, generated by $(c_3)^3=1$. The irreducible real representations are $\Irr(C_3)=\{1,E\}$ and satisfy the multiplication rule, $E\otimes E=2+E$. Here, $1$ is the 1D trivial representation, $c_3\mapsto 1$, and $E$ is the irreducible 2D representation, $c_3\mapsto\exp\{\frac{2\pi}{3}(\begin{smallmatrix}0&1\\-1&0\end{smallmatrix})\}$. The four-dimensional representation, $E\otimes E$, is equivalent to $1\oplus 1\oplus E$, this is evident by the eigenvalues which satisfy $(e^{\frac{2\pi i}{3}}+e^{-\frac{2\pi i}{3}})^2=2+e^{\frac{2\pi i}{3}}+e^{-\frac{2\pi i}{3}}$; see Ref.~\onlinecite{bradley2010mathematical} for a detailed discussion.
It is often overlooked that such a structure can be extended to a multiplication of quantum states.

Any two quantum groundstates of a system may be thought of as vector bundles over the BZ; this endows them with a multiplication given by the graded tensor product of the bundles. Particularly, since $\KR_G(\pt)\subset\KR_G(X)$ for any space, $X$, this implies that elements of all equivariant $\KK$-theories considered so far may be multiplied by $\KR_G(\pt)$, i.e.,
\begin{equation}\label{eq:KRX-mult}
\KR_G^{p,q}(\pt)\times\KR_G^{p',q'}(X)\to\KR_G^{p+p',q+q'}(X).
\end{equation}
Crucially, this product structure mixes different AZ classes in an additive manner.
Nonetheless, for $X=\pt$ and $p,p',q,q'=0$ it reduces to the multiplicative structure of $\Rep(G)$ discussed above. Note, that a similar module-structure for $\KU$ was discussed by Shiozaki, Sato, and Gomi in Ref.~\onlinecite{Shiozaki2017Topological}.

The multiplication on $\KK$-theory stems from an inherent multiplicative structure on the spectra $\KR$ and $\KR_G$, which, by definition, make these into ring-spectra. The $\AItoK$ map may thus be considered as a module-homomorphism, i.e., $\AItoK(u\cdot v)=u\cdot\AItoK(v)$. This has two important consequences that reveal the structure of the $\AItoK$ map.

First, it provides us with a simple expression for the most general structure of the map from an AI of a real irreducible representation $u\in\KR(\pt)\subset\AI_G$ to each real irreducible graded representation, $\rho$, in Eq.~\eqref{eq:CCiso},
\begin{equation}
\left[\AItoK(u)\right]_\rho = v_\rho\cdot u.
\end{equation}
Hence, for all AZ classes, the single element $v_\rho\in\trKR^{-p_\rho,-q_\rho}(\bSs^d)$ which is independent of the AZ class, determines the $\AItoK$ map. This element is itself determined by (at most) one integer corresponding to the classification of $d$ spatial dimensions at level $q=p_\rho-q_\rho~(\mathrm{mod}~8)$ in Table~\ref{tab:per}.
Mathematically, this immediately follows from the fundamental identity, $\Hom_\KR(\KR,\KR)=\KR$.
The analyses of complex and quaternionic representations are similar yet slightly more complicated. In particular, since any representation has a complex conjugate representation, this at most doubles the number of integers; see Appendix~\ref{sec:G-K-spec}.

Second, the module-homomorphism structure of the $\AItoK$ map implies that the topological classification of all AIs in any representation, $u\in\AI_G$, is determined by the classification, $\AItoK(1)$, of the fundamental AI corresponding to the trivial representation,
\begin{equation}\label{eq:ai_via_1}
    \AItoK(u)=\AItoK(u\cdot 1)=u\cdot\AItoK(1).
\end{equation}
We thus conclude that the multiplicative structure enables us to determine the topological classification of any AI of any orbital by the fundamental AI of an $s$-orbital at the same Wyckoff position. The latter is readily computed using Shiozaki's formula. This is how we obtain all the results presented in Table~\ref{tab:mainres} and Appendix~\ref{app:tables}.

\subsubsection{Immediate implications for symmetry indicators}\label{sec:no-SI}
As a final remark, the same reasoning as in Sec.~\ref{sec:no-AI}, that enabled us to exclude the existence of AIs at AZ classes AIII, DIII, and CI, applies for $\SIfull$ as well. The band labels correspond to individual HSM points, $\hsm\in\BZ$, and thus in gapped systems, they are also classified by the $\KK$-theory of a point,
\begin{equation}
\BS\simeq\frac{\bigoplus_\hsm\KR_{G_\hsm}(\pt)}{(\text{compatibility relations})},
\end{equation}
where the compatibility relations relate the invariants along high-symmetry lines, planes, and volumes in the BZ. Regardless of these compatibility relations, one sees that $\BS=0$ for AZ classes AIII, DIII, and CI for all Wyckoff positions of any space-group symmetry. This implies that $\SI=\BS/\AI=0$ and hence: \emph{In AZ symmetry classes \emph{AIII}, \emph{DIII}, and \emph{CI}, all non-trivial topological phases with anomalous surface states of any weakly interacting fermionic crystalline system in any spatial dimension cannot be detected by symmetry indicators}~\footnotemark[\thefnnumber]; see Table~\ref{tab:mainres}. Any non-trivial $\SIfull$ must indicate a gapless state and not a CTISC.

\section{Detailed derivation}\label{sec:detailed}
The topological classification tables in Table~\ref{tab:mainres} and Appendix~\ref{app:tables} were obtained by implementing the methods outlined in Sec.~\ref{sec:main_results} and described in Appendix~\ref{sec:math_background}. Our methods were implemented as a GAP4~\cite{GAP4} language algorithm. However, we believe it is highly beneficial for the reader to see some of the finer details.
First, in Sec.~\ref{sec:mult-TISC} we provide the basics of the hidden multiplicative structure within the periodic table of TISC. Then, in Sec.~\ref{sec:example}, we study a ``hands-on" pedagogical example, which portrays the use of our methods from top to bottom, i.e., choosing a particular space-group and fully deriving its complete classification of topologically distinct groundstates and topological phases with anomalous surface states. Finally, in Sec.~\ref{sec:hamilton}, we use this example and demonstrate how to construct model Hamiltonians for the numerous topological phases in Table~\ref{tab:mainres}.

\subsection{Multiplication tables of TISC}\label{sec:mult-TISC}

As discussed in Sec.~\ref{sec:mult}, the multiplicative structure of the $\KK$-theory and its inherent ring-spectrum, is at the core of our understanding of anomalous surface states.
The first hint of multiplicativity was already provided by $\KR_G^{0,0}(\pt)\simeq\Rep(G)$.
However, before one is ready to fully tackle a computational example, a more explicit sense of the ring-structure would be advantageous.

Therefore, we hereby discuss the (multiplicative) rings,
\begin{align}
\KR^{*,*}(\pt)&\defeq\bigoplus\nolimits_{p,q}\KR^{p,q}(\pt), \nonumber\\
\KC^{*,*}(\pt)&\defeq\bigoplus\nolimits_{p,q}\KC^{p,q}(\pt).
\end{align}
These rings and the maps between them, convey hidden aspects of the inherent ring-spectra and of the periodic table of TISC.

\begin{table}[t]
\centering
\caption{The additive structure on complex and real $\KK$-theory; cf.~the $d=0$ column of Table~\ref{tab:per}. In order to construct $\KR^{p,q}(\pt)$ at any bi-degree, one utilizes Bott-periodicity, $\KR^{-q-8\ell,0}(\pt)\simeq \beta^\ell \KR^{-q,0}(\pt)$ and $\KR^{-q-\ell,-\ell}(\pt)\simeq \mu^\ell \KR^{-q,0}(\pt)$. The same structure holds for $\KC$ with $\cc(\mu)=\mu$ and $\rr(\mu)=2\mu$. Trivial entries, i.e., $\rr(0)=0$ and $\cc(0)=0$, are omitted.}
\label{tab:KCKR}
\renewcommand{\arraystretch}{1.2}
\begin{tabularx}{\linewidth}{Xllll}
\hline\hline
$q$ & $\KC^{-q,0}(\pt)$ & Realification & $\KR^{-q,0}(\pt)$ & Complexification \\
& $\KU^{-q}(\pt)$ & & $\KO^{-q}(\pt)$ & \\
\hline
$0$ & $\ZZ$ & $\rr(1)=2$ & $\ZZ$ & $\cc(1)=1$ \\
$1$ & $0$ && $\ZZ_2\cdot\eta$ & $\cc(\eta)=0$ \\
$2$ & $\ZZ\cdot\xi$ & $\rr(\xi)=\eta^2$ & $\ZZ_2\cdot\eta^2$ & $\cc(\eta^2)=0$  \\
$3$ & $0$ && $0$ &\\
$4$ & $\ZZ\cdot\xi^2$ & $\rr(\xi^2)=\alpha$ & $\ZZ\cdot\alpha$ & $\cc(\alpha)=2\xi^2$  \\
$5$ & $0$ && $0$ &\\
$6$ & $\ZZ\cdot\xi^3$ & $\rr(\xi^3)=0$ & $0$ &\\
$7$ & $0$ && $0$ &\\
\hline
$8$ & $\ZZ\cdot\xi^4$ & $\rr(\xi^4)=2\beta$ & $\ZZ\cdot\beta$ & $\cc(\beta)=\xi^4$ \\
\hline\hline
\end{tabularx}
\renewcommand{\arraystretch}{1}
\end{table}

\begin{table}[t]
\centering
\caption{(left) The multiplication table of $\KC^{*,*}(\pt)$. (right) The multiplication table of $\KR^{*,*}(\pt)$. Any element may also be multiplied by $\mu^\ell$ for all $\ell\in\ZZ$. See Table~\ref{tab:KCKR}.}
\label{tab:KRmult}
\renewcommand{\arraystretch}{1.3}
\begin{tabular}{l|ll}
\hline\hline
$\bullet$ & $1$ & $\xi^j$ \\
\hline
$1$ & $1$ & $\xi^j$ \\
$\xi^i$ & $\xi^i$ & $\xi^{i+j}$ \\
\hline\hline
\end{tabular}
\qquad
\begin{tabular}{c|cccccccc}
\hline\hline
$\bullet$ & $1$ & $2$ & $\eta$ & $\eta^2$ & $\alpha$ & $\quad$ & $\beta^j$ \\
\hline
$1$ & $1$ & $2$ & $\eta$ & $\eta^2$ & $\alpha$ && $\beta^j$ \\
$2$ & $2$ & $4$ & $0$ & $0$ & $2\alpha$ && $2\beta^j$ \\
$\eta$ & $\eta$ & $0$ & $\eta^2$ & $0$ & $0$ && $\eta\beta^j$ \\
$\eta^2$ & $\eta^2$ & $0$ & $0$ & $0$ & $0$ && $\eta^2\beta^j$ \\ 
$\alpha$ & $\alpha$ & $2\alpha$ & $0$ & $0$ & $4\beta$ && $\alpha\beta^j$ \\
\\
$\beta^i$ & $\beta^i$ & $2\beta^i$ & $\eta\beta^i$ & $\eta^2\beta^i$ & $\alpha\beta^i$ && $\beta^{i+j}$ \\
\hline\hline
\end{tabular}
\renewcommand{\arraystretch}{1}
\end{table}

Using, $\trKR^{-q,0}(\bSs^d)=\KR^{-q,-d}(\pt)$, as discussed in Sec.~\ref{sec:P1}, these rings consist of all the different $\ZZ$ and $\ZZ_2$ invariants of the periodic table of TISC in Table~\ref{tab:per}.
In order to discern these different $\ZZ$ and $\ZZ_2$ invariants, we label their generators by $1,\eta,\eta^2,\alpha$, as summarized in Table~\ref{tab:KCKR}.
The twofold and eightfold Bott-periodicities are captured by the invertible elements, $\xi$ and $\beta$, respectively; the diagonal Bott-periodicity is captured by the invertible element, $\mu$.

For example,
the strong time-reversal invariant TI of real AZ class AII with $q=4$ and $d=3$ may be located in Table~\ref{tab:per} by starting at the $\ZZ_2\cdot\eta$ invariant of $q=1$ and $d=0$, and moving 3 diagonal steps down; it is thus represented by $\eta\mu^3$.
Similarly, the IQH effect of complex AZ class A with $q=0$ and $d=2$ may be located in Table~\ref{tab:per} by starting at the complex $\ZZ$ invariant of $q=d=0$, moving 2 diagonal steps down and applying the twofold Bott periodicity up; it is thus represented by $\xi^{-1}\mu^2$.

The multiplication table of $\KR^{*,*}(\pt)$ is presented in Table~\ref{tab:KRmult}. This multiplication table has some familiar features, such as $2\eta=0$ and $2\eta^2=0$ which encapsulate the $\ZZ_2$ nature of the corresponding topological phases, such as strong and weak time-reversal invariant TIs and Majorana bound states. Moreover, this multiplicative structure also has some intriguing consequences on CTISC.

The simplest example for such a direct consequence is the centrosymmetric space-group, $\mathrm{P\bar{1}}$, where the point-group, $G\simeq\ZZ_2$, acts by inversion, i.e., $\vk\mapsto-\vk$. The strong CTISC component of its $\AItoK$ map is given by
\begin{equation}
\KR^{-q,0}(\pt)\oto{\textstyle{v\mapsto\beta^{-1}\alpha\cdot v}}\KR^{4-q,0}(\pt),
\end{equation}
where,
\begin{align}\label{eq:KR-P1}
\KR^{-q,0}(\pt)\subset\KR^{-q,0}_{\ZZ_2}(\pt) &\subset\AI^{-q,0}, \nonumber\\
\KR^{4-q,0}(\pt)\simeq\trKR^{-q,0}_{\ZZ_2}(\bSs^3) &\subset\KR^{-q,0}_{\ZZ_2}(\BZ).
\end{align}
The algorithm for the derivation of the above equations is explained in detail in Sec.~\ref{sec:example}; we wish to shed light on some of the consequences.
Particularly, for AZ class AII we have $q=4$ and hence,
\begin{equation}\label{eq:alpha-to-4}
\KR^{-4,0}(\pt)\simeq\ZZ\cdot\alpha\oto{\textstyle{\alpha\mapsto 4}}\ZZ\simeq\KR^{0,0}(\pt),
\end{equation}
where we have set $v=\alpha$ and used $\beta^{-1}\alpha\cdot\alpha=4$, see Table~\ref{tab:KRmult}.
Eq.~\eqref{eq:alpha-to-4} implies that stacking four nontrivial crystalline-TI groundstates is topologically equivalent to an AI. This is the celebrated $\ZZ_4$ invariant~\cite{po2017symmetry,fang2017rotation,bradlyn2017topological} of space-group $\mathrm{P\bar{1}}$, which may be re-interpreted as a direct manifestation of the multiplication table, Table~\ref{tab:KRmult}.

We reserve the analyses of other examples to the following sections and Appendix~\ref{sec:math_background}. Nevertheless, we would like to emphasize a vital property of the multiplicative structure which stems from the inherent spectra: There are natural maps of realification, $\rr$, and complexification, $\cc$, between the spectra $\KR$ and $\KC$. Crucially, since the $\AItoK$ map is not just a $\KK$-theory map, but rather a map between the inherent spectra, \emph{any} $\AItoK$ map between any real or complex $\KK$-theories must be given by compositions of $\rr$, $\cc$, and multiplication by elements of the $\KK$-theories.
The manifestations of the realification and complexification maps on the $\KK$-theories is provided in Table~\ref{tab:KCKR}. In particular, $\rr(1)=2$ stems from $\CC\simeq\RR\oplus i\RR$ and $\cc(1)=1$ stems from $\CC\otimes\RR\simeq\CC$. For a complete list of possible maps see Table~\ref{tab:KR-hom} in Appendix~\ref{sec:math_background}.

\subsection{A pedagogical example: space-group \texorpdfstring{$\mathrm{P\bar{4}}$}{P-4}}\label{sec:example}

\subsubsection{Geometry}

\begin{figure}[t]
\centering
\includegraphics[width=\linewidth]{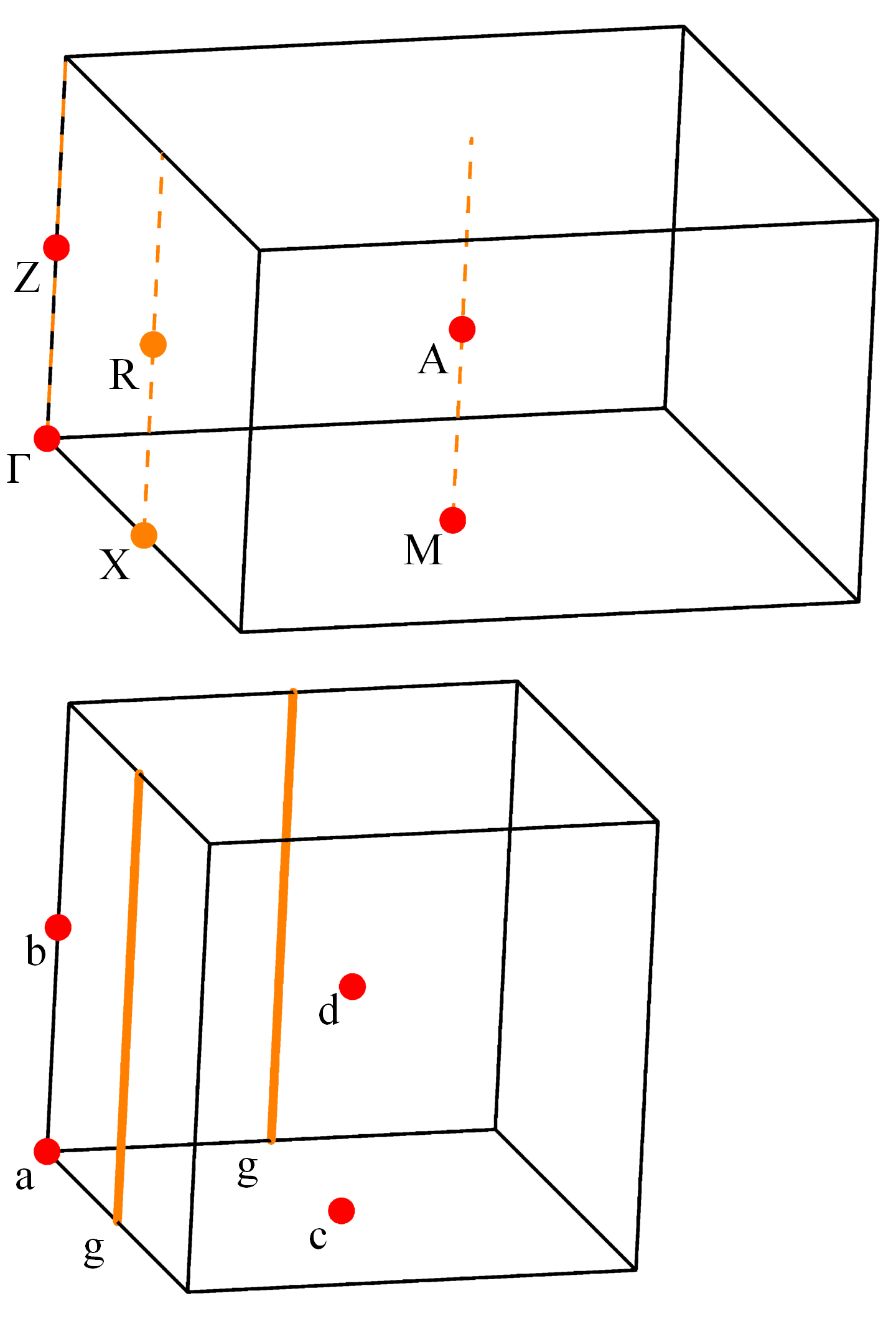}
\caption{(top) The Brillouin zone of space-group $\mathrm{P\bar{4}}$. The high-symmetry momenta, $\Gamma,Z,X,R,M,A$-points, at the centers of the equivariant cells are marked, cf.~Fig.~\ref{fig:CWP1}. The high-symmetry lines are depicted by the dashed lines. (bottom) The primitive unit-cell of space-group $\mathrm{P\bar{4}}$. The Wyckoff positions, $a,b,c,d,g$, reciprocal to the high-symmetry momenta are marked. Note that the $g$ Wyckoff position is reciprocal to both the $X$-point and the $R$-point HSM.}
\label{fig:P-4_HSM}
\end{figure}

\begin{table}[t]
\centering
\caption{The Wyckoff positions, $\wyck$, and high-symmetry momenta, $\hsm$, of space-group $\mathrm{P\bar{4}}$. The little group stabilizer, $G_\wyck=G_\hsm$, is given in Sch\"{o}nflies notation; the double point-group, $\twistcover{G}_{\wyck}=\twistcover{G}_{\hsm}$, is given as an abstract group. See Fig.~\ref{fig:P-4_HSM}.}
\label{tab:P-4_WyckHSM}
\renewcommand{\arraystretch}{1.3}
\begin{tabular}{cccccc}
\hline\hline
\multicolumn{2}{l}{\multirow{2}{*}{Wyckoff positions ($\wyck$)}} & \multicolumn{2}{l}{\multirow{2}{*}{\makecell[l]{High-symmetry \\ momenta ($\hsm$)}}} & \multicolumn{2}{c}{Point-group} \\
\cline{5-6}
\multicolumn{2}{l}{} & \multicolumn{2}{l}{} & Sch\"{o}nflies & $\twistcover{G}_{\wyck}=\twistcover{G}_{\hsm}$ \\
\hline
$a$ & $(0,0,0)$ & $\Gamma$ & $(0,0,0)$ & $S_4$ & $\ZZ_8$ \\
$b$ & $(0,0,\frac{1}{2})$ & $Z$ & $(0,0,\pi)$ & $S_4$ & $\ZZ_8$ \\
\multirow{2}{*}{$g$} & \multirow{2}{*}{$(\frac{1}{2},0,x),(0,\frac{1}{2},x)$} & $X$ & $(\pi,0,0)$ & $C_2$ & $\ZZ_4$ \\
&& $R$ & $(\pi,0,\pi)$ & $C_2$ & $\ZZ_4$ \\
$c$ & $(\frac{1}{2},\frac{1}{2},0)$ & $M$ & $(\pi,\pi,0)$ & $S_4$ & $\ZZ_8$ \\
$d$ & $(\frac{1}{2},\frac{1}{2},\frac{1}{2})$ & $A$ & $(\pi,\pi,\pi)$ & $S_4$ & $\ZZ_8$ \\
\hline\hline
\end{tabular}
\renewcommand{\arraystretch}{1}
\end{table}

\begin{table*}[t]
\centering
\caption{The complete $\KK$-theory classification of topologically distinct groundstates as well as the complete classification of topological phases with anomalous surface states, $\Surf=\KK/\AI$, of space-group $\mathrm{P\bar{4}}$. Each entry of the table is of the form $\KK\to\Surf$ with $\KK=\trKR^{-q,0}_{\twistcover{G}_\hsm}(\bSs^{d_\hsm})$ for AZ class, $q$, and HSM, $\hsm$.
The ``~$\cdot$~" symbol indicates a trivial classification, i.e., $0\to0$.}\label{tab:P-4_res}
\renewcommand{\arraystretch}{1.2}
\begin{tabularx}{\textwidth}{lrcXrcXrcXrcXrcXrcXrcl}
\hline\hline
HSM && $\Gamma$ &&& $Z$ &&& $X$ &&& $R$ &&& $M$ &&& $A$ && \multicolumn{3}{l}{\multirow{2}{*}{Total classification}} \\
\cline{1-1}
AZ class & &&& &&& &&& &&& &&& &&& \multicolumn{3}{l}{}\\
\hline
A & $\ZZ^{4}$ & $\to$ & $0$ & $\ZZ^{2}$ & $\to$ & $0$ & $\ZZ$ & $\to$ & $0$ && $\cdot$ && $\ZZ^{4}$ & $\to$ & $\ZZ$ & $\ZZ^{2}$ & $\to$ & $\ZZ_{2}$ & 
$\ZZ^{13}$ & $\to$ & $\ZZ\times\ZZ_2$ \\
AIII && $\cdot$ &&& $\cdot$ &&& $\cdot$ && $\ZZ$ & $\to$ & $\ZZ$ && $\cdot$ &&& $\cdot$ && 
$\ZZ$ & $\to$ & $\ZZ$ \\
\hline
AI & $\ZZ^{2}$ & $\to$ & $0$ & $\ZZ$ & $\to$ & $0$ && $\cdot$ &&& $\cdot$ && $\ZZ$ & $\to$ & $0$ & $\ZZ$ & $\to$ & $0$ &
$\ZZ^{5}$ & $\to$ & $0$ \\
BDI && $\cdot$ &&& $\cdot$ &&& $\cdot$ &&& $\cdot$ &&& $\cdot$ &&& $\cdot$ &&& 
$\cdot$ &\\
D & $\ZZ^{2}$ & $\to$ & $0$ & $\ZZ$ & $\to$ & $0$ & $\ZZ_{\hphantom{2}}$ & $\to$ & $\ZZ_{2}$ && $\cdot$ && $\ZZ^{3}$ & $\to$ & $\ZZ\times\ZZ_{2}$ & $\ZZ$ & $\to$ & $\ZZ_{2}$ & 
$\ZZ^{8}$ & $\to$ & $\ZZ\times\ZZ_2^3$ \\
DIII && $\cdot$ &&& $\cdot$ && $\ZZ_{2}$ & $\to$ & $\ZZ_{2}$ & $\ZZ_{\hphantom{2}}$ & $\to$ & $\ZZ$ & $\ZZ_{2}^{2}$ & $\to$ & $\ZZ_{2}^{2}$ & $\ZZ_{2}$ & $\to$ & $\ZZ_{2}$ & 
$\ZZ\times\ZZ_2^4$ & $\to$ & $\ZZ\times\ZZ_2^4$ \\
AII & $\ZZ^{2}$ & $\to$ & $0$ & $\ZZ$ & $\to$ & $0$ & $\ZZ_{2}$ & $\to$ & $0$ & $\ZZ_{2}$ & $\to$ & $\ZZ_{2}$ & $\ZZ\times\ZZ_{2}^{2}$ & $\to$ & $\ZZ_{2}$ & $\ZZ\times\ZZ_{2}$ & $\to$ & $\ZZ_{4}$ & 
$\ZZ^{5}\times\ZZ_2^5$ & $\to$ & $\ZZ_2^2\times\ZZ_4$ \\
CII && $\cdot$ &&& $\cdot$ &&& $\cdot$ && $\ZZ_{2}$ & $\to$ & $\ZZ_{2}$ && $\cdot$ && $\ZZ_{2}$ & $\to$ & $\ZZ_{2}$ & $\ZZ_2^2$ & $\to$ & $\ZZ_2^2$ \\
C & $\ZZ^{2}$ & $\to$ & $0$ & $\ZZ$ & $\to$ & $0$ & $\ZZ_{\hphantom{2}}$ & $\to$ & $0$ && $\cdot$ && $\ZZ^{3}$ & $\to$ & $\ZZ$ & $\ZZ\times\ZZ_{2}$ & $\to$ & $\ZZ_{2}$ & 
$\ZZ^8\times\ZZ_2$ & $\to$ & $\ZZ\times\ZZ_2$ \\
CI && $\cdot$ &&& $\cdot$ &&& $\cdot$ && $\ZZ_{\hphantom{2}}$ & $\to$ & $\ZZ$ && $\cdot$ &&& $\cdot$ && 
$\ZZ$ & $\to$ & $\ZZ$ \\
\hline\hline
\end{tabularx}
\renewcommand{\arraystretch}{1}
\end{table*}

Consider a crystalline material of space-group $\mathrm{P\bar{4}}$ with tetragonal-disphenoidal $S_4$ point-group symmetry. The primitive unit-cell may be chosen as a parallelepiped with edges along its primitive lattice vectors.
Here, this forms a square cuboid depicted in Fig.~\ref{fig:P-4_HSM}~(bottom). Similarly, the BZ may also be chosen as a parallelepiped with edges along its primitive reciprocal lattice vectors, $\lv_1,\lv_2,\lv_3$. This forms the square cuboid depicted in Fig.~\ref{fig:P-4_HSM}~(top).

The tetragonal-disphenoidal $S_4$ point-group symmetry is an abstract fourfold $G\simeq\ZZ_4$ symmetry which acts on the primitive reciprocal lattice vectors by
\begin{equation}\label{eq:S4action}
\begin{pmatrix}\lv_1 \\ \lv_2 \\ \lv_3\end{pmatrix}\xmapsto{\hat{s}_4}
\begin{pmatrix}
0 & 1 & 0 \\
-1 & 0 & 0 \\
0 & 0 & -1 \\
\end{pmatrix}
\begin{pmatrix}\lv_1 \\ \lv_2 \\ \lv_3\end{pmatrix},
\end{equation}
with a similar action on the primitive lattice vectors.
However, since we study spinful-electrons, one must take into account the (projective) spin-representation $(\hat{s}_4)^4=-1$. The easiest way to deal with this inconvenience is to consider $\hat{s}_4$ as a generator of the double point-group $\twistcover{G}\simeq\ZZ_8$ eightfold symmetry with $(\hat{s}_4)^8=1$. We shall use this description from here on.

The building blocks of the topological classification correspond to the six HSM, $\hsm$, with $\pi$-integer coordinates, $\Gamma,Z,X,R,M,A$. Each of them is stabilized by its little group $\twistcover{G}_\hsm$, these are listed in Table~\ref{tab:P-4_WyckHSM} and depicted in Fig.~\ref{fig:P-4_HSM}~(top). The building blocks of the AIs correspond to all isolated Wyckoff positions, $\wyck$. These are, $a,b,c,d,g$, and contain all points reciprocal to the HSM; see Table~\ref{tab:P-4_WyckHSM} and Fig.~\ref{fig:P-4_HSM}~(bottom). We use the naming conventions of Ref.~\onlinecite{bradley2010mathematical} for the HSM in the BZ and the naming conventions of Ref.~\onlinecite{BilbaoCrystallographicServerIDatabasesandcrystallographiccomputingprograms} for Wyckoff positions in the unit-cell.

\subsubsection{Overview of the full classification results}
The complete $\KK$-theory classification, $\KR_{\twistcover{G}}(\BZ)$, of topologically distinct groundstates as well as the the complete classification of topological phases with anomalous surface states, $\Surf=\KK/\AI$, of space-group $\mathrm{P\bar{4}}$ is presented in Table~\ref{tab:P-4_res}. In Sec.~\ref{sec:main_results} we focused on the complete classification of anomalous surface states (see the $\mathrm{P\bar{4}}$ entries in Table~\ref{tab:mainres}), these appear as the $\Surf$ in the last column of Table~\ref{tab:P-4_res}. In the following sections we derive these results in detail.

\subsubsection{Topologically distinct groundstates}
The complete $\KK$-theory classification, $\KR_{\twistcover{G}}^{-q,0}(\bTt^3)$, of topologically distinct groundstates in AZ symmetry class, $q$, is derived from our free spectra decomposition, Eq.~\eqref{eq:torus-G-decomposition}, and is given by our HSM decomposition, Eq.~\eqref{eq:KTtoKS}, i.e.,
\begin{multline}\label{eq:P-4_KTtoKS}
\KR_{\twistcover{G}}(\bTt^3)\simeq\KR_{\twistcover{G}_\Gamma}(\pt)\oplus\rKR_{\twistcover{G}_Z}(\bSs^1)\oplus\rKR_{\twistcover{G}_X}(\bSs^1) \\
\oplus\rKR_{\twistcover{G}_R}(\bSs^2)\oplus\rKR_{\twistcover{G}_M}(\bSs^2)\oplus\rKR_{\twistcover{G}_A}(\bSs^3).
\end{multline}
Each of these summands further decomposes according to the isomorphism of Ref.~\onlinecite{Cornfeld2019Classification}; see Eq.~\eqref{eq:CCiso}. This decomposition contains one summand for each irreducible $\ZZ_2$-graded representation of $\twistcover{G}_\hsm$.

\paragraph{Representation theory of \texorpdfstring{$\ZZ_8$}{Z/8}}--
Since $\twistcover{G}\simeq\ZZ_8$, we shall use the standard \emph{ungraded} complex representations of $\ZZ_8$ in order to label the $\ZZ_2$-graded representations in a prescribed manner. We denote the fundamental 1D complex representation of $\ZZ_8$ by $t_8$, such that
\begin{equation}
\hat{s}_4\xmapsto{t_8} e^{\frac{2\pi i}{8}}.
\end{equation}
Any representation of $\ZZ_8$ is a polynomial with integer coefficients of $t_8$ subject to the relation $(t_8)^8=1$, such that each coefficient corresponds to the multiplicity of an irreducible representation, $(t_8)^n$, and every representation has a complex conjugate representation $\overline{(t_8)^n}=(t_8)^{8-n}$. For example, the 3D geometric representation in Eq.~\eqref{eq:S4action} is a real representation, $t_8^2+t_8^6+t_8^4$, which is a direct sum of the $\frac{2\pi}{4}$-rotation 2D representation, $t_8^2+t_8^6$, and the 1D sign representation, $t_8^4$, i.e.,
\begin{align}
t_8^2+t_8^6 \colon &&& \hat{s}_4\mapsto
\begin{pmatrix}
0 & 1 \\
-1 & 0 \\
\end{pmatrix}
\sim
\begin{pmatrix}
i & 0 \\
0 & -i \\
\end{pmatrix}, \nonumber\\
t_8^4 \colon &&& \hat{s}_4\mapsto -1.
\end{align}
We use an analogous construction for the subgroup, $C_2\subset S_4$, with $\hat{c}_2^{\phantom{1}}=\hat{s}_4^2$, such that $t_4=\Res_{\ZZ_4}^{\ZZ_8}(t_8)$.

\paragraph{\texorpdfstring{$\ZZ_2$}{Z/2}-graded representations}--
The $\ZZ_2$-grading itself, i.e., a group homomorphism, $\twistcover{G}_\hsm\to\ZZ_2$, is determined by the determinant of the $\Ort(d_\hsm)$ geometric action of $\twistcover{G}_\hsm$ on the primitive reciprocal lattice vectors,
\begin{equation}\label{eq:Z2-grading-by-det}
\twistcover{G}_\hsm\to\Ort(d_\hsm)\oto{\det}\{\pm 1\}\simeq\ZZ_2.
\end{equation}
For example, the action in Eq.~\eqref{eq:S4action} yields the $\ZZ_2$-grading, $\hat{s}_4^{\vphantom{1}},\hat{s}_4^3,\hat{s}_4^5,\hat{s}_4^7\xmapsto{\det} -1$ and $1,\hat{s}_4^2,\hat{s}_4^4,\hat{s}_4^6\xmapsto{\det} 1$.

In general, there are three types of possible gradings; we dub these, type-0, type-1, and type-2, as we now explain.

Type-0: This is the simplest type corresponding to the $\Gamma,M$-points, where the $\ZZ_2$-grading is trivial. In this case, the summand $\rKR_{\twistcover{G}_\hsm}(\bSs^{d_\hsm})$ decomposes according to irreducible \emph{ungraded} real representations of $\twistcover{G}_\hsm$. For example, consider the $M$-point, which is stabilized by the $S_4$ little group acting on $\lv_1,\lv_2$,
\begin{equation}
\begin{pmatrix}\lv_1 \\ \lv_2\end{pmatrix}\xmapsto{\hat{s}_4}
\begin{pmatrix}
0 & 1 \\
-1 & 0 \\
\end{pmatrix}
\begin{pmatrix}\lv_1 \\ \lv_2\end{pmatrix},
\end{equation}
such that the actions of all elements of $\twistcover{G}_M$ have positive determinant. We thus have,
\begin{align}
&\rKR_{\twistcover{G}_M}^{p,q}(\bSs^2)\simeq\trKR^{p,q}(\bSs^2)_{1}\oplus\trKC^{p,q}(\bSs^2)_{t_8^{\vphantom{1}}+t_8^7} \nonumber\\
&\oplus\trKC^{p,q}(\bSs^2)_{t_8^2+t_8^6}\oplus\trKC^{p,q}(\bSs^2)_{t_8^5+t_8^3}\oplus\trKR^{p,q}(\bSs^2)_{t_8^4}.
\end{align}
Here, the complex $\KK$-theory summands correspond to irreducible real representations of complex type.

Type-1: This is the type corresponding to the $Z,X,R,A$-points, where the $\ZZ_2$-grading is non-trivial. In this case, there is a canonical sign representation $\signrep$, which stems from the $\ZZ_2$-grading; the summand $\rKR_{\twistcover{G}_\hsm}(\bSs^{d_\hsm})$ decomposes according to $\ZZ_2$-graded representations of $\twistcover{G}_\hsm$, each of which has even and odd parts, $\rho^0$ and $\rho^1$, with $\rho^0=\signrep\otimes\rho^1$. We encode this structure by the virtual representation, $[\rho^0 -\rho^1]$. For example, consider the $R$-point, which is stabilized by the $C_2$ little group acting on $\lv_1,\lv_3$,
\begin{equation}
\begin{pmatrix}\lv_1 \\ \lv_3\end{pmatrix}\xmapsto{\hat{c}_2^{\vphantom{1}}=\hat{s}_4^2}
\begin{pmatrix}
-1 & 0 \\
0 & 1 \\
\end{pmatrix}
\begin{pmatrix}\lv_1 \\ \lv_3\end{pmatrix}.
\end{equation}
The $\ZZ_2$-grading, $\hat{c}_2^{\vphantom{1}},\hat{c}_2^3\xmapsto{\det} -1$ and $1,\hat{c}_2^2\xmapsto{\det} 1$, provides us with a sign representation $\signrep=t_4^2$, and we thus have,
\begin{equation}
\rKR_{\twistcover{G}_R}^{p,q}(\bSs^2)\simeq\rKR^{p,q-1}(\bSs^2)_{1-t_4^2}\oplus\rKR^{p-1,q}(\bSs^2)_{t_4^{\vphantom{1}}-t_4^3},
\end{equation}
where the shifts in the $\KK$-theory degree, i.e., $(0,-1)$ and $(-1,0)$, are determined by the isomorphism of Ref.~\cite{Cornfeld2019Classification}; see Eq.~\eqref{eq:CCiso}.

Type-2: This type is only applicable for even-dimensional irreducible representations of groups with a non-trivial $\ZZ_2$-grading. These are absent in cyclic groups but present in dihedral and other groups. In this case, these representations yield summands corresponding to (virtual) representations of the \emph{even part} of $\twistcover{G}_\hsm$ with respect to the $\ZZ_2$-grading. Examples are given in Ref.~\onlinecite{Cornfeld2019Classification} and are not present in the classification of the current pedagogical example.

\begin{table*}[t]
\centering
\caption{The decomposition of the real equivariant $\KK$-theory classification, $\KR_{\twistcover{G}}^{p,q}(\bTt^3)\simeq \bigoplus_{\hsm}\trKR^{p,q}_{\twistcover{G}_\hsm}(\bSs^{d_\hsm})$, of space-group $\mathrm{P\bar{4}}$ into non-$G$-equivariant $\KK$-theory components.}
\label{tab:P-4_full_KR}
\renewcommand{\arraystretch}{1.8}
\begin{tabularx}{\linewidth}{rXcXlXcXl}
\hline\hline
$\rKR_{\twistcover{G}_\hsm}^{p,q}(\bSs^{d_\hsm})$ 
&& $\simeq$ &&
Spinless contributions
&& $\oplus$ &&
Spinful contributions
\\
\hline
$\KR_{\twistcover{G}_\Gamma}^{p,q}(\pt)$ 
&& $\simeq$ &&
$\KR^{p,q}(\pt)_{1}\oplus\KR^{p,q}(\pt)_{t_8^4}\oplus\KC^{p,q}(\pt)_{t_8^2+t_8^6}$
&& $\oplus$ &&
$\KC^{p,q}(\pt)_{t_8^{\vphantom{1}}+t_8^7}\oplus\KC^{p,q}(\pt)_{t_8^5+t_8^3}$
\\
$\rKR_{\twistcover{G}_Z}^{p,q}(\bSs^1)$ 
&& $\simeq$ &&
$\trKR^{p,q-1}(\bSs^1)_{1-t_8^4}\oplus\trKR^{p-1,q}(\bSs^1)_{t_8^2-t_8^6}$
&& $\oplus$ &&
$\trKC^{p-1,q}(\bSs^1)_{t_8^{\vphantom{1}}-t_8^5+t_8^7-t_8^3}$
\\
$\rKR_{\twistcover{G}_X}^{p,q}(\bSs^1)$ 
&& $\simeq$ &&
$\trKR^{p-1,q}(\bSs^1)_{t_4^{\vphantom{1}}-t_4^3}$
&& $\oplus$ &&
$\trKR^{p,q-1}(\bSs^1)_{1-t_4^2}$
\\
$\rKR_{\twistcover{G}_R}^{p,q}(\bSs^2)$ 
&& $\simeq$ &&
$\trKR^{p-1,q}(\bSs^2)_{t_4^{\vphantom{1}}-t_4^3}$
&& $\oplus$ &&
$\trKR^{p,q-1}(\bSs^2)_{1-t_4^2}$
\\
$\rKR_{\twistcover{G}_M}^{p,q}(\bSs^2)$ 
&& $\simeq$ &&
$\trKC^{p,q}(\bSs^2)_{t_8^{\vphantom{1}}+t_8^7}\oplus\trKC^{p,q}(\bSs^2)_{t_8^5+t_8^3}$
&& $\oplus$ &&
$\trKR^{p,q}(\bSs^2)_{1}\oplus\trKR^{p,q}(\bSs^2)_{t_8^4}\oplus\trKC^{p,q}(\bSs^2)_{t_8^2+t_8^6}$
\\
$\rKR_{\twistcover{G}_A}^{p,q}(\bSs^3)$ 
&& $\simeq$ &&
$\trKC^{p-1,q}(\bSs^3)_{t_8^{\vphantom{1}}-t_8^5+t_8^7-t_8^3}$
&& $\oplus$ &&
$\trKR^{p,q-1}(\bSs^3)_{1-t_8^4}\oplus\trKR^{p-1,q}(\bSs^3)_{t_8^2-t_8^6}$
\\
\hline\hline
\end{tabularx}
\renewcommand{\arraystretch}{1}
\end{table*}

\begin{table}[!t]
\centering
\caption{The decomposition of the complex equivariant $\KK$-theory classification, $\KC_{\twistcover{G}}^{p,q}(\bTt^3)\simeq \bigoplus_{\hsm}\trKC^{p,q}_{\twistcover{G}_\hsm}(\bSs^{d_\hsm})$, of space-group $\mathrm{P\bar{4}}$ into non-$G$-equivariant $\KK$-theory components.}
\label{tab:P-4_full_KC}
\renewcommand{\arraystretch}{2}
\begin{tabularx}{\linewidth}{rXcXlXcXl}
\hline\hline
$\rKC_{\twistcover{G}_\hsm}^{p,q}(\bSs^{d_\hsm})$ 
&& $\simeq$ &&
Spinless
&& $\oplus$ &&
Spinful
\\
\hline
$\KC_{\twistcover{G}_\Gamma}^{p,q}(\pt)$ 
&& $\simeq$ &&
$\bigoplus\limits_{\mathrlap{\rho=1,t_8^2,t_8^4,t_8^6}}\KC^{p,q}(\pt)_{\rho}$
&& $\oplus$ &&
$\bigoplus\limits_{\mathrlap{\rho=t_8^{\vphantom{1}},t_8^3,t_8^5,t_8^7}}\KC^{p,q}(\pt)_{\rho}$
\\
$\rKC_{\twistcover{G}_Z}^{p,q}(\bSs^1)$ 
&& $\simeq$ &&
$\bigoplus\limits_{\mathrlap{\rho=[1-t_8^4],[t_8^2-t_8^6]}}\trKC^{p-1,q}(\bSs^1)_{\rho}$
&& $\oplus$ &&
$\bigoplus\limits_{\mathrlap{\rho=[t_8^{\vphantom{1}}-t_8^5],[t_8^7-t_8^3]}}\trKC^{p-1,q}(\bSs^1)_{\rho}$
\\
$\rKC_{\twistcover{G}_X}^{p,q}(\bSs^1)$ 
&& $\simeq$ &&
$\trKC^{p-1,q}(\bSs^1)_{t_4^{\vphantom{1}}-t_4^3}$
&& $\oplus$ &&
$\trKC^{p-1,q}(\bSs^1)_{1-t_4^2}$
\\
$\rKC_{\twistcover{G}_R}^{p,q}(\bSs^2)$ 
&& $\simeq$ &&
$\trKC^{p-1,q}(\bSs^2)_{t_4^{\vphantom{1}}-t_4^3}$
&& $\oplus$ &&
$\trKC^{p-1,q}(\bSs^2)_{1-t_4^2}$
\\
$\rKC_{\twistcover{G}_M}^{p,q}(\bSs^2)$ 
&& $\simeq$ &&
$\bigoplus\limits_{\mathrlap{\rho=t_8^{\vphantom{1}},t_8^3,t_8^5,t_8^7}}\trKC^{p,q}(\bSs^2)_{\rho}$
&& $\oplus$ &&
$\bigoplus\limits_{\mathrlap{\rho=1,t_8^2,t_8^4,t_8^6}}\trKC^{p,q}(\bSs^2)_{\rho}$
\\
$\rKC_{\twistcover{G}_A}^{p,q}(\bSs^3)$ 
&& $\simeq$ &&
$\bigoplus\limits_{\mathrlap{\rho=[t_8^{\vphantom{1}}-t_8^5],[t_8^7-t_8^3]}}\trKC^{p-1,q}(\bSs^3)_{\rho}$
&& $\oplus$ &&
$\bigoplus\limits_{\mathrlap{\rho=[1-t_8^4],[t_8^2-t_8^6]}}\trKC^{p-1,q}(\bSs^3)_{\rho}$
\\
\hline\hline
\end{tabularx}
\renewcommand{\arraystretch}{1}
\end{table}

\begin{table*}[t]
\centering
\caption{The decomposition of real AZ classes AIs, $\AI^{p,q}\simeq\bigoplus_\wyck\KR_{\twistcover{G}_\wyck}^{p,q}(\pt)$, at Wyckoff position, $\wyck$, of space-group $\mathrm{P\bar{4}}$ into non-$G$-equivariant $\KK$-theory components.}
\label{tab:P-4_full_KR_AI}
\renewcommand{\arraystretch}{1.8}
\begin{tabularx}{\linewidth}{rXcXlXcXl}
\hline\hline
$\KR_{\twistcover{G}_\wyck}^{p,q}(\pt)$ 
&& $\simeq$ &&
Spinless contributions
&& $\oplus$ &&
Spinful contributions
\\
\hline
$\KR_{\twistcover{G}_{a,b,c,d}}^{p,q}(\pt)$ 
&& $\simeq$ &&
$\KR^{p,q}(\pt)_{1}\oplus\KR^{p,q}(\pt)_{t_8^4}\oplus\KC^{p,q}(\pt)_{t_8^2+t_8^6}$
&& $\oplus$ &&
$\KC^{p,q}(\pt)_{t_8^{\vphantom{1}}+t_8^7}\oplus\KC^{p,q}(\pt)_{t_8^5+t_8^3}$
\\
$\KR_{\twistcover{G}_g}^{p,q}(\pt)$ 
&& $\simeq$ &&
$\KR^{p,q}(\pt)_{1}\oplus\KR^{p,q}(\pt)_{t_4^2}$
&& $\oplus$ &&
$\KC^{p,q}(\pt)_{t_4^{\vphantom{1}}+t_4^3}$
\\
\hline\hline
\end{tabularx}
\renewcommand{\arraystretch}{1}
\end{table*}

\begin{table}[!t]
\centering
\caption{The decomposition of complex AZ classes AIs, $\AI^{p,q}\simeq\bigoplus_\wyck\KC_{\twistcover{G}_\wyck}^{p,q}(\pt)$, at Wyckoff position, $\wyck$, of space-group $\mathrm{P\bar{4}}$ into non-$G$-equivariant $\KK$-theory components.}
\label{tab:P-4_full_KC_AI}
\renewcommand{\arraystretch}{2}
\begin{tabularx}{\linewidth}{rXcXlXcXl}
\hline\hline
$\KC_{\twistcover{G}_\wyck}^{p,q}(\pt)$ 
&& $\simeq$ &&
Spinless
&& $\oplus$ &&
Spinful
\\
\hline
$\KC_{\twistcover{G}_{a,b,c,d}}^{p,q}(\pt)$ 
&& $\simeq$ &&
$\bigoplus\limits_{\mathrlap{\rho=1,t_8^2,t_8^4,t_8^6}}\KC^{p,q}(\pt)_{\rho}$
&& $\oplus$ &&
$\bigoplus\limits_{\mathrlap{\rho=t_8^{\vphantom{1}},t_8^3,t_8^5,t_8^7}}\KC^{p,q}(\pt)_{\rho}$
\\
$\KC_{\twistcover{G}_g}^{p,q}(\pt)$ 
&& $\simeq$ &&
$\bigoplus\limits_{\mathrlap{\rho=1,t_4^2}}\KC^{p,q}(\pt)_{\rho}$
&& $\oplus$ &&
$\bigoplus\limits_{\mathrlap{\rho=t_4^{\vphantom{1}},t_4^3}}\KC^{p,q}(\pt)_{\rho}$
\\
\hline\hline
\end{tabularx}
\renewcommand{\arraystretch}{1}
\end{table}

\paragraph{\texorpdfstring{$\KK$}{K}-theory classification results}--
By repeating this analyses for all HSM we obtain, in Table~\ref{tab:P-4_full_KR}, the complete decomposition of the equivariant $\KK$-theory classification, $\KR_{\twistcover{G}}^{p,q}(\bTt^3)$, into non-$G$-equivariant $\KK$-theory components for each irreducible $\ZZ_2$-graded representation.
Note, that since we are using the double point group, $\twistcover{G}$, we get contributions corresponding both to spinless and spinful electrons~\footnote{In general, one may have to use a further, HSM dependent, double cover in order to split the $\KK$-theory, see Appendix~\ref{sec:spins-and-reps}; this is fortunately not the case for space-group $\mathrm{P\bar{4}}$.}. In order to distinguish these contributions, in Table~\ref{tab:P-4_full_KR}, we place the spinless summands on the left and the spinful summands on the right.
All summands may be easily read from the invariants in the periodic table of TISC (see Table~\ref{tab:per}), using Bott periodicity~\cite{atiyah1964clifford,bott1969lectures,kitaev2009periodic},
\begin{equation}\label{eq:Bott-pq}
\trKR^{p,q}(\bSs^d)=\KR^{p,q-d}(\pt)\simeq\pi_0(\sR_{q-p-d}).
\end{equation}
The spinful contributions for all AZ classes are explicitly presented in Table~\ref{tab:P-4_res}.

The complex AZ classes, A and AIII, are classified by complex $\KK$-theory which decomposes according to irreducible complex representations. In particular, every irreducible real representation of complex type splits into two irreducible complex representations, e.g., $t_8^5+t_8^3$ splits to $t_8^5$ and $t_8^3$. This is explicitly presented in Table~\ref{tab:P-4_full_KC}.

\subsubsection{Anomalous surface states}\label{sec:P-4_ASS}
The complete classification of topological phases with anomalous surface states in AZ symmetry class, $q$, is given by topologically distinct groundstates which are not related to one another by atomic insulators, i.e., $\Surf^{-q,0}=\KR^{-q,0}_{\twistcover{G}}(\bTt^3)/\AI^{-q,0}$. In order to obtain these, one must quotient out the groundstates corresponding to AIs, i.e., the image of the $\AItoK$ map, $\AI^{-q,0}\oto{\AItoK}\KR^{-q,0}_{\twistcover{G}}(\bTt^3)$.
As discussed in Sec.~\ref{sec:AI}, the AIs themselves are described by the $\KK$-theory of a point for each isolated Wyckoff position,
\begin{align}\label{eq:AIsum}
\AI = \ & \KR_{\twistcover{G}_a}(\pt)\oplus\KR_{\twistcover{G}_b}(\pt)\oplus\KR_{\twistcover{G}_g}(\pt) \nonumber\\
&\oplus\KR_{\twistcover{G}_c}(\pt)\oplus\KR_{\twistcover{G}_d}(\pt),
\end{align}
see Fig.~\ref{fig:P-4_HSM} and Table~\ref{tab:P-4_WyckHSM}. Note, that the $g$ Wyckoff position is a line of \emph{equivalent} points and thus also classified by a single equivariant $\KK$-theory component. Similar to Tables~\ref{tab:P-4_full_KR} and \ref{tab:P-4_full_KC}, each of the Wyckoff position contributions is decomposed into non-$G$-equivariant components for each irreducible (ungraded) representation. This is presented in Tables~\ref{tab:P-4_full_KR_AI} and \ref{tab:P-4_full_KC_AI}.

\paragraph{Structure of the \texorpdfstring{$\AItoK$}{AI} map}--
The $\AItoK$ map is a $\KK$-theory map. Specifically, it is a homomorphism between the modules, $\AI^{*,*}$ and $\KR_{\twistcover{G}}^{*,*}(\bTt^3)$, over the ring, $\KR_{\twistcover{G}}^{*,*}(\pt)$, i.e., a map which respects the multiplicative structure of $\KK$-theory, Eq.~\eqref{eq:KRX-mult}. As discussed in Sec.~\ref{sec:mult}, such maps satisfy,
\begin{equation}
\AItoK^{p,q}(u)=u\cdot\AItoK^{0,0}(1),
\end{equation}
for all $u\in\AI^{p,q}$.
Since $1\in\KR_{\twistcover{G}}^{0,0}(\pt)$ is a spinless representation, it is now clear why we have not omitted these contributions in Tables~\ref{tab:P-4_full_KR}-\ref{tab:P-4_full_KC_AI}.

The $\AItoK^{*,*}$ map is thus completely determined by $\AItoK^{0,0}(1)$.
Furthermore, it follows from Eq.~\eqref{eq:KTtoKS} and Eq.~\eqref{eq:AIsum} that the $\AItoK$ map can be displayed as a block-matrix, $[\AItoK]_{\hsm\wyck}=[\AItoK(u_\wyck)]_\hsm$, whose entries are the contributions of $u_\wyck\in\KR_{\twistcover{G}_\wyck}(\pt)\subset\AI$ to $\rKR_{\twistcover{G}_\hsm}(\bSs^{d_\hsm})\subset\KR_{\twistcover{G}_\hsm}(\bTt^3)$. For each $\hsm$ and $\wyck$, the block's entries correspond to the irreducible $\ZZ_2$-graded representations of $\twistcover{G}_\hsm$ and the irreducible representations of $\twistcover{G}_\wyck$ as given in Tables \ref{tab:P-4_full_KR} and \ref{tab:P-4_full_KR_AI}.

We find that the matrix $[\AItoK^{0,0}(1)]_{\hsm\wyck}=[\AItoK^{0,0}(1_\wyck)]_{\hsm}$ for space-group $\mathrm{P\bar{4}}$ is given by
\begin{equation}\label{eq:ai1-block}
\begin{smallarray}{rccccccc}
\vphantom{{}_{\big|}}&& a & b & g & c & d\\
\Gamma & \multirow{6}{*}{$\left(\vphantom{\substack{\Bigg|\\\Bigg|}}\right.$} & 1 & 1 & 1+t_8^4 & 1 & 1 & \multirow{6}{*}{$\left.\vphantom{\substack{\Bigg|\\\Bigg|}}\right)$} \\
Z && 0 & [t_8^2-t_8^6] & 0 & 0 & [t_8^2-t_8^6] \\
X && 0 & 0 & [t_4^{\vphantom{1}}-t_4^3] & [t_4^{\vphantom{1}}-t_4^3] & [t_4^{\vphantom{1}}-t_4^3] \\
R && 0 & 0 & 0 & 0 & 0 \\
M && 0 & 0 & 0 & [t_8^{\vphantom{1}}+t_8^7]\xi^3 & [t_8^{\vphantom{1}}+t_8^7]\xi^3 \\
A && 0 & 0 & 0 & 0 & [t_8^{\vphantom{1}}-t_8^5+t_8^7-t_8^3]\xi^3
\end{smallarray},
\end{equation}
where $1$ is the identity element, $1\in\KR^{0,0}(\pt)_{1}$. Note, that $\trKR^{p-p_\rho,q-q_\rho}(\bSs^d)_{\rho=[\rho^0-\rho^1]}$ is a shifted copy of $\KR^{p,q}(\pt)$, therefore, we denote its canonical generator by
\begin{equation}
[\rho^0 -\rho^1]\defeq 1\in\trKR^{0,d}(\bSs^d)_{\rho^0 -\rho^1}\simeq\KR^{0,0}(\pt).
\end{equation}
We use a similar construction for $\KC$ components.
Note, that none of the maps discussed in Sec.~\ref{sec:mult-TISC} are affected by the invertible elements, $\beta$, $\mu$, and $\xi^4$. Therefore, here and henceforth, we  set $\mu=\beta=\xi^4=1$ for all elements of $\trKR^{p,q}(\bSs^d)=\KR^{p,q-d}(\pt)$, see, e.g., Eq.~\eqref{eq:xi3-example}.

Let us explain how to obtain the matrix in Eq.~\eqref{eq:ai1-block}.
We begin by noting three generic properties:

First, one notices that $[\AItoK]_{\hsm\wyck}$ is an upper triangular block-matrix. The component $\rKR_{\twistcover{G}_\hsm}(\bSs^{d_\hsm})$ does not capture groundstates which are topologically equivalent to states on the boundaries of the cell centered at $\hsm$, see Fig.~\ref{fig:CWP1}. This provides us with a partial order where every HSM is only contributed from `greater' Wyckoff positions,
\begin{equation}
\xymatrix@R=-5pt@C=20pt{
& {M,c} \ar@{.}[rr]|-{\displaystyle{>}} && {X,g} \ar@{.}[rrd]|-{\displaystyle{>}} && \\
{A,d} \ar@{.}[ru]|-{\displaystyle{>}} \ar@{.}[rrd]|-{\displaystyle{>}} &&&&& {\Gamma,a}. \\
&& {R,g} \ar@{.}[rr]|-{\displaystyle{>}} \ar@{.}[ruu]|-{\displaystyle{>}} && {Z,b} \ar@{.}[ru]|-{\displaystyle{>}} &
}
\end{equation}
We say that $\hsm_1>\hsm_0$ if all coordinate values of $\hsm_1$, as given in Table~\ref{tab:P-4_WyckHSM}, are greater than those of $\hsm_0$. Hence, an AI at a Wyckoff position, $\wyck_1$, reciprocal to $\hsm_1$, only contributes to $\rKR_{\twistcover{G}_{\hsm_0}}(\bSs^{d_{\hsm_0}})$ if $\hsm_1\ge\hsm_0$.

Second, one notices that all entries left to the diagonal are multiples (within the representation ring) of the diagonal entry. Consider a particular diagonal entry with $\hsm_0$ reciprocal to $\wyck_0$. The contribution to $\rKR_{\twistcover{G}_{\hsm_0}}(\bSs^{d_{\hsm_0}})$ of an AI at any $\wyck_1>\wyck_0$ is given by restricting the groundstate to the sub-torus centered at $\hsm_0$. This gives an AI with the same $\vk$-dependence as that at $\wyck_0$ and with a representation determined by an appropriate restriction and/or induction between $\twistcover{G}_{\wyck_1}$ and $\twistcover{G}_{\wyck_0}$. For example, consider an AI at the $g$ Wyckoff position with a trivial representation of $\twistcover{G}\simeq\ZZ_4$. It corresponds to two $s$-orbitals at $\vx=(\frac{1}{2},0,x),(0,\frac{1}{2},x)$, which are interchanged by $\hat{s}_4$. Therefore, as a representation of $\hat{G}_\Gamma\simeq\ZZ_8$, its contribution to $\KR^{0,0}_{\twistcover{G}_\Gamma}(\pt)\simeq\Rep(\ZZ_8)$ is $\Ind_{\ZZ_4}^{\ZZ_8}(1)=1+t_8^4$, which is a two-dimensional representation corresponding to both orbitals combined.

Third, one notices that no Wyckoff position contributes to the $R$-point. This is a consequence of the high-symmetry line that connects the $R$-point with the $X$-point; see Fig.~\ref{fig:P-4_HSM}. Any AI groundstate supported on the $g$ Wyckoff position, $\vx=(\frac{1}{2},0,x),(0,\frac{1}{2},x)$, may be continuously deformed to $\vx=(\frac{1}{2},0,0),(0,\frac{1}{2},0)$ which does not contribute to the $R$-point. This is also a generic property.

An immediate conclusion of these three properties is that the $[\AItoK]_{\hsm\wyck}$ block-matrix may be block-diagonalized using column operations [valued in $\Rep(\twistcover{G})$]. This implies that the image of the $\AI$ map is identical to the image of the block-diagonal entries. These entries correspond to contributions of AIs at the center of the point-group and thus may be evaluated using Shiozaki's formula~\cite{shiozaki2019classification}, Eq.~\eqref{eq:shiozaki}.

\paragraph{Application of Shiozaki's formula}--
In order to interpret Eq.~\eqref{eq:shiozaki} in terms of our decomposition, we utilize the Atiyah-Bott-Shapiro construction~\cite{atiyah1964clifford,lawson2016spin} which expresses elements of the real $\KK$-theory in terms of $\ZZ_2$-graded representations of Clifford algebras, $\Cl_{p,q}$. In essence, this amounts to lifting the geometric action, $G_\hsm\to\Ort(d_\hsm)$ [e.g., Eq.~\eqref{eq:S4action}], to a ``geometric-algebra" action $\spincover{G}_\hsm\to\Cl_{d_\hsm,0}$ and decompose this action into irreducible $\ZZ_2$-graded representations.
We leave the mathematical details to Appendix~\ref{sec:math-AI} and provide here a ``quick and dirty" method to obtain the diagonal entries of $[\AItoK^{0,0}(1)]_{\hsm\wyck}$. 

The gist of the ``quick and dirty" method is as follows. We first use the geometric action, $G_\hsm\to\Ort(d_\hsm)$, i.e.,
\begin{align}
\lv_i\xmapsto{g}\sum\nolimits_j [O_{g}]_{ij}\lv_j, && O_{g_1}O_{g_2}=O_{g_1\cdot g_2},
\end{align}
and restrict our attention to the image, $\minor{G}_\hsm$, of $G_\hsm$ in $\Ort(d_\hsm)$, i.e., such that $G_\hsm\twoheadrightarrow\minor{G}_\hsm\subset\Ort(d_\hsm)$.
We then construct a double cover, $\minor{\spincover{G}}_\hsm$, of $\minor{G}_\hsm$, which is faithfully represented by $\SUnit(2)$, which itself is the double cover of $\SOrt(3)$, i.e., such that $\minor{\spincover{G}}_\hsm\subset\SUnit(2)$~\footnote{Formally speaking, one must construct a cover $\minor{\spincover{G}}_\hsm\subset\Pin_-(d_\hsm)\subset\Cl_{d_\hsm,0}$. However, since $\Pin_-(3)\simeq\SUnit(2)\times\ZZ_2$, $\Pin_-(2)\subset\SUnit(2)$, and $\Pin_-(1)\simeq\ZZ_4\subset\SUnit(2)$, we find that the $\SUnit(2)$ description suffices for all $d_\hsm$.}. The unitary matrices $U_g\in\SUnit(2)$ are constructed as to satisfy,
\begin{align}\label{eq:lift2U2}
U_{g}^{\vphantom{1}}\sigma_i U_{g}^{-1}=\sum\nolimits_j\det(O_{g})[O_{g}]_{ij}\sigma_j, && U_{g_1}U_{g_2}=U_{g_1\cdot g_2},
\end{align}
where $\sigma_{1,2,3}$ are the Pauli matrices. Finally, the diagonal entry of the $\AItoK^{0,0}(1)$ matrix [Eq.~\eqref{eq:ai1-block}] is simply found by decomposing the above unitary matrix representation into irreducible ungraded representations of $\twistcover{G}_\hsm$.

Note, that the faithful representation, Eq.~\eqref{eq:lift2U2}, of $\minor{\spincover{G}}_\hsm$ provides us with a generically non-faithful representation of $\spincover{G}_\hsm\twoheadrightarrow\minor{\spincover{G}}_\hsm$. Here, $\spincover{G}_\hsm$ is a double cover of $G_\hsm$ which might be different than the physical double point-group, $\twistcover{G}_\hsm$. Fortunately, for all HSM of space-group $\mathrm{P\bar{4}}$, one has $\spincover{G}_\hsm\simeq\twistcover{G}_\hsm$  (the generic case is treated in Appendix~\ref{sec:spins-and-reps}). 

One might worry that some data is lost when restricting the $\Ort(d_\hsm)$ action to $\SOrt(d_\hsm)$ in Eq.~\eqref{eq:lift2U2} thus forgetting the $\ZZ_2$-grading by the determinant [Eq.~\eqref{eq:Z2-grading-by-det}].
However, since $\KR^{0,0}_{\twistcover{G}_\hsm}(\bSs^{d_\hsm})$ is already decomposed into $\ZZ_2$-graded representations (see Table.~\ref{tab:P-4_full_KR}), we at most encounter a global sign ambiguity for each row of $[\AItoK^{0,0}(1)]_{\hsm\wyck}$; this does not alter the quotient by the image of the $\AItoK$ map. One may always get rid of this inconsequential ambiguity by following Appendix~\ref{sec:math-AI}.

Let us provide a couple of explicit examples, which would make our method clearer:

First, consider the contribution of an AI at the $c$ Wyckoff position to the $M$-point.
The $G_M\to \Ort(2)$ action is
\begin{equation}
\begin{pmatrix}\lv_1 \\ \lv_2\end{pmatrix}\xmapsto{s_4}
\begin{pmatrix}
0 & 1 \\
-1 & 0 \\
\end{pmatrix}
\begin{pmatrix}\lv_1 \\ \lv_2\end{pmatrix}.
\end{equation}
This is a faithful action, so $\minor{G}_M=G_M\simeq\ZZ_4$. We construct a faithful unitary matrix representation of $\minor{\spincover{G}}_M=\twistcover{G}_M\simeq\ZZ_8$, given by,
\begin{align}
U_{\hat{s}_4}= e^{-\frac{2\pi i}{8}\sigma_z}, && U_{\hat{s}_4}^{\vphantom{1}}\begin{pmatrix}\sigma_1 \\ \sigma_2\end{pmatrix} U_{\hat{s}_4}^{-1}=\begin{pmatrix}
0 & 1 \\
-1 & 0 \\
\end{pmatrix}
\begin{pmatrix}\sigma_1 \\ \sigma_2\end{pmatrix}.
\end{align}
This is clearly the $t_8^{\vphantom{1}}+t_8^7$ representation, which leads to $[\AItoK^{0,0}(1_c)]_{M}=[t_8^{\vphantom{1}}+t_8^7]\xi^3$ in Eq.~\eqref{eq:ai1-block}. The $\xi^3$ factor stems from
\begin{equation}\label{eq:xi3-example}
\trKC^{0,0}(\bSs^2)_{t_8^{\vphantom{1}}+t_8^7}=\KC^{0,-2}(\pt)\simeq\KC^{-6,0}(\pt)=\ZZ\cdot\xi^3,
\end{equation}
where we have used $\mu=\xi^4=1$ as above.

Second, consider the contribution of an AI at the $b$ Wyckoff position to the $Z$-point.
The $G_Z\to \Ort(1)$ action is
\begin{equation}
\lv_3\xmapsto{s_4}(-1)
\lv_3.
\end{equation}
This is not a faithful action, so we restrict to $\minor{G}_Z\simeq\ZZ_2$. We construct a faithful unitary matrix representation of $\minor{\spincover{G}}_Z\simeq\ZZ_4$, given by,
\begin{align}
U_{\hat{s}_4}= e^{\frac{2\pi i}{4}\sigma_3}, && U_{\hat{s}_4}^{\vphantom{1}}\sigma_3 U_{\hat{s}_4}^{-1}=-(-1)\sigma_3.
\end{align}
Here, in order to spare the reader of excess notation, we identified the generator of $\minor{\spincover{G}}_Z$ with the $\hat{s}_4$ generator of $\ZZ_8\simeq\twistcover{G}_Z\twoheadrightarrow\minor{\spincover{G}}_Z$.
This $\SUnit(2)$ action is clearly the $t_8^2+t_8^6$ ungraded representation, which leads to the $\ZZ_2$-graded representation, $[\AItoK^{0,0}(1_b)]_{Z}=[t_8^2-t_8^6]$, in Eq.~\eqref{eq:ai1-block}. This is where the inconsequential global sign ambiguity comes about.

All the other diagonal entries of $[\AItoK^{0,0}(1)]_{\hsm\wyck}$, given in Eq.~\eqref{eq:ai1-block}, are similarly derived.

\paragraph{Application of \texorpdfstring{$\KK$}{K}-theory multiplication}--
Once $\AItoK^{0,0}(1)$ is attained, one wishes to compute $\AItoK^{-q,0}(u)=u\cdot\AItoK^{0,0}(1)\in\KR^{-q,0}_{\twistcover{G}}(\bTt^3)$, for all $u\in\AI^{-q,0}$ and all AZ symmetry classes, $q$. This provides the image of the $\AItoK$ map, that is, the full classification of AI groundstates and thus $\Surf^{-q,0}=\KR^{-q,0}_{\twistcover{G}}(\bTt^3)/\AI^{-q,0}$.
It thus suffices to find the images of the generators of $\AI^{-q,0}$, as given in Table~\ref{tab:P-4_full_KR_AI}, corresponding to the irreducible representations of $\twistcover{G}_\wyck$.

We thus focus on each block-diagonal entry of $[\AItoK]_{\hsm\wyck}$ and expand that block into a matrix $[\AItoK]_{{\rho_\hsm}{\rho_\wyck}}=[\AItoK(u_{\rho_\wyck})]_{{\rho_\hsm}}$ whose entries are the contributions of the AI, $u_{\rho_\wyck}\in\KR_{\twistcover{G}_\wyck}^{p,q}(\pt)$, to the $\rho_\hsm$-component of $\KR_{\twistcover{G}_\hsm}^{p,q}(\bSs^{d_\hsm})$.
Each ${\rho_\hsm}$ and ${\rho_\wyck}$ correspond to irreducible $\ZZ_2$-graded representations of $\twistcover{G}_\hsm$ and the irreducible representations of $\twistcover{G}_\wyck$ as given in Tables \ref{tab:P-4_full_KR} and \ref{tab:P-4_full_KR_AI}. We find that the blocks, $[\AItoK]_{{\rho_\hsm} {\rho_\wyck}}=[\AItoK(u_{\rho_\wyck})]_{{\rho_\hsm}}$, for space-group $\mathrm{P\bar{4}}$ are as follows: $[\AItoK]_{\Gamma a}$ is always the identity matrix, 
\begin{equation}\label{eq:ai-Ga}
\begin{array}{rcccc}
\vphantom{{}_{\big|}}&& [t_8^{\vphantom{1}}+t_8^7]_a & [t_8^5+t_8^3]_a \\
{[t_8^{\vphantom{1}}+t_8^7]_\Gamma} & \multirow{2}{*}{\bigg(} & u & 0 & \multirow{2}{*}{\bigg)} \\
{[t_8^5+t_8^3]_\Gamma} && 0 & u
\end{array},
\end{equation}
$[\AItoK]_{Rg}$ is the zero matrix, and the rest are given by
\begin{equation}
\begin{array}{rcccc}
\vphantom{{}_{\big|}}&& [t_8^{\vphantom{1}}+t_8^7]_b & [t_8^5+t_8^3]_b \\
{[t_8^{\vphantom{1}}-t_8^5+t_8^7-t_8^3]_Z} & \Big( & -u & u & \Big)
\end{array},
\end{equation}
\begin{equation}
\begin{array}{rc}
\vphantom{{}_{\big|}}& [t_4^{\vphantom{1}}+t_4^3]_g\\
{[1-t_4^2]_X} & \Big(-\rr(\xi^3\cdot u)\Big)
\end{array},
\end{equation}
\begin{equation}
\begin{array}{rcccc}
\vphantom{{}_{\big|}}&& [t_8^{\vphantom{1}}+t_8^7]_c & [t_8^5+t_8^3]_c\\
{[1]_M} & \multirow{3}{*}{\Bigg(} & \rr(\xi^3\cdot u) & 0 & \multirow{3}{*}{\Bigg)}\\
{[t_8^4]_M} && 0 & \rr(\xi^3\cdot u)\\
{[t_8^2+t_8^6]_M} && \xi^3\cdot u & \xi^3\cdot u
\end{array},
\end{equation}
\begin{equation}\label{eq:ai-Ad}
\begin{array}{rcccc}
\vphantom{{}_{\big|}}&& [t_8^{\vphantom{1}}+t_8^7]_d & [t_8^5+t_8^3]_d\\
{[1-t_8^4]_A} & \multirow{2}{*}{\bigg(} & \rr(\xi^2\cdot u) & -\rr(\xi^2\cdot u) & \multirow{2}{*}{\bigg)}\\
{[t_8^2-t_8^6]_A} && \rr(\xi^3\cdot u) & -\rr(\xi^3\cdot u)
\end{array}.
\end{equation}
Here, we have only presented the results for the spinful components in Tables \ref{tab:P-4_full_KR} and \ref{tab:P-4_full_KR_AI}; the spinless results are analogous.

In order to obtain these results, we utilize the multiplicative structure of the equivariant $\KK$-theory. In particular, one finds that the inherent equivariant spectra constrain the $\AItoK$ map to such an extent that it is uniquely determined by the multiplicative structure of the \emph{ungraded} complex representations, as we immediately demonstrate using an explicit example.

Consider $[\AItoK]_{Ad}=[\AItoK(u_{\rho_d})]_{A}$. Let us find the image of
\begin{equation}
u\in\KC^{-q,0}(\pt)_{t_8^{\vphantom{1}}+t_8^7}\subset\KR^{-q,0}_{\twistcover{G}_d}(\pt),
\end{equation}
within
\begin{multline}
\KR^{-q-4,0}(\pt)_{1-t_8^4}\oplus\KR^{-q-6,0}(\pt)_{t_8^2-t_8^6} \\
\simeq\rKR^{-q,-1}(\bSs^3)_{1-t_8^4}\oplus\rKR^{-q-1,0}(\bSs^3)_{t_8^2-t_8^6} \\
\subset\KR^{-q,0}_{\twistcover{G}_A}(\bTt^3),
\end{multline}
cf.~Eq.~\eqref{eq:xi3-example}.
Following the discussion in Sec.~\ref{sec:mult-TISC}, the only possible $\KK$-theory map, stemming from a module-spectra homomorphism of type $\KC\to\KR\oplus\KR$, is given by the realification map,
\begin{equation}\label{eq:n1n2-def}
[t_8^{\vphantom{1}}+t_8^7]_d u \xmapsto{\AItoK} [1-t_8^4]_A n'\rr(\xi^2\cdot u)+[t_8^2-t_8^6]_A n''\rr(\xi^3\cdot u),
\end{equation}
where $n',n''\in\ZZ$ are integers. Moreover, from the relevant entry of Eq.~\eqref{eq:ai1-block}, we know that
\begin{equation}
[1]_d \xmapsto{\AItoK} [t_8^{\vphantom{1}}-t_8^5+t_8^7-t_8^3]_A\xi^3.
\end{equation}
In order to extract the integers, $n'$ and $n''$, we observe that Eq.~\eqref{eq:n1n2-def} translates to the complex $\KK$-theory in Tables \ref{tab:P-4_full_KC} and \ref{tab:P-4_full_KC_AI} as follows,
\begin{align}
[t_8^{\vphantom{1}}]_d u &\xmapsto{\AItoK} [1-t_8^4]_A n'\xi^2\cdot u+[t_8^2-t_8^6]_A n''\xi^3\cdot u, \nonumber\\
[t_8^7]_d u &\xmapsto{\AItoK} [1-t_8^4]_A n'\xi^2\cdot u-[t_8^2-t_8^6]_A n''\xi^3\cdot u.
\end{align}
The complex $\KK$-theory maps are determined by the multiplication rules of the \emph{ungraded} complex representations of $\ZZ_8$. Specifically, we recall that $\AItoK(u) =u\cdot\AItoK(1)$, and thus by setting $\xi=1$, we may utilize
\begin{equation}
t_8^{\vphantom{1}}\otimes(t_8^{\vphantom{1}}-t_8^5+t_8^7-t_8^3)=(t_8^2-t_8^6)+(1-t_8^4),
\end{equation}
to conclude that $n'=n''=1$. All other entries of Eqs.~\eqref{eq:ai-Ga}-\eqref{eq:ai-Ad} are similarly obtained using the correspondences between the real and complex $\KK$-theories which are provided in Table~\ref{tab:KR-hom} in Appendix~\ref{sec:math_background}.

Note, that since $\rr(u)\neq\eta$ for all $u\in\KC^{-q,0}(\pt)$, the only type of K-theory maps undetermined by the complex representation theory is $v\mapsto n\eta\cdot v$. Nevertheless, by $2\eta=0$, one has $n\eta\cdot v=(n\tmod 2)\eta\cdot v$, such that this map is completely determined by the dimensionality (modulo $2$) of the representation.

\paragraph{Classification of anomalous surface states}--
Once in full possession of $\AItoK(u)$, for all $u\in\AI^{-q,0}$, it is but a simple manner of picking a particular AZ symmetry class, $q$, and quotienting-out the image of the $\AItoK$ map, $\AI^{-q,0}\oto{\AItoK}\KR^{-q,0}_{\twistcover{G}}(\bTt^3)$, in order to finally obtain all anomalous surface states $\Surf^{-q,0}=\KR^{-q,0}_{\twistcover{G}}(\bTt^3)/\AI^{-q,0}$.

As discussed above, the complete classification results are presented in Table~\ref{tab:P-4_res}. In order to demonstrate the abelian calculations leading to it, let us focus on the $A$-point HSM in AZ symmetry class AII of $q=4$, where
\begin{equation}
\KR^{-4,0}_{\twistcover{G}_d}(\pt)\oto{\AItoK}\trKR^{-4,0}_{\twistcover{G}_A}(\bSs^{3}).
\end{equation}

For any particular AZ class, $\AItoK^{-q,0}$ is an abelian-group homomorphism. For our case, it takes the form,
\begin{equation}
\ZZ\times\ZZ\oto{\AItoK}\ZZ\times\ZZ_2.
\end{equation}
This is obtained using Tables \ref{tab:P-4_full_KR}, \ref{tab:P-4_full_KR_AI}, and Table~\ref{tab:KR-hom} in Appendix~\ref{sec:math_background}, by focusing on spinful-electrons; specifically,
\begin{align}
\KC^{-4,0}(\pt)_{t_8^{\vphantom{1}}+t_8^7}&\simeq\ZZ\cdot\xi^2, & \ZZ&\simeq\trKR^{-4,-1}(\bSs^3)_{1-t_8^4}, \nonumber\\
\KC^{-4,0}(\pt)_{t_8^5+t_8^3}&\simeq\ZZ\cdot\xi^2, & \ZZ_2\cdot\eta^2&\simeq\trKR^{-5,0}(\bSs^3)_{t_8^2-t_8^6},
\end{align}
cf.~Eq.~\eqref{eq:xi3-example}.
Using our explicit results in Eq.~\eqref{eq:ai-Ad}, and setting $u=\xi^2=\xi^{-2}$ (by $\xi^4=1$ as discussed above), we find that the $[\AItoK]_{Ad}$ matrix is given by
\begin{equation}
[\AItoK]_{\rho_A \rho_d}=
\begin{pmatrix}
\rr(\xi^2\cdot \xi^{-2}) & -\rr(\xi^2\cdot \xi^{-2}) \\
\rr(\xi^3\cdot \xi^{-2}) & -\rr(\xi^3\cdot \xi^{-2})
\end{pmatrix}
=
\begin{pmatrix}
2 & -2 \\
\eta^2 & -\eta^2
\end{pmatrix}.
\end{equation}
Here, we have evaluated the realification map as given in Table~\ref{tab:KCKR}.

In general, in order to find the quotient by the image of an integer-valued matrix, i.e., its cokernel, one brings it to its Smith normal form, where the elementary divisors are the diagonal entries. The $[\AItoK]_{\hsm\wyck}$ matrix is $\KR^{*,*}(\pt)$-valued; nevertheless, we may always manually impose $2\eta=2\eta^2=0$ and bring it to a pure integral formulation. In our case, this amounts to adding an extra $\left(\substack{0\\2\eta^2}\right)$-column. We find,
\begin{equation}
\begin{pmatrix}
2 & -2 \\
\eta^2 & -\eta^2
\end{pmatrix}
\mapsto
\begin{pmatrix}
2 & -2 & 0\\
1 & -1 & 2
\end{pmatrix}
\xmapsto[\text{Smith}]{}
\begin{pmatrix}
1 & 0 & 0\\
0 & 4 & 0
\end{pmatrix}.
\end{equation}
Finally, since $\ZZ_1\times\ZZ_4\simeq\ZZ_4$, we conclude that
\begin{equation}\label{eq:P-4_Z4}
[\Surf^{-4,0}]_{A}=\frac{\ZZ\times\ZZ_2}{\AItoK(\ZZ\times\ZZ)}\simeq\ZZ_4.
\end{equation}
In particular, we find that if we stack four copies of the elementary topologically distinct groundstate of the ${[1-t_8^4]_A}$ representation we get a state which is topologically equivalent to an AI. All other invariants in Table~\ref{tab:P-4_res} or obtained via analogous calculations.

\subsubsection{Summary}\label{sec:P-4_summary}
This completes our ``top to bottom" derivation of the full classification of  topologically distinct groundstates and topological phases with anomalous surface states for space-group $\mathrm{P\bar{4}}$. We have constructed the following data:
\begin{itemize}
\item[(i)] The identification of the HSM with $\pi$-integer primitive coordinates, their reciprocal Wyckoff positions, and their stabilizer double point-groups (little groups); see Table~\ref{tab:P-4_WyckHSM}.
\item[(ii)] The decomposition of the full $\KK$-theory classification into components corresponding to each HSM via our equivariant spectra paradigm; see Eq.~\eqref{eq:P-4_KTtoKS}.
\item[(iii)] The decomposition of the $\KK$-theory component corresponding to each HSM into $\ZZ_2$-graded representations; see Tables \ref{tab:P-4_full_KR} and \ref{tab:P-4_full_KC}. 

This yields the complete classification of topologically distinct groundstates; see Table~\ref{tab:P-4_res}.
\item[(iv)] The identification of all AIs at each Wyckoff position for all AZ classes via ungraded representations; see Tables \ref{tab:P-4_full_KR_AI} and \ref{tab:P-4_full_KC_AI}.
\item[(v)] The determination of the topologically distinct groundstate within the $\KK$-theory classification corresponding to the (spinless) trivial AI at each Wyckoff position; see Eq.~\eqref{eq:ai1-block}.
\item[(vi)] The conclusion of the full image of all (spinful) AIs at all Wyckoff positions within the $\KK$-theory groundstate classification for all AZ classes at once via the multiplicative structure; see Eqs.~\eqref{eq:ai-Ga}-\eqref{eq:ai-Ad}.
\item[(vii)] Computation of the resulting quotients; see, e.g., Eq.~\eqref{eq:P-4_Z4}.

This yields the complete classification of topological phases with anomalous surface states; see Table~\ref{tab:P-4_res}.
\end{itemize}
Such calculations were implemented as a GAP4~\cite{GAP4} language algorithm and yielded all the topological classification tables in Table~\ref{tab:mainres} and Appendix~\ref{app:tables}. A much deeper mathematical perspective is provided in Appendix~\ref{sec:math_background} but the gist of all crucial steps have been explicitly demonstrated.

Before concluding our paper in Sec.~\ref{sec:discussion}, we depart from the theoretical algebraic perspective taken so far, which have been extremely useful in obtaining quantitative results, and use our test-case of space-group $\mathrm{P\bar{4}}$ to provide the reader with tight-binding Hamiltonians manifesting our predicted anomalous surface states.

\subsection{Model Hamiltonians}\label{sec:hamilton} 

\begin{figure}[t]
\centering
\includegraphics[width=\linewidth]{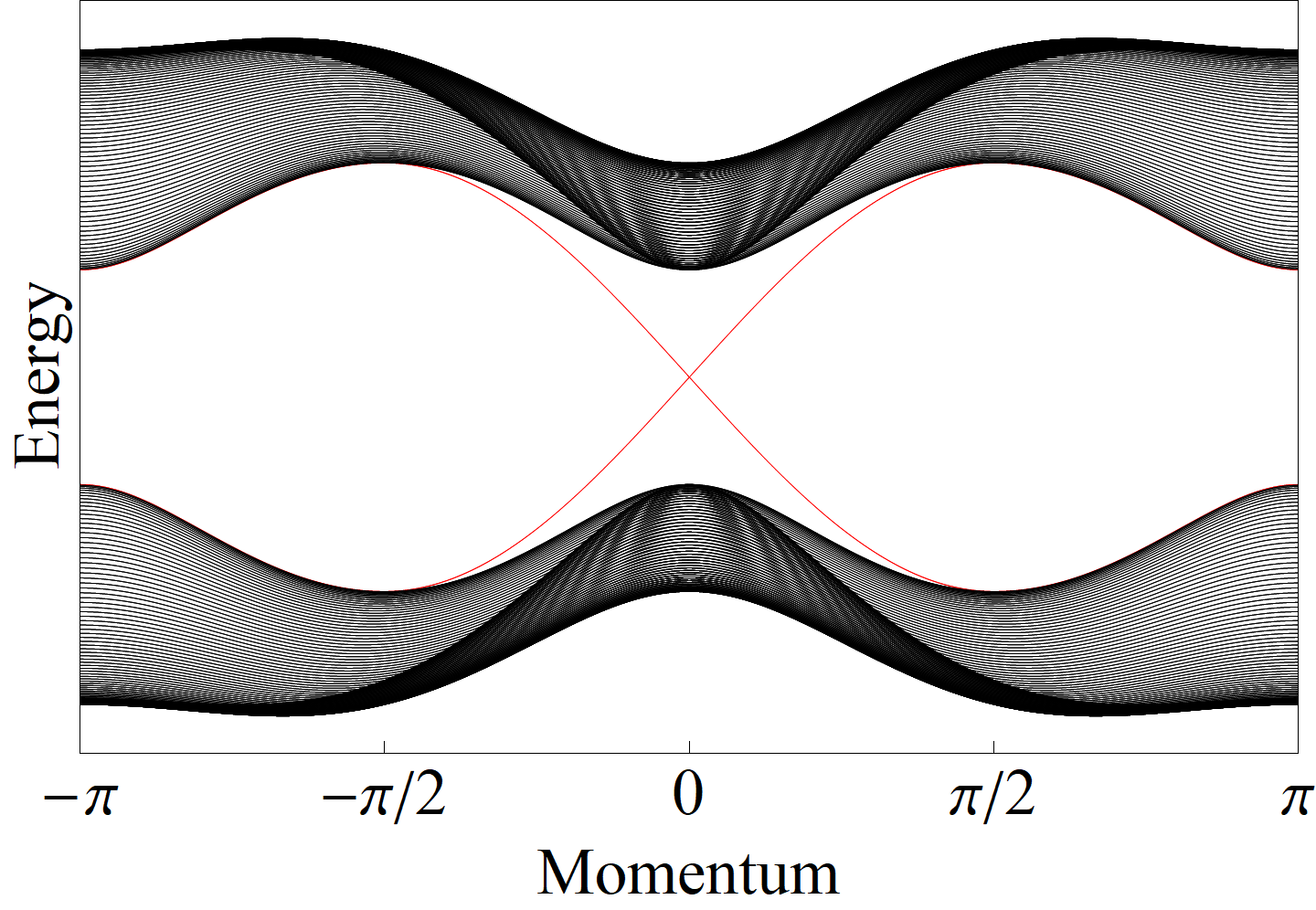}
\caption{The band structure of a model Hamiltonian [Eq.~\eqref{eq:Hamiltonian_P-4_DIII}] in AZ class DIII for the $\ZZ_2$ bulk topological phase at the $A$-point of space-group $\mathrm{P\bar{4}}$. We take 50 unit-cells with open boundary conditions along the $z$-axis and plot the energy levels $\epsilon_n(k_x)$ at $k_y=0$.}
\label{fig:P-4_bands}
\end{figure}

The constructive isomorphism in Eq.~\eqref{eq:CCiso} provides explicit model Hamiltonians for all classes of topologically distinct groundstates and hence for all topological phases with anomalous surface states; see Ref.~\onlinecite{Cornfeld2019Classification}.
Let us demonstrate this for the $\ZZ_2$ invariant of the $A$-point HSM at AZ symmetry class DIII; see Table~\ref{tab:P-4_res}. This is depicted in Fig.~\ref{fig:P-4_bands}.

AZ class DIII consists of time-reversal invariant chiral superconductors with broken spin rotation $\SUnit(2)$ symmetry (e.g., via spin-orbit coupling). We thus construct a Dirac-like tight-binding Bogoliubov de Gennes Hamiltonian
which respects the anti-unitary symmetries, $\TR^2=-1$ and $\PH^2=1$ (see Fig.~\ref{fig:bott}), as well as the $S_4$ spinful roto-reflection symmetry, $(\hat{s}_4)^4=-1$, i.e.,

\begin{align}
\TR H(\vk) \TR^{-1} &=H(-\vk), \nonumber\\
\PH H(\vk) \PH^{-1} &=-H(-\vk), \nonumber\\
\hat{s}_4^{\vphantom{1}}H(k_x,k_y,k_z)\hat{s}_4^{-1} &=H(k_y,-k_x,-k_z).
\end{align}
In momentum-space, it is given by
\begin{multline}\label{eq:Hamiltonian_P-4_DIII}
H(\vk)= \Delta\left(\gamma_x\sin k_x+\gamma_y\sin k_y\right)+\Delta'\gamma_z\sin k_z \\
+\gamma_0\left(\mu-t\left(\cos k_x+\cos k_y\right)-t'\cos k_z\right),
\end{multline}
where $\gamma_x,\gamma_y,\gamma_z,\gamma_0$ are Dirac matrices and $\Delta,\Delta',t,t',\mu$ are real parameters.

The isomorphism [Eq.~\eqref{eq:CCiso}] guarantees the existence of a matrix representation for $\hat{s}_4$ such that the eight operator combinations, $\gamma_x, \gamma_y, \gamma_z, i\gamma_0, \TR, i\TR, i\TR\PH, \hat{s}_4\gamma_z e^{\frac{\pi}{4}\gamma_x\gamma_y}$, are all mutually anti-commuting and square to $\pm1$. They thus form an ungraded presentation of the Clifford algebra, $\Cl_{4,4}\simeq\Mat{16}(\RR)$. The fundamental real representation of this algebra is 16-dimensional; it is isomorphic to the following 8-dimensional complex representation:
\begin{align}
\TR &= i\sigma_2\otimes\sigmaO\otimes\sigmaO\mathcal{K}, \nonumber\\
\PH &= \phantom{i}\sigmaO\otimes\sigma_1\otimes\sigmaO\mathcal{K}, \nonumber\\
\gamma_0 &= \phantom{i}\sigma_2\otimes\sigma_2\otimes\sigmaO, \nonumber\\
\gamma_x &= \phantom{i}\sigma_1\otimes\sigmaO\otimes\sigma_1, \nonumber\\
\gamma_y &= \phantom{i}\sigma_1\otimes\sigmaO\otimes\sigma_3, \nonumber\\
\gamma_z &= \phantom{i}\sigma_3\otimes\sigmaO\otimes\sigmaO, \nonumber\\
\hat{s}_4 &=i\sigma_2\otimes\sigmaO\otimes e^{\frac{\pi}{4}i\sigma_2}.
\end{align}
Here, $\mathcal{K}$ is the complex conjugation operator, $\sigma_{1,2,3}$ are the Pauli matrices, and $\sigmaO$ is the $2\times 2$ identity matrix.

With this choice of matrices, the Hamiltonian [Eq.~\eqref{eq:Hamiltonian_P-4_DIII}] exhibits topologically protected anomalous surface states and topological phase transitions for various values of $\mu$. These surface states are clearly visible in Fig.~\ref{fig:P-4_bands} where we aesthetically set $t'=t$, $\Delta'=\Delta=2t$, and $\mu=2t$. The Dirac cone is fourfold degenerate and is topologically protected by $\TR$, $\PH$, and $\hat{s}_4$.

As discussed in Sec.~\ref{sec:mult}, the topology of this chiral superconductor cannot be captured by the symmetry indicators, as there are no symmetry indicators for gapped systems in AZ class DIII. Moreover, it is straightforward to check that it has a zero winding number and hence also eludes the $\ZZ$ winding number invariant of 3D chiral TSCs in AZ class DIII; see Table~\ref{tab:per}. This is also evident without explicit calculation by noting that the only possible homomorphism from the $\ZZ_2$ classification of CTISC we found to the $\ZZ$ winding number of TISC, $\ZZ_2\to\ZZ$, is the zero homomorphism.

Finally, we emphasize that there is nothing unique about the choice of AZ class DIII, and the same procedure of constructing Dirac matrices and tight-binding Hamiltonians may be applied to all ten AZ symmetry classes.

\section{Discussion and outlook} \label{sec:discussion}
\subsection{Comparison with other works}\label{sec:comp}

\subsubsection{Topologically distinct groundstates}
We have obtained unified results for the classification of topologically distinct groundstates for the full tenfold-way of the two complex and eight real AZ non-spatial symmetry classes; see Fig.~\ref{fig:bott}.
Nevertheless, there have been numerous works that obtained results in specific AZ classes.

In Ref.~\onlinecite{Shiozaki2017Topological}, Shiozaki, Sato, and Gomi utilized long exact sequences of complex $\KK$-theory, e.g., Mayer-Vietoris and Gysin, to classify all wallpaper-groups in the two complex AZ classes, A and AIII. Our results are in complete agreement with their results; see Appendix~\ref{sec:layer-and-rod}.

In Ref.~\onlinecite{Kruthoff2017Topological}, Kruthoff et~al.~utilized band-structure combinatorics and Segal's formula~\cite{hirzebruch1990euler} for complex $\KK$-theory in order to classify all wallpaper-groups and some space-groups in AZ class A. Our results are in complete agreement with their results; see Appendix~\ref{app:tables}.

In Ref.~\onlinecite{shiozaki2018atiyah,shiozaki2018generalized}, Shiozaki et~al.~have explicitly calculated the Atiyah-Hirzebruch spectral sequence for all space-group symmetries for the two complex AZ classes as well as several other examples for the real AZ classes. In general, the $E_\infty$-page of the Atiyah-Hirzebruch spectral sequence approximates the $\KK$-theory.
Our $\KK$-theory results consistently fit their spectral sequence results.

\subsubsection{Topological phases with anomalous surface states}
One of our main results, see Table~\ref{tab:mainres}, is the unified classification of topological phases with anomalous surface states for the full tenfold-way of all AZ symmetry classes, A, AIII, AI, BDI, D, DIII, AII, CII, C, and CI; see Fig.~\ref{fig:bott}.

However, over the past couple of years, two independent works have presented a classification of anomalous surface states for AZ class AII using different methods: In Ref.~\onlinecite{Khalaf2018Symmetry}, Khalaf, Po, Vishwanath, and Watanabe have studied surface Hamiltonians and stacked (doubled) strong TIs. In Ref.~\onlinecite{Song2019Topological}, Song, Huang, Qi, Fang, and Hermele have studied topological crystals using group~cohomology. Our results are in complete agreement with the results of both research groups.

Moreover, in Ref.~\onlinecite{shiozaki2019classification}, Shiozaki studied the surface states of CTISC with magnetic point-group symmetry; see Sec.~\ref{sec:mag}. These correspond to the contribution of the volume cell (cf.~the $R$-point cell in Fig.~\ref{fig:CWP1}), analogous to strong TISC in the non-crystalline case. These results for magnetic point-group symmetries may be compared with our results for AZ classes, A, AI, BDI, D, DIII, and AII, and are in complete agreement.

\subsubsection{Symmetry indicators}

The majority of SI studies have naturally focused on Dyson's threefold-way of AZ classes A, AI, and AII~\cite{Dyson1962Threefold,zirnbauer2010symmetry}; see, e.g., Refs.~\onlinecite{bradlyn2017topological,po2017symmetry}. Nevertheless, an extension to the other AZ classes has been successfully achieved; see, e.g., Refs.~\onlinecite{Ono2018Unified,Geier2020Symmetry,Ono2020Refined,ono2020Z2}.

However, as discussed in Sec.~\ref{sec:over}, the anomalous surface states do not necessarily have to be indicated by the SI; see Fig.~\ref{fig:main-diag}. Moreover, it is possible for SI to indicate a \emph{gapless} state such as a (semi-)metal. Nonetheless, we find that whenever the SI indicate \emph{gapped} states (i.e., CTISC) we find them to be quotients of our $\KK$-theory classification as required by Eq.~\eqref{eq:main-diag}.

\subsection{Extensions}\label{sec:ext}
\subsubsection{strongly-interacting fermions \& more spectra}\label{sec:cobord}

Many of the theoretical results derived here [e.g., Eq.~\eqref{eq:torus-G-decomposition}] are not limited in their applicability only to $\KR$-spectra but, in fact, may be either directly applied or easily generalized to many spectra corresponding to other classification problems beyond weakly-interacting fermions~\cite{kitaev2011toward,kitaev2013topological,kitaev2015homotopy,freed2014short,freed2016reflection,kapustin2014symmetry,Kapustin2015Topological,Chen2013Symmetry,Gu2014Symmetry,shiozaki2018generalized,Xiong2018Minimalist}.

In particular, a direct non-trivial extension of this work may be in the study of invertible symmetry protected topological phases of \emph{strongly-interacting} fermions which are classified by real cobordisms~\cite{freed2016reflection}.

There is an intriguing sense in which one may gradually interpolate between $\KK$-theory and real cobordisms, by mean of the chromatic filtration~\cite{barthel2019chromatic}. Accordingly, it is natural to hypothesize that the  topological phases of fermions with higher-order interactions (e.g., quartic interactions), are classified by cohomology theories of higher chromatic heights. We unfortunately as yet have no evidence for such a relation, other than the two extreme cases.

\subsubsection{Magnetic space-groups \& topological superconductivity}\label{sec:mag}

In this paper we have limited our scope to symmetry actions where the spatial symmetries all commute with the anti-unitary non-spatial symmetries, $\TR$ and $\PH$, see Fig.~\ref{fig:bott}. However, although more complicated, our analysis is generalizable to treat cases where these symmetries intertwine~\footnote{See, e.g., Table~X in Ref.~\onlinecite{Cornfeld2019Classification}.}.

One manifestation is when considering the intertwining of time-reversal symmetry, $\TR$, and spatial symmetries~\cite{lifshitz2005magnetic}; these are the so-called ``magnetic space-groups", see Refs.~\onlinecite{Zhang2015Topological,Watanabe2018Structure,okuma2018topological,shiozaki2019classification,ono2020Z2} for related classification works.

Recently, in Refs.~\onlinecite{Geier2020Symmetry,Ono2020Refined,ono2020Z2}, Geier et~al.~and Ono et~al.~have studied the SI of superconducting systems where they had also treated the possibility of intertwining the particle-hole anti-symmetry, $\PH$, and the spatial symmetries. Since $\TR$ and $\PH$ are both manifestations of $\ZT$ with different actions on the underlying $\KU$ spectrum, both intertwinings should be akinly treated in our paradigm; see discussion in Sec.~\ref{sec:intro_sym}.

\subsubsection{Higher order topological insulators and superconductors}
One of the most intriguing properties of CTISC phases is the existence of higher-order surface-states which are supported on the lower-dimensional edges or corners of a crystalline material~\cite{parameswaran2017topological,Benalcazar2017Quantized,Benalcazar2017Electric,Song2017d,Langbehn2017Reflection,Schindler2018Higher,schindler2018bismuth,xu2017topological,Shapourian2018Topological,lin2017topological,Ezawa2018Higher,Khalaf2018Higher,Geier2018Second,trifunovic2018higher,fang2017rotation,okuma2018topological}.

In Refs.~\onlinecite{trifunovic2018higher,Geier2018Second}, Trifunovic, Brouwer, and Geier have presented a theoretical formulation for the classification of such HOTISC using a filtration of the $\KK$-theory. In Ref.~\onlinecite{shiozaki2018generalized}, Shiozaki, Xiong, and Gomi have shown that this filtration naturally fits within the Atiyah-Hirzebruch spectral sequence. We expect that the knowledge of the full $\KK$-theory classification provided in this paper would be used to resolve the sequence and obtain a complete classification of HOTISC phenomena.

\subsubsection{Hexagonal space-groups}

While the general machinery used in this paper is relevant also for the hexagonal space-groups, one key ingredient breaks down. Namely, the free $G$-equivariant spectrum of a torus with a hexagonal action does not decompose into a direct sum according to its cells, and so the $G$-equivariant $\KK$-theory does not reduce to equivariant spheres. Nevertheless, there are still advantages in the equivariant spectra paradigm. 

If $\bTt^d$ is a torus with an action of a group $G$, one can always equivariantly construct $\bTt^d$ out of cells~\cite{shiozaki2018atiyah,Song2019Topological}; cf.~Fig.~\ref{fig:CWP1}. For example, consider layer-group $\mathrm{p6mm}$, where $C_{6v}$ acts on the 2D torus, $\bTt^2$. This resulting $G$-equivariant torus may be decomposed into two equilateral triangles, permuted transitively by the $C_{6v}$-action. While the $G$-equivariant attaching data of the triangles to their boundary no longer stably trivializes, the part of the data determining the $G$-equivariant $\KK$-theory of the torus is computable and depends only on a small number of parameters. Hence, we expect that our methods will prove advantageous in the classification of topological phases for hexagonal symmetry as well. 

\subsection{Summary}\label{sec:summ}

In this paper, we have utilized the mathematical equivariant spectra paradigm to obtain explicit quantitative results for the classification of topologically distinct groundstates as well as topological phases with anomalous surface states of crystalline topological insulators and superconductors.
This is done in a unified manner, which captures the full tenfold-way of Altland-Zirnbauer non-spatial symmetry classes; see Fig.~\ref{fig:bott}.
In Table~\ref{tab:mainres}, we have focused on the full classification of key 3D space-groups, but our analysis naturally captures 2D layer-groups and 1D rod-groups as well; see Appendix~\ref{app:tables}. These full classification results are \emph{exhaustive} and thus extend beyond the symmetry indicators of band topology.

We have established both the theoretical and the computational benefits of the equivariant spectra paradigm:

First, we have successfully utilized the modern mathematical framework of equivariant spectra to obtain a geometric equivariant spectra formulation of crystalline systems; see Eq.~\eqref{eq:torus-G-decomposition}. These infrastructural spectra are independent of the specific classification of topological phases we set out to complete.

Next, we emphasized the $\KK$-theory multiplication which physically translates to the hidden multiplicative structure within the periodic table of topological insulators and superconductors, mixing the different AZ symmetry classes. This fundamental multiplicative structure is independent of any crystalline symmetry. Nevertheless, this generalizes to the crystalline case and unveils further relations between different CTISC phases; see Sec.~\ref{sec:mult}.

Furthermore, by treating the atomic insulators and superconductors within the same equivariant spectra paradigm, we provide a deeper understanding of the AIs' classification and thus of the anomalous surface states. This understanding translates to an efficient computational approach [see Sec.~\ref{sec:P-4_summary}] which allowed us to attain results for the full tenfold-way of AZ symmetry classes in a unified manner.

To conclude, as discussed above, we showed that our results consistently broaden the existing knowledge of CTISC phases and that the paradigm we have established holds the potential to lead the path towards the discovery and the understanding of other diverse topological phenomena.

\section*{Acknowledgments}
We are grateful for illuminating discussions with \'{A}.~Nagy, N.~Okuma, K.~Shiozaki, R.~Thorngren, E.~Berg, D.~Clausen, and T.~Schlank. E.~C.~acknowledges support from the Deutsche Forschungsgemeinschaft (DFG, German Research Foundation) – project grant 277101999 – within the CRC network TRR 183. S.~C.~was supported by the Adams Fellowship Program of the Israel Academy of Sciences and Humanities.

\appendix

\section{Classification tables}\label{app:tables}
\begin{table}[t]
\centering
\caption{The layout of Table~\ref{tab:app}.}
\begin{tabular}{cccc}
\hline\hline
Space-group & $\cdots$ & HSM & $\cdots$ \\
\hline
$\vdots$ \\
AZ class & & $\KK\to\Surf$ &\\
$\vdots$\\
\hline\hline
\end{tabular}
\label{tab:layout}
\end{table}

In this appendix, we present the full classification tables which provide detailed information that complements the main results presented in Table~\ref{tab:mainres}.

In Table~\ref{tab:app} (ahead) we present the complete K-theory classification of topologically distinct groundstates for the signed-permutation representation space-groups as well as the complete classification of topological phases with anomalous surface states, $\Surf=\KK/\AI$. Each entry of the table is laid-out as in Table~\ref{tab:layout} which provides both classifications for each AZ symmetry class and each of the high-symmetry momenta at the center of the equivariant cell component of the BZ torus; see Fig.~\ref{fig:CWP1}. The complete classification for each AZ class is a direct sum of the abelian groups for all tabulated HSM. We use the naming convention of Ref.~\onlinecite{bradley2010mathematical} for the HSM in the BZ.

The abelian groups presented in Table~\ref{tab:app} may be easily used to obtain the full classification of many layer-groups and rod-groups as well. Since any layer-group/rod-group is a sub-group of some space-group, it is straightforward to extract their topological classification. This is discussed in Sec.~\ref{sec:layer-and-rod}.

\newcommand{\captionAppendixTables}{For each space-group in Table~\ref{tab:mainres}, we provide the complete classification of topologically distinct groundstates as well as the complete classification of topological phases with anomalous surface states. Each entry of the table is of the form $\KK\to\Surf$; see discussion in Appendix~\ref{app:tables}. The ``~$\cdot$~" symbol indicates a trivial classification, i.e., $0\to0$.}

\renewcommand{\arraystretch}{0.89}
\begin{table}[H]
\centering
\caption{\captionAppendixTables}
\label{tab:app}
{\scriptsize
\begin{tabularx}{\linewidth}{XXXXXXXXX}
\hline\hline
$\mathrm{P1}\vphantom{\dot{|}}$ & $\Gamma$ & $Z$ & $Y$ & $X$ & $T$ & $U$ & $S$ & $R$ \\
\hline
A & $\ZZ\!\!\shortrightarrow\!\!0$ & $\cdot$ & $\cdot$ & $\cdot$ & $\ZZ\!\!\shortrightarrow\!\!\ZZ$ & $\ZZ\!\!\shortrightarrow\!\!\ZZ$ & $\ZZ\!\!\shortrightarrow\!\!\ZZ$ & $\cdot$ \\
AIII & $\cdot$ & $\ZZ\!\!\shortrightarrow\!\!\ZZ$ & $\ZZ\!\!\shortrightarrow\!\!\ZZ$ & $\ZZ\!\!\shortrightarrow\!\!\ZZ$ & $\cdot$ & $\cdot$ & $\cdot$ & $\ZZ\!\!\shortrightarrow\!\!\ZZ$ \\
\hline
AI & $\ZZ\!\!\shortrightarrow\!\!0$ & $\cdot$ & $\cdot$ & $\cdot$ & $\cdot$ & $\cdot$ & $\cdot$ & $\cdot$ \\
BDI & $\ZZ_{2}\!\!\shortrightarrow\!\!0$ & $\ZZ\!\!\shortrightarrow\!\!\ZZ$ & $\ZZ\!\!\shortrightarrow\!\!\ZZ$ & $\ZZ\!\!\shortrightarrow\!\!\ZZ$ & $\cdot$ & $\cdot$ & $\cdot$ & $\cdot$ \\
D & $\ZZ_{2}\!\!\shortrightarrow\!\!0$ & $\ZZ_{2}\!\!\shortrightarrow\!\!\ZZ_{2}$ & $\ZZ_{2}\!\!\shortrightarrow\!\!\ZZ_{2}$ & $\ZZ_{2}\!\!\shortrightarrow\!\!\ZZ_{2}$ & $\ZZ\!\!\shortrightarrow\!\!\ZZ$ & $\ZZ\!\!\shortrightarrow\!\!\ZZ$ & $\ZZ\!\!\shortrightarrow\!\!\ZZ$ & $\cdot$ \\
DIII & $\cdot$ & $\ZZ_{2}\!\!\shortrightarrow\!\!\ZZ_{2}$ & $\ZZ_{2}\!\!\shortrightarrow\!\!\ZZ_{2}$ & $\ZZ_{2}\!\!\shortrightarrow\!\!\ZZ_{2}$ & $\ZZ_{2}\!\!\shortrightarrow\!\!\ZZ_{2}$ & $\ZZ_{2}\!\!\shortrightarrow\!\!\ZZ_{2}$ & $\ZZ_{2}\!\!\shortrightarrow\!\!\ZZ_{2}$ & $\ZZ\!\!\shortrightarrow\!\!\ZZ$ \\
AII & $\ZZ\!\!\shortrightarrow\!\!0$ & $\cdot$ & $\cdot$ & $\cdot$ & $\ZZ_{2}\!\!\shortrightarrow\!\!\ZZ_{2}$ & $\ZZ_{2}\!\!\shortrightarrow\!\!\ZZ_{2}$ & $\ZZ_{2}\!\!\shortrightarrow\!\!\ZZ_{2}$ & $\ZZ_{2}\!\!\shortrightarrow\!\!\ZZ_{2}$ \\
\hline\hline
\end{tabularx}}
\end{table}

\addtocounter{table}{-1}

\begin{table}[H]
\centering
\caption{\textit{(Continued)}}
{\scriptsize\begin{tabularx}{\linewidth}{XXXXXXXXX}
\hline\hline
$\mathrm{P\bar{1}}\vphantom{\dot{|}}$ & $\Gamma$ & $Z$ & $Y$ & $X$ & $T$ & $U$ & $S$ & $R$ \\
\hline
A & $\ZZ^{2}\!\!\shortrightarrow\!\!0$ & $\ZZ\!\!\shortrightarrow\!\!0$ & $\ZZ\!\!\shortrightarrow\!\!0$ & $\ZZ\!\!\shortrightarrow\!\!0$ & $\ZZ^{2}\!\!\shortrightarrow\!\!\ZZ$ & $\ZZ^{2}\!\!\shortrightarrow\!\!\ZZ$ & $\ZZ^{2}\!\!\shortrightarrow\!\!\ZZ$ & $\ZZ\!\!\shortrightarrow\!\!\ZZ_{2}$ \\
AIII & $\cdot$ & $\cdot$ & $\cdot$ & $\cdot$ & $\cdot$ & $\cdot$ & $\cdot$ & $\cdot$ \\
\hline
AI & $\ZZ^{2}\!\!\shortrightarrow\!\!0$ & $\ZZ\!\!\shortrightarrow\!\!0$ & $\ZZ\!\!\shortrightarrow\!\!0$ & $\ZZ\!\!\shortrightarrow\!\!0$ & $\ZZ\!\!\shortrightarrow\!\!0$ & $\ZZ\!\!\shortrightarrow\!\!0$ & $\ZZ\!\!\shortrightarrow\!\!0$ & $\ZZ\!\!\shortrightarrow\!\!0$ \\
BDI & $\ZZ_{2}^{2}\!\!\shortrightarrow\!\!0$ & $\ZZ_{2}\!\!\shortrightarrow\!\!0$ & $\ZZ_{2}\!\!\shortrightarrow\!\!0$ & $\ZZ_{2}\!\!\shortrightarrow\!\!0$ & $\cdot$ & $\cdot$ & $\cdot$ & $\cdot$ \\
D & $\ZZ_{2}^{2}\!\!\shortrightarrow\!\!0$ & $\ZZ_{2}\!\!\shortrightarrow\!\!0$ & $\ZZ_{2}\!\!\shortrightarrow\!\!0$ & $\ZZ_{2}\!\!\shortrightarrow\!\!0$ & $\ZZ\!\!\shortrightarrow\!\!\ZZ$ & $\ZZ\!\!\shortrightarrow\!\!\ZZ$ & $\ZZ\!\!\shortrightarrow\!\!\ZZ$ & $\cdot$ \\
DIII & $\cdot$ & $\cdot$ & $\cdot$ & $\cdot$ & $\cdot$ & $\cdot$ & $\cdot$ & $\cdot$ \\
AII & $\ZZ^{2}\!\!\shortrightarrow\!\!0$ & $\ZZ\!\!\shortrightarrow\!\!0$ & $\ZZ\!\!\shortrightarrow\!\!0$ & $\ZZ\!\!\shortrightarrow\!\!0$ & $\ZZ\!\!\shortrightarrow\!\!\ZZ_{2}$ & $\ZZ\!\!\shortrightarrow\!\!\ZZ_{2}$ & $\ZZ\!\!\shortrightarrow\!\!\ZZ_{2}$ & $\ZZ\!\!\shortrightarrow\!\!\ZZ_{4}$ \\
\hline\hline
\end{tabularx}}
\end{table}

\addtocounter{table}{-1}

\begin{table}[H]
\centering
\caption{\textit{(Continued)}}
{\scriptsize\begin{tabularx}{\linewidth}{Xllllllll}
\hline\hline
$\mathrm{P2}\vphantom{\dot{|}}$ & $\Gamma$ & $B$ & $Z$ & $Y$ & $D$ & $A$ & $C$ & $E$ \\
\hline
A & $\ZZ^{2}\!\!\shortrightarrow\!\!0$ & $\ZZ\!\!\shortrightarrow\!\!0$ & $\cdot$ & $\ZZ\!\!\shortrightarrow\!\!0$ & $\cdot$ & $\ZZ^{2}\!\!\shortrightarrow\!\!\ZZ$ & $\cdot$ & $\cdot$ \\
AIII & $\cdot$ & $\cdot$ & $\ZZ^{2}\!\!\shortrightarrow\!\!\ZZ^{2}$ & $\cdot$ & $\ZZ\!\!\shortrightarrow\!\!\ZZ$ & $\cdot$ & $\ZZ\!\!\shortrightarrow\!\!\ZZ$ & $\ZZ^{2}\!\!\shortrightarrow\!\!\ZZ^{2}$ \\
\hline
AI & $\ZZ\!\!\shortrightarrow\!\!0$ & $\cdot$ & $\cdot$ & $\cdot$ & $\cdot$ & $\cdot$ & $\cdot$ & $\cdot$ \\
BDI & $\cdot$ & $\cdot$ & $\ZZ\!\!\shortrightarrow\!\!\ZZ$ & $\cdot$ & $\cdot$ & $\cdot$ & $\cdot$ & $\cdot$ \\
D & $\ZZ\!\!\shortrightarrow\!\!0$ & $\ZZ\!\!\shortrightarrow\!\!\ZZ_{2}$ & $\cdot$ & $\ZZ\!\!\shortrightarrow\!\!\ZZ_{2}$ & $\cdot$ & $\ZZ^{2}\!\!\shortrightarrow\!\!\ZZ\!\times\!\ZZ_{2}$ & $\cdot$ & $\cdot$ \\
DIII & $\cdot$ & $\ZZ_{2}\!\!\shortrightarrow\!\!\ZZ_{2}$ & $\ZZ\!\!\shortrightarrow\!\!\ZZ$ & $\ZZ_{2}\!\!\shortrightarrow\!\!\ZZ_{2}$ & $\ZZ\!\!\shortrightarrow\!\!\ZZ$ & $\ZZ_{2}^{2}\!\!\shortrightarrow\!\!\ZZ_{2}^{2}$ & $\ZZ\!\!\shortrightarrow\!\!\ZZ$ & $\ZZ^{2}\!\!\shortrightarrow\!\!\ZZ^{2}$ \\
AII & $\ZZ\!\!\shortrightarrow\!\!0$ & $\ZZ_{2}\!\!\shortrightarrow\!\!0$ & $\cdot$ & $\ZZ_{2}\!\!\shortrightarrow\!\!0$ & $\ZZ_{2}\!\!\shortrightarrow\!\!\ZZ_{2}$ & $\ZZ_{2}^{2}\!\!\shortrightarrow\!\!\ZZ_{2}$ & $\ZZ_{2}\!\!\shortrightarrow\!\!\ZZ_{2}$ & $\ZZ_{2}^{2}\!\!\shortrightarrow\!\!\ZZ_{2}^{2}$ \\
\hline\hline
\end{tabularx}}
\end{table}

\addtocounter{table}{-1}

\begin{table}[H]
\centering
\caption{\textit{(Continued)}}
{\scriptsize\begin{tabularx}{\linewidth}{XXXXXXX}
\hline\hline
$\mathrm{C2}\vphantom{\dot{|}}$ & $\Gamma$ & $A$ & $V$ & $L$ & $Z$ & $M$ \\
\hline
A & $\ZZ^{2}\!\!\shortrightarrow\!\!0$ & $\ZZ\!\!\shortrightarrow\!\!0$ & $\cdot$ & $\ZZ\!\!\shortrightarrow\!\!\ZZ$ & $\cdot$ & $\cdot$ \\
AIII & $\cdot$ & $\cdot$ & $\ZZ\!\!\shortrightarrow\!\!\ZZ$ & $\cdot$ & $\ZZ\!\!\shortrightarrow\!\!\ZZ$ & $\ZZ^{2}\!\!\shortrightarrow\!\!\ZZ^{2}$ \\
\hline
AI & $\ZZ\!\!\shortrightarrow\!\!0$ & $\cdot$ & $\cdot$ & $\cdot$ & $\cdot$ & $\cdot$ \\
BDI & $\cdot$ & $\cdot$ & $\ZZ\!\!\shortrightarrow\!\!\ZZ$ & $\cdot$ & $\cdot$ & $\cdot$ \\
D & $\ZZ\!\!\shortrightarrow\!\!0$ & $\ZZ\!\!\shortrightarrow\!\!\ZZ_{2}$ & $\ZZ_{2}\!\!\shortrightarrow\!\!\ZZ_{2}$ & $\ZZ\!\!\shortrightarrow\!\!\ZZ$ & $\cdot$ & $\cdot$ \\
DIII & $\cdot$ & $\ZZ_{2}\!\!\shortrightarrow\!\!\ZZ_{2}$ & $\ZZ_{2}\!\!\shortrightarrow\!\!\ZZ_{2}$ & $\ZZ_{2}\!\!\shortrightarrow\!\!\ZZ_{2}$ & $\ZZ\!\!\shortrightarrow\!\!\ZZ$ & $\ZZ^{2}\!\!\shortrightarrow\!\!\ZZ^{2}$ \\
AII & $\ZZ\!\!\shortrightarrow\!\!0$ & $\ZZ_{2}\!\!\shortrightarrow\!\!0$ & $\cdot$ & $\ZZ_{2}\!\!\shortrightarrow\!\!\ZZ_{2}$ & $\ZZ_{2}\!\!\shortrightarrow\!\!\ZZ_{2}$ & $\ZZ_{2}^{2}\!\!\shortrightarrow\!\!\ZZ_{2}^{2}$ \\
CII & $\cdot$ & $\cdot$ & $\ZZ\!\!\shortrightarrow\!\!\ZZ$ & $\cdot$ & $\ZZ_{2}\!\!\shortrightarrow\!\!\ZZ_{2}$ & $\ZZ_{2}^{2}\!\!\shortrightarrow\!\!\ZZ_{2}^{2}$ \\
C & $\ZZ\!\!\shortrightarrow\!\!0$ & $\ZZ\!\!\shortrightarrow\!\!0$ & $\cdot$ & $\ZZ\!\!\shortrightarrow\!\!\ZZ$ & $\cdot$ & $\cdot$ \\
CI & $\cdot$ & $\cdot$ & $\cdot$ & $\cdot$ & $\ZZ\!\!\shortrightarrow\!\!\ZZ$ & $\ZZ^{2}\!\!\shortrightarrow\!\!\ZZ^{2}$ \\
\hline\hline
\end{tabularx}}
\end{table}

\addtocounter{table}{-1}

\begin{table}[H]
\centering
\caption{\textit{(Continued)}}
{\scriptsize\begin{tabularx}{\linewidth}{XXXXXXXXX}
\hline\hline
$\mathrm{Pm}\vphantom{\dot{|}}$ & $\Gamma$ & $B$ & $Z$ & $Y$ & $D$ & $A$ & $C$ & $E$ \\
\hline
A & $\ZZ^{2}\!\!\shortrightarrow\!\!0$ & $\cdot$ & $\ZZ\!\!\shortrightarrow\!\!0$ & $\cdot$ & $\cdot$ & $\ZZ^{2}\!\!\shortrightarrow\!\!\ZZ^{2}$ & $\cdot$ & $\ZZ\!\!\shortrightarrow\!\!\ZZ$ \\
AIII & $\cdot$ & $\ZZ^{2}\!\!\shortrightarrow\!\!\ZZ^{2}$ & $\cdot$ & $\ZZ^{2}\!\!\shortrightarrow\!\!\ZZ^{2}$ & $\ZZ\!\!\shortrightarrow\!\!\ZZ$ & $\cdot$ & $\ZZ\!\!\shortrightarrow\!\!\ZZ$ & $\cdot$ \\
\hline
AI & $\ZZ\!\!\shortrightarrow\!\!0$ & $\cdot$ & $\cdot$ & $\cdot$ & $\cdot$ & $\ZZ\!\!\shortrightarrow\!\!\ZZ$ & $\cdot$ & $\ZZ\!\!\shortrightarrow\!\!\ZZ$ \\
BDI & $\cdot$ & $\ZZ\!\!\shortrightarrow\!\!\ZZ$ & $\cdot$ & $\ZZ\!\!\shortrightarrow\!\!\ZZ$ & $\cdot$ & $\cdot$ & $\cdot$ & $\cdot$ \\
D & $\ZZ\!\!\shortrightarrow\!\!0$ & $\cdot$ & $\ZZ\!\!\shortrightarrow\!\!\ZZ_{2}$ & $\cdot$ & $\cdot$ & $\ZZ\!\!\shortrightarrow\!\!\ZZ$ & $\cdot$ & $\cdot$ \\
DIII & $\cdot$ & $\ZZ\!\!\shortrightarrow\!\!\ZZ$ & $\ZZ_{2}\!\!\shortrightarrow\!\!\ZZ_{2}$ & $\ZZ\!\!\shortrightarrow\!\!\ZZ$ & $\ZZ\!\!\shortrightarrow\!\!\ZZ$ & $\cdot$ & $\ZZ\!\!\shortrightarrow\!\!\ZZ$ & $\cdot$ \\
AII & $\ZZ\!\!\shortrightarrow\!\!0$ & $\cdot$ & $\ZZ_{2}\!\!\shortrightarrow\!\!0$ & $\cdot$ & $\ZZ_{2}\!\!\shortrightarrow\!\!\ZZ_{2}$ & $\ZZ\!\!\shortrightarrow\!\!\ZZ$ & $\ZZ_{2}\!\!\shortrightarrow\!\!\ZZ_{2}$ & $\ZZ\!\!\shortrightarrow\!\!\ZZ$ \\
\hline\hline
\end{tabularx}}
\end{table}

\addtocounter{table}{-1}

\begin{table}[H]
\centering
\caption{\textit{(Continued)}}
{\scriptsize\begin{tabularx}{\linewidth}{XXXXXXX}
\hline\hline
$\mathrm{Cm}\vphantom{\dot{|}}$ & $\Gamma$ & $A$ & $V$ & $L$ & $Z$ & $M$ \\
\hline
A & $\ZZ^{2}\!\!\shortrightarrow\!\!0$ & $\cdot$ & $\cdot$ & $\ZZ\!\!\shortrightarrow\!\!\ZZ$ & $\cdot$ & $\ZZ\!\!\shortrightarrow\!\!\ZZ$ \\
AIII & $\cdot$ & $\ZZ^{2}\!\!\shortrightarrow\!\!\ZZ^{2}$ & $\ZZ\!\!\shortrightarrow\!\!\ZZ$ & $\cdot$ & $\ZZ\!\!\shortrightarrow\!\!\ZZ$ & $\cdot$ \\
\hline
AI & $\ZZ\!\!\shortrightarrow\!\!0$ & $\cdot$ & $\cdot$ & $\cdot$ & $\cdot$ & $\ZZ\!\!\shortrightarrow\!\!\ZZ$ \\
BDI & $\cdot$ & $\ZZ\!\!\shortrightarrow\!\!\ZZ$ & $\ZZ\!\!\shortrightarrow\!\!\ZZ$ & $\cdot$ & $\cdot$ & $\cdot$ \\
D & $\ZZ\!\!\shortrightarrow\!\!0$ & $\cdot$ & $\ZZ_{2}\!\!\shortrightarrow\!\!\ZZ_{2}$ & $\ZZ\!\!\shortrightarrow\!\!\ZZ$ & $\cdot$ & $\cdot$ \\
DIII & $\cdot$ & $\ZZ\!\!\shortrightarrow\!\!\ZZ$ & $\ZZ_{2}\!\!\shortrightarrow\!\!\ZZ_{2}$ & $\ZZ_{2}\!\!\shortrightarrow\!\!\ZZ_{2}$ & $\ZZ\!\!\shortrightarrow\!\!\ZZ$ & $\cdot$ \\
AII & $\ZZ\!\!\shortrightarrow\!\!0$ & $\cdot$ & $\cdot$ & $\ZZ_{2}\!\!\shortrightarrow\!\!\ZZ_{2}$ & $\ZZ_{2}\!\!\shortrightarrow\!\!\ZZ_{2}$ & $\ZZ\!\!\shortrightarrow\!\!\ZZ$ \\
CII & $\cdot$ & $\ZZ\!\!\shortrightarrow\!\!\ZZ$ & $\ZZ\!\!\shortrightarrow\!\!\ZZ$ & $\cdot$ & $\ZZ_{2}\!\!\shortrightarrow\!\!\ZZ_{2}$ & $\ZZ_{2}\!\!\shortrightarrow\!\!\ZZ_{2}$ \\
C & $\ZZ\!\!\shortrightarrow\!\!0$ & $\cdot$ & $\cdot$ & $\ZZ\!\!\shortrightarrow\!\!\ZZ$ & $\cdot$ & $\ZZ_{2}\!\!\shortrightarrow\!\!\ZZ_{2}$ \\
CI & $\cdot$ & $\ZZ\!\!\shortrightarrow\!\!\ZZ$ & $\cdot$ & $\cdot$ & $\ZZ\!\!\shortrightarrow\!\!\ZZ$ & $\cdot$ \\
\hline\hline
\end{tabularx}}
\end{table}

\addtocounter{table}{-1}

\begin{table}[H]
\centering
\caption{\textit{(Continued)}}
{\scriptsize\begin{tabularx}{\linewidth}{Xllllllll}
\hline\hline
$\mathrm{P2/m}\vphantom{\dot{|}}$ & $\Gamma$ & $B$ & $Z$ & $Y$ & $D$ & $A$ & $C$ & $E$ \\
\hline
A & $\ZZ^{4}\!\!\shortrightarrow\!\!0$ & $\ZZ^{2}\!\!\shortrightarrow\!\!0$ & $\ZZ^{2}\!\!\shortrightarrow\!\!0$ & $\ZZ^{2}\!\!\shortrightarrow\!\!0$ & $\ZZ\!\!\shortrightarrow\!\!0$ & $\ZZ^{4}\!\!\shortrightarrow\!\!\ZZ^{2}$ & $\ZZ\!\!\shortrightarrow\!\!0$ & $\ZZ^{2}\!\!\shortrightarrow\!\!\ZZ$ \\
AIII & $\cdot$ & $\cdot$ & $\cdot$ & $\cdot$ & $\cdot$ & $\cdot$ & $\cdot$ & $\cdot$ \\
\hline
AI & $\ZZ^{2}\!\!\shortrightarrow\!\!0$ & $\ZZ\!\!\shortrightarrow\!\!0$ & $\ZZ\!\!\shortrightarrow\!\!0$ & $\ZZ\!\!\shortrightarrow\!\!0$ & $\ZZ\!\!\shortrightarrow\!\!0$ & $\ZZ^{2}\!\!\shortrightarrow\!\!\ZZ$ & $\ZZ\!\!\shortrightarrow\!\!0$ & $\ZZ^{2}\!\!\shortrightarrow\!\!\ZZ$ \\
BDI & $\cdot$ & $\cdot$ & $\cdot$ & $\cdot$ & $\cdot$ & $\cdot$ & $\cdot$ & $\cdot$ \\
D & $\ZZ^{2}\!\!\shortrightarrow\!\!0$ & $\ZZ\!\!\shortrightarrow\!\!0$ & $\ZZ\!\!\shortrightarrow\!\!0$ & $\ZZ\!\!\shortrightarrow\!\!0$ & $\cdot$ & $\ZZ^{2}\!\!\shortrightarrow\!\!\ZZ$ & $\cdot$ & $\cdot$ \\
DIII & $\cdot$ & $\cdot$ & $\cdot$ & $\cdot$ & $\cdot$ & $\cdot$ & $\cdot$ & $\cdot$ \\
AII & $\ZZ^{2}\!\!\shortrightarrow\!\!0$ & $\ZZ\!\!\shortrightarrow\!\!0$ & $\ZZ\!\!\shortrightarrow\!\!0$ & $\ZZ\!\!\shortrightarrow\!\!0$ & $\ZZ\!\!\shortrightarrow\!\!\ZZ_{2}$ & $\ZZ^{2}\!\!\shortrightarrow\!\!\ZZ$ & $\ZZ\!\!\shortrightarrow\!\!\ZZ_{2}$ & $\ZZ^{2}\!\!\shortrightarrow\!\!\ZZ\!\times\!\ZZ_{2}$ \\
\hline\hline
\end{tabularx}}
\end{table}

\addtocounter{table}{-1}

\begin{table}[H]
\centering
\caption{\textit{(Continued)}}
{\scriptsize\begin{tabularx}{\linewidth}{XXXXXXX}
\hline\hline
$\mathrm{C2/m}\vphantom{\dot{|}}$ & $\Gamma$ & $A$ & $V$ & $L$ & $Z$ & $M$ \\
\hline
A & $\ZZ^{4}\!\!\shortrightarrow\!\!0$ & $\ZZ^{2}\!\!\shortrightarrow\!\!0$ & $\ZZ\!\!\shortrightarrow\!\!0$ & $\ZZ^{2}\!\!\shortrightarrow\!\!\ZZ$ & $\ZZ\!\!\shortrightarrow\!\!0$ & $\ZZ^{2}\!\!\shortrightarrow\!\!\ZZ$ \\
AIII & $\cdot$ & $\cdot$ & $\cdot$ & $\cdot$ & $\cdot$ & $\cdot$ \\
\hline
AI & $\ZZ^{2}\!\!\shortrightarrow\!\!0$ & $\ZZ\!\!\shortrightarrow\!\!0$ & $\ZZ\!\!\shortrightarrow\!\!0$ & $\ZZ\!\!\shortrightarrow\!\!0$ & $\ZZ\!\!\shortrightarrow\!\!0$ & $\ZZ^{2}\!\!\shortrightarrow\!\!\ZZ$ \\
BDI & $\cdot$ & $\cdot$ & $\ZZ_{2}\!\!\shortrightarrow\!\!0$ & $\cdot$ & $\cdot$ & $\cdot$ \\
D & $\ZZ^{2}\!\!\shortrightarrow\!\!0$ & $\ZZ\!\!\shortrightarrow\!\!0$ & $\ZZ_{2}\!\!\shortrightarrow\!\!0$ & $\ZZ\!\!\shortrightarrow\!\!\ZZ$ & $\cdot$ & $\cdot$ \\
DIII & $\cdot$ & $\cdot$ & $\cdot$ & $\cdot$ & $\cdot$ & $\cdot$ \\
AII & $\ZZ^{2}\!\!\shortrightarrow\!\!0$ & $\ZZ\!\!\shortrightarrow\!\!0$ & $\ZZ\!\!\shortrightarrow\!\!0$ & $\ZZ\!\!\shortrightarrow\!\!\ZZ_{2}$ & $\ZZ\!\!\shortrightarrow\!\!\ZZ_{2}$ & $\ZZ^{2}\!\!\shortrightarrow\!\!\ZZ\!\times\!\ZZ_{2}$ \\
CII & $\cdot$ & $\cdot$ & $\cdot$ & $\cdot$ & $\ZZ_{2}\!\!\shortrightarrow\!\!\ZZ_{2}$ & $\ZZ_{2}^{2}\!\!\shortrightarrow\!\!\ZZ_{2}^{2}$ \\
C & $\ZZ^{2}\!\!\shortrightarrow\!\!0$ & $\ZZ\!\!\shortrightarrow\!\!0$ & $\cdot$ & $\ZZ\!\!\shortrightarrow\!\!\ZZ$ & $\ZZ_{2}\!\!\shortrightarrow\!\!0$ & $\ZZ_{2}^{2}\!\!\shortrightarrow\!\!\ZZ_{2}$ \\
CI & $\cdot$ & $\cdot$ & $\cdot$ & $\cdot$ & $\cdot$ & $\cdot$ \\
\hline\hline
\end{tabularx}}
\end{table}

\addtocounter{table}{-1}

\begin{table}[H]
\centering
\caption{\textit{(Continued)}}
{\scriptsize\begin{tabularx}{\linewidth}{XXXXXXXXX}
\hline\hline
$\mathrm{P222}\vphantom{\dot{|}}$ & $\Gamma$ & $Z$ & $Y$ & $X$ & $T$ & $U$ & $S$ & $R$ \\
\hline
A & $\ZZ\!\!\shortrightarrow\!\!0$ & $\cdot$ & $\cdot$ & $\cdot$ & $\cdot$ & $\cdot$ & $\cdot$ & $\cdot$ \\
AIII & $\cdot$ & $\ZZ\!\!\shortrightarrow\!\!\ZZ$ & $\ZZ\!\!\shortrightarrow\!\!\ZZ$ & $\ZZ\!\!\shortrightarrow\!\!\ZZ$ & $\ZZ^{2}\!\!\shortrightarrow\!\!\ZZ^{2}$ & $\ZZ^{2}\!\!\shortrightarrow\!\!\ZZ^{2}$ & $\ZZ^{2}\!\!\shortrightarrow\!\!\ZZ^{2}$ & $\ZZ^{4}\!\!\shortrightarrow\!\!\ZZ^{4}$ \\
\hline
AI & $\ZZ\!\!\shortrightarrow\!\!0$ & $\cdot$ & $\cdot$ & $\cdot$ & $\cdot$ & $\cdot$ & $\cdot$ & $\cdot$ \\
BDI & $\cdot$ & $\cdot$ & $\cdot$ & $\cdot$ & $\cdot$ & $\cdot$ & $\cdot$ & $\cdot$ \\
D & $\cdot$ & $\cdot$ & $\cdot$ & $\cdot$ & $\cdot$ & $\cdot$ & $\cdot$ & $\cdot$ \\
DIII & $\cdot$ & $\ZZ\!\!\shortrightarrow\!\!\ZZ$ & $\ZZ\!\!\shortrightarrow\!\!\ZZ$ & $\ZZ\!\!\shortrightarrow\!\!\ZZ$ & $\ZZ^{2}\!\!\shortrightarrow\!\!\ZZ^{2}$ & $\ZZ^{2}\!\!\shortrightarrow\!\!\ZZ^{2}$ & $\ZZ^{2}\!\!\shortrightarrow\!\!\ZZ^{2}$ & $\ZZ^{4}\!\!\shortrightarrow\!\!\ZZ^{4}$ \\
AII & $\ZZ\!\!\shortrightarrow\!\!0$ & $\ZZ_{2}\!\!\shortrightarrow\!\!0$ & $\ZZ_{2}\!\!\shortrightarrow\!\!0$ & $\ZZ_{2}\!\!\shortrightarrow\!\!0$ & $\ZZ_{2}^{2}\!\!\shortrightarrow\!\!\ZZ_{2}$ & $\ZZ_{2}^{2}\!\!\shortrightarrow\!\!\ZZ_{2}$ & $\ZZ_{2}^{2}\!\!\shortrightarrow\!\!\ZZ_{2}$ & $\ZZ_{2}^{4}\!\!\shortrightarrow\!\!\ZZ_{2}^{3}$ \\
\hline\hline
\end{tabularx}}
\end{table}

\addtocounter{table}{-1}

\begin{table}[H]
\centering
\caption{\textit{(Continued)}}
{\scriptsize\begin{tabularx}{\linewidth}{XXXXXXX}
\hline\hline
$\mathrm{C222}\vphantom{\dot{|}}$ & $\Gamma$ & $Z$ & $S$ & $R$ & $Y$ & $T$ \\
\hline
A & $\ZZ\!\!\shortrightarrow\!\!0$ & $\cdot$ & $\ZZ\!\!\shortrightarrow\!\!0$ & $\cdot$ & $\cdot$ & $\cdot$ \\
AIII & $\cdot$ & $\ZZ\!\!\shortrightarrow\!\!\ZZ$ & $\cdot$ & $\ZZ\!\!\shortrightarrow\!\!\ZZ$ & $\ZZ^{2}\!\!\shortrightarrow\!\!\ZZ^{2}$ & $\ZZ^{4}\!\!\shortrightarrow\!\!\ZZ^{4}$ \\
\hline
AI & $\ZZ\!\!\shortrightarrow\!\!0$ & $\cdot$ & $\cdot$ & $\cdot$ & $\cdot$ & $\cdot$ \\
BDI & $\cdot$ & $\cdot$ & $\cdot$ & $\cdot$ & $\cdot$ & $\cdot$ \\
D & $\cdot$ & $\cdot$ & $\ZZ\!\!\shortrightarrow\!\!\ZZ_{2}$ & $\cdot$ & $\cdot$ & $\cdot$ \\
DIII & $\cdot$ & $\ZZ\!\!\shortrightarrow\!\!\ZZ$ & $\ZZ_{2}\!\!\shortrightarrow\!\!\ZZ_{2}$ & $\ZZ\!\!\shortrightarrow\!\!\ZZ$ & $\ZZ^{2}\!\!\shortrightarrow\!\!\ZZ^{2}$ & $\ZZ^{4}\!\!\shortrightarrow\!\!\ZZ^{4}$ \\
AII & $\ZZ\!\!\shortrightarrow\!\!0$ & $\ZZ_{2}\!\!\shortrightarrow\!\!0$ & $\ZZ_{2}\!\!\shortrightarrow\!\!0$ & $\ZZ_{2}\!\!\shortrightarrow\!\!\ZZ_{2}$ & $\ZZ_{2}^{2}\!\!\shortrightarrow\!\!\ZZ_{2}$ & $\ZZ_{2}^{4}\!\!\shortrightarrow\!\!\ZZ_{2}^{3}$ \\
CII & $\ZZ_{2}\!\!\shortrightarrow\!\!0$ & $\ZZ_{2}\!\!\shortrightarrow\!\!0$ & $\cdot$ & $\ZZ_{2}\!\!\shortrightarrow\!\!\ZZ_{2}$ & $\ZZ_{2}^{2}\!\!\shortrightarrow\!\!\ZZ_{2}$ & $\ZZ_{2}^{4}\!\!\shortrightarrow\!\!\ZZ_{2}^{3}$ \\
C & $\ZZ_{2}\!\!\shortrightarrow\!\!0$ & $\cdot$ & $\ZZ\!\!\shortrightarrow\!\!0$ & $\cdot$ & $\cdot$ & $\cdot$ \\
CI & $\cdot$ & $\ZZ\!\!\shortrightarrow\!\!\ZZ$ & $\cdot$ & $\ZZ\!\!\shortrightarrow\!\!\ZZ$ & $\ZZ^{2}\!\!\shortrightarrow\!\!\ZZ^{2}$ & $\ZZ^{4}\!\!\shortrightarrow\!\!\ZZ^{4}$ \\
\hline\hline
\end{tabularx}}
\end{table}

\addtocounter{table}{-1}

\begin{table}[H]
\centering
\caption{\textit{(Continued)}}
{\scriptsize\begin{tabularx}{\linewidth}{XXXXXXXXX}
\hline\hline
$\mathrm{Pmm2}\vphantom{\dot{|}}$ & $\Gamma$ & $Z$ & $Y$ & $X$ & $T$ & $U$ & $S$ & $R$ \\
\hline
A & $\ZZ\!\!\shortrightarrow\!\!0$ & $\cdot$ & $\cdot$ & $\cdot$ & $\ZZ\!\!\shortrightarrow\!\!\ZZ$ & $\ZZ\!\!\shortrightarrow\!\!\ZZ$ & $\cdot$ & $\ZZ^{2}\!\!\shortrightarrow\!\!\ZZ^{2}$ \\
AIII & $\cdot$ & $\ZZ\!\!\shortrightarrow\!\!\ZZ$ & $\ZZ\!\!\shortrightarrow\!\!\ZZ$ & $\ZZ\!\!\shortrightarrow\!\!\ZZ$ & $\cdot$ & $\cdot$ & $\ZZ^{2}\!\!\shortrightarrow\!\!\ZZ^{2}$ & $\cdot$ \\
\hline
AI & $\ZZ\!\!\shortrightarrow\!\!0$ & $\cdot$ & $\cdot$ & $\cdot$ & $\ZZ\!\!\shortrightarrow\!\!\ZZ$ & $\ZZ\!\!\shortrightarrow\!\!\ZZ$ & $\cdot$ & $\ZZ^{2}\!\!\shortrightarrow\!\!\ZZ^{2}$ \\
BDI & $\cdot$ & $\ZZ\!\!\shortrightarrow\!\!\ZZ$ & $\cdot$ & $\cdot$ & $\cdot$ & $\cdot$ & $\cdot$ & $\cdot$ \\
D & $\cdot$ & $\cdot$ & $\cdot$ & $\cdot$ & $\cdot$ & $\cdot$ & $\cdot$ & $\cdot$ \\
DIII & $\cdot$ & $\cdot$ & $\ZZ\!\!\shortrightarrow\!\!\ZZ$ & $\ZZ\!\!\shortrightarrow\!\!\ZZ$ & $\cdot$ & $\cdot$ & $\ZZ^{2}\!\!\shortrightarrow\!\!\ZZ^{2}$ & $\cdot$ \\
AII & $\ZZ\!\!\shortrightarrow\!\!0$ & $\cdot$ & $\ZZ_{2}\!\!\shortrightarrow\!\!0$ & $\ZZ_{2}\!\!\shortrightarrow\!\!0$ & $\ZZ\!\!\shortrightarrow\!\!\ZZ$ & $\ZZ\!\!\shortrightarrow\!\!\ZZ$ & $\ZZ_{2}^{2}\!\!\shortrightarrow\!\!\ZZ_{2}$ & $\ZZ^{2}\!\!\shortrightarrow\!\!\ZZ^{2}$ \\
\hline\hline
\end{tabularx}}
\end{table}

\addtocounter{table}{-1}

\begin{table}[H]
\centering
\caption{\textit{(Continued)}}
{\scriptsize\begin{tabularx}{\linewidth}{XXXXXXX}
\hline\hline
$\mathrm{Cmm2}\vphantom{\dot{|}}$ & $\Gamma$ & $Z$ & $S$ & $R$ & $Y$ & $T$ \\
\hline
A & $\ZZ\!\!\shortrightarrow\!\!0$ & $\cdot$ & $\ZZ\!\!\shortrightarrow\!\!0$ & $\cdot$ & $\cdot$ & $\ZZ^{2}\!\!\shortrightarrow\!\!\ZZ^{2}$ \\
AIII & $\cdot$ & $\ZZ\!\!\shortrightarrow\!\!\ZZ$ & $\cdot$ & $\ZZ\!\!\shortrightarrow\!\!\ZZ$ & $\ZZ^{2}\!\!\shortrightarrow\!\!\ZZ^{2}$ & $\cdot$ \\
\hline
AI & $\ZZ\!\!\shortrightarrow\!\!0$ & $\cdot$ & $\cdot$ & $\cdot$ & $\cdot$ & $\ZZ^{2}\!\!\shortrightarrow\!\!\ZZ^{2}$ \\
BDI & $\cdot$ & $\ZZ\!\!\shortrightarrow\!\!\ZZ$ & $\cdot$ & $\cdot$ & $\cdot$ & $\cdot$ \\
D & $\cdot$ & $\cdot$ & $\ZZ\!\!\shortrightarrow\!\!\ZZ_{2}$ & $\cdot$ & $\cdot$ & $\cdot$ \\
DIII & $\cdot$ & $\cdot$ & $\ZZ_{2}\!\!\shortrightarrow\!\!\ZZ_{2}$ & $\ZZ\!\!\shortrightarrow\!\!\ZZ$ & $\ZZ^{2}\!\!\shortrightarrow\!\!\ZZ^{2}$ & $\cdot$ \\
AII & $\ZZ\!\!\shortrightarrow\!\!0$ & $\cdot$ & $\ZZ_{2}\!\!\shortrightarrow\!\!0$ & $\ZZ_{2}\!\!\shortrightarrow\!\!\ZZ_{2}$ & $\ZZ_{2}^{2}\!\!\shortrightarrow\!\!\ZZ_{2}$ & $\ZZ^{2}\!\!\shortrightarrow\!\!\ZZ^{2}$ \\
CII & $\ZZ_{2}\!\!\shortrightarrow\!\!0$ & $\ZZ\!\!\shortrightarrow\!\!\ZZ$ & $\cdot$ & $\ZZ_{2}\!\!\shortrightarrow\!\!\ZZ_{2}$ & $\ZZ_{2}^{2}\!\!\shortrightarrow\!\!\ZZ_{2}$ & $\ZZ_{2}^{2}\!\!\shortrightarrow\!\!\ZZ_{2}^{2}$ \\
C & $\ZZ_{2}\!\!\shortrightarrow\!\!0$ & $\ZZ_{2}\!\!\shortrightarrow\!\!\ZZ_{2}$ & $\ZZ\!\!\shortrightarrow\!\!0$ & $\cdot$ & $\cdot$ & $\ZZ_{2}^{2}\!\!\shortrightarrow\!\!\ZZ_{2}^{2}$ \\
CI & $\cdot$ & $\ZZ_{2}\!\!\shortrightarrow\!\!\ZZ_{2}$ & $\cdot$ & $\ZZ\!\!\shortrightarrow\!\!\ZZ$ & $\ZZ^{2}\!\!\shortrightarrow\!\!\ZZ^{2}$ & $\cdot$ \\
\hline\hline
\end{tabularx}}
\end{table}

\addtocounter{table}{-1}

\begin{table}[H]
\centering
\caption{\textit{(Continued)}}
{\scriptsize\begin{tabularx}{\linewidth}{XXXXXXX}
\hline\hline
$\mathrm{Amm2}\vphantom{\dot{|}}$ & $\Gamma$ & $Z$ & $S$ & $R$ & $Y$ & $T$ \\
\hline
A & $\ZZ\!\!\shortrightarrow\!\!0$ & $\cdot$ & $\cdot$ & $\cdot$ & $\ZZ\!\!\shortrightarrow\!\!\ZZ$ & $\ZZ^{2}\!\!\shortrightarrow\!\!\ZZ^{2}$ \\
AIII & $\cdot$ & $\ZZ\!\!\shortrightarrow\!\!\ZZ$ & $\ZZ^{2}\!\!\shortrightarrow\!\!\ZZ^{2}$ & $\ZZ\!\!\shortrightarrow\!\!\ZZ$ & $\cdot$ & $\cdot$ \\
\hline
AI & $\ZZ\!\!\shortrightarrow\!\!0$ & $\cdot$ & $\cdot$ & $\cdot$ & $\ZZ\!\!\shortrightarrow\!\!\ZZ$ & $\ZZ^{2}\!\!\shortrightarrow\!\!\ZZ^{2}$ \\
BDI & $\cdot$ & $\cdot$ & $\ZZ\!\!\shortrightarrow\!\!\ZZ$ & $\cdot$ & $\cdot$ & $\cdot$ \\
D & $\cdot$ & $\cdot$ & $\cdot$ & $\cdot$ & $\cdot$ & $\cdot$ \\
DIII & $\cdot$ & $\ZZ\!\!\shortrightarrow\!\!\ZZ$ & $\ZZ\!\!\shortrightarrow\!\!\ZZ$ & $\ZZ\!\!\shortrightarrow\!\!\ZZ$ & $\cdot$ & $\cdot$ \\
AII & $\ZZ\!\!\shortrightarrow\!\!0$ & $\ZZ_{2}\!\!\shortrightarrow\!\!0$ & $\cdot$ & $\ZZ_{2}\!\!\shortrightarrow\!\!\ZZ_{2}$ & $\ZZ\!\!\shortrightarrow\!\!\ZZ$ & $\ZZ^{2}\!\!\shortrightarrow\!\!\ZZ^{2}$ \\
CII & $\ZZ_{2}\!\!\shortrightarrow\!\!0$ & $\ZZ_{2}\!\!\shortrightarrow\!\!0$ & $\ZZ\!\!\shortrightarrow\!\!\ZZ$ & $\ZZ_{2}\!\!\shortrightarrow\!\!\ZZ_{2}$ & $\ZZ_{2}\!\!\shortrightarrow\!\!\ZZ_{2}$ & $\ZZ_{2}^{2}\!\!\shortrightarrow\!\!\ZZ_{2}^{2}$ \\
C & $\ZZ_{2}\!\!\shortrightarrow\!\!0$ & $\cdot$ & $\cdot$ & $\cdot$ & $\ZZ_{2}\!\!\shortrightarrow\!\!\ZZ_{2}$ & $\ZZ_{2}^{2}\!\!\shortrightarrow\!\!\ZZ_{2}^{2}$ \\
CI & $\cdot$ & $\ZZ\!\!\shortrightarrow\!\!\ZZ$ & $\ZZ\!\!\shortrightarrow\!\!\ZZ$ & $\ZZ\!\!\shortrightarrow\!\!\ZZ$ & $\cdot$ & $\cdot$ \\
\hline\hline
\end{tabularx}}
\end{table}

\addtocounter{table}{-1}

\begin{table}[H]
\centering
\caption{\textit{(Continued)}}
{\scriptsize\begin{tabularx}{\linewidth}{lXXXXXXXX}
\hline\hline
$\mathrm{Pmmm}\vphantom{\dot{|}}$ & $\Gamma$ & $Z$ & $Y$ & $X$ & $T$ & $U$ & $S$ & $R$ \\
\hline
A & $\ZZ^{2}\!\!\shortrightarrow\!\!0$ & $\ZZ\!\!\shortrightarrow\!\!0$ & $\ZZ\!\!\shortrightarrow\!\!0$ & $\ZZ\!\!\shortrightarrow\!\!0$ & $\ZZ^{2}\!\!\shortrightarrow\!\!\ZZ$ & $\ZZ^{2}\!\!\shortrightarrow\!\!\ZZ$ & $\ZZ^{2}\!\!\shortrightarrow\!\!\ZZ$ & $\ZZ^{4}\!\!\shortrightarrow\!\!\ZZ^{3}$ \\
AIII & $\cdot$ & $\cdot$ & $\cdot$ & $\cdot$ & $\cdot$ & $\cdot$ & $\cdot$ & $\cdot$ \\
\hline
AI & $\ZZ^{2}\!\!\shortrightarrow\!\!0$ & $\ZZ\!\!\shortrightarrow\!\!0$ & $\ZZ\!\!\shortrightarrow\!\!0$ & $\ZZ\!\!\shortrightarrow\!\!0$ & $\ZZ^{2}\!\!\shortrightarrow\!\!\ZZ$ & $\ZZ^{2}\!\!\shortrightarrow\!\!\ZZ$ & $\ZZ^{2}\!\!\shortrightarrow\!\!\ZZ$ & $\ZZ^{4}\!\!\shortrightarrow\!\!\ZZ^{3}$ \\
BDI & $\cdot$ & $\cdot$ & $\cdot$ & $\cdot$ & $\cdot$ & $\cdot$ & $\cdot$ & $\cdot$ \\
D & $\cdot$ & $\cdot$ & $\cdot$ & $\cdot$ & $\cdot$ & $\cdot$ & $\cdot$ & $\cdot$ \\
DIII & $\cdot$ & $\cdot$ & $\cdot$ & $\cdot$ & $\cdot$ & $\cdot$ & $\cdot$ & $\cdot$ \\
AII & $\ZZ^{2}\!\!\shortrightarrow\!\!0$ & $\ZZ\!\!\shortrightarrow\!\!0$ & $\ZZ\!\!\shortrightarrow\!\!0$ & $\ZZ\!\!\shortrightarrow\!\!0$ & $\ZZ^{2}\!\!\shortrightarrow\!\!\ZZ$ & $\ZZ^{2}\!\!\shortrightarrow\!\!\ZZ$ & $\ZZ^{2}\!\!\shortrightarrow\!\!\ZZ$ & $\ZZ^{4}\!\!\shortrightarrow\!\!\ZZ^{3}$ \\
\hline\hline
\end{tabularx}}
\end{table}

\addtocounter{table}{-1}

\begin{table}[H]
\centering
\caption{\textit{(Continued)}}
{\scriptsize\begin{tabularx}{\linewidth}{XXXXXXX}
\hline\hline
$\mathrm{Cmmm}\vphantom{\dot{|}}$ & $\Gamma$ & $Z$ & $S$ & $R$ & $Y$ & $T$ \\
\hline
A & $\ZZ^{2}\!\!\shortrightarrow\!\!0$ & $\ZZ\!\!\shortrightarrow\!\!0$ & $\ZZ^{2}\!\!\shortrightarrow\!\!0$ & $\ZZ\!\!\shortrightarrow\!\!0$ & $\ZZ^{2}\!\!\shortrightarrow\!\!\ZZ$ & $\ZZ^{4}\!\!\shortrightarrow\!\!\ZZ^{3}$ \\
AIII & $\cdot$ & $\cdot$ & $\cdot$ & $\cdot$ & $\cdot$ & $\cdot$ \\
\hline
AI & $\ZZ^{2}\!\!\shortrightarrow\!\!0$ & $\ZZ\!\!\shortrightarrow\!\!0$ & $\ZZ\!\!\shortrightarrow\!\!0$ & $\ZZ\!\!\shortrightarrow\!\!0$ & $\ZZ^{2}\!\!\shortrightarrow\!\!\ZZ$ & $\ZZ^{4}\!\!\shortrightarrow\!\!\ZZ^{3}$ \\
BDI & $\cdot$ & $\cdot$ & $\cdot$ & $\cdot$ & $\cdot$ & $\cdot$ \\
D & $\cdot$ & $\cdot$ & $\ZZ\!\!\shortrightarrow\!\!0$ & $\cdot$ & $\cdot$ & $\cdot$ \\
DIII & $\cdot$ & $\cdot$ & $\cdot$ & $\cdot$ & $\cdot$ & $\cdot$ \\
AII & $\ZZ^{2}\!\!\shortrightarrow\!\!0$ & $\ZZ\!\!\shortrightarrow\!\!0$ & $\ZZ\!\!\shortrightarrow\!\!0$ & $\ZZ\!\!\shortrightarrow\!\!\ZZ_{2}$ & $\ZZ^{2}\!\!\shortrightarrow\!\!\ZZ$ & $\ZZ^{4}\!\!\shortrightarrow\!\!\ZZ^{3}$ \\
CII & $\ZZ_{2}^{2}\!\!\shortrightarrow\!\!0$ & $\ZZ_{2}\!\!\shortrightarrow\!\!0$ & $\cdot$ & $\ZZ_{2}\!\!\shortrightarrow\!\!\ZZ_{2}$ & $\ZZ_{2}^{2}\!\!\shortrightarrow\!\!\ZZ_{2}$ & $\ZZ_{2}^{4}\!\!\shortrightarrow\!\!\ZZ_{2}^{3}$ \\
C & $\ZZ_{2}^{2}\!\!\shortrightarrow\!\!0$ & $\ZZ_{2}\!\!\shortrightarrow\!\!0$ & $\ZZ\!\!\shortrightarrow\!\!0$ & $\ZZ_{2}\!\!\shortrightarrow\!\!0$ & $\ZZ_{2}^{2}\!\!\shortrightarrow\!\!\ZZ_{2}$ & $\ZZ_{2}^{4}\!\!\shortrightarrow\!\!\ZZ_{2}^{3}$ \\
CI & $\cdot$ & $\cdot$ & $\cdot$ & $\cdot$ & $\cdot$ & $\cdot$ \\
\hline\hline
\end{tabularx}}
\end{table}

\addtocounter{table}{-1}

\begin{table}[H]
\centering
\caption{\textit{(Continued)}}
{\scriptsize\begin{tabularx}{\linewidth}{XXXXXll}
\hline\hline
$\mathrm{P4}\vphantom{\dot{|}}$ & $\Gamma$ & $Z$ & $X$ & $R$ & $M$ & $A$ \\
\hline
A & $\ZZ^{4}\!\!\shortrightarrow\!\!0$ & $\cdot$ & $\ZZ\!\!\shortrightarrow\!\!0$ & $\cdot$ & $\ZZ^{4}\!\!\shortrightarrow\!\!\ZZ$ & $\cdot$ \\
AIII & $\cdot$ & $\ZZ^{4}\!\!\shortrightarrow\!\!\ZZ^{4}$ & $\cdot$ & $\ZZ\!\!\shortrightarrow\!\!\ZZ$ & $\cdot$ & $\ZZ^{4}\!\!\shortrightarrow\!\!\ZZ^{4}$ \\
\hline
AI & $\ZZ^{2}\!\!\shortrightarrow\!\!0$ & $\cdot$ & $\cdot$ & $\cdot$ & $\ZZ\!\!\shortrightarrow\!\!0$ & $\cdot$ \\
BDI & $\cdot$ & $\ZZ^{2}\!\!\shortrightarrow\!\!\ZZ^{2}$ & $\cdot$ & $\cdot$ & $\cdot$ & $\ZZ\!\!\shortrightarrow\!\!\ZZ$ \\
D & $\ZZ^{2}\!\!\shortrightarrow\!\!0$ & $\cdot$ & $\ZZ\!\!\shortrightarrow\!\!\ZZ_{2}$ & $\cdot$ & $\ZZ^{3}\!\!\shortrightarrow\!\!\ZZ\!\times\!\ZZ_{2}$ & $\cdot$ \\
DIII & $\cdot$ & $\ZZ^{2}\!\!\shortrightarrow\!\!\ZZ^{2}$ & $\ZZ_{2}\!\!\shortrightarrow\!\!\ZZ_{2}$ & $\ZZ\!\!\shortrightarrow\!\!\ZZ$ & $\ZZ_{2}^{2}\!\!\shortrightarrow\!\!\ZZ_{2}^{2}$ & $\ZZ^{3}\!\!\shortrightarrow\!\!\ZZ^{3}$ \\
AII & $\ZZ^{2}\!\!\shortrightarrow\!\!0$ & $\cdot$ & $\ZZ_{2}\!\!\shortrightarrow\!\!0$ & $\ZZ_{2}\!\!\shortrightarrow\!\!\ZZ_{2}$ & $\ZZ\!\times\!\ZZ_{2}^{2}\!\!\shortrightarrow\!\!\ZZ_{2}$ & $\ZZ_{2}^{2}\!\!\shortrightarrow\!\!\ZZ_{2}^{2}$ \\
CII & $\cdot$ & $\ZZ^{2}\!\!\shortrightarrow\!\!\ZZ^{2}$ & $\cdot$ & $\ZZ_{2}\!\!\shortrightarrow\!\!\ZZ_{2}$ & $\cdot$ & $\ZZ\!\times\!\ZZ_{2}^{2}\!\!\shortrightarrow\!\!\ZZ\!\times\!\ZZ_{2}^{2}$ \\
C & $\ZZ^{2}\!\!\shortrightarrow\!\!0$ & $\cdot$ & $\ZZ\!\!\shortrightarrow\!\!0$ & $\cdot$ & $\ZZ^{3}\!\!\shortrightarrow\!\!\ZZ$ & $\cdot$ \\
CI & $\cdot$ & $\ZZ^{2}\!\!\shortrightarrow\!\!\ZZ^{2}$ & $\cdot$ & $\ZZ\!\!\shortrightarrow\!\!\ZZ$ & $\cdot$ & $\ZZ^{3}\!\!\shortrightarrow\!\!\ZZ^{3}$ \\
\hline\hline
\end{tabularx}}
\end{table}

\addtocounter{table}{-1}

\begin{table}[H]
\centering
\caption{\textit{(Continued)}}
{\scriptsize\begin{tabularx}{\linewidth}{XXXXXXX}
\hline\hline
$\mathrm{P\bar{4}}\vphantom{\dot{|}}$ & $\Gamma$ & $Z$ & $X$ & $R$ & $M$ & $A$ \\
\hline
A & $\ZZ^{4}\!\!\shortrightarrow\!\!0$ & $\ZZ^{2}\!\!\shortrightarrow\!\!0$ & $\ZZ\!\!\shortrightarrow\!\!0$ & $\cdot$ & $\ZZ^{4}\!\!\shortrightarrow\!\!\ZZ$ & $\ZZ^{2}\!\!\shortrightarrow\!\!\ZZ_{2}$ \\
AIII & $\cdot$ & $\cdot$ & $\cdot$ & $\ZZ\!\!\shortrightarrow\!\!\ZZ$ & $\cdot$ & $\cdot$ \\
\hline
AI & $\ZZ^{2}\!\!\shortrightarrow\!\!0$ & $\ZZ\!\!\shortrightarrow\!\!0$ & $\cdot$ & $\cdot$ & $\ZZ\!\!\shortrightarrow\!\!0$ & $\ZZ\!\!\shortrightarrow\!\!0$ \\
BDI & $\cdot$ & $\cdot$ & $\cdot$ & $\cdot$ & $\cdot$ & $\cdot$ \\
D & $\ZZ^{2}\!\!\shortrightarrow\!\!0$ & $\ZZ\!\!\shortrightarrow\!\!0$ & $\ZZ\!\!\shortrightarrow\!\!\ZZ_{2}$ & $\cdot$ & $\ZZ^{3}\!\!\shortrightarrow\!\!\ZZ\!\times\!\ZZ_{2}$ & $\ZZ\!\!\shortrightarrow\!\!\ZZ_{2}$ \\
DIII & $\cdot$ & $\cdot$ & $\ZZ_{2}\!\!\shortrightarrow\!\!\ZZ_{2}$ & $\ZZ\!\!\shortrightarrow\!\!\ZZ$ & $\ZZ_{2}^{2}\!\!\shortrightarrow\!\!\ZZ_{2}^{2}$ & $\ZZ_{2}\!\!\shortrightarrow\!\!\ZZ_{2}$ \\
AII & $\ZZ^{2}\!\!\shortrightarrow\!\!0$ & $\ZZ\!\!\shortrightarrow\!\!0$ & $\ZZ_{2}\!\!\shortrightarrow\!\!0$ & $\ZZ_{2}\!\!\shortrightarrow\!\!\ZZ_{2}$ & $\ZZ\!\times\!\ZZ_{2}^{2}\!\!\shortrightarrow\!\!\ZZ_{2}$ & $\ZZ\!\times\!\ZZ_{2}\!\!\shortrightarrow\!\!\ZZ_{4}$ \\
CII & $\cdot$ & $\cdot$ & $\cdot$ & $\ZZ_{2}\!\!\shortrightarrow\!\!\ZZ_{2}$ & $\cdot$ & $\ZZ_{2}\!\!\shortrightarrow\!\!\ZZ_{2}$ \\
C & $\ZZ^{2}\!\!\shortrightarrow\!\!0$ & $\ZZ\!\!\shortrightarrow\!\!0$ & $\ZZ\!\!\shortrightarrow\!\!0$ & $\cdot$ & $\ZZ^{3}\!\!\shortrightarrow\!\!\ZZ$ & $\ZZ\!\times\!\ZZ_{2}\!\!\shortrightarrow\!\!\ZZ_{2}$ \\
CI & $\cdot$ & $\cdot$ & $\cdot$ & $\ZZ\!\!\shortrightarrow\!\!\ZZ$ & $\cdot$ & $\cdot$ \\
\hline\hline
\end{tabularx}}
\end{table}

\addtocounter{table}{-1}

\begin{table}[H]
\centering
\caption{\textit{(Continued)}}
{\scriptsize\begin{tabularx}{\linewidth}{XXXXXXX}
\hline\hline
$\mathrm{P4/m}\vphantom{\dot{|}}$ & $\Gamma$ & $Z$ & $X$ & $R$ & $M$ & $A$ \\
\hline
A & $\ZZ^{8}\!\!\shortrightarrow\!\!0$ & $\ZZ^{4}\!\!\shortrightarrow\!\!0$ & $\ZZ^{2}\!\!\shortrightarrow\!\!0$ & $\ZZ\!\!\shortrightarrow\!\!0$ & $\ZZ^{8}\!\!\shortrightarrow\!\!\ZZ^{2}$ & $\ZZ^{4}\!\!\shortrightarrow\!\!\ZZ$ \\
AIII & $\cdot$ & $\cdot$ & $\cdot$ & $\cdot$ & $\cdot$ & $\cdot$ \\
\hline
AI & $\ZZ^{4}\!\!\shortrightarrow\!\!0$ & $\ZZ^{2}\!\!\shortrightarrow\!\!0$ & $\ZZ\!\!\shortrightarrow\!\!0$ & $\ZZ\!\!\shortrightarrow\!\!0$ & $\ZZ^{4}\!\!\shortrightarrow\!\!\ZZ$ & $\ZZ^{3}\!\!\shortrightarrow\!\!\ZZ$ \\
BDI & $\cdot$ & $\cdot$ & $\cdot$ & $\cdot$ & $\cdot$ & $\cdot$ \\
D & $\ZZ^{4}\!\!\shortrightarrow\!\!0$ & $\ZZ^{2}\!\!\shortrightarrow\!\!0$ & $\ZZ\!\!\shortrightarrow\!\!0$ & $\cdot$ & $\ZZ^{4}\!\!\shortrightarrow\!\!\ZZ$ & $\ZZ\!\!\shortrightarrow\!\!0$ \\
DIII & $\cdot$ & $\cdot$ & $\cdot$ & $\cdot$ & $\cdot$ & $\cdot$ \\
AII & $\ZZ^{4}\!\!\shortrightarrow\!\!0$ & $\ZZ^{2}\!\!\shortrightarrow\!\!0$ & $\ZZ\!\!\shortrightarrow\!\!0$ & $\ZZ\!\!\shortrightarrow\!\!\ZZ_{2}$ & $\ZZ^{4}\!\!\shortrightarrow\!\!\ZZ$ & $\ZZ^{3}\!\!\shortrightarrow\!\!\ZZ\!\times\!\ZZ_{2}$ \\
CII & $\cdot$ & $\cdot$ & $\cdot$ & $\ZZ_{2}\!\!\shortrightarrow\!\!\ZZ_{2}$ & $\cdot$ & $\ZZ_{2}^{2}\!\!\shortrightarrow\!\!\ZZ_{2}^{2}$ \\
C & $\ZZ^{4}\!\!\shortrightarrow\!\!0$ & $\ZZ^{2}\!\!\shortrightarrow\!\!0$ & $\ZZ\!\!\shortrightarrow\!\!0$ & $\ZZ_{2}\!\!\shortrightarrow\!\!0$ & $\ZZ^{4}\!\!\shortrightarrow\!\!\ZZ$ & $\ZZ\!\times\!\ZZ_{2}^{2}\!\!\shortrightarrow\!\!\ZZ_{2}$ \\
CI & $\cdot$ & $\cdot$ & $\cdot$ & $\cdot$ & $\cdot$ & $\cdot$ \\
\hline\hline
\end{tabularx}}
\end{table}

\addtocounter{table}{-1}

\begin{table}[H]
\centering
\caption{\textit{(Continued)}}
{\scriptsize\begin{tabularx}{\linewidth}{XXXXXXX}
\hline\hline
$\mathrm{P422}\vphantom{\dot{|}}$ & $\Gamma$ & $Z$ & $X$ & $R$ & $M$ & $A$ \\
\hline
A & $\ZZ^{2}\!\!\shortrightarrow\!\!0$ & $\cdot$ & $\cdot$ & $\cdot$ & $\ZZ\!\!\shortrightarrow\!\!0$ & $\cdot$ \\
AIII & $\cdot$ & $\ZZ^{2}\!\!\shortrightarrow\!\!\ZZ^{2}$ & $\ZZ\!\!\shortrightarrow\!\!\ZZ$ & $\ZZ^{2}\!\!\shortrightarrow\!\!\ZZ^{2}$ & $\ZZ^{2}\!\!\shortrightarrow\!\!\ZZ^{2}$ & $\ZZ^{5}\!\!\shortrightarrow\!\!\ZZ^{5}$ \\
\hline
AI & $\ZZ^{2}\!\!\shortrightarrow\!\!0$ & $\cdot$ & $\cdot$ & $\cdot$ & $\ZZ\!\!\shortrightarrow\!\!0$ & $\cdot$ \\
BDI & $\cdot$ & $\cdot$ & $\cdot$ & $\cdot$ & $\cdot$ & $\cdot$ \\
D & $\cdot$ & $\cdot$ & $\cdot$ & $\cdot$ & $\cdot$ & $\cdot$ \\
DIII & $\cdot$ & $\ZZ^{2}\!\!\shortrightarrow\!\!\ZZ^{2}$ & $\ZZ\!\!\shortrightarrow\!\!\ZZ$ & $\ZZ^{2}\!\!\shortrightarrow\!\!\ZZ^{2}$ & $\ZZ^{2}\!\!\shortrightarrow\!\!\ZZ^{2}$ & $\ZZ^{5}\!\!\shortrightarrow\!\!\ZZ^{5}$ \\
AII & $\ZZ^{2}\!\!\shortrightarrow\!\!0$ & $\ZZ_{2}^{2}\!\!\shortrightarrow\!\!0$ & $\ZZ_{2}\!\!\shortrightarrow\!\!0$ & $\ZZ_{2}^{2}\!\!\shortrightarrow\!\!\ZZ_{2}$ & $\ZZ\!\times\!\ZZ_{2}^{2}\!\!\shortrightarrow\!\!\ZZ_{2}$ & $\ZZ_{2}^{5}\!\!\shortrightarrow\!\!\ZZ_{2}^{3}$ \\
CII & $\ZZ_{2}^{2}\!\!\shortrightarrow\!\!0$ & $\ZZ_{2}^{2}\!\!\shortrightarrow\!\!0$ & $\ZZ_{2}\!\!\shortrightarrow\!\!0$ & $\ZZ_{2}^{2}\!\!\shortrightarrow\!\!\ZZ_{2}$ & $\ZZ_{2}^{3}\!\!\shortrightarrow\!\!\ZZ_{2}$ & $\ZZ_{2}^{5}\!\!\shortrightarrow\!\!\ZZ_{2}^{3}$ \\
C & $\ZZ_{2}^{2}\!\!\shortrightarrow\!\!0$ & $\cdot$ & $\cdot$ & $\cdot$ & $\ZZ_{2}\!\!\shortrightarrow\!\!0$ & $\cdot$ \\
CI & $\cdot$ & $\ZZ^{2}\!\!\shortrightarrow\!\!\ZZ^{2}$ & $\ZZ\!\!\shortrightarrow\!\!\ZZ$ & $\ZZ^{2}\!\!\shortrightarrow\!\!\ZZ^{2}$ & $\ZZ^{2}\!\!\shortrightarrow\!\!\ZZ^{2}$ & $\ZZ^{5}\!\!\shortrightarrow\!\!\ZZ^{5}$ \\
\hline\hline
\end{tabularx}}
\end{table}

\addtocounter{table}{-1}

\begin{table}[H]
\centering
\caption{\textit{(Continued)}}
{\scriptsize\begin{tabularx}{\linewidth}{XXXXXll}
\hline\hline
$\mathrm{P4mm}\vphantom{\dot{|}}$ & $\Gamma$ & $Z$ & $X$ & $R$ & $M$ & $A$ \\
\hline
A & $\ZZ^{2}\!\!\shortrightarrow\!\!0$ & $\cdot$ & $\cdot$ & $\ZZ\!\!\shortrightarrow\!\!\ZZ$ & $\ZZ\!\!\shortrightarrow\!\!0$ & $\ZZ^{2}\!\!\shortrightarrow\!\!\ZZ^{2}$ \\
AIII & $\cdot$ & $\ZZ^{2}\!\!\shortrightarrow\!\!\ZZ^{2}$ & $\ZZ\!\!\shortrightarrow\!\!\ZZ$ & $\cdot$ & $\ZZ^{2}\!\!\shortrightarrow\!\!\ZZ^{2}$ & $\ZZ\!\!\shortrightarrow\!\!\ZZ$ \\
\hline
AI & $\ZZ^{2}\!\!\shortrightarrow\!\!0$ & $\cdot$ & $\cdot$ & $\ZZ\!\!\shortrightarrow\!\!\ZZ$ & $\ZZ\!\!\shortrightarrow\!\!0$ & $\ZZ^{2}\!\!\shortrightarrow\!\!\ZZ^{2}$ \\
BDI & $\cdot$ & $\ZZ^{2}\!\!\shortrightarrow\!\!\ZZ^{2}$ & $\cdot$ & $\cdot$ & $\cdot$ & $\ZZ\!\!\shortrightarrow\!\!\ZZ$ \\
D & $\cdot$ & $\cdot$ & $\cdot$ & $\cdot$ & $\cdot$ & $\cdot$ \\
DIII & $\cdot$ & $\cdot$ & $\ZZ\!\!\shortrightarrow\!\!\ZZ$ & $\cdot$ & $\ZZ^{2}\!\!\shortrightarrow\!\!\ZZ^{2}$ & $\cdot$ \\
AII & $\ZZ^{2}\!\!\shortrightarrow\!\!0$ & $\cdot$ & $\ZZ_{2}\!\!\shortrightarrow\!\!0$ & $\ZZ\!\!\shortrightarrow\!\!\ZZ$ & $\ZZ\!\times\!\ZZ_{2}^{2}\!\!\shortrightarrow\!\!\ZZ_{2}$ & $\ZZ^{2}\!\!\shortrightarrow\!\!\ZZ^{2}$ \\
CII & $\ZZ_{2}^{2}\!\!\shortrightarrow\!\!0$ & $\ZZ^{2}\!\!\shortrightarrow\!\!\ZZ^{2}$ & $\ZZ_{2}\!\!\shortrightarrow\!\!0$ & $\ZZ_{2}\!\!\shortrightarrow\!\!\ZZ_{2}$ & $\ZZ_{2}^{3}\!\!\shortrightarrow\!\!\ZZ_{2}$ & $\ZZ\!\times\!\ZZ_{2}^{2}\!\!\shortrightarrow\!\!\ZZ\!\times\!\ZZ_{2}^{2}$ \\
C & $\ZZ_{2}^{2}\!\!\shortrightarrow\!\!0$ & $\ZZ_{2}^{2}\!\!\shortrightarrow\!\!\ZZ_{2}^{2}$ & $\cdot$ & $\ZZ_{2}\!\!\shortrightarrow\!\!\ZZ_{2}$ & $\ZZ_{2}\!\!\shortrightarrow\!\!0$ & $\ZZ_{2}^{3}\!\!\shortrightarrow\!\!\ZZ_{2}^{3}$ \\
CI & $\cdot$ & $\ZZ_{2}^{2}\!\!\shortrightarrow\!\!\ZZ_{2}^{2}$ & $\ZZ\!\!\shortrightarrow\!\!\ZZ$ & $\cdot$ & $\ZZ^{2}\!\!\shortrightarrow\!\!\ZZ^{2}$ & $\ZZ_{2}\!\!\shortrightarrow\!\!\ZZ_{2}$ \\
\hline\hline
\end{tabularx}}
\end{table}

\addtocounter{table}{-1}

\begin{table}[H]
\centering
\caption{\textit{(Continued)}}
{\scriptsize\begin{tabularx}{\linewidth}{XXXXXll}
\hline\hline
$\mathrm{P\bar{4}2m}\vphantom{\dot{|}}$ & $\Gamma$ & $Z$ & $X$ & $R$ & $M$ & $A$ \\
\hline
A & $\ZZ^{2}\!\!\shortrightarrow\!\!0$ & $\ZZ\!\!\shortrightarrow\!\!0$ & $\cdot$ & $\cdot$ & $\ZZ\!\!\shortrightarrow\!\!0$ & $\ZZ^{2}\!\!\shortrightarrow\!\!\ZZ$ \\
AIII & $\cdot$ & $\cdot$ & $\ZZ\!\!\shortrightarrow\!\!\ZZ$ & $\ZZ^{2}\!\!\shortrightarrow\!\!\ZZ^{2}$ & $\ZZ^{2}\!\!\shortrightarrow\!\!\ZZ^{2}$ & $\ZZ\!\!\shortrightarrow\!\!\ZZ$ \\
\hline
AI & $\ZZ^{2}\!\!\shortrightarrow\!\!0$ & $\ZZ\!\!\shortrightarrow\!\!0$ & $\cdot$ & $\cdot$ & $\ZZ\!\!\shortrightarrow\!\!0$ & $\ZZ^{2}\!\!\shortrightarrow\!\!\ZZ$ \\
BDI & $\cdot$ & $\cdot$ & $\cdot$ & $\cdot$ & $\cdot$ & $\cdot$ \\
D & $\cdot$ & $\cdot$ & $\cdot$ & $\cdot$ & $\cdot$ & $\cdot$ \\
DIII & $\cdot$ & $\cdot$ & $\ZZ\!\!\shortrightarrow\!\!\ZZ$ & $\ZZ^{2}\!\!\shortrightarrow\!\!\ZZ^{2}$ & $\ZZ^{2}\!\!\shortrightarrow\!\!\ZZ^{2}$ & $\ZZ\!\!\shortrightarrow\!\!\ZZ$ \\
AII & $\ZZ^{2}\!\!\shortrightarrow\!\!0$ & $\ZZ\!\!\shortrightarrow\!\!0$ & $\ZZ_{2}\!\!\shortrightarrow\!\!0$ & $\ZZ_{2}^{2}\!\!\shortrightarrow\!\!\ZZ_{2}$ & $\ZZ\!\times\!\ZZ_{2}^{2}\!\!\shortrightarrow\!\!\ZZ_{2}$ & $\ZZ^{2}\!\times\!\ZZ_{2}\!\!\shortrightarrow\!\!\ZZ\!\times\!\ZZ_{2}$ \\
CII & $\ZZ_{2}^{2}\!\!\shortrightarrow\!\!0$ & $\ZZ_{2}\!\!\shortrightarrow\!\!0$ & $\ZZ_{2}\!\!\shortrightarrow\!\!0$ & $\ZZ_{2}^{2}\!\!\shortrightarrow\!\!\ZZ_{2}$ & $\ZZ_{2}^{3}\!\!\shortrightarrow\!\!\ZZ_{2}$ & $\ZZ_{2}^{3}\!\!\shortrightarrow\!\!\ZZ_{2}^{2}$ \\
C & $\ZZ_{2}^{2}\!\!\shortrightarrow\!\!0$ & $\ZZ_{2}\!\!\shortrightarrow\!\!0$ & $\cdot$ & $\cdot$ & $\ZZ_{2}\!\!\shortrightarrow\!\!0$ & $\ZZ_{2}^{2}\!\!\shortrightarrow\!\!\ZZ_{2}$ \\
CI & $\cdot$ & $\cdot$ & $\ZZ\!\!\shortrightarrow\!\!\ZZ$ & $\ZZ^{2}\!\!\shortrightarrow\!\!\ZZ^{2}$ & $\ZZ^{2}\!\!\shortrightarrow\!\!\ZZ^{2}$ & $\ZZ\!\!\shortrightarrow\!\!\ZZ$ \\
\hline\hline
\end{tabularx}}
\end{table}

\addtocounter{table}{-1}

\begin{table}[H]
\centering
\caption{\textit{(Continued)}}
{\scriptsize\begin{tabularx}{\linewidth}{XXXXXll}
\hline\hline
$\mathrm{P\bar{4}m2}\vphantom{\dot{|}}$ & $\Gamma$ & $Z$ & $X$ & $R$ & $M$ & $A$ \\
\hline
A & $\ZZ^{2}\!\!\shortrightarrow\!\!0$ & $\ZZ\!\!\shortrightarrow\!\!0$ & $\cdot$ & $\ZZ\!\!\shortrightarrow\!\!\ZZ$ & $\ZZ\!\!\shortrightarrow\!\!0$ & $\ZZ^{2}\!\!\shortrightarrow\!\!\ZZ$ \\
AIII & $\cdot$ & $\cdot$ & $\ZZ\!\!\shortrightarrow\!\!\ZZ$ & $\cdot$ & $\ZZ^{2}\!\!\shortrightarrow\!\!\ZZ^{2}$ & $\ZZ\!\!\shortrightarrow\!\!\ZZ$ \\
\hline
AI & $\ZZ^{2}\!\!\shortrightarrow\!\!0$ & $\ZZ\!\!\shortrightarrow\!\!0$ & $\cdot$ & $\ZZ\!\!\shortrightarrow\!\!\ZZ$ & $\ZZ\!\!\shortrightarrow\!\!0$ & $\ZZ^{2}\!\!\shortrightarrow\!\!\ZZ$ \\
BDI & $\cdot$ & $\cdot$ & $\cdot$ & $\cdot$ & $\cdot$ & $\cdot$ \\
D & $\cdot$ & $\cdot$ & $\cdot$ & $\cdot$ & $\cdot$ & $\cdot$ \\
DIII & $\cdot$ & $\cdot$ & $\ZZ\!\!\shortrightarrow\!\!\ZZ$ & $\cdot$ & $\ZZ^{2}\!\!\shortrightarrow\!\!\ZZ^{2}$ & $\ZZ\!\!\shortrightarrow\!\!\ZZ$ \\
AII & $\ZZ^{2}\!\!\shortrightarrow\!\!0$ & $\ZZ\!\!\shortrightarrow\!\!0$ & $\ZZ_{2}\!\!\shortrightarrow\!\!0$ & $\ZZ\!\!\shortrightarrow\!\!\ZZ$ & $\ZZ\!\times\!\ZZ_{2}^{2}\!\!\shortrightarrow\!\!\ZZ_{2}$ & $\ZZ^{2}\!\times\!\ZZ_{2}\!\!\shortrightarrow\!\!\ZZ\!\times\!\ZZ_{2}$ \\
CII & $\ZZ_{2}^{2}\!\!\shortrightarrow\!\!0$ & $\ZZ_{2}\!\!\shortrightarrow\!\!0$ & $\ZZ_{2}\!\!\shortrightarrow\!\!0$ & $\ZZ_{2}\!\!\shortrightarrow\!\!\ZZ_{2}$ & $\ZZ_{2}^{3}\!\!\shortrightarrow\!\!\ZZ_{2}$ & $\ZZ_{2}^{3}\!\!\shortrightarrow\!\!\ZZ_{2}^{2}$ \\
C & $\ZZ_{2}^{2}\!\!\shortrightarrow\!\!0$ & $\ZZ_{2}\!\!\shortrightarrow\!\!0$ & $\cdot$ & $\ZZ_{2}\!\!\shortrightarrow\!\!\ZZ_{2}$ & $\ZZ_{2}\!\!\shortrightarrow\!\!0$ & $\ZZ_{2}^{2}\!\!\shortrightarrow\!\!\ZZ_{2}$ \\
CI & $\cdot$ & $\cdot$ & $\ZZ\!\!\shortrightarrow\!\!\ZZ$ & $\cdot$ & $\ZZ^{2}\!\!\shortrightarrow\!\!\ZZ^{2}$ & $\ZZ\!\!\shortrightarrow\!\!\ZZ$ \\
\hline\hline
\end{tabularx}}
\end{table}

\addtocounter{table}{-1}

\begin{table}[H]
\centering
\caption{\textit{(Continued)}}
{\scriptsize\begin{tabularx}{\linewidth}{lXXXXXX}
\hline\hline
$\mathrm{P4/mmm}\vphantom{\dot{|}}$ & $\Gamma$ & $Z$ & $X$ & $R$ & $M$ & $A$ \\
\hline
A & $\ZZ^{4}\!\!\shortrightarrow\!\!0$ & $\ZZ^{2}\!\!\shortrightarrow\!\!0$ & $\ZZ\!\!\shortrightarrow\!\!0$ & $\ZZ^{2}\!\!\shortrightarrow\!\!\ZZ$ & $\ZZ^{4}\!\!\shortrightarrow\!\!\ZZ$ & $\ZZ^{5}\!\!\shortrightarrow\!\!\ZZ^{3}$ \\
AIII & $\cdot$ & $\cdot$ & $\cdot$ & $\cdot$ & $\cdot$ & $\cdot$ \\
\hline
AI & $\ZZ^{4}\!\!\shortrightarrow\!\!0$ & $\ZZ^{2}\!\!\shortrightarrow\!\!0$ & $\ZZ\!\!\shortrightarrow\!\!0$ & $\ZZ^{2}\!\!\shortrightarrow\!\!\ZZ$ & $\ZZ^{4}\!\!\shortrightarrow\!\!\ZZ$ & $\ZZ^{5}\!\!\shortrightarrow\!\!\ZZ^{3}$ \\
BDI & $\cdot$ & $\cdot$ & $\cdot$ & $\cdot$ & $\cdot$ & $\cdot$ \\
D & $\cdot$ & $\cdot$ & $\cdot$ & $\cdot$ & $\cdot$ & $\cdot$ \\
DIII & $\cdot$ & $\cdot$ & $\cdot$ & $\cdot$ & $\cdot$ & $\cdot$ \\
AII & $\ZZ^{4}\!\!\shortrightarrow\!\!0$ & $\ZZ^{2}\!\!\shortrightarrow\!\!0$ & $\ZZ\!\!\shortrightarrow\!\!0$ & $\ZZ^{2}\!\!\shortrightarrow\!\!\ZZ$ & $\ZZ^{4}\!\!\shortrightarrow\!\!\ZZ$ & $\ZZ^{5}\!\!\shortrightarrow\!\!\ZZ^{3}$ \\
CII & $\ZZ_{2}^{4}\!\!\shortrightarrow\!\!0$ & $\ZZ_{2}^{2}\!\!\shortrightarrow\!\!0$ & $\ZZ_{2}\!\!\shortrightarrow\!\!0$ & $\ZZ_{2}^{2}\!\!\shortrightarrow\!\!\ZZ_{2}$ & $\ZZ_{2}^{4}\!\!\shortrightarrow\!\!\ZZ_{2}$ & $\ZZ_{2}^{5}\!\!\shortrightarrow\!\!\ZZ_{2}^{3}$ \\
C & $\ZZ_{2}^{4}\!\!\shortrightarrow\!\!0$ & $\ZZ_{2}^{2}\!\!\shortrightarrow\!\!0$ & $\ZZ_{2}\!\!\shortrightarrow\!\!0$ & $\ZZ_{2}^{2}\!\!\shortrightarrow\!\!\ZZ_{2}$ & $\ZZ_{2}^{4}\!\!\shortrightarrow\!\!\ZZ_{2}$ & $\ZZ_{2}^{5}\!\!\shortrightarrow\!\!\ZZ_{2}^{3}$ \\
CI & $\cdot$ & $\cdot$ & $\cdot$ & $\cdot$ & $\cdot$ & $\cdot$ \\
\hline\hline
\end{tabularx}}
\end{table}

\addtocounter{table}{-1}

\begin{table}[H]
\centering
\caption{\textit{(Continued)}}
{\scriptsize\begin{tabularx}{\linewidth}{XXXXX}
\hline\hline
$\mathrm{R3}\vphantom{\dot{|}}$ & $\Gamma$ & $L$ & $F$ & $Z$ \\
\hline
A & $\ZZ^{3}\!\!\shortrightarrow\!\!0$ & $\cdot$ & $\ZZ\!\!\shortrightarrow\!\!\ZZ$ & $\cdot$ \\
AIII & $\cdot$ & $\ZZ\!\!\shortrightarrow\!\!\ZZ$ & $\cdot$ & $\ZZ^{3}\!\!\shortrightarrow\!\!\ZZ^{3}$ \\
\hline
AI & $\ZZ^{2}\!\!\shortrightarrow\!\!0$ & $\cdot$ & $\cdot$ & $\cdot$ \\
BDI & $\ZZ_{2}\!\!\shortrightarrow\!\!0$ & $\ZZ\!\!\shortrightarrow\!\!\ZZ$ & $\cdot$ & $\ZZ\!\!\shortrightarrow\!\!\ZZ$ \\
D & $\ZZ\!\times\!\ZZ_{2}\!\!\shortrightarrow\!\!0$ & $\ZZ_{2}\!\!\shortrightarrow\!\!\ZZ_{2}$ & $\ZZ\!\!\shortrightarrow\!\!\ZZ$ & $\cdot$ \\
DIII & $\cdot$ & $\ZZ_{2}\!\!\shortrightarrow\!\!\ZZ_{2}$ & $\ZZ_{2}\!\!\shortrightarrow\!\!\ZZ_{2}$ & $\ZZ^{2}\!\!\shortrightarrow\!\!\ZZ^{2}$ \\
AII & $\ZZ^{2}\!\!\shortrightarrow\!\!0$ & $\cdot$ & $\ZZ_{2}\!\!\shortrightarrow\!\!\ZZ_{2}$ & $\ZZ_{2}\!\!\shortrightarrow\!\!\ZZ_{2}$ \\
CII & $\cdot$ & $\ZZ\!\!\shortrightarrow\!\!\ZZ$ & $\cdot$ & $\ZZ\!\times\!\ZZ_{2}\!\!\shortrightarrow\!\!\ZZ\!\times\!\ZZ_{2}$ \\
C & $\ZZ\!\!\shortrightarrow\!\!0$ & $\cdot$ & $\ZZ\!\!\shortrightarrow\!\!\ZZ$ & $\cdot$ \\
CI & $\cdot$ & $\cdot$ & $\cdot$ & $\ZZ^{2}\!\!\shortrightarrow\!\!\ZZ^{2}$ \\
\hline\hline
\end{tabularx}}
\end{table}

\addtocounter{table}{-1}

\begin{table}[H]
\centering
\caption{\textit{(Continued)}}
{\scriptsize\begin{tabularx}{\linewidth}{XXXXX}
\hline\hline
$\mathrm{R\bar{3}}\vphantom{\dot{|}}$ & $\Gamma$ & $L$ & $F$ & $Z$ \\
\hline
A & $\ZZ^{6}\!\!\shortrightarrow\!\!0$ & $\ZZ\!\!\shortrightarrow\!\!0$ & $\ZZ^{2}\!\!\shortrightarrow\!\!\ZZ$ & $\ZZ^{3}\!\!\shortrightarrow\!\!\ZZ_{2}$ \\
AIII & $\cdot$ & $\cdot$ & $\cdot$ & $\cdot$ \\
\hline
AI & $\ZZ^{4}\!\!\shortrightarrow\!\!0$ & $\ZZ\!\!\shortrightarrow\!\!0$ & $\ZZ\!\!\shortrightarrow\!\!0$ & $\ZZ^{2}\!\!\shortrightarrow\!\!0$ \\
BDI & $\ZZ_{2}^{2}\!\!\shortrightarrow\!\!0$ & $\ZZ_{2}\!\!\shortrightarrow\!\!0$ & $\cdot$ & $\cdot$ \\
D & $\ZZ^{2}\!\times\!\ZZ_{2}^{2}\!\!\shortrightarrow\!\!0$ & $\ZZ_{2}\!\!\shortrightarrow\!\!0$ & $\ZZ\!\!\shortrightarrow\!\!\ZZ$ & $\ZZ\!\!\shortrightarrow\!\!0$ \\
DIII & $\cdot$ & $\cdot$ & $\cdot$ & $\cdot$ \\
AII & $\ZZ^{4}\!\!\shortrightarrow\!\!0$ & $\ZZ\!\!\shortrightarrow\!\!0$ & $\ZZ\!\!\shortrightarrow\!\!\ZZ_{2}$ & $\ZZ^{2}\!\!\shortrightarrow\!\!\ZZ_{4}$ \\
CII & $\cdot$ & $\cdot$ & $\cdot$ & $\ZZ_{2}\!\!\shortrightarrow\!\!\ZZ_{2}$ \\
C & $\ZZ^{2}\!\!\shortrightarrow\!\!0$ & $\cdot$ & $\ZZ\!\!\shortrightarrow\!\!\ZZ$ & $\ZZ\!\times\!\ZZ_{2}\!\!\shortrightarrow\!\!\ZZ_{2}$ \\
CI & $\cdot$ & $\cdot$ & $\cdot$ & $\cdot$ \\
\hline\hline
\end{tabularx}}
\end{table}

\addtocounter{table}{-1}

\begin{table}[H]
\centering
\caption{\textit{(Continued)}}
{\scriptsize\begin{tabularx}{\linewidth}{XXXXX}
\hline\hline
$\mathrm{R32}\vphantom{\dot{|}}$ & $\Gamma$ & $L$ & $F$ & $Z$ \\
\hline
A & $\ZZ^{3}\!\!\shortrightarrow\!\!0$ & $\ZZ\!\!\shortrightarrow\!\!0$ & $\cdot$ & $\cdot$ \\
AIII & $\cdot$ & $\cdot$ & $\ZZ\!\!\shortrightarrow\!\!\ZZ$ & $\ZZ^{3}\!\!\shortrightarrow\!\!\ZZ^{3}$ \\
\hline
AI & $\ZZ^{2}\!\!\shortrightarrow\!\!0$ & $\cdot$ & $\cdot$ & $\cdot$ \\
BDI & $\cdot$ & $\cdot$ & $\cdot$ & $\cdot$ \\
D & $\ZZ\!\!\shortrightarrow\!\!0$ & $\ZZ\!\!\shortrightarrow\!\!\ZZ_{2}$ & $\cdot$ & $\cdot$ \\
DIII & $\cdot$ & $\ZZ_{2}\!\!\shortrightarrow\!\!\ZZ_{2}$ & $\ZZ\!\!\shortrightarrow\!\!\ZZ$ & $\ZZ^{3}\!\!\shortrightarrow\!\!\ZZ^{3}$ \\
AII & $\ZZ^{2}\!\!\shortrightarrow\!\!0$ & $\ZZ_{2}\!\!\shortrightarrow\!\!0$ & $\ZZ_{2}\!\!\shortrightarrow\!\!\ZZ_{2}$ & $\ZZ_{2}^{3}\!\!\shortrightarrow\!\!\ZZ_{2}^{2}$ \\
CII & $\ZZ_{2}\!\!\shortrightarrow\!\!0$ & $\cdot$ & $\ZZ_{2}\!\!\shortrightarrow\!\!\ZZ_{2}$ & $\ZZ_{2}^{3}\!\!\shortrightarrow\!\!\ZZ_{2}^{2}$ \\
C & $\ZZ\!\times\!\ZZ_{2}\!\!\shortrightarrow\!\!0$ & $\ZZ\!\!\shortrightarrow\!\!0$ & $\cdot$ & $\cdot$ \\
CI & $\cdot$ & $\cdot$ & $\ZZ\!\!\shortrightarrow\!\!\ZZ$ & $\ZZ^{3}\!\!\shortrightarrow\!\!\ZZ^{3}$ \\
\hline\hline
\end{tabularx}}
\end{table}

\addtocounter{table}{-1}

\begin{table}[H]
\centering
\caption{\textit{(Continued)}}
{\scriptsize\begin{tabularx}{\linewidth}{XXXXX}
\hline\hline
$\mathrm{R3m}\vphantom{\dot{|}}$ & $\Gamma$ & $L$ & $F$ & $Z$ \\
\hline
A & $\ZZ^{3}\!\!\shortrightarrow\!\!0$ & $\cdot$ & $\cdot$ & $\ZZ\!\!\shortrightarrow\!\!\ZZ$ \\
AIII & $\cdot$ & $\ZZ^{2}\!\!\shortrightarrow\!\!\ZZ^{2}$ & $\ZZ\!\!\shortrightarrow\!\!\ZZ$ & $\ZZ\!\!\shortrightarrow\!\!\ZZ$ \\
\hline
AI & $\ZZ^{2}\!\!\shortrightarrow\!\!0$ & $\cdot$ & $\cdot$ & $\ZZ\!\!\shortrightarrow\!\!\ZZ$ \\
BDI & $\cdot$ & $\ZZ\!\!\shortrightarrow\!\!\ZZ$ & $\cdot$ & $\ZZ\!\!\shortrightarrow\!\!\ZZ$ \\
D & $\ZZ\!\!\shortrightarrow\!\!0$ & $\cdot$ & $\cdot$ & $\cdot$ \\
DIII & $\cdot$ & $\ZZ\!\!\shortrightarrow\!\!\ZZ$ & $\ZZ\!\!\shortrightarrow\!\!\ZZ$ & $\cdot$ \\
AII & $\ZZ^{2}\!\!\shortrightarrow\!\!0$ & $\cdot$ & $\ZZ_{2}\!\!\shortrightarrow\!\!\ZZ_{2}$ & $\ZZ\!\!\shortrightarrow\!\!\ZZ$ \\
CII & $\ZZ_{2}\!\!\shortrightarrow\!\!0$ & $\ZZ\!\!\shortrightarrow\!\!\ZZ$ & $\ZZ_{2}\!\!\shortrightarrow\!\!\ZZ_{2}$ & $\ZZ\!\times\!\ZZ_{2}\!\!\shortrightarrow\!\!\ZZ\!\times\!\ZZ_{2}$ \\
C & $\ZZ\!\times\!\ZZ_{2}\!\!\shortrightarrow\!\!0$ & $\cdot$ & $\cdot$ & $\ZZ_{2}^{2}\!\!\shortrightarrow\!\!\ZZ_{2}^{2}$ \\
CI & $\cdot$ & $\ZZ\!\!\shortrightarrow\!\!\ZZ$ & $\ZZ\!\!\shortrightarrow\!\!\ZZ$ & $\ZZ_{2}\!\!\shortrightarrow\!\!\ZZ_{2}$ \\
\hline\hline
\end{tabularx}}
\end{table}

\addtocounter{table}{-1}

\begin{table}[H]
\centering
\caption{\textit{(Continued)}}
{\scriptsize\begin{tabularx}{\linewidth}{XXXXX}
\hline\hline
$\mathrm{R\bar{3}m}\vphantom{\dot{|}}$ & $\Gamma$ & $L$ & $F$ & $Z$ \\
\hline
A & $\ZZ^{6}\!\!\shortrightarrow\!\!0$ & $\ZZ^{2}\!\!\shortrightarrow\!\!0$ & $\ZZ\!\!\shortrightarrow\!\!0$ & $\ZZ^{3}\!\!\shortrightarrow\!\!\ZZ$ \\
AIII & $\cdot$ & $\cdot$ & $\cdot$ & $\cdot$ \\
\hline
AI & $\ZZ^{4}\!\!\shortrightarrow\!\!0$ & $\ZZ\!\!\shortrightarrow\!\!0$ & $\ZZ\!\!\shortrightarrow\!\!0$ & $\ZZ^{3}\!\!\shortrightarrow\!\!\ZZ$ \\
BDI & $\cdot$ & $\cdot$ & $\cdot$ & $\cdot$ \\
D & $\ZZ^{2}\!\!\shortrightarrow\!\!0$ & $\ZZ\!\!\shortrightarrow\!\!0$ & $\cdot$ & $\cdot$ \\
DIII & $\cdot$ & $\cdot$ & $\cdot$ & $\cdot$ \\
AII & $\ZZ^{4}\!\!\shortrightarrow\!\!0$ & $\ZZ\!\!\shortrightarrow\!\!0$ & $\ZZ\!\!\shortrightarrow\!\!\ZZ_{2}$ & $\ZZ^{3}\!\!\shortrightarrow\!\!\ZZ\!\times\!\ZZ_{2}$ \\
CII & $\ZZ_{2}^{2}\!\!\shortrightarrow\!\!0$ & $\cdot$ & $\ZZ_{2}\!\!\shortrightarrow\!\!\ZZ_{2}$ & $\ZZ_{2}^{3}\!\!\shortrightarrow\!\!\ZZ_{2}^{2}$ \\
C & $\ZZ^{2}\!\times\!\ZZ_{2}^{2}\!\!\shortrightarrow\!\!0$ & $\ZZ\!\!\shortrightarrow\!\!0$ & $\ZZ_{2}\!\!\shortrightarrow\!\!0$ & $\ZZ_{2}^{3}\!\!\shortrightarrow\!\!\ZZ_{2}$ \\
CI & $\cdot$ & $\cdot$ & $\cdot$ & $\cdot$ \\
\hline\hline
\end{tabularx}}
\end{table}

\addtocounter{table}{-1}

\begin{table}[H]
\centering
\caption{\textit{(Continued)}}
{\scriptsize\begin{tabularx}{\linewidth}{XXXXX}
\hline\hline
$\mathrm{P23}\vphantom{\dot{|}}$ & $\Gamma$ & $X$ & $M$ & $R$ \\
\hline
A & $\ZZ^{3}\!\!\shortrightarrow\!\!0$ & $\cdot$ & $\cdot$ & $\cdot$ \\
AIII & $\cdot$ & $\ZZ\!\!\shortrightarrow\!\!\ZZ$ & $\ZZ^{2}\!\!\shortrightarrow\!\!\ZZ^{2}$ & $\ZZ^{4}\!\!\shortrightarrow\!\!\ZZ^{4}$ \\
\hline
AI & $\ZZ^{2}\!\!\shortrightarrow\!\!0$ & $\cdot$ & $\cdot$ & $\cdot$ \\
BDI & $\cdot$ & $\cdot$ & $\cdot$ & $\ZZ\!\!\shortrightarrow\!\!\ZZ$ \\
D & $\ZZ\!\!\shortrightarrow\!\!0$ & $\cdot$ & $\cdot$ & $\cdot$ \\
DIII & $\cdot$ & $\ZZ\!\!\shortrightarrow\!\!\ZZ$ & $\ZZ^{2}\!\!\shortrightarrow\!\!\ZZ^{2}$ & $\ZZ^{3}\!\!\shortrightarrow\!\!\ZZ^{3}$ \\
AII & $\ZZ^{2}\!\!\shortrightarrow\!\!0$ & $\ZZ_{2}\!\!\shortrightarrow\!\!0$ & $\ZZ_{2}^{2}\!\!\shortrightarrow\!\!\ZZ_{2}$ & $\ZZ_{2}^{2}\!\!\shortrightarrow\!\!\ZZ_{2}$ \\
CII & $\ZZ_{2}\!\!\shortrightarrow\!\!0$ & $\ZZ_{2}\!\!\shortrightarrow\!\!0$ & $\ZZ_{2}^{2}\!\!\shortrightarrow\!\!\ZZ_{2}$ & $\ZZ\!\times\!\ZZ_{2}^{2}\!\!\shortrightarrow\!\!\ZZ\!\times\!\ZZ_{2}$ \\
C & $\ZZ\!\times\!\ZZ_{2}\!\!\shortrightarrow\!\!0$ & $\cdot$ & $\cdot$ & $\cdot$ \\
CI & $\cdot$ & $\ZZ\!\!\shortrightarrow\!\!\ZZ$ & $\ZZ^{2}\!\!\shortrightarrow\!\!\ZZ^{2}$ & $\ZZ^{3}\!\!\shortrightarrow\!\!\ZZ^{3}$ \\
\hline\hline
\end{tabularx}%
}
\end{table}
\renewcommand{\arraystretch}{1}

\begin{table*}[t]
\centering

\caption{Obtaining the complete topological classification of layer-group (LG) \#51 $\mathrm{p\bar{4}}$ and rod-group (RG) \#27 $\mathrm{p\bar{4}}$ from the invariants classifying space-group (SG) \#81 $\mathrm{P\bar{4}}$ which are provided in Table~\ref{tab:app}.
The ``~$\cdot$~" symbol indicates a trivial classification, i.e., $0\to0$.}\label{tab:P-4_layer_and_rod}

\renewcommand{\arraystretch}{1.15}
%
\renewcommand{\arraystretch}{1}
\end{table*}

\subsection{Layer-groups \& rod-groups}\label{sec:layer-and-rod}

The BZ of either a layer-group or a rod-group system is correspondingly either a 2D or a 1D sub-torus of some embedding 3D space-group. Therefore its topological classification is a direct sum of the topological invariants corresponding only to the HSM of the sub-torus.

As an example, consider space-group $\mathrm{P\bar{4}}$ discussed in Sec.~\ref{sec:example}. It's topological classification is obtained as a direct sum over the $\Gamma,Z,X,R,M,A$-points. The layer-group $\mathrm{p\bar{4}}$ is a subgroup of $\mathrm{P\bar{4}}$ and its 2D BZ contains only the $\Gamma,X,M$-points; see Fig.~\ref{fig:P-4_HSM}. The direct sum of the topological invariants of these HSM, which are tabulated in Table~\ref{tab:app}, provide the complete classification for layer-group $\mathrm{p\bar{4}}$. Similarly, the rod-group $\mathrm{p\bar{4}}$ is a subgroup of $\mathrm{P\bar{4}}$ and its 1D BZ contains only the $\Gamma$-point and the $Z$-point; see Fig.~\ref{fig:P-4_HSM}. The direct sum of the topological invariants of these HSM provide the complete classification for rod-group $\mathrm{p\bar{4}}$. This is explicitly demonstrated in Table~\ref{tab:P-4_layer_and_rod}.

\section{Mathematical background}\label{sec:math_background}
The main objects of study in this paper is the $G$-equivariant real $\KK$-theory $\KR^{p,q}_G(\bTt)$ of a torus $\bTt$ and the image of the $\AI$-map $\AItoK^{p,q} \colon \AI^{p,q} \to \KR^{p,q}_G(\bTt)$ in it; see Sec.~\ref{sec:surf}. For every value of $(p,q)$, the map $\AItoK^{p,q}$ is a map of abelian groups, and it is not so easy to compute these maps independently. One of the insights that we present is that these maps are highly interlaced. They are all different manifestations of a single map of objects, which depend on a much smaller number of parameters then a tuple of maps of abelian groups for every value of $(p,q)\in \ZZ^2$. To explain this common origin and dependencies between the various degrees, let us first give a ``shadow" of it on the level of cohomology groups. 

As discussed in Sec.~\ref{sec:mult}, the graded ring $\KR^{*,*}(\pt)$, given explicitly in Eq.~\eqref{eq:formula-for-KR-of-point} ahead, acts on $\AI^{*,*}$ and $\KR_G^{*,*}(\bTt)$, turning them into \emph{bi-graded modules} over the bi-graded ring $\KR^{*,*}(\pt)$. In other words, if $v\in \AI^{p,q}$ and $u\in \KR^{p',q'}(\pt)$ then 
\begin{equation}
\AItoK(u\cdot v)=u\cdot\AItoK(v) \in \KR_G^{p+p',q+q'}(\bTt).    
\end{equation}
This fact already provides many relations between the maps, $\AItoK^{p,q}$, for various $(p,q)$-s. As a first trivial application, using the multiplication by $\mu$ and $\beta$, we see that the map depends only on its $(p,0)$ component for $0\le p\le 7$, recovering the classical observation that the computation of the entire $\AItoK$ map reduces to those degrees. 

However, and especially for summands of $\AI^{*,*}$ of the form $\KC^{*,*}(\pt)$, this is \emph{not enough} to entirely determine the map $\AItoK$. Specifically, the module $\KC^{*,*}(\pt)$ is not a \emph{free} module over $\KR^{*,*}(\pt)$, and is not generated by a single element, but rather it is minimally generated by $1,\xi,\xi^2,$ and $\xi^3$. 
This deficiency naturally leads to the question of whether every map of $\KR^{*,*}(\pt)$-modules, i.e., a map that respects the multiplication by classes of the real Bott-periodicity, can appear as a matrix entry in a matrix representation of $\AItoK^{*,*}$. For example, one can ask whether every map $\KC^{*,*}(\pt)\to \KR^{*,*}(\pt)$ can appear in the computation. It turns out that this is not the case. For example, the map given by
\begin{equation}
\xi^{4k+m}\mu^\ell\mapsto\beta^k\mu^\ell\times\begin{cases} 2 & m=0, \\
0 & m=1,  \\ \alpha & m = 2, \\ 0 & m = 3, \end{cases}
\end{equation}
cannot appear as such a matrix entry, even though it is entirely consistent with the $\KR^{*,*}$-module structure. The reason for this phenomenon is that the map $\AItoK$ must come from a map of modules over the ring-spectrum $\KR$ itself, not just the cohomology ring $\KR^{*,*}(\pt)$. Namely, the map $\AItoK$ intertwines the operation of tensoring with a real vector space on the geometric level, before passing to homotopy groups. To make this idea precise, one needs to explain in what sense, and where, the object $\KR$ is a ring, the objects $\AI$ and $\KR_G(\bTt)$ are modules over $\KR$ and  $\AItoK\colon\AI\to \KR_G(\bTt)$ is a map of modules over $\KR$. A natural context to make these notions precise is that of \emph{$\ZT$-equivariant spectra}. It turns out that thinking about the classification problem in terms of equivariant spectra has many other advantages that we shall later explain. 

The rest of this appendix is organized as follows:
In Sec.~\ref{sec:rings}, we recall the classical theory of rings, modules, and representations of groups, both in the usual and the $\ZZ_2$-graded settings. In Sec.~\ref{sec:KRKC-theory}, we discuss real and complex $\KK$-theory, their multiplicative structure, and the relations between them. In Sec.~\ref{sec:spectra}, we consider the notions of spectra and $G$-equivariant spectra. We discuss the tensor product of spectra and the resulting theory of commutative rings and modules internally to spectra. We then specialize to the case of $G$-equivariant real $\KK$-theory and explain how to define it as a commutative ring in $G$-equivariant spectra. 
In sec.~\ref{sec:signed-permutation}, we study the $G$-equivariant $\KK$-theory of the torus associated with a special type of equivariant lattices which we call signed permutation representations. We show that the $G$-equivariant spectrum generated from such a torus splits as a direct sum of representation spheres, hence reducing our calculations to the case of spheres.   
In Sec.~\ref{sec:spins-and-reps}, we study the $G$-equivariant $\KK$-theory of representation spheres endowed with a Pin-structure. We show how a Pin-structure on the representation allows us to compute the $G$-equivariant $\KK$-theory of the representation sphere as a real $\KK$-theory module using only representation-theoretic considerations.  
Finally, in Sec.~\ref{sec:math-AI}, we use all the machinery introduced in the previous sections, and Shiozaki's formula, to give a uniform classification, in representation-theoretic terms, of the atomic insulators for crystals with signed permutation representation symmetry.

\subsection{Rings, representations, \& gradings}\label{sec:rings}

In the next sections, we shall discuss some ``brave new algebra", i.e., algebra done in the category of spectra. Hence, as a warm-up, we recall some classical notions from classical algebra.

\subsubsection{Rings \& modules}

Recall that a \emph{ring} is a set $R$ with two binary associative operations $+$ and $\cdot$, such that $+$ makes $R$ into an additive abelian group and the distributive laws hold: $(a+b)c = ac + bc$ and $a(b+c) = ab + ac$. We also demand the existence of a unit for the multiplication, $1\in R$, such that $1a=a1=a$. A typical example is the ring of real $n$ by $n$ matrices, $\Mat{n}(\RR)$ with addition and matrix multiplication. Note, that we do not require the multiplication to be commutative. If it does, i.e., \emph{if} $ab=ba$, we say that $R$ is a commutative ring. A typical example is the commutative ring $\ZZ$ of integers with addition and multiplication.

If $R$ is a ring, we can form the ring of polynomials, 
$R[x]=\{\sum_i r_i x^i : r_i\in R\}$, with its natural addition and multiplication of polynomials. More generally, if $R$ is a commutative ring and $f_1(x_1,..,x_\ell),\ldots,f_m(x_1,\ldots,x_\ell)$ are polynomials with coefficients in $R$, we can form the ring, 
\begin{equation}
\label{eq:polynomial-ring}
\frac{R[x_1,\ldots,x_\ell]}{(f_1(x_1,\ldots,x_\ell),\ldots,f_m(x_1,\ldots,x_\ell))},
\end{equation}
given by adding formal variables $x_i$ subject only to the relations, $f_j(x_1,\ldots,x_\ell)=0$, for $j=1,\ldots,m$. This will be a common way for us to present the rings involved in our computations. For example, in this convention, we have $\CC \simeq \RR[x]/(x^2 +1)$. 
In fact, the ring in Eq.~\eqref{eq:polynomial-ring} is an \emph{$R$-algebra}, i.e., endowed with a ring map from $R$. More precisely, it is the $R$-algebra with generators $x_i$ and relations $f_j(x_1,\ldots,x_\ell)$.

A \emph{module} over a ring $R$ is an abelian group $M$ on which $R$ acts linearly. More precisely, we have a binary operation, $(r,m)\mapsto rm$, which satisfies $(rr')\cdot m=r\cdot (r'\cdot m)$, $(r+r')\cdot m=r\cdot m + r'\cdot m$ and $r\cdot (m+m')=r\cdot m + r\cdot m'$. Finally, we demand that $1\cdot m = m$. Equivalently, if $M$ is an abelian group, the set of homomorphisms, $\Hom_\Ab(M,M)$, of abelian groups from $M$ to $M$ admits a ring structure, for which the addition and multiplication are given by $(f+g)(m)=f(m)+g(m)$ and $(fg)(m)=f(g(m))$, the multiplicative unit is the identity map $\Id_M$. An $R$-module structure on $M$ is just a ring homomorphism $R\to \Hom_{\Ab}(M,M)$. The collection of modules over $R$ organizes into a \emph{category} $\Mod_R(\Ab)$ of $R$-modules in abelian groups. This means that for every two $R$-modules $M,M'$ there is a set of $R$-linear maps, $\Hom_{R}(M,M')$, i.e., group homomorphisms for which $f(r\cdot m)=r\cdot f(m)$. In fact, $\Hom_{R}(M,M')$ has a canonical addition turning it into an abelian group. If $R$ is moreover \emph{commutative}, then $\Hom_{R}(M,M')$ admits an $R$-module structure, via 
 $(r \cdot f) (m) = r\cdot f(m) = f(r\cdot m)$, and we always consider $\Hom_{R}(M,M')$ as endowed with this module structure.   
In the case where $M'=M$ we also denote $\Hom_R(M,M)$ by $\End_R(M)$.

\subsubsection{Group rings \& representations}
If $G$ is a group and $R$ is a ring with a $G$-action given by
$r\xmapsto{g}g(r)\in R$,
one can produce the \emph{twisted} group ring, $R[G]$, given by formal sums, $\sum_{i} r_i g_i$, with addition, $rg+r'g=(r+r')g$,
and multiplication,
\begin{equation}\label{eq:grop-ring_mult}
rg\cdot r'g'=(r\cdot g(r'))(gg').
\end{equation}
In the special case where $G$ acts trivially, so that $g(r)=r$, we obtain the (non-twisted) \emph{group ring}, $R[G]$. 

The seemingly arbitrary construction of the twisted group ring is characterized up to Morita equivalence via a very simple property of the category of modules. Since the ingredients involved in the characterization play a crucial role in the passage from groups and algebras to spectra and ring spectra, we shall discuss it before moving on to graded objects.  
Let $\Ab^G$ denote the category of abelian groups with $G$-action. We can think of $R$ as a ring in $\Ab^G$, i.e., a ring with a $G$-action. More precisely, groups with $G$-action have a \emph{tensor product}, $(M,N)\mapsto M\otimes N$, given by the usual tensor product with the $G$-action $g(m\otimes n)=g(m)\otimes g(n)$. Then, the action of $G$ on the ring $R$ is encoded via a multiplication, $R\otimes R \to R$, where both the source and target are in $\Ab^G$, and the tensor product is the tensor product in $\Ab^G$.  Since $R$ is a ring in $\Ab^G$, it makes sense to talk about modules over $R$ internally to $\Ab^G$. We denote this category by $\Mod_R(\Ab^G)$. Explicitly, this is an object, $M\in \Ab^G$, together with action map, $R\otimes M\to M$, satisfying the usual module identities. 
We have an equivalence of categories 
\begin{equation}
    \Mod_{R[G]}(\Ab)\simeq \Mod_{R}(\Ab^{G}).
\end{equation}
In the special case where $R=\RR$ or $R=\CC$ with the trivial $G$-action, we recover the usual group rings $\RR[G]$ and $\CC[G]$. The categories $\Mod_{\RR[G]}(\Ab)$ and $\Mod_{\CC[G]}(\Ab)$ are the categories of real and complex representations of $G$ respectively. For a finite group $G$, the algebra $\CC[G]$ has a very simple description. If $\rho_1,\ldots,\rho_\ell$ are the different irreducible complex representations of $G$, and if $\dim(\rho_i)=d_i$, then 
\begin{equation}
\label{eq:Group_algebra_isomorphism_complex}
    \CC[G]\simeq \sum\nolimits_i \Mat{d_i}(\CC). 
\end{equation} 
To describe this isomorphism in a canonical way, we shall discuss the notion of \emph{group representation} in more detail. If $G$ is a finite group, a (complex) \emph{representation} of $G$ is a complex vector space $\cvs$ and a group homomorphism, $\rho\colon G\to \GL(\cvs)$, where $\GL(\cvs)$ is the group of invertible linear maps from $\cvs$ to $\cvs$. It is always possible to choose a Hermitian metric on $\cvs$ which is preserved by the $G$-action. By choosing such a metric and an orthonormal basis to $\cvs$, we can depict a $d$-dimensional representation as a group homomorphism $\rho\colon G\to \Unit(d)$. 

A representation, $\cvs$, of $G$ is \emph{irreducible} if there is no $G$-invariant decomposition, $\cvs\simeq \cvs_1 \oplus \cvs_2$. Let $\rho_1,\ldots,\rho_\ell$ be the set of pairwise non-isomorphic irreducible representations of $G$. 
The map $\rho_i$, given in some orthonormal basis, by $\rho_i\colon G \to \Unit(d_i) \subseteq \Mat{{d_i}}(\CC)$, extends by linearity to a map, $\CC[G] \to \Mat{{d_i}}(\CC)$. The isomorphism in Eq.~\eqref{eq:Group_algebra_isomorphism_complex} is then the direct sum of the these extended maps. 

In the real case, the situation is slightly more complicated, as real irreducible representations naturally split into 3 kinds. To describe those, we shall consider some standard linear algebra constructions related to real and complex vector spaces. 

A real vector space, $\rvs$, has a \emph{complexification}, $\cc(\rvs) \defeq \rvs \otimes_\RR \CC$. Similarly, a complex vector space, $\cvs$, has a \emph{realification}, $\rr(\cvs) \simeq \cvs$, considered as a real vector space by forgetting the multiplication with complex scalars. Accordingly, a real representation, $\tau \colon G\to \GL(\rvs)$, can be complexified, so that
\begin{equation}
\label{eq:Complexification-of-representation}
\xymatrix{
G \ar[r]_-\tau \ar@/^1pc/[rr]^{\cc(\tau)} & \GL(\rvs) \ar@{}[r]|-{\displaystyle\subset} & \GL(\cc(\rvs)),
} 
\end{equation}
and a complex representation, $\rho \colon G\to \GL(\cvs)$, can be realified via
\begin{equation}
\xymatrix{
G \ar[r]_-\rho \ar@/^1pc/[rr]^{\rr(\rho)} & \GL(\cvs) \ar@{}[r]|-{\displaystyle\subset} & \GL(\rr(\cvs)).
}
\end{equation}
Finally, a complex vector space $\cvs$ has a complex conjugate $\conj{\cvs}$, which is the same real vector space with the scalar multiplication defined by $a\cdot v \defeq \conj{a}v$. Moreover, this construction is \emph{functorial}, i.e., a complex-linear map, $A\colon U_1\to U_2$, induces a linear map, $\conj{A}\colon \wideconj{U_1}\to \wideconj{U_2}$. Accordingly, a complex representation, $\rho\colon G\to \GL(\cvs)$, has a complex conjugate, 
\begin{equation}
\xymatrix{
G \ar[r]_-\rho \ar@/^1pc/[rr]^{\conj{\rho}} & \GL(\cvs) \ar@{}[r]|-{\displaystyle\simeq} & \GL(\conj{\cvs}).
}
\end{equation}
In coordinates, the map $\conj{\rho}\colon G\to \Unit(d)$ is the composition $G\xrightarrow{\rho} \Unit(d) \xrightarrow{A\mapsto \conj{A}} \Unit(d)$.
The operations, $\cc$ and $\rr$, satisfy the identities, $\cc (\rr(\cvs))\simeq \cvs\oplus \conj{\cvs}$ and $\rr (\cc (\rvs)) \simeq \rvs \oplus \rvs$.

Let $\tau\colon G\to \GL(\rvs)$ be an irreducible real representation. The representation, $\rho \defeq \cc(\tau)$, is equipped with an involutive isomorphism, $\ConjOp\colon \conj{\rho} \xrightarrow{\sim} \rho$, coming from the complex conjugation on $\CC$. 
In fact, a real representation can be identified with a complex representation endowed with such an involutive isomorphism with the complex conjugate.
If $\cc(\tau)$ decomposes as a direct sum of two different irreducible representations, $\cc(\tau) \simeq \rho_1 \oplus \rho_2$, then $\End_G(\tau)\simeq \CC$, and $\tau$ contributes a summand of the form $\Mat{d}(\CC)$ to $\RR[G]$. We say in this case that $\tau$ is of \emph{complex type}. Moreover, in this case $\rho_2 \simeq \conj{\rho}_1$ and $\rr(\rho_1)\simeq \rr(\rho_2)\simeq \tau$.  

If $\rho = \cc(\tau)$ is an irreducible complex representation, then the involutive isomorphism, $\ConjOp\colon \conj{\rho} \arsim \rho$, endows $\rho$ with a real structure, and so $\tau$ contributes a summand of the form $\Mat{d}(\RR)$ to $\RR[G]$, with $d=\dim(\tau)$. In this case, we say that $\tau$ is of \emph{real type}, and one has $\rr(\rho)\simeq \tau \oplus \tau$.

Finally, it is possible that $\cc(\tau) \simeq \rho \oplus \rho$ for a single complex irreducible representation, $\rho$. In this case we have an isomorphism, $\ConjOp\colon \conj{\rho} \arsim \rho$, but we can not choose $\ConjOp$ to be involutive. Rather, the composition, $\rho \oto{\conj{\ConjOp}} \conj{\rho} \oto{\ConjOp} \rho$, can be chosen to be the scalar linear map, $-\Id_\rho$. Then, the representation $\tau$ contributes a copy of $\Mat{d/4}(\HH)$ to $\RR[G]$, where $d=\dim(\tau)$ and $\HH$ is the quaternion algebra, i.e., the $\RR$-algebra generated by $i$ and $j$ with $i^2=-1$, $j^2=-1$, and $ij=-ji$. We say that $\tau$ is of \emph{quaternionic type}; one has $\rr(\rho) \simeq \tau$ and $\cc(\tau) \simeq \rho \oplus \rho$. 

To conclude, we have an isomorphism, 
\begin{multline}\label{eq:Group_algebra_isomorphism_real}
\RR[G]\simeq
\bigoplus_{\tau \text{ real}} \Mat{d_\tau}(\RR)  \oplus\bigoplus_{\tau \text{ complex}}  \Mat{\frac{d_\tau}{2}}(\CC) \\ \oplus\bigoplus_{\tau \text{ quaternionic}} \Mat{\frac{d_\tau}{4}}(\HH).
\end{multline}
 
Representations of different groups are related. If $H\subseteq G$ is a subgroup and $\tau$ is a real, or complex, representation of $G$, we can regard it as a representation of $H$, which we denote by $\Res^G_H(\tau)$. In the other direction, if $\rho$ is a representation of $H$, we can form the induced representation, $\Ind_H^G(\rho)\defeq \Hom_{\RR[H]}(\RR[G],\rho).$
The induction and restriction functors are related via the adjunction relation, 
\begin{equation}
\Hom_G(\tau, \Ind_H^G(\rho)) \simeq \Hom_H(\Res_H^G(\tau),\rho).
\end{equation}

Representation theory provides us with interesting examples of rings. 
If $G$ is a group, we denote by $\CRep(G)$ the \emph{Grothendieck ring} of complex representations of $G$. Namely, elements of $\CRep(G)$ are formal integral combinations, $\sum_i m_i [\rho_i]$, of complex representations, subject to the relation, $[\rho \oplus \rho']=[\rho]+[\rho']$. The product of two elements in the ring is given by the tensor product of representations, $[\rho][\rho']\defeq [\rho \otimes \rho']$. Specifically, if $\rho_1 \colon G\to \GL(\cvs_1)$ and $\rho_2 \colon G\to \GL(\cvs_2)$ then $\rho_1 \otimes \rho_2 \colon G\to \GL(\cvs_1 \otimes \cvs_2)$ is given by $(\rho_1 \otimes \rho_2)(g) \defeq \rho_1(g)\otimes \rho_2(g)$.   

Similarly, we denote by $\RRep(G)$ the Grothendieck ring of real representations of $G$. There is a ring homomorphism, $\cc \colon \RRep(G)\to \CRep(G)$, turning $\CRep(G)$ into an $\RRep(G)$-algebra. In the opposite direction, we have a map, $\rr \colon \CRep(G)\to \RRep(G)$, of $\RRep(G)$-modules. Namely, for $u\in \CRep(G)$ and $v\in \RRep(G)$ we have 
\begin{equation}
\label{eq:projection_formula_rep_rings}
    \rr(\cc(v)u)=v\rr(u).
\end{equation}
Moreover, since the complexification is a ring homomorphism, one has
\begin{equation}
    \cc(vv')=\cc(v)\cc(v').
\end{equation}
Importantly, the realification is \emph{not} a ring homomorphism but rather a module homomorphism and hence no analogous relation exists for $\rr$, cf.~Eq.~\eqref{eq:projection_formula_rep_rings}.
Furthermore, the ring $\CRep(G)$ admits an involution, $u\mapsto \conj{u}$, reflecting the complex conjugation of complex representations. The operations, $\cc(v)$, $\rr(u)$ and $\conj{u}$, satisfy the relations,
\begin{align}
\label{eq:relations_complexification_realification_conjugation_rep_ring}
\rr(\cc(v)) &= 2v, \nonumber\\
\cc(\rr(u)) &= u + \conj{u}, \nonumber\\ 
\rr(\conj{u}) &= \rr(u).
\end{align}

\subsubsection{\texorpdfstring{$\ZZ_2$}{Z/2}-gradings} 
Computations within the $G$-equivariant $\KK$-theory of spaces requires not only representations and rings but also with their $\ZZ_2$-graded versions. 

A $\ZZ_2$-graded abelian group is an abelian group, $A$, with a decomposition, $A= A^0 \oplus A^1$. We call $A^0$ the even part of $A$ and $A^1$ the odd part of $A$. We let $\grAb$ denote the category of $\ZZ_2$-graded abelian groups. 
For $A\in \grAb$, we say that $a\in A$ is \emph{homogeneous} if either $a\in A^0$ or $a\in A^1$. In this case the parity of $a$ is simply
\begin{equation}
|a| = \begin{cases} 0 & a\in A^0, \\ 1 & a\in A^1.\end{cases}
\end{equation}
We also have a tensor product, $A\gotimes B$, on $\grAb$ given by
\begin{align}
(A\gotimes B)^0 &= (A^0 \otimes B^0) \oplus (A^1 \otimes B^1), \nonumber\\
(A\gotimes B)^1 &= (A^0 \otimes B^1) \oplus (A^1 \otimes B^0).
\end{align} 
Accordingly, we can define $\ZZ_2$-graded rings and modules over such. A $\ZZ_2$-graded ring is a $\ZZ_2$-graded abelian group $R$ endowed with an associative and distributive multiplication, $R\gotimes R \to R$, as well as a unit, $1\in R^0$. We say that $R$ is \emph{super-commutative} if $ab = (-1)^{|a||b|} ba$ for homogeneous $a,b\in R$. 
We denote the category of $\ZZ_2$-graded modules over $R$ by $\Mod_R(\grAb)$. Specifically, an object, $M\in \Mod_R(\grAb)$, is a $\ZZ_2$-graded abelian group $M$ endowed with an associative and distributive action, $R\gotimes M \to M$. 

One way to produce, not necessarily super-commutative, $\ZZ_2$-graded rings, is as group rings of $\ZZ_2$-graded groups. Here, a $\ZZ_2$-graded group is a group, $G$, endowed with a homomorphism, $\parity \colon G\to \ZZ_2$. We then set $G^0 \defeq \parity^{-1}(0)$ and $G^1 \defeq \parity^{-1}(1)$, such that $G^0$ is the even subgroup and $G^1$ is the odd part which is \emph{not} a group. 
If $G$ is a $\ZZ_2$-graded group, a $\ZZ_2$-graded (complex or real) representation of $G$ is a (complex or real) representation, $\rho \colon G\to \GL(\cvs)$, for a $\ZZ_2$-graded vector space $\cvs$, such that $|\rho(g)(u)|=(-1)^{\parity(g)}|u|$ for every $g\in G$ and homogeneous $u\in \cvs$.  

The $\ZZ_2$-graded complex representations of $G$ are encoded in the $\ZZ_2$-graded group ring, $\CC^{\parity}[G]$, which is $\ZZ_2$-graded via $\CC^{\parity}[G]^0 \defeq \CC[G^0]$ and $\CC^{\parity}[G]^1 \defeq\CC[G^1]$ (note that the latter is \emph{not} a group ring). Similarly, we have $\RR^{\parity}[G]^0\defeq\RR[G^0]$ and $\RR^{\parity}[G]^1 \defeq \RR[G^1]$. 
The $\ZZ_2$-graded group ring, $\CC^{\parity}[G]$, factors into a sum of simple $\ZZ_2$-graded algebras according to the classification of the irreducible complex $\ZZ_2$-graded representations. In order to provide this factorization, we recall the notion of Clifford algebras. 

\subsubsection{The Clifford algebras}

For $p\in \NN$, we denote by $\CCl_p$ the Clifford algebra on $p$ generators. Namely, $\CCl_p$ is the $\ZZ_2$-graded algebra over $\CC$ generated by the \emph{odd} elements, $\cliff{1},\ldots,\cliff{p}$, with $\cliff{i}^2 = -1$, and $\cliff{i} \cliff{j} = -\cliff{j} \cliff{i}$ for $i\ne j$.
Similarly, for $p,q\in \NN$, the real Clifford algebras, $\Cl_{p,q}$, are generated by the elements, $\cliffM{1},\ldots,\cliffM{p},\cliffP{p+1},\ldots,\cliffP{p+q}$, subject to the relations, 
\begin{align}\label{eq:clifford-def}
\cliffM{i}^2 &= -1, & i &=1,\ldots,p, \nonumber\\
\cliffP{i}^2 &= +1, & i &=p+1,\ldots,p+q, \nonumber\\
\cliff{i}\cliff{j}&=-\cliff{j}\cliff{i}, & i &\ne j,
\end{align}
and such that that the generators are all \emph{odd}, i.e., $|\cliff{i}|=1$. Particularly, one has $\CCl_{p+q}=\CC\otimes_\RR \Cl_{p,q}$. The first few real Clifford algebras are presented in Table~\ref{tab:real-Clifford}.

\begin{table}[t]
\centering
\caption{Real Clifford algebras, $\Cl_{p,q}$. The even part of each algebra is isomorphic to $\Cl_{p-1,q}$ found one row above it. The first few algebras are explicitly displayed, all others are obtained via Bott-periodicity, $\Cl_{p+1,q+1}\simeq\Mat{2}(\Cl_{p,q})$ and $\Cl_{p+4,q}\simeq\Cl_{p,q+4}\simeq\Mat{2}(\HH)\otimes\Cl_{p,q}$.}
\label{tab:real-Clifford}
\begin{tabularx}{\linewidth}{lXXXXXr}
\hline\hline
$p \backslash q\quad$ &  0 & 1 & 2 & 3 & 4 & $\cdots$\\
\hline
& & $\RR$ & $\CC$  & $\HH$  & $\HH\oplus\HH$\\
\hline
0 & $\RR$ & $\RR\oplus\RR$ & $\Mat{2}(\RR)$ & $\Mat{2}(\CC)$ & $\Mat{2}(\HH)$\\
1 & $\CC$ & $\Mat{2}(\RR)$ & $\Mat{2}^{\oplus 2}(\RR)$ & $\Mat{4}(\RR)$ & \\
2 & $\HH$ & $\Mat{2}(\CC)$ & $\Mat{4}(\RR)$ & & \\
3 & $\HH\oplus\HH$  & $\Mat{2}(\HH)$ & & & \\
4 & $\Mat{2}(\HH)$ & & & & \\
$\vdots$ \\
\hline\hline
\end{tabularx}
\end{table}

Let $G$ be a group which is $\ZZ_2$-graded by $\parity\colon G\to \ZZ_2$. The $\ZZ_2$-graded algebra $\CC^{\parity}[G]$ decomposes as a direct sum of $\ZZ_2$-graded algebras of the form $\Mat{d}(\CCl_p)$. Similarly, The $\ZZ_2$-graded algebra $\RR^{\parity}[G]$ decomposes as a direct sum of $\ZZ_2$-graded algebras of the forms, $\Mat{d}(\CCl_p)$ and $\Mat{d}(\Cl_{p,q})$. This decomposition is tied with the vital notion of $\ZZ_2$-graded representations which we now explain.

\subsubsection{\texorpdfstring{$\ZZ_2$}{Z/2}-graded representations}
\label{sec:Z_2-graded}

Let us begin with complex representations. First, if $\parity$ is trivial, so that $G^0=G$, then $\CC^{\parity}[G]$ is entirely even, and the decomposition coincides with the one in Eq.~\eqref{eq:Group_algebra_isomorphism_complex}. 

Otherwise, the construction, $\cvs\mapsto \cvs^0$, induces an identification~\cite{vershik2008new},
\begin{equation}
\label{eq:Identification_graded_reps_reps_of_even_part}
\Mod_{\CC^{\parity}[G]}(\grAb)\simeq \Mod_{\CC[G^0]}(\Ab),  
\end{equation}
with inverse given by $\cvs\mapsto \cvs\otimes_{\CC[G^0]}\CC[G] \simeq \Ind_{G_0}^G(U)$. 
Hence, irreducible $\ZZ_2$-graded representations of $G$ are in direct correspondence with irreducible representations of $G^0$. Let us fix an arbitrary odd element, $g\in G^1$. If $\tau$ is an irreducible representation of $G^0$, so that $\rho = \Ind_{G^0}^G(\tau)$ is an irreducible $\ZZ_2$-graded representation of $G$, we can use $g$ to twist the representation $\tau$ and obtain a new representation, $\tau^{g}$, such that $\tau^{g}(h)\defeq \tau(ghg^{-1})$. The relation between $\tau$ and $\tau^g$ determines the contribution of $\tau$ to $\CC^\parity[G]$. As with the classification of real representations, there are two types of irreducible representations of $G^0$. 

If $\tau^g\simeq \tau$, we necessarily have $\Ind_{G^0}^G(\tau) \simeq \rho_1 \oplus \rho_2$ for two different irreducible representations, $\rho_1$ and $\rho_2$, of $G$. In this case, the $\ZZ_2$-graded representation $\rho$ contributes a copy of $\Mat{d}(\CCl_1)$ to $\CC^{\parity}[G]$ for $d=\dim{\tau}$. The underlying ungraded $\CC$-algebra is then isomorphic to $\Mat{d}(\CC)\oplus \Mat{d}(\CC)$ and corresponds to the summands of $\CC[G]$ coming from $\rho_1$ and $\rho_2$. 

If $\tau$ and $\tau^g$ are different irreducible representations, then $\Ind_{G^0}^G(\tau)$ is irreducible as an ungraded representation. In this case, as ungraded representations we have $\Ind_{G^0}^G(\tau)\simeq \Ind_{G^0}^G(\tau^g)\simeq \rho$ and so, there are two non-equivalent $\ZZ_2$-gradings on $\rho$. In this case, the pair $(\tau,\tau^g)$ contributes a copy of $\Mat{d}(\CCl_2)$ to $\CC^{\parity}[G]$. The underlying ungraded algebra is $\Mat{2d}(\CC)$, which is the summand of $\CC[G]$ corresponding to $\rho$.  

To conclude, if $\parity\colon G\to \ZZ_2$ is non-trivial, we have
\begin{equation}
\label{eq:Decomposition_of_group_algebra_complex_Z2_graded}
\CC^{\parity}[G]\simeq \bigoplus_{\tau\neq\tau^g} \Mat{d_\tau}(\CCl_1) \oplus \bigoplus_{\tau = \tau^g} \Mat{d_\tau}(\CCl_2).
\end{equation}

Moving on to real representations, one finds that the decomposition of a $\ZZ_2$-graded group ring $\RR^{\parity}[G]$ into real Clifford algebras is more complicated as it involves both the complex conjugation and the twisting by $g\in G^1$. First, if $G^0=G$, then $\RR^{\parity}[G]$ is entirely even, and the decomposition coincides with the one in Eq.~\eqref{eq:Group_algebra_isomorphism_real}. Otherwise, there are ten different types of real $\ZZ_2$-graded representations which are summarized in Table~\ref{tab:real-Clifford-Irr}. These correspond to the ten AZ symmetry classes.

\begin{table}[t]
\centering
\caption{The ten irreducible $\ZZ_2$-graded representations and their corresponding matrix algebras. For each irreducible ungraded representation $\tau$ of $G^0$, there is a $\ZZ_2$-graded representation, $\rho = \Ind_{G^0}^G(\tau)$, of $G$. It is determined by the real, complex, and quaternionic structures on $\tau$ and $\rho$, and on whether $\tau$ and $\tau^g$ are isomorphic or not for $g\in G^1$.}\label{tab:real-Clifford-Irr}
\begin{tabularx}{\linewidth}{llXXX}
\hline\hline
$\tau\backslash\rho$ & & $\RR$-type & $\CC$-type & $\HH$-type\\
\hline
\multirow{2}{*}{$\RR$-type} & $\tau=\tau^g$ & $\tMat{d_\tau}(\Cl_{0,1})$ & $\tMat{d_\tau}(\Cl_{1,0})$ & \\
& $\tau\neq\tau^g$ & $\tMat{d_\tau}(\Cl_{1,1})$ & &\\
\multirow{2}{*}{$\CC$-type} & $\tau=\tau^g$ & & $\ttMat{\frac{d_\tau}{2}}(\CCl_{1})$\\
& $\tau\neq\tau^g$ & $\ttMat{\frac{d_\tau}{2}}(\Cl_{0,2})$ & $\ttMat{\frac{d_\tau}{2}}(\CCl_{2})$ & $\ttMat{\frac{d_\tau}{2}}(\Cl_{2,0})$\\
\multirow{2}{*}{$\HH$-type} & $\tau=\tau^g$ & & $\ttMat{\frac{d_\tau}{4}}(\Cl_{0,3})$ & $\ttMat{\frac{d_\tau}{4}}(\Cl_{3,0})$ \\
& $\tau\neq\tau^g$ & & & $\ttMat{\frac{d_\tau}{4}}(\Cl_{4,0})$\\
\hline\hline
\end{tabularx}
\end{table}

\subsection{Real and complex \texorpdfstring{$\KK$}{K}-theory}\label{sec:KRKC-theory}

\subsubsection{The ring structure on \texorpdfstring{$\KU$}{KU}}
Complex $\KK$-theory is a classical generalized cohomology theory introduced by Atiyha and Hirzebruch~\cite{atiyah1961vector}, see also Ref.~\onlinecite{bott1969lectures}. Recall that a \emph{vector bundle} on a topological space $X$ is a choice of a finite dimensional complex vector space $\cvs_x$ for every $x\in X$, depending continuously, in an appropriate sense, on $x\in X$. We let $\KU^0(X)$ be the abelian group generated by all complex vector bundle on $X$, subject to the relation, $[\cvs_1]+[\cvs_2]= [\cvs_1\oplus \cvs_2]$. Every element in $\KU^0(X)$ can be written as a formal difference, $[\cvs_1]-[\cvs_2]$, which we refer to as a \emph{virtual bundle}, i.e., an hypothetical object that, if added to the vector bundle $\cvs_2$, yields the honest vector bundle $\cvs_1$. 

The functoriality of $\KU^0$ is described via the pullback of vector bundles, i.e., if $f\colon X\to Y$ is a continuous map and $\cvs$ is a complex vector bundle on $Y$, one can form the vector bundle, $(f^*U)_x = U_{f(x)}$. We hence obtain a homomorphism of abelian groups, $f^*\colon \KU^0(Y)\to \KU^0(X)$.

It turns out that $\KU^0(X)=\pi_0(\Map(X,\sC_0))$, and so one can extend $\KU$ to a cohomology theory by setting 
$\KU^{-q}(X)=\pi_0(\Map(X,\sC_q))$, see discussion in Sec.~\ref{sec:intro_class}. 
While the definition based on maps to the classifying space $\sC_q$ exhibits the additive structure on $\KK$-theory through loop concatenation in $\sC_{q}\simeq \Omega \sC_{q-1}$, the cohomology theory $\KU$ also admits a multiplicative structure.  
The tensor product of vector bundles, given by $(\cvs_1\otimes \cvs_2)_x \defeq (\cvs_1)_x \otimes (\cvs_2)_x$ endows $\KU^0$ with a multiplicative structure, and this in turn extend to all other degrees, turning $\KU$ into a \emph{multiplicative} generalized cohomology theory. 
Namely, we have a multiplication, $\KU^p(X)\otimes \KU^q(X) \to \KU^{p+q}(X)$, for every space $X$, which is also super-commutative, i.e., $uv = (-1)^{pq} vu$ for $u\in \KU^p(X)$ and $v\in \KU^{q}(X)$. This multiplication endows  
$\KU^*(X)\defeq \bigoplus_q \KU^q(X)$ with a super-commutative graded ring structure.
In the case where $X=\pt$, the ring structure on $\KU^*(\pt)$ reveals a tight relationship between the different levels of $\KK$-theory. In fact, there is a class $\xi \in \KU^{-2}(\pt) = \trKU^0(\Ss^2)$, which is invertible in the ring $\KU^*(\pt)$ and such that 
$\KU^{2\ell}(\pt)=\ZZ\cdot \xi^{-\ell}$. 
One summarizes this situation by the following shorthand notation, 
\begin{align}
&\KU^*(\pt)=\ZZ[\xi,\xi^{-1}], &|\xi|=-2.
\end{align}
Here, the elements in the brackets represent generators of the ring $\KU^*(\pt)$ and $|\xi|$ stands for the degree of $\xi$ in the cohomology ring, e.g., if $|u|=q$ then $u\in \KU^{q}(\pt)$. The complex Bott-periodicity is now just multiplication by $\xi^{-1}$; see Table~\ref{tab:KCKR}.

\subsubsection{Complex conjugation \& real \texorpdfstring{$\KK$}{K}-theory}
The cohomology theory $\KU$ admits a canonical action of $\ZT$, through its compatible action on the $\sC_q$-s. At degree $q=0$, this action corresponds to the conjugation of complex vector bundles. Namely, a complex vector bundle $\cvs$ has a \emph{complex conjugate} $\conj{\cvs}$, given by $\conj{U}_x \defeq \wideconj{U_x}$. On $\KU^*(\pt)$, this action is given by 
\begin{equation}
    \conj{\xi}^k=(-1)^k\xi^k.
\end{equation}
Using the ring structure, we see that this formula is \emph{imposed} by the simple fact that $\conj{\xi}=-\xi$, along with the observation that $\cvs\mapsto\conj{\cvs}$ is a ring homomorphism: $\wideconj{\cvs_1\otimes \cvs_2}\simeq \wideconj{\cvs_1}\otimes \wideconj{\cvs_2}$.   

The action of $\ZT$ through complex conjugation on complex vector bundles endows the cohomology theory $\KU$ with a $\ZT$-equivariant structure, i.e., turns it into a cohomology theory defined on topological spaces endowed with a $\ZT$-action. The resulting  $\ZT$-equivariant cohomology theory, introduced by Atiyah in Ref.~\onlinecite{atiyah1966k}, is denoted by $\KR$. Being $\ZT$-equivariant, the cohomology theory $\KR$ is naturally \emph{bi-graded}; we can form the $(p,q)$-cohomology, $\KR^{p,q}(X)$, for a space $X$ with a $\ZT$-action. This bi-grading is induced from the two natural based circles with $\ZT$-action: The one with the trivial action, denoted simply $\Ss^1$, and the one with the action by reflection along an axis, denoted $\bSs^1$. Namely, we have 
$\KR^{p,q}(\Ss^{d} \wedge \bSs^{d'} \wedge (X\sqcup \pt)) \simeq \KR^{p-d,q-d'}(X)$, where $X\sqcup \pt$ is the disjoint union of $X$ and a point. 

The bi-graded $\KR$-theory of the point is slightly more complicated than its non-equivariant counterpart $\KU$. Namely, $\KR^{*,*}(\pt)\defeq \bigoplus_{p,q}\KR^{p,q}(\pt)$ admits the following presentation, as a super-commutative graded ring~\cite{atiyah1964clifford,bott1969lectures}, by generators and relations:
\begin{equation}
\label{eq:formula-for-KR-of-point}
    \KR^{*,*}(\pt) =  \frac{\ZZ[\eta,\alpha,\beta,\beta^{-1},\mu,\mu^{-1}]}{(2\eta,\eta^3,\eta\alpha,\alpha^2-4\beta)},
\end{equation}
where the $(p,q)$-degrees of the generators are given by  
\begin{align}
|\eta| &=(-1,0), & |\alpha|&=(-4,0),  \nonumber\\ 
|\beta| &= (-8,0), & |\mu|&= (-1,-1), 
\end{align}
see Refs.~\onlinecite{Rognes2015Topological,dobson2007ko}.
These generators span the abelian groups summarized in Table~\ref{tab:KCKR} (cf.~Table~\ref{tab:per}) and the relations yield the multiplication rules summarized in Table~\ref{tab:KRmult}.

The group $\KR^{0,0}(X)$ for a space $X$ with a $\ZT$-action given by the involution, $\inv\colon X\to X$, can be interpreted as follows. It is the abelian group spanned by complex vector bundles $\cvs$ on $X$ and endowed with an involutive identification, $\ConjOp\colon \wideconj{\cvs_x} \arsim \cvs_{\inv(x)}$, which depends continuously on $x\in X$. The classes of complex vector bundles are subject to the relation $[\cvs]+[\cvs']=[\cvs\oplus \cvs']$. 
In particular, if $\inv$ acts trivially on $X$, we recover the group of \emph{real} vector bundles on $X$, hence we have an identification of $\KR^{p,0}(X)$ with the real, eight-fold periodic, $\KK$-theory of $X$, usually denoted $\KO^p(X)$.  

In fact, via this relation, one is ought to think of $\KO$ as being the \emph{fixed points} of $\KR$ by the $\ZT$-action. Similarly, we introduce a $\ZT$-equivariant cohomology theory $\KC$, by assigning a $\ZT$-space its $\KU$-cohomology, regardless of the $\ZT$-action. Then, we can view $\KU$ as the fixed points of $\KC$. The bi-graded $\KC$-cohomology of the point consist of infinitely many shifted copies of $\KU^*(\pt)$, namely, 
\begin{equation}
    \KC^{*,*}(\pt)\simeq \ZZ[\xi^{\pm 1},\mu^{\pm 1}].
\end{equation}

The above cohomology theories are related by some maps between them, reflecting operations on real and complex vector spaces. We have a \emph{complexification} map, $\cc\colon \KR\to \KC$, which forgets the identification of a vector bundle with its twisted complex conjugate. In the other direction, we have a (\emph{non-multiplicative}) \emph{realification} map, $\rr\colon \KC\to \KR$, which takes a vector bundle $\cvs$ to the sum, $\cvs\oplus \inv^*\conj{\cvs}$, on which the map $\ConjOp$ is given by
\begin{equation}
    \ConjOp \colon \wideconj{\cvs\oplus \inv^*\conj{\cvs}} \simeq \conj{\cvs} \oplus \inv^*\cvs \oto{\left(\begin{smallmatrix}0 & 1 \\ 1 & 0\end{smallmatrix}\right)}\inv^*\cvs \oplus \conj{\cvs} \simeq \inv^*(\cvs \oplus \inv^*\conj{\cvs}). 
\end{equation}
Note, that these operations descend to the fixed points and provide maps $\KO\rightleftarrows \KU$, but we shall only consider the $\ZT$-equivariant versions in this paper.

The effect of these maps on the generators is summarized in Table~\ref{tab:KCKR}. We shall explore the precise role and origin of these maps via the language of $\ZT$-spectra, introduced in the next section.

\subsection{Spectra and equivariant spectra}\label{sec:spectra} 

\subsubsection{The category of spectra}
To define a cohomology theory, such as $\KU$, one have to provide an infinite delooping of a pointed space $X_0$. Namely, to find a sequence of pointed spaces $X_i$ and homotopy equivalences, $\Omega X_i \simeq X_{i-1}$, which extrapolates the sequence $X_{-i} \defeq \Omega^i X_0$ to all integers. From such a sequence, we get a generalized cohomology theory by taking the $i$-th cohomology of $X'$ to be $\pi_0(\Map(X',X_i))$. A sequence of the form $\{X_i\}_{i\in \ZZ}$ as above is classically known as a \emph{spectrum}, and in this case $X_0$ is an \emph{infinite loop space}. To see that this construction provides a cohomology theory, one has to introduce an abelian group structure on $\pi_0(\Map(X',X_i))$. This structure is induced from the identification, 
\begin{equation}
    \pi_0(\Map(X',X_i))\simeq \pi_2(\Map(X',X_{i+2})),
\end{equation} 
via the canonical abelian group structure on $\pi_2$ of a space by concatenation of loops-of-loops. 

The above description of the addition in the cohomology theory $\KU$ is, however, unnecessarily complicated. We have defined the additive group structure on $\KU^0(X)$ via the direct sum operation on complex vector bundles. Thinking of this operation as a result of a loop concatenation in a delooping of the classifying space $\sC_0$ is a less straight-forward construction.  
One may think of the situation as the exact opposite. We can consider the existence of the addition of vector bundle as the \emph{reason}, rather than the \emph{consequence}, of a delooping to $\sC_0$~\footnote{The situation is analogous with the one already present in the theory of groups. If $G$ is an abstract group, one can a-posteriori endow the set $G$ with a multiplication given by loop concatenation in $\Omega K(G,1)\simeq G$, where $K(G,1)$ is the quotient of a contractible space by a free $G$-action. This indeed recovers the group structure of $G$, but should not be thought of as the \emph{reason} for its existence, which was given to us a-priori.}.  
This line of thought leads to a  model of spectra which is much more algebraic, as we shall now see.

As mentioned above, the notion of a spectrum is analogous to that of an abelian group. We can think of spectra as a version of abelian groups where one replaces the set of elements with a \emph{space} of elements, i.e., a nice enough topological space, well defined up to homotopy equivalence.  
For every two spaces, $X$ and $Y$, we can form the space of maps $\Map(X,Y)$, allowing continuous families of maps, $X\to Y$, depending on a continuous parameter. We denote the collection of spaces by $\spaces$. 
Such a structure, in which maps between objects form a space, is called an \emph{$\infty$-category}, see Ref.~\onlinecite{Lurie2009Higher}. We shall not get into the details of what that precisely means, but informally, this is a machinery designed to encode algebraic and geometric structures in which identities are satisfied only \emph{up to coherent homotopy}. We shall soon see examples of this phenomenon.  

It is now natural to ask what a commutative group in $\spaces$ should be. We start with the notion of a commutative monoid. 
Recall that a \emph{commutative monoid} is a set, $M$, with a commutative and associative binary operation, $+$, which admits a zero element, $0$. Similarly, one can define a commutative monoid in $\spaces$, to be a space $M$ endowed with a binary operation, $+$, which is associative and commutative up to coherent homotopy and with a zero object, $0$, which is neutral to the addition up to coherent homotopy. Let us denote the collection (in fact, $\infty$-category) of such commutative monoids by $\CMon(\spaces)$. To give a flavor of what that means,  we replace, for example, the demand that the associativity relation $(a+b)+c=a+(b+c)$ holds, with the structure of a \emph{specified} homotopy between the left-hand side and the right-hand side of this equation, varying continuously in $(a,b,c)\in M^3$. Similarly, we require a choice of a homotopy between the expressions, $a+b$ and $b+a$, depending continuously on $(a,b)\in M^2$, and a further hierarchy of higher homotopies connecting these basic ones.  

A (usual) commutative monoid $M$ is an abelian group, precisely when the map, $M\times M\to M\times M$, given by 
\begin{equation}
(a,b)\mapsto (a,a+b),
\end{equation}
is a bijection. Similarly, we say that a commutative monoid, $M\in \CMon(\spaces)$,  is \emph{group-like} if 
the map, $M\times M\to M\times M$, given by the same formula is a homotopy equivalence. It turns out that a group-like commutative monoid is the same as a spectrum with only non-negative homotopy groups. Namely, to specify a group-like commutative monoid, one has to specify a sequence of pointed spaces 
$\{X_i\}$ with chosen homotopy equivalences, $\Omega X_i\simeq X_{i-1}$, and such that the connectivity assumption, $\pi_i(X_j) =0$ for $i<j$, holds. This is a special case of a general principle called the \emph{May recognition principle}~\cite{may2006geometry}, saying concisely that
\begin{equation}
\begin{gathered}
\{\text{group-like commutative monoids}\}\\
\updownarrow\\
\{\text{Infinite loop spaces}\}.
\end{gathered}
\end{equation}
In other words, all group-like additions on a space are canonically given by concatenation of loops in a suitable delooping. Nevertheless, it is usually more natural to allow other types of group-like monoid structures, without explicitly identifying them with loop concatenation, as mentioned above.  

We shall be interested in spectra which have negative homotopy groups, such as $\KU$. To achieve this, we note that if $M\in \CMon(\spaces)$, then $\Omega M$ is automatically a group-like commutative monoid, and the loop concatenation and pointwise addition on $\Omega M$ coincide up to homotopy. Spectra are obtained from commutative monoids in $\spaces$ by formally inverting the loop operation,
\begin{equation}
\Sp \defeq \CMon(\spaces)[\Omega^{-1}].
\end{equation} 
Objects of $\Sp$ then correspond to sequences $\{X_i\}$ without the connectivity assumption.
The operation $\Omega^{-1}$ is denoted $\Sigma$, as it is related, but not to be confused with, the suspension operation on pointed spaces. 
If $E$ is a spectrum, we can now define its homotopy groups $\pi_i(E)$ for all integer values $i\in\ZZ$. If $i\ge 0$, we take $\pi_i(E)$ to be the $i$-th homotopy of the commutative monoid corresponding to $E$. For negative homotopies, we set $\pi_{-i}(E) = \pi_0 (\Sigma^i E)$.   

\subsubsection{Constructions in the category of spectra}\label{sec:constructions-in-spectra}
The analogy, $\Sp \approx \Ab$, suggests that constructions in abelian groups may be adapted to spectra. For example, an abelian group $A$ has an underlying set of elements, obtained by forgetting the addition. Similarly, a spectrum $E$ has an underlying space, $\Under{E}\in \spaces$. It has the property that $\pi_i(\Under{E}) = \pi_i(E)$ for $i\ge 0$.
For every two spectra, $E$ and $E'$, there is a mapping space, $\Map(E',E)\in \spaces$. In fact, similarly to the fact that $\Hom_\Ab(A,B)$ for abelian groups, $A$ and $B$, has a canonical abelian group structure, there is a \emph{mapping spectrum},
\begin{equation}\label{eq:mapping-spectrum}
E(E')\defeq\Hom_{\Sp}(E',E)\in \Sp,
\end{equation}
with an underlying space $\Map(E',E)$. 
 
The abelian group, $\ZZ$, has as an analog the \emph{sphere spectrum}, $\SSp$. It is the free spectrum on a point, just like $\ZZ$ is the free abelian group on a point. More generally, if $X$ is a set, we have the free abelian group, 
\begin{equation}
\textstyle\ZZ[X] \defeq \left\{\sum\nolimits_i n_i x_i : x_i \in X, n_i\in \ZZ\right\}.    
\end{equation}

Similarly, there is a \emph{free spectrum on a space}, $\SSp[X]$, characterized by the property,
\begin{equation} 
\Under{\Hom_{\Sp}(\SSp[X],Y)}\simeq \Map_{\spaces}(X,\Under{Y}).
\end{equation} 
For historical reasons, this operation is usually denoted $\Sigma^\infty$, but we use $\SSp[-]$ to emphasize the analogy with the free abelian group construction. 

For two abelian groups, $A$ and $B$, we can form their direct sum, $A\oplus B$. There is a similar operation for spectra, compatible with homotopy groups. Moreover, we have $\SSp[X\sqcup Y]\simeq \SSp[X]\oplus \SSp[Y]$. There is also a tensor product of abelian groups, $A\otimes B$, classifying bilinear maps from $A\times B$. Similarly, $\Sp$ has a symmetric monoidal structure, $\otimes$, which is uniquely characterized by the property that $\SSp[X]\otimes\SSp[Y]\simeq \SSp[X\times Y]$ for $X,Y\in \spaces$ and is compatible with the gluing of spectra in the appropriate sense; see section 4.8.2 of Ref.~\onlinecite{Lurie2017Higher}. This operation is classically known as the \emph{smash product} and usually denoted $E \wedge E'$ for spectra $E$ and $E'$. In particular, the tensor product of spectra distributes over the direct sum, 
\begin{equation}
(E'\oplus E'')\otimes E \simeq (E'\otimes E) \oplus (E''\otimes E).
\end{equation}

Having a symmetric monoidal structure $\otimes$, we can define commutative rings and modules over them, as in section 4.5 of Ref.~\onlinecite{Lurie2017Higher}. Namely, a commutative ring in $\Sp$ is a spectrum $R$ endowed with a multiplication map, $R\otimes R\to R$, and a unit, $\SSp\to R$, which are associative, commutative, and unital up to coherent homotopy. Similarly, a module over such a ring $R$ is a spectrum $M$ endowed with an action, $a\colon R\otimes M\to M$, which is associative up to coherent homotopy, and for which the unit of $R$ acts as the identity up to coherent homotopy.  

Spectra also give rise to cohomology theories. If $E$ is a spectrum and $X$ is a space, we can form the spectrum,
\begin{equation}
E(X) \defeq \Hom_{\Sp}(\SSp[X],E).
\end{equation}
Then, we obtain the cohomology by setting
\begin{equation} 
E^{-i}(X) \defeq \pi_{i}(E(X)).
\end{equation}
It can be shown that this definition give a cohomology theory for every spectrum $E$, hence relating spectra to cohomology theories, cf.~Eq.~\eqref{eq:Hom-BZ-KR}. Similarly, one can turn a spectrum $E$ into a homology theory by setting 
\begin{equation}
E_i(X) \defeq \pi_i (E\otimes \SSp[X]).
\end{equation}

\subsubsection{The \texorpdfstring{$\KK$}{K}-theory spectrum \& the monoid of vector spaces}

The cohomlogy theory $\KU$, determined by the infinite loop space $\{\sC_q\}$, has a multiplicative structure turning it into a multiplicative cohomology theory. Our aim now is to directly describe a construction of the spectrum $\KU$ as a commutative monoid and lift its multiplicative structure to $\Sp$.  

Let $\Vect_\CC$ denote the collection of complex vector spaces. We can view $\Vect_\CC$ as a space, whose points are complex vector spaces and the space of maps from $\cvs_1$ to $\cvs_2$ is the (possibly empty) space of invertible linear transformations from $\cvs_1$ to $\cvs_2$.

The direct sum operation of vector spaces turns $\Vect_\CC$ into a commutative monoid in $\spaces$. Of course, it is not group-like, as one can not subtract vector spaces. To turn it into a group-like commutative monoid, one formally adds inverses to the elements of $\Vect_\CC$. This operation, known as \emph{group completion}, is the homotopy-theoretic analog of the way $\KU^0(X)$ is obtained from the monoid of vector bundles on $X$ by adding inverses to the elements. 
We denote the group completion of a commutative $M$ by $M^\gp$. The spectrum $\Vect_\CC^\gp$ is then a version of $\KU$ with only non-negative homotopy groups. To get $\KU$, one then inverts the class, $\xi\in \pi_2(\Vect_\CC^\gp)$.
More precisely, the tensor product of vector spaces endows the group-like commutative monoid $\Vect_\CC^\gp$ with a commutative ring structure in $\Sp$ and $\pi_2(\Vect_\CC^\gp)\simeq \ZZ$ is generated by a class $\xi$. Then, formally adding  its inverse, we get $\KU = \Vect_\CC^\gp[\xi^{-1}]$, which is a commutative ring spectrum lifting the cohomology theory $\KU^*(X)$ to the spectral level; namely,
\begin{equation}
    \KU^{q}(X)\simeq \pi_{-q}(\KU(X)).
\end{equation}

\subsubsection{\texorpdfstring{$G$}{G}-Spectra} 
\label{sec:G-spectra}

Similarly to the spectra, we can define an $\infty$-category, $\Sp^G$, of spectra with a $G$-action. 
To do so, we mimic the line of thought presented in the previous section, but replacing spaces with $G$-spaces, i.e., spaces endowed with a $G$-action. Namely, let $\spaces^G$ be the $\infty$-category whose objects are CW-complexes with action of the finite group $G$, and whose mapping spaces $\Map_{\spaces^G}(X,Y)$ consist of maps, $f\colon X\to Y$, for which $f(g(x))=g(f(x))$ for every $g\in G$ and $x\in X$.  

As for the non-equivariant version, we can define (group-like) $G$\emph{-commutative monoids}. The reader is warned that these are not just commutative monoids in $\spaces^G$. Namely, to define such a monoid, one should specify summation maps not only for $I$-indexed families of elements of the monoid, whenever $I$ is a finite set of indices. Rather, we should have such a map for every finite set $I$ with, possibly non-trivial, $G$-action.
For example, if $G= \ZZ_2$, then a $G$-monoid, $E$, should have not only a $G$-equivariant summation map, $E\times E\to E$, but also a \emph{trace map}, into the $\ZZ_2$-fixed points, given informally by $a \mapsto a + \inv(a)$ for $\inv$ the generator of $\ZZ_2$. See Ref.~\onlinecite{barwick2020spectral} for further details.

To obtain the category $\Sp^G$, one inverts the formation of loops on $G$-commutative monoids: 
\begin{equation}
    \Sp^G \defeq \CMon_G[\Omega^{-1}].
\end{equation}
The $\infty$-category, $\Sp^G$, is endowed with a tensor product, $\otimes$, and moreover, for every pair, $E,E'\in \Sp^G$, there is a spectrum of homomorphisms, $\Hom_{G}(E,E')\in \Sp$.
There is a `\emph{free $G$-spectrum} construction, $\SSp[-]\colon \spaces^G\to \Sp^G$, analogous to the non-equivariant version, which similarly turns disjoint unions to direct sums and products to tensor products.   

The constructions of $\spaces^G$ and $\Sp^G$ for various groups, $G$, interact. Namely, if $X$ is a $G$-space and $H\subseteq G$, we can regard $X$ as an $H$-space by restriction of the $G$-action to $H$. We denote this operation by either $\Res^G_H(X)$ or $X|_H$. Conversely, if $X$ is an $H$-space we can turn it into a $G$-space in two different ways. On one hand, we can turn $X$ into a $G$-space via $X\mapsto G\times_H X$, informally described as the quotient of $G\times X$ by the relation $(gh,x)\sim (g,h(x))$, where $G$ acts via the first coordinate. This construction satisfies $\Map_G(G\times_H X,Y)\simeq \Map_H(X,Y|_H)$. On the other hand, we can consider the construction, $X\mapsto \Map_H(G,X)$, with $G$ acting on the source.  This construction satisfies
\begin{equation}
\label{eq:adjunction-for-wrong-induction}
    \Map_G(X,\Map_H(G,Y))\simeq \Map_H(X|_H,Y).
\end{equation}

Shifting our attention to spectra, the spectral analog of the functor, $G\times_H X$, is the functor, $\Ind_H^G\colon \Sp^H \to \Sp^G$. It satisfies 
\begin{equation}
\Ind_H^G(\SSp_H[X])\simeq \SSp[G\times_H X],
\end{equation}
as well as an adjunction relation,
\begin{equation}
    \Hom_G(\Ind_H^G(E'),E)\simeq \Hom_H(E',\Res_H^G(E)),
\end{equation}
similar to the induction and restriction of representations.

The spectral analog of the functor, $\Map_H(G,X)$, is more complicated, and called the \emph{Hopkins-Hill-Ravenel norm}; see Ref.~\onlinecite{hill2016nonexistence}. It is informally defined by 
$\Norm_H^G(E)= \bigotimes_{G/H} E$ for $E\in \Sp^H$, with the action of $G$ given by permuting the factors and acting on each of them simultaneously. 
The norm is related to $\Map_H(G,X)$ via 
\begin{equation}
\label{eq:Norm-of-free} 
\Norm_H^G(\SSp[X]) \simeq \SSp[\Map_H(G,X)].
\end{equation}

The Hopkins-Hill-Ravenel norm interacts with the direct sum and tensor product of $H$-spectra in a way which we shall now describe. Namely, first of all, being itself an iterated tensor product, it commutes with the tensor product, so that 
\begin{equation}
\Norm_H^G(E\otimes E')\simeq \Norm_H^G(E)\otimes \Norm_H^G(E'). 
\end{equation}
The interaction with the direct sum operation is more complicated. If $H\subseteq G$ is a subgroup, and $E_1,\ldots,E_m$ are $H$-spectra, we want to decompose $\Norm_H^{G}(\bigoplus_{i=1}^m E_i)$. To explicitly present this decomposition, more notation is needed. For a function, $f\colon G/H \to \{1,\ldots,m\}$, denote 
\begin{equation}
    G_f = \{ g\in G \colon f([gg']) = f([g']) \forall g'\in G\}.
\end{equation} 
The set, $G/H$, decomposes into $G_f$-orbits, and we denote the set of orbits by $G_f\backslash G/H$. We then have
\begin{multline}
\label{eq:Norm-of-sum-formula}
\Norm_H^G\left(\bigoplus\nolimits_{i=1}^m E_i\right) \\
\simeq \bigoplus\nolimits_{f} \Ind_{G_f}^G \left(
\bigotimes\nolimits_{[g]\in G_f\backslash G/H} \Norm_{G_{f}\cap gHg^{-1}}^{G_f}(E'_{[g]})\right),
\end{multline}
where the external sum run over all representatives of orbits of $G$ on the set of functions $f\colon G/H\to \{1,..,m\}$, and $E'_{[g]}$ is the $gHg^{-1}$-spectrum corresponding to $E_{f([g])}$ under the conjugation isomorphism, $gHg^{-1} \simeq H$.  
To illustrate what this formula means for two summands, we have
\begin{equation}
    \Norm_H^G(E_1\oplus E_2)\simeq \bigoplus_{A\sqcup B= G/H} E_1^{\otimes A} \otimes E_2^{\otimes B},
\end{equation}
with an appropriate $G$-action.

\begin{table*}[t]
\centering
\caption{All possible $\KR$-module homomorphisms. Any particular spectra-homomorphism descends to a $\KR^{*,*}(\pt)$-linear map, for which the images of all $v\in\KR^{p,q}(\pt)$ and $u\in\KC^{p,q}(\pt)$ in all bi-degrees ($p,q$) are determined by at most two elements, either $v'\in\KR^{*,*}(\pt)$ or $u',u''\in\KC^{*,*}(\pt)$. The last column contains the complexified $\KC^{*,*}(\pt)$-linear maps.}
\label{tab:KR-hom}
\renewcommand{\arraystretch}{1}
\begin{tabularx}{\linewidth}{XlXl}
\hline\hline
Spectra-homomorphism & Elements & $\KR$-theory map & $\KC$-theory map\\
\hline
$\Hom_{\KR}(\KR,\KR) \simeq \KR$ & $v'$ & $v \mapsto v'\cdot v$ & $u \mapsto \cc(v')\cdot u$\\ 
$\Hom_{\KR}(\KR,\KC) \simeq \KC$ & $u'$ & $v\mapsto u'\cdot \cc(v)$ & $u\mapsto (u'\cdot u) \oplus (\conj{u}'\cdot u)$\\ 
$\Hom_{\KR}(\KC,\KR) \simeq \KC$ & $u'$ & $u\mapsto\rr(u'\cdot u)$ & $u_1\oplus u_2 \mapsto u'\cdot u_1+\conj{u}'\cdot u_2$\\ 
$\Hom_{\KR}(\KC,\KC) \simeq \KC \oplus\KC$ & $u',u''$ & $u \mapsto u'\cdot u + u''\cdot\conj{u}$ & $u_1\oplus u_2 \mapsto (u'\cdot u_1 + u''\cdot u_2)\oplus(\conj{u}'\cdot u_1 + \conj{u}''\cdot u_2)$\\
\hline\hline
\end{tabularx}
\renewcommand{\arraystretch}{1}
\end{table*}

If $E\in \Sp^G$, then we can define $\pi_n(E) = \pi_0(\Hom_G(\SSp,\Omega^nE))$. More generally, if $\rho\colon G\to \Ort(d)$ is a real linear representation of $G$, we can define a $G$-spectrum, $\SSp^{\rho}$, as follows. Denote by  $\Ss^\rho \in \spaces^G$ the $G$-space obtained from $\RR^d$ via one point compactification, on which $G$ acts via $\rho$. We set $\SSp^\rho$ to be the fiber of the map, $\SSp[\Ss^\rho] \to \SSp$, obtained from the constant map $\Ss^\rho \to \pt$ by applying $\SSp[-]$. In particular, since $\Ss^\rho$ has a $G$-fixed point at the north pole of the one point compactification, we have 
\begin{equation}
\label{eq:Direct-sum-decomposition-for-equivariant-sphere} 
\SSp[\Ss^\rho] \simeq \SSp \oplus \SSp^\rho. 
\end{equation}
The construction, $\rho \mapsto \SSp^\rho$, nicely interacts with the tensor product and the norm of $G$-spectra. For example,
\begin{equation}
\label{eq:Tensor-product-representation-spheres} 
\SSp^{\rho_0} \otimes \SSp^{\rho_1} \simeq \SSp^{\rho_0 \oplus \rho_1}.  
\end{equation}
This allows us to define the $\RRep(G)$-graded homotopy groups,
\begin{equation}\label{eq:rho-homotopy}
\pi_\rho(E)\defeq \pi_0(\Hom_G(\SSp^\rho,E)).
\end{equation}
In fact, this makes sense also when $\rho$ is a virtual representation.
The behavior of the Hopkins-Hill-Ravenel norm on representation spheres is also relatively simple. Namely, we have 
\begin{equation}
\label{eq:norm-of-representation-sphere}
    \Norm_H^G(\SSp_H^\rho) \simeq \SSp^{\Ind_H^G(\rho)}, 
\end{equation}
where $\Ind_H^G$ stands for the induction of real representations.

\subsubsection{The \texorpdfstring{$G$}{G}-equivariant \texorpdfstring{$\KK$}{K}-theory spectrum}\label{sec:G-K-spec}

Having the language of $G$-spectra set up, we can discuss the spectral versions of equivariant $\KK$-theory. We start with $\KR$. The commutative monoid in spaces, $\Vect_\CC$, admits an involution, given by complex conjugation of complex vector spaces. 
This involution is obtained from a $\ZT$-equivariant commutative monoid structure on $\Vect_\CC$, for which the fixed points are $\Vect_\RR$ and the trace map, $\Vect_\CC \to \Vect_\RR$, is the realification map.

Applying an equivariant version of group completion, we get a $\ZT$-commutative ring in $\Sp^{\ZT}$. We get $\KR$ by inverting the classes $\beta$ and $\mu$. Since $\Rep(\ZT)\simeq \ZZ\oplus \ZZ$, the homotopy groups of $\KR$ assemble to a bi-graded commutative ring, $\KR^{p,q}(\pt)=\pi_{-p,-q}(\KR)$; cf. Eq.~\eqref{eq:rho-homotopy}. Hence, $\KR$ is a commutative ring in $\Sp^{\ZT}$, lifting the $\ZT$-equivariant cohomology theory, $X\mapsto\KR^{*,*}(X)$.

A more complicated, yet similar, construction provides the $G$-equivariant $K$-theory, $\KR_G$. It is a $G\times \ZT$-equivariant commutative ring spectrum obtained via $G$-group completion (and inverting certain classes) from a $G$-commutative monoid structure on $\Vect_\CC$, whose $H$-fixed points, for $H\subseteq G$, are the monoids in $\spaces$ consisting of $H$-representations. For $H_0 \subseteq H_1 \subseteq G$, the trace map associated with the surjection, $G/H_0 \to G/H_1$, for this $G$-commutative monoid, is given by induction of representations from $H_0$ to $H_1$.   

To get the analogous $\KC_G$, one applies the general induction and restriction machinery constructed in the previous section. Namely, 
\begin{equation}
    \KC_G \defeq \Ind^{G\times \ZT}_G \Res^{G\times \ZT}_G \KR_G.
\end{equation} In particular, the underlying spectrum of $\KR_G$ is $\KU$, while the underlying spectrum of $\KC_G$ is $\KU\oplus \KU$. The $\ZT$ action for $\KR_G$ corresponds to the conjugation action on $\KU$, while for $\KC_G$ it corresponds to the action, $(u,u')\mapsto (u',\conj{u})$, on $\KU \oplus \KU$.
The complexification map, $\cc\colon \KR_G\to \KC_G$, then corresponds to a map, $\KU \to \KU \oplus \KU$, given by $u\mapsto (u,\conj{u})$. Similarly, the realification map, $\rr\colon \KC_G\to \KR_G$, corresponds to a summation map, $\KU \oplus \KU \to \KU$, given by $(u,u')\mapsto u+\conj{u}'$. 
The reader might find these formulas similar to the realification and complexification maps on $\RR$ and $\CC$. Indeed, the relation between $\KC$ and $\KR$ is a categorification of the relation between $\CC$ and $\RR$, replacing real/complex numbers by real/complex vector spaces, addition by direct sum, and multiplication by tensor product. 

We shall now return to the main reason we are interested in these spectral level constructions, that is, the classification of maps between $\KR$-modules (see Table~\ref{tab:KR-hom}). 
Note, that the map, $\rr\colon \KC\to \KR$, is a map of $\KR$-modules. This is a consequence of the obvious fact from linear algebra that tensoring with a real vector space commutes with realification. Similarly, the map, $\cc\colon \KR \to \KC$, is a map of commutative rings in $\Sp^{\ZT}$. If $M,N$ are modules over $\KR$, we have a $\KR$-module of linear maps between them, $\Hom_{\KR}(M,N)\in \Mod_{\KR}(\Sp^{\ZT})$. Indeed, we can multiply a $\KR$-linear map with a scalar from $\KR$ to get a new $\KR$-linear map, just like we do for classical, non-equivariant rings.  In fact, 
\begin{align}
\Hom_{\KR}(\KR,\KR) &\simeq \KR, \nonumber\\
\Hom_{\KR}(\KR,\KC) &\simeq \KC, \nonumber\\
\Hom_{\KR}(\KC,\KR) &\simeq \KC, \nonumber\\
\Hom_{\KR}(\KC,\KC) &\simeq \KC \oplus\KC.
\end{align}
Note, that these identifications are of $\KR$-modules, and hence contain all the needed information on the maps of arbitrary $(p,q)$-degrees. For example, 
\begin{equation}
    \pi_0(\Hom_{\KR}(\KR,\Sigma^{p,q} \KR))\simeq \pi_0(\Sigma^{p,q}\KR) \simeq \KR^{p,q}(\pt).
\end{equation}
 
To get these formulas, we note that $\KR$ is free over itself, and so $\Hom_\KR(\KR,M)\simeq M\in \Mod_{\KR}(\Sp^{\ZT})$. 
To see that $\Hom_{\KR}(\KC,\KR)\simeq \KC$, we note that 
\begin{multline}
    \Hom_{\KR}(\KC,\KR)\simeq \Hom_{\KR}(\Ind_{\trivg}^{\ZT}(\KU),\KR) \\ 
    \simeq\Ind_{\trivg}^{\ZT}\Hom_{\KU}(\KU,\KU)\simeq \Ind_{\trivg}^{\ZT}\KU\simeq \KC. 
\end{multline}
Finally, by scalar extension, we have  
\begin{align}
\Hom_{\KR}(\KC,\KC) &\simeq \Hom_{\KC}(\KC\otimes_{\KR}\KC,\KC), \nonumber\\  
\KC\otimes_\KR \KC &\simeq \KC \oplus \KC,  
\end{align}
analogous to the fact that $\CC\otimes_\RR \CC \simeq \CC \oplus \CC$. 

While derived from fairly abstract considerations, it is not hard to translate those formulas into maps of cohomology theories. We consider them as \emph{constraints} on the possible maps of the associated $\ZT$-equivariant cohomology theories in presence of a genuine $\KR$-linear structure on the map, lifting the na\"{i}ve $\pi_{*,*}(\KR)$-modules structure. 
Specifically, a map, $\KR \to \KR$, of bi-degree ($p,q$) is given by multiplication by a class in $\pi_{p,q}(\KR) \simeq \KR^{-p,-q}(\pt)$. A map, $\KR \to \KC$, of any bi-degree is of the form $v\mapsto u'\cc(v)$ where $u'\in \pi_{*,*}(\KC)$.
A map, $\KC\to \KR$, is of the form $u\mapsto \rr(u'u)$ for $u'\in \pi_{*,*}(\KC)$. Finally, a map, $\KC\to \KC$, is of the form $u\mapsto u'u +u''\conj{u}$ for $u',u''\in \pi_{*,*}(\KC)$. Any of these maps may be complexified by tensoring with $\KC$. This produces $\pi_{*,*}(\KC)$-linear maps by the identifications,
\begin{align}
\KR &\otimes_\KR\KC \simeq \KC,  & \KC &\otimes_\KR\KC \simeq \KC\oplus\KC, \nonumber\\
v' &\otimes u \leftrightarrow \cc(v')u, & u' &\otimes u \leftrightarrow u'u\oplus\conj{u}'u.
\end{align}
The above is summarized in Table~\ref{tab:KR-hom}.

This is the precise meaning of our claim that the map $\AItoK$ in all bi-degrees depend on a small number of parameters: Being a map of $\KR$-modules, it is specified by a class, or at most two classes, in either complex or real $\KK$-theory per matrix coefficient, for all bi-degrees ($p,q$) combined.  

\subsection{Signed permutation representations}\label{sec:signed-permutation}
Each space-group symmetry has a normal subgroup of translations corresponding to a Bravais lattice. Its reciprocal lattice, $\Lambda$, defines the geometry of the BZ torus, $\BZ=\RR^3/\Lambda$, with an action of the point-group, $G$. Since $\bTSp{3}=\SSp[\BZ]$ plays an important role in the topological classification, we hereby describe the generic construction of $\TSp_\Lambda$ for any $G$-lattice $\Lambda$.

One way to produce periodic lattices with $G$-action is by starting with a linear representation, $\rho \colon G\to \Ort(d)$, and a basis, $\lv_1,\dots,\lv_d$, such that for every $i=1,\ldots,d$ there is $j=1,\ldots,d$ for which $\rho(g)\lv_i = \pm \lv_j$. We call such a lattice, a signed permutation lattice for $G$, and $\rho$ a \emph{signed permutation representation}. 
If $G$ acts on a lattice $\Lambda\subseteq \RR^d$, one may form the torus, $\Tt_\Lambda= \RR^d / \Lambda \in \spaces^G$. Applying $\SSp[-]$, we obtain a $G$-spectrum, $\TSp_\Lambda \defeq \SSp[T_\Lambda] \in \Sp^G$. 

Let $\rho\colon G\to \Aut(\Lambda)$ be a signed permutation representation, with basis, $\lv_1,\ldots,\lv_n$, for which $\rho(g)\lv_i = \pm \lv_j$ and let $\Lambda$ be the lattice generated by this basis.
We wish to describe the $G$-spectrum, $\TSp_\Lambda$. 
The construction, $\Lambda \mapsto \TSp_\Lambda$, satisfies 
\begin{equation}
    \TSp_{\Lambda \oplus \Lambda'}\simeq \SSp[T_{\Lambda}\times T_{\Lambda'}] \simeq \TSp_{\Lambda}\otimes \TSp_{\Lambda'}\in \Sp^G,
\end{equation} 
and so it essentially suffices to describe $\TSp_{\Lambda}$ for \emph{indecomposable} lattices $\Lambda$. 

In light of the previous discussion, we may assume that $\Lambda$ is indecomposable, so that $G$ permutes the lines spanned by the $\lv_i$-s transitively. Let $\Lambda_1 = \ZZ \cdot \lv_1 \subseteq \RR$ and $G_1$ be the the stabilizer of the line $\RR \cdot \lv_1$ in $\RR^d$.  
We have a $G_1$-equivariant projection, $\proj\colon \Tt_\Lambda \to \RR / \ZZ\cdot \lv_1$, given by $\proj(a_1\lv_1 + \ldots + a_n \lv_d) = a_1 \lv_1$.  By Eq.~\eqref{eq:adjunction-for-wrong-induction}, this projection determines a $G$-equivariant map, 
\begin{equation}
\label{eq:multiplicatively-induced-indecomposable-torus}
\Tt_\Lambda \to \Map_{G_1}(G,\Tt_{\Lambda_1}).
\end{equation}
The map in Eq.~\eqref{eq:multiplicatively-induced-indecomposable-torus} is a $G$-equivariant \emph{homeomorphism} and, in particular, a $G$-equivariant homotopy equivalence. 
Using  Eq.~\eqref{eq:Norm-of-free} and Eq.~\eqref{eq:multiplicatively-induced-indecomposable-torus}, we see that 
\begin{equation}
\TSp_\Lambda \simeq \SSp[\Map_{G_1}(G,\Tt_{\Lambda_1})]\simeq \Norm_{G_1}^G(\TSp_{\Lambda_1}).
\end{equation}
To finish the analysis, let $\rho_1\colon G_1 \to \Ort(1)$ denote the representation of $G_1$ on the line $\RR \cdot \lv_1$. We have a direct sum decomposition, 
\begin{equation}
    \TSp_{\Lambda_1} \simeq \SSp_{G_1}[\Ss^{\rho_1}] \simeq \SSp_{G_1}\oplus \SSp_{G_1}^{\rho_1} \in \Sp^G.
\end{equation} 
It follows that $\TSp_\Lambda \simeq \Norm_{G_1}^{G}(\SSp_{G_1}\oplus \SSp_{G_1}^{\rho_1})$. 

Combining Eq.~\eqref{eq:Norm-of-sum-formula},
Eq.~\eqref{eq:Tensor-product-representation-spheres}, and Eq.~\eqref{eq:norm-of-representation-sphere},  we can finally give the following description of $\TSp_{\Lambda}$ for arbitrary signed permutation lattices. Each orbit of $G$ on $\lv_1,\ldots,\lv_d$ may be represented by a set $I_k\subseteq \{1,\ldots,d\}$ such that $\bigsqcup_{k=1}^\ell I_k=\{1,\ldots,d\}$.
Let $G_k$ be the stabilizer of $I_k$ and $\rho_k\colon G\to \Ort(|I_k|)$ be the representation of $G_k$ on the linear span of $\{\lv_i\}_{i\in I_k}$. We have the \emph{stable equivariant splitting},
\begin{equation}
\label{eq:stable-splitting}
\TSp_\Lambda \simeq \bigoplus\nolimits_{k=1}^{\ell}\Ind_{G_k}^{G}(\SSp^{\rho_{k}}).
\end{equation}
Hence, the study of the $G$-equivariant $\KK$-theory of the torus of a signed permutations representation reduces to that of representation spheres, which is covered by Refs.~\onlinecite{Cornfeld2019Classification,KAROUBI2002Equivariant}.

\subsection{Spins and representations}\label{sec:spins-and-reps}

\subsubsection{Quadratic representations \& Pin-structures}\label{sec:pin-structures}

For a finite group $G$, the representations of $G$ come into play in the computation of the $G$-equivariant $\KK$-theory in two different, interacting ways. First, as seen in Eq.~\eqref{eq:stable-splitting}, the $G$-spectrum $\TSp_\Lambda$ decomposes as a sum of representation sphere spectra. Second, the $G$-equivariant $\KK$-theory spectrum itself encodes $G$-representations. Hence, we shall now review some of the features of the theory that come into play in both aspects and in the interaction between them. 

Recall that we can think of a real representation of $G$ as a map, $\rho\colon G\to \GL(\rvs)$, for a real vector space $\rvs$. We are mostly interested in representations which are orthogonal with respect to a symmetric bilinear form. Namely, let $Q = Q_{ij}$ be a symmetric bilinear form on $\rvs$. We denote by $\Ort(Q)$ the group of linear maps, $A\colon \rvs\to \rvs$, such that $AQA^\transpose = Q$, or, in terms of a basis, such that 
$\sum_{ij}A_{ki} Q_{ij} A_{\ell j} = Q_{k \ell}$. 
For example, if $Q=\Id_p \oplus - \Id_q$ then $\Ort(Q) = \Ort(p,q)$. 
By a \emph{quadratic representation} of $G$  we mean a linear space $\rvs$ endowed with a non-degenerate bilinear form $Q$ and a homomorphism, $\rho\colon G\to \Ort(Q)$. Note, that $\Ort(Q)=\Ort(-Q)$ and so a quadratic representation $\rho$ on  $(\rvs,Q)$ gives a quadratic representation $\rho^-$ on $(\rvs,-Q)$.

We recall the notion of a Pin-structure on a quadratic representation. For a quadratic form $Q$ on a linear space $\rvs$, let the \emph{Clifford algebra} of $Q$, denoted $\Cl_Q$, be the free algebra generated by $\rvs$, subject only to the relations, $v^2 + Q(v,v) = 0$, for $v\in \rvs$. 
In particular, if $Q = \Id_p \oplus -\Id_q$ then $\Cl_Q\simeq\Cl_{p,q}$.
We denote by $\Pin(Q)\subseteq \Cl_Q^\times$ the \emph{Pin group} of $Q$, i.e., the subgroup of  invertible elements in $\Cl_Q$ of the form, $\pinelm = v_1\cdot v_2\cdots v_\ell$, for $v_i\in \rvs$.
Recall that we have a canonical double cover, 
\begin{equation}
1\to \ZZ_2 \to \Pin(Q) \oto{\pi} \Ort(Q) \to 1,  
\end{equation}
where the map, $\pi \colon\Pin(Q) \to \Ort(Q)$, takes $v\in \rvs$ to the reflection along the hyper-plane orthogonal to $v$. 
A \emph{Pin-structure} on a quadratic representation, $\rho\colon G\to \Ort(Q)$, is a homomorphic lift, $\spincover{\rho}\colon G\to \Pin(Q)$ of $\rho$, i.e., such that the diagram, 
\begin{equation}
\label{eq:pin_structure_diagram}
\xymatrix{
G \ar_-{\spincover{\rho}}[r] \ar@/^1pc/[rr]^{\rho} & \Pin(Q) \ar_{\pi}[r] & \Ort(Q),
}    
\end{equation}
commutes.
We warn that this notion depends heavily on $Q$. In particular, depending on whether $Q$ is positive or negative definite, it coincides with the classical notion of $\Pin_+$ and $\Pin_-$-structures respectively. 
Note, moreover, that a $\Pin$-structure on a representation does not always exist.
Even when it exists, a $\Pin$-structure might not be unique; however, this will not affect our computations. 

If $\rho$ admits no $\Pin$-structure, one can always construct a double cover, $\ZZ_2\to\spincover{G} \oto{\pi} G$, such that the composition, $\spincover{G}\stackrel{\pi}{\to} G \stackrel{\rho}{\to}\Ort(Q)$, admits a $\Pin$-structure. For example, if $\rho$ is injective, we can take $\spincover{G}$ to be the preimage of $G$ in $\Pin(Q)$ under $\pi$. In practice, we shall always assume that a $\Pin$-structure exists on our representations, replacing $G$ with $\spincover{G}$ if necessary. This replacement have the effect of adding some irrelevant components to the $\KK$-theory that later will have to be identified and ignored; this technicality poses no practical troubles.

Another advantage of passing to such double covers, apart from allowing for a $\Pin$-structure, is a convenient description of spinful fermions captured by twisted $\KK$-theory~\cite{freed2013twisted,Shiozaki2017Topological}. The intrinsic spin-structure of spinful fermions, such as electrons, imply that they transform under the so-called \emph{double point-group} corresponding to the double cover,
\begin{equation}
    1\to \ZZ_2 \to \twistcover{G}\to G \to 1.
\end{equation} 

In general, recall that, if $\omega$ is a $\ZZ_2$-valued $2$-cocycle on $G$ then one can form the $\omega$-twisted $\KK$-theory, $\KR_G^{\omega}$, classifying $\omega$-twisted  vector bundles~\cite{freed2013twisted,Shiozaki2017Topological}. 
Here, by a $2$-cocycle we mean a function, $\omega\colon G\times G\to \ZZ_2$, satisfying
\begin{equation}
\omega(g_2,g_3) + \omega(g_1,g_2g_3) + \omega(g_1g_2,g_3)+ \omega(g_1,g_2) = 0,
\end{equation}
for every $g_1,g_2,g_3\in G$.

The $2$-cocycle $\omega$ determines the double cover, $\twistcover{G}$.
Accordingly, for a $G$-space $X$ we have a splitting,
\begin{equation}\label{eq:twist-split}
\KR_{\twistcover{G}} (X)\simeq \KR_G(X)\oplus \KR_{G}^{\omega}(X).
\end{equation}
Hence, replacing $G$ by $\twistcover{G}$, we may reduce the computation of twisted $\KK$-theory to the computation of untwisted $\KK$-theory. For spinful fermions, all the formulas presented in Sec.~\ref{sec:over} and Sec.~\ref{sec:main_results} apply for $\KR_{\twistcover{G}}$; the results in Table~\ref{tab:mainres} are easily obtained by dropping the abelian groups stemming from the unphysical $\KR_G(X)$ summand in Eq.~\eqref{eq:twist-split}. This is explicitly demonstrated in Sec.~\ref{sec:example}.

\subsubsection{The Clifford group-algebra of a quadratic representation}\label{sec:change-vars}
In Ref.~\onlinecite{Cornfeld2019Classification}, the $G$-equivariant $\KK$-theory of a representation sphere $\SSp^\rho$ was described in terms of the $\ZZ_2$-graded algebra $\Cl_Q[G]$, i.e., the algebra generated from $G$ and $\Cl_Q$
subject to the relation, $gvg^{-1}=\rho(g)(v)$. To exploit this description, the authors had to decompose $\Cl_Q[G]$ into a sum of Clifford algebras. the result becomes cleaner given Pin-structure on the representation $\rho$. Namely, for $\ZZ_2$-graded algebras, $A$ and $B$, let $A\gotimes B$ be their $\ZZ_2$-graded tensor product, i.e., the  algebra generated from $A$ and $B$ subject to the relation, $ab = (-1)^{|a||b|}ba$. Then, if $\rho\colon G\to \Ort(Q)$ is a quadratic representation which admits a Pin-structure, we have 
\begin{equation}
\label{eq:separation-of-variables}
\Cl_Q[G]\simeq \Cl_Q\gotimes \RR^{\repgr{\rho}}[G].    
\end{equation}
Note that, in particular, the algebra $\Cl_Q[G]$ depends only on $\repgr{\rho}$.

This ``separation of variables" results from a change of coordinates~\cite{KAROUBI2002Equivariant,Cornfeld2019Classification}. 
Given a real representation $\rho$ of $G$, one obtains a $\ZZ_2$-grading, $\repgr{\rho}\colon G\to \ZZ_2$, on $G$, for which $g\xmapsto{\repgr{\rho}}\det(\rho(g))$, i.e., the parity of $g\in G$ is sign of the determinant.
Let $\spincover{\rho}\colon G\to \Pin(Q)$ be a Pin-structure on $\rho$. For every $g\in G$, set
\begin{equation}
\newg{g}\defeq g\spincover{\rho}(g)^{-1}.
\end{equation}
We claim that these elements generate a copy of $\RR^{\repgr{\rho}}[G]$ in $\Cl_Q[G]$ which super-commutes with $\Cl_Q$. 
To see this, first note that the conjugation action of $\Pin(Q)$ on itself, is given by $\pinelm v \pinelm^{-1} = (-1)^{|\pinelm|} \pi(\pinelm)v$ for $v\in \rvs$ and $\pinelm \in \Pin(Q)$, see Eq.~\eqref{eq:pin_structure_diagram}. Hence, for every $v\in \rvs$, we have 
\begin{equation}
\newg{g}v=(-1)^{\repgr{\rho}(g)}v\newg{g}.
\end{equation}
Similarly, the elements $\newg{g}$ satisfy the same multiplication table as that of $G$: 
\begin{align}
\newg{g}_1\newg{g}_2 &= 
g_1 \spincover{\rho}(g_1)^{-1} g_2 \spincover{\rho}(g_2)^{-1} = 
g_1 \spincover{\rho}(g_1)^{-1} g_2 \spincover{\rho}(g_2)^{-1} \nonumber\\ 
&=g_1 g_2 \spincover{\rho}(g_2)^{-1}\spincover{\rho}(g_1)^{-1}\spincover{\rho}(g_2)\spincover{\rho}(g_2)^{-1} \nonumber\\
&= g_1 g_2 \spincover{\rho}(g_1 g_2)^{-1} 
= \newg{(g_1g_2)}.
\end{align}
These two identities prove that the generators, $v\in V$ and $\newg{g}$, together generate an algebra isomorphic to $\Cl_Q \gotimes \RR^{\repgr{\rho}}[G]$.  

\subsubsection{Diagrammatic formulation}\label{sec:monster-diag}

Let us make things more explicit by focusing on a 3D physical system. On one hand, for any $G\subset \Ort(3)$ we may always construct the double point-group, $\twistcover{G}\subset\Pin_{-}(3)$, that encodes the physical twist corresponding to the intrinsic spin-structure of spinful fermions. On the other hand, any subgroup, $G_\hsm\subseteq G$, which stabilizes a $d$-dimensional sub-torus, acts on it via $G_\hsm\twoheadrightarrow\minor{G}_\hsm\subset\Ort(d)$.
It therefore has a \emph{different} double cover, $\minor{\spincover{G}}_\hsm\subset\Pin_{-}(d)$, required for the separation of variables Eq.~\eqref{eq:separation-of-variables}. Nevertheless, one can always replace $G_\hsm$ with its quadruple-cover, $\quadcover{G}_\hsm$, which admits \emph{both} Pin-structures:
\begin{equation}\label{eq:monster-diag}
\xymatrix@!=14pt{
& & \ZZ_2 \ar@{^{(}->}[dr] \ar@{=}[dl] & \quadcover{G}^0_\hsm\vphantom{\big|} \ar@{^{(}->}[d] \ar@{.>}[dddll] & \ZZ_2 \ar@{_{(}->}[dl] \ar@{=}[dr] & & \\
& \ZZ_2 \ar@{^{(}->}[dr] 
\ar@{=}[d] & & \quadcover{G}_\hsm \ar[d] \ar@{>>}[dl] \ar@{>>}[dr] & & \ZZ_2 \ar@{_{(}->}[dl] 
\ar@{=}[r] & \ZZ_2 \ar@{_{(}->}[dl] \ar@{=}[dd] \\
\ZZ_2 \ar@{=}[d] \ar@{^{(}->}[dr] & \ZZ_2 \ar@{^{(}->}[dr] \ar@{=}[dl] & \spincover{G}_\hsm \ar@{>>}[dr] 
\ar@{>>}[d] & \ZZ_2 \ar@{:}[dddll] & \twistcover{G}_\hsm \ar@{>>}[dl] 
\ar@{^{(}->}[r] & \twistcover{G}\vphantom{|} \ar@{>>}[dl] \ar@{^{(}->}[dd] & & \\
\ZZ_2 \ar@{^{(}->}[dr] & \SPin(d) \ar@{^{(}->}[d] \ar@{>>}[dr] & \minor{\spincover{G}}_\hsm \ar@{>>}[dr] \ar@{_{(}->}[dl] & G_\hsm\vphantom{\big|} \ar[ddl] 
\ar@{^{(}->}[ddr] 
\ar@{>>}[d] \ar@{^{(}->}[r] & G\vphantom{|} \ar@{^{(}->}[dd] & & \ZZ_2 \ar@{_{(}->}[dl] \\
& \Pin_-(d) \ar@{>>}[d] \ar@{>>}[dr] & \SOrt(d) \ar@{^{(}->}[d] & \minor{G}_\hsm \ar@{_{(}->}[dl] & & \Pin_-(3) \ar@{>>}[dl] & \\
& \ZZ_2 \ar@{=}[dr] & \Ort(d) \ar^-{\det}@{>>}[d] & & \Ort(3) & & \\
& & \ZZ_2 & & & &
}
\end{equation}
In particular, for any $G_\hsm\subseteq G$ with geometric actions via $\Ort(3)$ and $\Ort(d)$, one can always explicitly construct (e.g., as a GAP4~\cite{GAP4} language algorithm) all the above groups and homomorphisms such that the diagram commutes and all sequences along straight lines are short (or long) exact sequences.

\subsubsection{The equivariant \texorpdfstring{$\KK$}{K}-theory of representation spheres}\label{sec:K-of-rep-spheres}
For a quadratic representation, $\rho \colon G\to \Ort(Q)$, we associate a $G\times \ZT$-equivariant spectrum $\SSp^{\rho}$, generalizing the construction of the representation sphere from Sec.~\ref{sec:G-spectra}.  

Let $\rho\colon G\to \Ort(Q)$ be a quadratic representation of $G$.
We can always decompose $Q$ as a $G$-invariant direct sum $Q= Q_1 \oplus Q_2$ such that $Q_1$ is positive definite and $Q_2$ is negative definite. We then set 
\begin{equation}
    \SSp^{\rho} = \SSp^{\rho_1}\otimes \bar{\SSp}^{\rho_2}\in \Sp^{G\times \ZT}.  
\end{equation}
The uniqueness of the decomposition, $Q= Q_1 \oplus Q_2$, up to homotopy, shows that this association is well-defined. 
Our aim now is to give a description of $\KR_G(\SSp^\rho)\in \Mod_{\KR}(\Sp^{\ZT})$ for a quadratic representation $\rho$ in elementary terms.  

Assume from now on that $\rho$ admits a Pin-structure, $\spincover{\rho}\colon G\to \Pin(Q)$.  As we have seen in Sec.~\ref{sec:Z_2-graded}, we have a decomposition of $\RR^{\repgr{\rho}}[G]$ into a sum of matrix algebras over Clifford algebras of the form
\begin{equation}
    \RR^{\repgr{\rho}}[G]\simeq \bigoplus_\iR \Mat{d_\iR}(\Cl_{p_\iR,q_\iR}) \oplus \bigoplus_\iC\Mat{d_\iC}(\CCl_{p_\iC}).
\end{equation}  
If $Q$ is of signature $(p,q)$ then the above decomposition, together with the separation of variables [Eq.~\eqref{eq:separation-of-variables}], implies that 
\begin{multline}
    \Cl_Q[G]\simeq \bigoplus\nolimits_\iR \Mat{d_i}(\Cl_{p+p_\iR ,q+q_\iR}) \\
    \oplus \bigoplus\nolimits_{\iC}\Mat{\iC}(\CCl_{p + q + p_\iR}).
\end{multline}
This determines a decomposition of the $G$-equivariant $\KK$-theory spectrum: 
\begin{multline}
    \KR_G(\SSp^\rho)\\ 
    \simeq \bigoplus\nolimits_\iR \Sigma^{-p-p_\iR, -q-q_\iR}\KR \oplus \bigoplus\nolimits_\iC \Sigma^{- p -p_\iC, - q}\KC.
\end{multline}
It thus provides a lift of Eq.~\eqref{eq:CCiso} to the $\infty$-category of modules over $\KR$. 

While the computation above describes $\KR_G(\SSp^\rho)$ completely as a $\KR$-module, we shall need in our computation of the $\AItoK$ map, a slightly more canonical description of its homotopy groups. By doing it for all representations, $\rho$, together, it suffices to describe the 0-th homotopy group of $\KR_G(\SSp^\rho)$. 

Let $\gRRep(\Cl_Q[G])$ be the $\ZZ_2$-graded representation ring of $\Cl_Q[G]$, so that an element of $\gRRep(\Cl_Q[G])$ is a formal integral combination of $\ZZ_2$-graded representations. Given a $\ZZ_2$-graded representation, $\rvs= (\rvs^0,\rvs^1)$, of $\Cl_Q[G]$, we can construct a $G$-equivariant vector bundle, $W$, on $S^\rho$, as follows: We take the constant $G$-bundle, $\rvs^0$, on the upper hemisphere of $S^\rho$, and take the constant $G$-bundle, $\rvs^1$, on the lower hemisphere. We then glue these two $G$-bundles along the equator of $S^\rho$ in the following way. Every point on the equator of $S^\rho$ can be seen as a unit vector in the representation space of $\rho$, and hence as an odd element, $\cliff{}\in\Cl_Q$. We identify $\rvs^0$ and $\rvs^1$ at $\cliff{}$ using the module structure of $\rvs$ over $\Cl_Q$, which in particular provides a linear isomorphism, $\cliff{}\colon \rvs^0 \arsim \rvs^1$. 

By sending the class of $[\rvs] \in \gRRep(\Cl_Q[G])$  to the class $[W] - \dim(W) \in \rKR_G(S^\rho)$ one obtains a map, 
\begin{equation}
    \gRRep(\Cl_Q[G])\to \pi_0(\rKR_G(S^\rho))\simeq \pi_0(\KR_G(\SSp^\rho)),
\end{equation} 
which is the $G$-analog of the Atiyah-Bott-Shapiro map~\cite{atiyah1964clifford}. In fact, the resulting map factors through the quotient of $\gRRep(\Cl_Q[G])$ by the image of the restriction map, and induces an isomorphism,
\begin{gather}
\label{eq:ABS_equivariant}
\pi_0(\KR_G(\SSp^\rho))\simeq
    \frac{\gRRep(\Cl_Q[G])}{\im(\Res)}, \\
\Res\colon \gRRep(\Cl_{1,0}\gotimes\Cl_Q[G]) \to \gRRep(\Cl_Q[G]).\label{eq:clifford-restriction}
\end{gather}
This isomorphism is functorial with respect to maps of representations. A morphism, $f\colon \rho_0\to \rho_1$, of quadratic representations, with corresponding quadratic forms, $Q_0$ and $Q_1$, induces a map, $\SSp^{\rho_0} \to \SSp^{\rho_1}$, which in turn gives a map,
$f^*\colon \KR_G(\SSp^{\rho_1}) \to \KR_G(\SSp^{\rho_0})$.
The map induced from $f^*$ on the 0-th homotopy group translates via Eq.~\eqref{eq:ABS_equivariant} to the restriction map, $\gRRep{\Cl_{Q_1}[G]} \to \gRRep{\Cl_{Q_0}[G]}$, along the algebra homomorphism, $\Cl_{Q_0}[G]\to \Cl_{Q_1}[G]$ induced from $f$.

\subsection{Atomic insulators}\label{sec:math-AI}
As discussed in Sec.~\ref{sec:intro_BS}, besides the determination of the $G$-equivariant $\KK$-theory of the BZ torus $\bTt^d$, or equivalently, of the free $G\times \ZT$-spectrum $\bTSp{d}$ generated from it, we are also interested in the identification of the AI-bundles on the BZ. Namely, we wish to compute the map, 
\begin{equation}
    \AItoK^{p,q} \colon  \AI_G^{p,q}\to \KR^{p,q}(\bTSp{3}).
\end{equation}
We shall now explain how this map is computed uniformly in $(p,q)$ for sign permutation representations.

\subsubsection{Definition of AI-bundles}
Let us first recall the definition of the AI-bundle associated with a representation of the little group $G_\wyck$ for a Wyckoff position $\wyck$. For such a data, 
We define the \emph{fundamental AI-bundle} associated with $\wyck$ as follows. Let $\Relatt$ denote the lattice of unit cell origins in the real (euclidean) space, $\Euc{d}$. For every $g\in G_\wyck$, we have by definition that $g(\wyck) - \wyck$ belongs to $\Relatt$. Hence, the function,
\begin{equation}\label{eq:nu-def}
\AIfunc{\wyck}{g}{\vk}\defeq\exp\{i(g(\wyck) - \wyck)\cdot\vk\},
\end{equation}
is well defined on the BZ torus, $\bTt^d$. 

Consider the trivial one-dimensional vector bundle $\rvs$ on the BZ, with a nowhere vanishing global section, $\ket{\psi(\vk)}$. We endow $\rvs$ with the $G$-equivariant structure determined by the condition, 
\begin{equation}\label{eq:g_on_psi}
    g\ket{\psi(\vk)} = \AIfunc{\wyck}{g}{\vk}\ket{\psi(g(\vk))}.
\end{equation}
This way, we obtain a $G_\wyck$-equivariant vector bundle on the BZ torus, whose underlying non-equivariant vector bundle is the trivial one-dimensional bundle. This bundle is represented by a class, $\fundAI{\wyck}\in \KR_{G_\wyck}^{0,0}(\bTt^d)$. 
Recall that $\KR_{G_\wyck}(\bTt^d)$ is a module over $\KR_{G_\wyck}(\pt)$, so that one may multiply the class $\fundAI{\wyck}$ with classes in $\KR_{G_\wyck}^{p,q}(\pt)$. 
The map, $\AItoK$, is now given by the formula,
\begin{equation}
\label{eq:fundamental-AI-def}
\AItoK_{\wyck}(v) = \Ind_{G_\wyck}^{G}(v \cdot  \fundAI{\wyck})\in \KR_G^{p,q}(\bTt^d),
\end{equation}
for $v\in \KR_{G_\wyck}^{p,q}(\pt)$. 
We warn the reader that here, as in the previous sections, we assume that the representation of $G$ on the euclidean space $\Euc{d}$ is endowed with a Pin-structure, replacing $G$ with a further cover, $\twistcover{G}$, if necessary; see Sec.~\ref{sec:pin-structures}. Hence, we eliminated the twisting of the $\KK$-theory we consider, and work with classes of usual, untwisted, representations of $G_\wyck$. 

\subsubsection{Reduction to representation spheres}
For a general system, in order to compute the map, $\AItoK$, one needs to consider all possible Wyckoff positions in $\Euc{d}$. It turns out, that for sign permutation representations, there is a canonical choice of Wyckoff positions that exhaust the AI-bundles for every choice of $G$. Specifically, consider the set, $\uniwyck$, of vectors, $\vx\in \Euc{d}$, for which all components, $(\wyckcoord_1,\ldots,\wyckcoord_d)$, of $\wyck=\wyckcoord_1\dlv_1+\ldots+\wyckcoord_d\dlv_d$, in \emph{primitive lattice vector} coordinates, are either $0$ or $\frac{1}{2}$.
Note, that in this convention, a Wyckoff position corresponding to a high-symmetry line/plane/volume in the BZ will be represented by a discrete set of points.
Nevertheless, we show that the AI-bundles associated with the Wyckoff positions in $\uniwyck$ generate the entire subgroup of  AI-bundles in $\KR_G^{p,q}(\bTt^d)$. Namely, the image of $\AItoK$ on $\bigoplus_{\wyck \in \Euc{d} / \Relatt} \KR^{p,q}_{G_{\wyck}}(\pt)$ and on the subgroup $\bigoplus_{\wyck\in \uniwyck} \KR^{p,q}_{G_{\wyck}}(\pt)$ coincide. 

To achieve this, we use some relations satisfied by the map, $\AItoK$, with respect to different Wyckoff positions. Suppose we are given a 1-parameter family, $\wyck(t)$, of Wyckoff positions, for which $G_{\wyck(t)}$ is constant for $0\le t<1$ and $G_{\wyck(0)}\subseteq G_{\wyck(1)}$. 
Then, we have
\begin{equation}
\fundAI{\wyck(0)} = \Res^{G_{\wyck(1)}}_{G_{\wyck(0)}} (\fundAI{\wyck(1)}).     
\end{equation}
Let $v\in \KR_{G_{\wyck(0)}}^{p,q}(\pt)$ and set $v' = \Ind_{G_{\wyck(0)}}^{G_{\wyck(1)}}(v)$. The compatibility of the restriction, induction, and product in $\KK$-theory yields
\begin{align}
\label{eq:wyck-dependancies}
\AItoK_{\wyck(0)}(v) &= \Ind_{G_{\wyck(0)}}^G(\fundAI{\wyck(0)}\cdot v) \nonumber\\ 
&=\Ind_{G_{\wyck(1)}}^G (\Ind_{G_{\wyck(0)}}^{G_{\wyck(1)}}(\Res^{G_{\wyck(1)}}_{G_{\wyck(0)}}(\fundAI{\wyck(1)})\cdot v)) \nonumber\\ 
&= \Ind_{G_{\wyck(1)}}^G (\fundAI{\wyck(1)}\cdot\Ind_{G_{\wyck(0)}}^{G_{\wyck(1)}} (v)) = 
\AItoK_{\wyck(1)}(v').
\end{align}
In particular, we can compute the class in $\KK$-theory of every AI-bundle associated with the Wyckoff position $\wyck(0)$ using a corresponding class coming from $\wyck(1)$. 
It follows that, to compute the image of $\AItoK$, it suffices to consider points which have a maximal little group in their neighborhood. In the case of a signed permutation representation, it is straightforward to check that $\uniwyck$ contains all such points in $\Euc{d}/\Relatt$. 

In view of the discussion above, it makes sense to set the source of $\AItoK$ to be the sum, $\bigoplus_{\wyck\in \uniwyck} \KR_{G_\wyck}^{p,q}(\pt)$. Hence, from now on we adapt the convention that 
\begin{equation}
    \AI^{p,q} \defeq \bigoplus_{\wyck\in \uniwyck} \KR_{G_\wyck}^{p,q}(\pt).
\end{equation}

Let $\HSM$ denote the subset of high symmetry momenta located at the centers of the cells of the BZ torus; cf.~Fig.~\ref{fig:CWP1}. There is a bijection, $\wyck \mapsto \hsm=\wtf{\wyck}$, from $\uniwyck$ to $\HSM$, sending the vector $\wyck = x_1\dlv_1+\ldots+x_d\dlv_d$ in primitive lattice vector coordinates to the vector $\hsm=\hsmcoord_1\lv_1+\ldots+\hsmcoord_d\lv_d$ in terms of the reciprocal basis, such that $(\wyckcoord_1,\ldots,\wyckcoord_d)=(\frac{\hsmcoord_1}{2\pi},\ldots,\frac{\hsmcoord_d}{2\pi})$.
Since $\uniwyck$ and $\HSM$ are reciprocal sets, we may now write $\AItoK$ as a ``square matrix": 
\begin{equation}
    \AItoK^{p,q}\colon \smashoperator[r]{\bigoplus_{\wyck \in \uniwyck}} \KR_{G_\wyck}^{p,q}(\pt)\to \KR_G^{p,q}(\bTt^d) \simeq\smashoperator{\bigoplus_{\hsm \in \HSM}} \KR_{G_\hsm}^{p,q}(\SSp^{\rho_{\hsm}}),
\end{equation}
where $\rho_\hsm$ denotes the representation  corresponding to the cell with center $\hsm$.
Let us denote by $\AItoK_{\hsm\wyck}$ the $(\hsm,\wyck)$ component of this matrix. Namely, $\AItoK_{\hsm\wyck}$ is the composition, 
\begin{equation}
    \AItoK_{\hsm\wyck}^{p,q} \colon \KR_{G_\wyck}^{p,q}(\pt)\hookrightarrow \AI_G^{p,q} \oto{\AItoK} \KR_G^{p,q}(\bTt^d) \twoheadrightarrow \KR_{G_\hsm}^{p,q}(\SSp^{\rho_\hsm}).
\end{equation}
The map, $\AItoK$, in not diagonal, i.e., in general, $\AItoK_{\hsm\wyck} \ne 0$ even if $\hsm \ne \wtf{\wyck}$.
However, we shall now show that the quotient of $\KR_G^{p,q}(\bTt)$ by the image of of the map, $\AItoK$, does not change if we replace the matrix by its diagonal.
To obtain this, we shall use some dependencies between the different matrix entries of $\AItoK$.

For every $\hsm \in \HSM$, let $\bTt_\hsm$ denote the coordinate torus containing $\hsm$ as its center. We have a $G_\hsm$-equivariant projection map, $\proj_\hsm \colon  \bTt^d  \to \bTt_\hsm$. 
A key observation regarding the AI-bundles is that they are functorial with respect to sub-tori. Namely, for $\wyck\in \uniwyck$, the position vector, $g(\wyck) - \wyck$, is orthogonal to the momenta in the cell with center $\wtf{\wyck}$. Hence, we have  
\begin{equation}
    \AIfunc{\wyck}{g}{\vk} = \exp\{i(g(\wyck) - \wyck) \cdot \vk\} = 
\exp\{i(g(\wyck) - \wyck) \cdot \proj_\wyck(\vk)\},
\end{equation} 
and consequently, $\fundAI{\wyck}$ is in the image of the map, $\proj_{\wtf{\wyck}}^*\colon \KR_G^{p,q}(\bTt_{\wtf{\wyck}})\to \KR_G^{p,q}(\bTt^{d})$. More precisely, the pullback morphism, $\proj_{\wtf{\wyck}}^*$, takes the fundamental AI-bundle of the Wyckoff position $\wyck$ on the smaller torus, $\bTt_{\wtf{\wyck}}$, to the fundamental AI-bundle corresponding to the same Wyckoff position $\wyck$ for the BZ-torus, $\bTt^d$.
If $v\in \KR_{G_\hsm}^{p,q}(\bTt_\hsm)$, we can decompose $\proj_\hsm^*(v)$ as a sum of components associated with the cells of the BZ torus. In fact, an element of the form $\proj_\hsm^*(v)$ has non-zero components only for the cells which are contained in the closure of the cell centered at $\hsm$.   
It follows, that the matrix entries, $\AItoK_{\hsm\wyck}$, are non-zero only when $\wtf{\wyck} \ge \hsm$ coordinate-wise, i.e., $\wtf{\wyckcoord}_i\ge\hsmcoord^{\vphantom{*}}_i$ for all $i=1,\ldots,d$.

We now analyse the case where $\wtf{\wyck}\ge \hsm$.
The fundamental AI-bundle associated with $\wyck$ is constructed using the function $\AIfunc{\wyck}{g}{\vk}$, see Eq.~\eqref{eq:nu-def}. 
Let $\wtf{\wyck}_0=\hsm_0$ denote a particular reciprocal pair.
We wish to find the $\hsm_0$-component of the fundamental AI-bundle for all $\wyck_1\ge\wyck_0$. Every $\wyck_1\ge\wyck_0$ can be written as $\wyck_1 = \wyck_0 + \vx_\perp$, where $\vx_\perp$ is orthogonal to the torus with center $\hsm_0$.  For every $g\in G_{\hsm_0}$ and every $\vk\in \bTt_{\hsm_0}$, since $\vk \cdot \vx_\perp = 0$, one has
\begin{align}
\AIfunc{\wyck_1}{g}{\hsm}&=\exp\{i(g(\wyck_1) -\wyck_1)\cdot\vk\} \nonumber\\
&= \exp\{i(g(\wyck_0 + \vx_\perp) -\wyck_0 - \vx_\perp)\cdot\vk\} \nonumber\\
&=\exp\{i(g(\wyck_0) - \wyck_0)\cdot \vk + i(g(\vx_\perp) -   \vx_\perp)\cdot\vk\} \nonumber\\ 
&=\exp\{i(g(\wyck_0) - \wyck_0)\cdot \vk\}=\AIfunc{\wyck_0}{g}{\hsm},
\end{align}
which is the function used to construct the fundamental AI-bundle of the smaller torus, $\bTt_{\hsm_0}$. It follows that, up to appropriate inductions and restrictions, the entry, $\AItoK_{ \hsm_0\wyck_1}$, of the matrix $\AItoK$ is a multiple of the entry, $\AItoK_{\hsm_0\wyck_0}$, i.e., of the corresponding diagonal entry. This readily implies that the upper triangular matrix, $\AItoK_{\hsm\wyck}$, can be diagonalized using invertible column operations. Hence, we have an isomorphism of the cokernel of $\AItoK$ and that of its diagonalized version, $\bigoplus_\wyck \AItoK_{\wtf{\wyck}\wyck}$.       

\subsubsection{The spectral lift of atomic insulators}

We now show that the collection of abelian groups, $\AI_G^{p,q}$, and the collection of maps, $\AItoK^{p,q}$, are all the $(p,q)$-pieces of a single map of $\KR$-modules. 
We can view $\KR_G$ as a ring in $\Sp^{G\times \ZT}$, and we shall construct a $\KR_G$-module, $\AI_G$, together with a map of modules, $\AItoK\colon \AI_G\to \KR_G(\bTt)$, which specialize to the desired lifts to $\KR$-modules by forgetting the $G$-action.  

We start with the groups $\AI_G^{p,q}$. Let $\Gwyck$ denote the set of points in $\Euc{d}/\Relatt$ with half-integral entries in reciprocal primitive lattice vector coordinates so that $\uniwyck$ is the set of representatives for each orbit of a point in $\Gwyck$ under the $G$-action on $\Gwyck$. 
We can view $\Gwyck$ as a discrete object of the $\infty$-category $\spaces^{G\times \ZT}$ by letting $\ZT$ act trivially. 
This entices the definition, 
\begin{equation}
    \AI_G \defeq
    \KR_G\otimes \SSp[\Gwyck] \in \Mod_{\KR_G}(\Sp^{G\times \ZT}).
\end{equation}
To see that $\pi_{-p,-q}(\AI_G) = \AI_G^{p,q}$, with respect to the previous definition of the AI groups, we note that $\SSp[\Gwyck]\simeq \bigoplus_{\wyck \in \uniwyck} \Ind_{G_\wyck}^{G}(\SSp)$. Hence, by tensoring with $\KR_G$, we get 
\begin{equation}
\AI_G \simeq \bigoplus_{\wyck \in \uniwyck} \Ind_{G_\wyck}^{G}(\SSp)\otimes \KR_G \simeq \bigoplus_{\wyck \in \uniwyck} \Ind_{G_\wyck}^G(\KR_G).     
\end{equation}
Taking bi-graded homotopy groups we find,
\begin{multline}
    \pi_{-p,-q}(\AI_G) \simeq  \bigoplus_{\wyck\in \uniwyck} \pi_{-p,-q}(\Ind_{G_\wyck}^G(\KR_{G_\wyck})) \\ \simeq \bigoplus_{\wyck\in \uniwyck} \KR_{G_\wyck}^{p,q}(\pt). 
\end{multline}
Hence, $\AI_G$ is a lift of the bi-graded abelian group of AI-bundles. 

We now turn to the maps $\AItoK^{p,q}$. To define a lift of them to a map of modules over $\KR_G$, we need to specify a map,  
$\AItoK\colon \KR_G\otimes \SSp[\Gwyck] \to \KR_G(\bTt^d)$. The left-hand side is the free $\KR_G$-module on the finite $G\times \ZT$-set, $\Gwyck$. Hence, such a map is determined uniquely by a class in 
$\KR_G^{0,0}(\Gwyck\times \bTt^d) \simeq \bigoplus_{\wyck \in \uniwyck} \KR_{G_\wyck}^{0,0}(\pt)$. Consider the function, $(\vx,g,\vk)\mapsto \AIfunc{\wyck}{g}{\vk}$, defined on $\Gwyck\times G \times \bTt^d$. We can associate with it, as in the construction of the fundamental AI-bundles [Eq.~\eqref{eq:g_on_psi}], a $G$-equivariant vector bundle on $\Gwyck\times \bTt^d$. This bundle yields the fundamental AI-bundles by restricting it to $\uniwyck$, which is the set of orbits of $G$ on $\Gwyck$. This provides the desired class in $\KR_G^{0,0}(\Gwyck\times\bTt^d)$. 

\subsubsection{Representation-theoretic interpretation}

We end this appendix by describing the fundamental AI-class, $\fundAI{\wyck}$, explicitly. Using such description, with the $\KR$-module lifts from the previous section, one may compute the map, $\AItoK$, uniformly in the bi-degrees. Based on the reduction of $\AItoK$ to its diagonal terms by column operations, we may consider only the diagonal entries of the matrix $\AItoK$. Fortunately, an elegant computation scheme was given by Shiozaki in Ref.~\onlinecite{shiozaki2019classification}. 

Let us focus, from now on, on a single Wyckoff position $\wyck$ and its associated high symmetry momentum $\hsm = \wtf{\wyck}$. Particularly, for simplicity's sake, we take $\wyck=(\frac{1}{2},\ldots,\frac{1}{2})$, where $G=G_\hsm=G_\wyck$, other Wyckoff positions are treated completely analogously.

Let $\rho\colon G \to \Ort(Q)$ denote the quadratic representation of $G$ on the cell with center $\hsm$ with respect to a negative definite $G$-invariant quadratic form. We then wish to compute the $\hsm$-component of $\fundAI{\wyck}$, i.e., a class in $\KR_G^{0,0}(\SSp^{\rho})$. By Eq.~\eqref{eq:ABS_equivariant}, such a class can be specified by a $\ZZ_2$-graded representation of the $\ZZ_2$-graded algebra $\Cl_{Q}[G]$.   

The representation sphere, $\bSs^\rho$, has opposite quadratic representation to $\rho$, i.e., it is the same linear representation with quadratic structure given by $-Q$.
The algebra, $\Cl_{Q\oplus -Q}\simeq \Cl_Q \gotimes \Cl_{-Q}$, has a natural action on $\Cl_Q$ via the left and right multiplication of $\Cl_Q$. This action is compatible with the action of $G$, turning $\Cl_Q$ into a $\Cl_{Q\oplus -Q}[G]$-module. 
In fact, using the Atiyah-Bott-Shapiro construction, Eq.~\eqref{eq:ABS_equivariant}, the isomorphism, $\KR_G(\pt)\arsim {\KR_G(\SSp^\rho\otimes\bar{\SSp}^{\rho})}$, is given by $V\mapsto V\otimes \Cl_Q \in \gRRep(\Cl_{Q \oplus -Q}[G])$,
for $V\in \RRep(G)$. 
In particular, the image of $1\in \KR_G^{0,0}(\pt)$ is represented by $\Cl_Q$ as a $\ZZ_2$-graded representation of $\Cl_{Q \oplus -Q}[G]$. Hence, following the discussion below Eq.~\eqref{eq:clifford-restriction}, if $i\colon \Ss^\rho \hookrightarrow \Ss^{\rho}\wedge\bSs^{\rho}$ is the linear embedding, then the composition, 
\begin{equation}
    \KR_G(\pt) \arsim \KR_G(\SSp^\rho \otimes \bar{\SSp}^{\rho}) \oto{i^*} \KR_G(\SSp^\rho), 
\end{equation}
takes the generator, $1\in \KR_G(\pt)$, to the representation, $\Cl_Q$, of $\Cl_Q[G]$, cf.~Eq.~\eqref{eq:shiozaki}.
The result of Shiozaki~\cite{shiozaki2019classification}, regarding the fundamental AI-bundle, can now be stated as follows: The $\hsm$-component of the fundamental AI-class, $\fundAI{\wyck}$, is representable by the $\Cl_{Q}[G]$-module, $\Cl_{Q}$. 

As a module, $\Cl_{Q}$
is spanned by the monomials, $(\cliff{i_1} \cdots\cliff{i_k})$, for $i=1,\ldots,\dim(Q)$, see Eq.~\eqref{eq:clifford-def}. The algebra $\Cl_{Q}$ acts on itself by multiplication from the left. Similarly, the group $G$ acts on $\Cl_{Q}$ via its action on the quadratic representation space of $\rho$ and the functoriality of the construction of the Clifford algebra. These two compatible actions combine to endow $\Cl_{Q}$ with a structure of a $\Cl_{Q}[G]$-module. 
Concretely,
\begin{equation}
(\cliff{} g)\cdot(\cliff{i_1} \cdots\cliff{i_k}) \defeq \cliff{}\cdot g(\cliff{i_1})\cdots g(\cliff{i_k}), 
\end{equation}
for $\cliff{},\cliff{i_1},\ldots,\cliff{i_k}
\in \Cl_Q$ and $g\in G$, cf.~Eq.~\eqref{eq:grop-ring_mult}.

This explicit description allows us to represent the fundamental AI-bundle as a direct sum of irreducible $\ZZ_2$-graded representations of $\Cl_Q[G]$, and hence completely identify its class in $\pi_0(\KR_G(\SSp^\rho))$ via the $G$-equivariant isomorphism in Eq.~\eqref{eq:ABS_equivariant}.
Once identified, one may easily quotient out the AIs from the $\KK$-theory classification.

\section{Isomorphism}\label{app:iso}
In this appendix, we recapitulate the essentials of the isomorphism Eq.~\eqref{eq:CCiso}. This isomorphism was originally presented in Ref.~\onlinecite{Cornfeld2019Classification} by E.~C.~and A.~Chapman as part of the classification of Dirac Hamiltonians invariant under any point-group symmetry, $G$. In Appendix~\ref{sec:spins-and-reps} we have presented the proof of this isomorphism in the $\KK$-theoretic context used in this paper. Here, we provide a succinct description and some explicit examples.

Before presenting the isomorphism, we alert the reader that a working knowledge of Clifford algebras and $\ZZ_2$-graded representation theory is required for the understanding of this appendix (this was not necessary for the understanding of the main text). All the relevant mathematical background is provided in Appendix~\ref{sec:rings}; we also refer the reader to Ref.~\onlinecite{Cornfeld2019Classification} where an extensive and detailed introduction to these subjects is given.

\subsection{Description of the isomorphism}
The aim of the isomorphism is to decompose the spin-twisted $G$-equivariant reduced $\KK$-theory of a sphere, $\trKR_G^{\omega;p,q}(\bSs^d)$, into direct summands of non-equivariant $\KK$-theories, $\trKR^{p-p_\rho,q-q_\rho}(\bSs^d)$ and $\trKC^{p-p_\rho,q}(\bSs^d)$, for some explicitly determined representations and degrees, $\rho$ and $p_\rho,q_\rho$.

As discussed in Sec.~\ref{sec:K-of-rep-spheres}, the $\KK$-theory is completely determined by the $\ZZ_2$-graded real representation ring,
\begin{equation}\label{eq:KR-by-rep}
\rKR_G^{-p,-q}(\bSs^d)\simeq\frac{\gRRep(\Cl_{p,q}\gotimes\Cl_{0,d}[G])}{\gRRep(\Cl_{p+1,q}\gotimes\Cl_{0,d}[G])}.
\end{equation}
Here, $\Cl_{0,d}[G]$ is the algebra generated from $G$ and $\cliffP{1},\ldots,\cliffP{d}\in\Cl_{0,d}$ such that $G$ acts geometrically, i.e., 
\begin{equation}
g\cliffP{i}g^{-1}=\sum\nolimits_j[O_g]_{ij}\cliffP{j},
\end{equation}
where $O_g\in\Ort(d)$ is the geometric action of the point-group. Moreover, if $G$ admits a $\Pin_-(d)$ structure, then 
\begin{equation}\label{eq:Cl-split}
\Cl_{0,d}[G]\simeq\Cl_{0,d}\gotimes\RR^{\repgr{}}[G],
\end{equation}
for a proof by change of variables see Sec.~\ref{sec:change-vars}.
Here, $\RR^{\repgr{}}[G]$ is the $\ZZ_2$-graded real group ring of $G$, graded by the the determinant of the point-group geometric action,
\begin{equation}
\xymatrix{
G \ar_-{g\mapsto O_g}[r] \ar@/^1.25pc/[rrr]^-{\repgr{}} & \Ort(d) \ar_-{\det}[r] & \{\pm 1\} \ar@{}[r]|-{\displaystyle\simeq} & \ZZ_2.
}  
\end{equation}
The $\ZZ_2$-graded real group ring decomposes into direct summands of matrix algebras over Clifford algebras, 
\begin{equation}\label{eq:RG-split}
    \RR^{\repgr{}}[G]\simeq \bigoplus_\iR \Mat{d_\iR}(\Cl_{p_\iR,q_\iR}) \oplus \bigoplus_\iC\Mat{d_\iC}(\CCl_{p_\iC}).
\end{equation}
Here, each summand corresponds to an irreducible $\ZZ_2$-graded representation; these may be readily determined by Table~\ref{tab:real-Clifford-Irr}.

Finally, by combining Eq.~\eqref{eq:KR-by-rep}, \eqref{eq:Cl-split}, and \eqref{eq:RG-split}, we obtain Eq.~\eqref{eq:CCiso},
\begin{equation}
\rKR_G^{p,q}(\bSs^d)\!\simeq\!\bigoplus_\iR\rKR^{p-p_\iR,q-q_\iR}(\bSs^d) \oplus \bigoplus_\iC\rKC^{p-p_\iC,q}(\bSs^d).
\end{equation}
Here, we have utilized $\Cl_{p,q}\gotimes\Cl_{p',q'}\simeq\Cl_{p+p',q+q'}$ as well as the Morita equivalence, $\gRep(\Mat{n}(A))\simeq\gRep(A)$.

Formally speaking, there is an exception when $\repgr{}=0$, i.e., when all elements of $G$ are of determinant $+1$. In that case,
\begin{multline}
\RR[G]\simeq
\bigoplus_{\rho \text{ real}} \Mat{d_\rho}(\RR)  \oplus\bigoplus_{\rho \text{ complex}}  \Mat{\frac{d_\rho}{2}}(\CC) \\ \oplus\bigoplus_{\rho \text{ quaternionic}} \Mat{\frac{d_\rho}{4}}(\HH).
\end{multline}
Nevertheless, by noting that $\RR\simeq\Cl_{0,0}$, $\CC\simeq\CCl_{0}$ and $\Mat{2}(\HH)\simeq\Cl_{4,0}\simeq\Cl_{0,4}$, this case may be treated using the same framework.

\subsubsection{Spins \& twistings}
Before moving on to the examples, let us be explicit about spins and twistings in the $\KK$-theory. First, spinful fermions have an intrinsic physical spin-structure of 3D space and the action of a point-group $G$ is hence twisted by it. This can be easily accounted for by considering the double point-group, $\twistcover{G}\twoheadrightarrow G$, and disposing the spinless contributions at the end,
\begin{equation}
\RR^{\repgr{}}[\twistcover{G}]\simeq\RR^{\repgr{}}[G]\oplus\RR^{\repgr{}}[G]^{\omega_{\Pin_-(3)}}.
\end{equation}
However, the attentive reader would notice that the Clifford algebras, and hence, the $\KK$-theory, are oblivious to the 3D physical spin-structure and require a $d$-dimensional spin-structure. In particular,
\begin{equation}\label{eq:Cl-spin}
\Cl_{0,d}[G]^{\omega_{\Pin_-(3)}}\simeq\Cl_{0,d}\gotimes\RR^{\repgr{}}[G]^{\omega_{\Pin_-(3)}+\omega_{\Pin_-(d)}},
\end{equation}
cf.~Eq.~\eqref{eq:Cl-split}. This may easily be solved by taking a \emph{further} double-cover, $\quadcover{G}\twoheadrightarrow\twistcover{G}\twoheadrightarrow G$, which admits both spin-structures,
\begin{multline}
\RR^{\repgr{}}[\quadcover{G}]\simeq\RR^{\repgr{}}[G]
\oplus\RR^{\repgr{}}[G]^{\omega_{\Pin_-(3)}}
\oplus\RR^{\repgr{}}[G]^{\omega_{\Pin_-(d)}}\\
\oplus\RR^{\repgr{}}[G]^{\omega_{\Pin_-(3)}+\omega_{\Pin_-(d)}}.
\end{multline}
This ``quadruple" cover is uniquely defined by Eq.~\eqref{eq:monster-diag}. It enables one to carry out all computations in terms of \emph{non-twisted} representations of $\quadcover{G}$, as described above.

\subsection{Examples}
We hereby present three pedagogical examples of increasing intricacy; many more examples are provided in Ref.~\onlinecite{Cornfeld2019Classification}. 

\subsubsection{3D point-group $C_i$}
This is one of the simplest possible examples. The point-group, $G\simeq\ZZ_2$, is generated by $I^2=1$ and acts by $O_I=\left(\begin{smallmatrix}-1&0&0\\0&-1&0\\0&0&-1\end{smallmatrix}\right)$, hence $\det(O_I)=-1$. In this case, there is no need for taking any covers, and one simply finds,
\begin{align}
&\RR[G]\simeq[\RR]_1\oplus[\RR]_{t_2},\qquad\RR^{\repgr{}}[G]\simeq[\Cl_{0,1}]_{1-t_2},\nonumber\\
&\rKR^{p,q}_{G}(\bSs^3)\simeq\rKR^{p,q-1}(\bSs^3)_{1-t_2},
\end{align}
cf.~Eq.~\eqref{eq:KR-P1}, where we have used Bott periodicity, $\trKR^{p,q}(\bSs^d)\simeq\KR^{p-q+d,0}(\pt)$.
Here, $1$ is the trivial representation and $t_2$ is the sign representation of $\ZZ_2$, i.e., $I\mapsto -1$.

\subsubsection{3D point-group $S_4$}
This is the example discussed in Sec.~\ref{sec:example}. The double point-group, $\twistcover{G}\simeq\ZZ_8$, is generated by $(\hat{s}_4)^4=-1$ and acts by Eq.~\eqref{eq:S4action}.
In this case (since $d=3$) the double point-group suffices, and one finds,
\begin{align}
\RR[\twistcover{G}]&\simeq[\RR]_1\oplus[\RR]_{t_8^4}\oplus[\CC]_{t_8^2+t_8^6}\oplus[\CC]_{t_8^{\vphantom{1}}+t_8^7}\oplus[\CC]_{t_8^3+t_8^5},\nonumber\\
\RR^{\repgr{}}[\twistcover{G}]&\simeq[\Cl_{0,1}]_{1-t_8^4}\oplus[\Cl_{1,0}]_{t_8^2-t_8^6}\oplus[\CCl_{1}]_{t_8^{\vphantom{1}}-t_8^5+t_8^7-t_8^3},
\end{align}
where $1$ is the trivial representation and $t_8$ is the fundamental complex representation of $\ZZ_8$, i.e., $\hat{s}_4\mapsto e^{\frac{2\pi i}{8}}$.
Notice, that since $\omega$ is $\ZZ_2$-valued (see Sec.~\ref{sec:pin-structures}) and since $d=3$, we have $\omega_{\Pin_-(d)}+\omega_{\Pin_-(3)}=0$. Therefore, by Eq.~\eqref{eq:Cl-spin}, the spinless representations of $\RR^{\repgr{}}[\twistcover{G}]$ span the spinful $\KK$-theory, i.e.,
\begin{equation}
\rKR_{G}^{\omega;p,q}(\bSs^3)\simeq\rKR^{p,q-1}(\bSs^3)_{1-t_8^4}\oplus\rKR^{p-1,q}(\bSs^3)_{t_8^2-t_8^6},
\end{equation}
cf.~Table~\ref{tab:P-4_full_KR}.

\subsubsection{1D (rod) point-group $C_{2h}$}
The double point-group, $\twistcover{G}\simeq\ZZ_2\times\ZZ_4$, is generated by inversions, $\hat{I}^2=1$, and $\pi$-rotations around the crystal axis, $(\hat{c}_2)^2=-1$.
In this case, the inversion symmetry acts as a mirror on the 1D system, hence a further cover is required for a $\Pin_-(1)$-structure; namely, $\quadcover{G}\simeq\ZZ_4\times\ZZ_4$ is generated by $\quadcover{I}^2=-1$ and $\quadcover{c}_2=\hat{c}_2$. One thus finds,
\begin{align}
\RR[\quadcover{G}]&\simeq\nonumber\\
&[\RR]_{1} &&\oplus[\RR]_{t_4^2}  &&\oplus[\CC]_{t_4^{\vphantom{1}}+t_4^3}\nonumber\\
\oplus &[\RR]_{t_4'^2} &&\oplus[\RR]_{t_4^2t_4'^2} &&\oplus[\CC]_{(t_4^{\vphantom{1}}+t_4^3)t_4'^2}\nonumber\\
\oplus &[\CC]_{t_4'^{\vphantom{1}}+t_4'^3}\!\!\!\!\!\! &&\oplus[\CC]_{t_4^2(t_4'^{\vphantom{1}}+t_4'^3)}\!\!\!\!\!\! &&\oplus[\CC]_{t_4^{\vphantom{1}}t_4'^{\vphantom{1}}+t_4^3t_4'^3}\oplus[\CC]_{t_4^3t_4'^{\vphantom{1}}+t_4^{\vphantom{1}}t_4'^3}.
\end{align}
Here, $t_4$ and $t_4'$ are the fundamental complex representations of $\ZZ_4\times\ZZ_4$, i.e., $\quadcover{I}\mapsto i$ and $\quadcover{c}_2\mapsto i$, respectively. The $\ZZ_2$-grading is determined by $O_{I}=(-1)$ and $O_{c_2}=(+1)$, such that,
\begin{align}
\RR^{\repgr{}}[\quadcover{G}]&\simeq [\Cl_{0,1}]_{1-t_4^2}\oplus[\Cl_{1,0}]_{t_4^{\vphantom{1}}-t_4^3}\nonumber\\
&\oplus[\Cl_{0,1}]_{(1-t_4^2)t_4'^2}\oplus[\Cl_{1,0}]_{(t_4^{\vphantom{1}}-t_4^3)t_4'^2}\nonumber\\
&\oplus[\CCl_{1}]_{(1-t_4^2)(t_4'^{\vphantom{1}}+t_4'^3)}\oplus[\CCl_{1}]_{(t_4^{\vphantom{1}}-t_4^3)(t_4'^{\vphantom{1}}-t_4'^3)}.
\end{align}
By Eq.~\eqref{eq:Cl-spin}, the spinful $\KK$-theory is spanned by the doubly-twisted representation of $\RR^{\repgr{}}[\quadcover{G}]$, i.e.,
\begin{equation}
\rKR_{G}^{\omega;p,q}(\bSs^1)\simeq\rKC^{p-1,q}(\bSs^1)_{(t_4^{\vphantom{1}}-t_4^3)(t_4'^{\vphantom{1}}-t_4'^3)}.
\end{equation}
This provides the $\ZZ$ invariants of the cell centered at the $Z$-point in space-group $\mathrm{P2/m}$; see Table~\ref{tab:app}.

\end{document}